\documentclass[acmtog]{acmart}
\pdfoutput=1
\renewcommand\footnotetextcopyrightpermission[1]{} % removes footnote with conference information in first column
\usepackage{booktabs} % For formal tables
\usepackage{amsmath}
\usepackage{amssymb}
\usepackage{amsfonts}
\usepackage{mathtools}
\usepackage{gensymb}
\usepackage{mathrsfs}
\usepackage{nicefrac}
\usepackage{xspace}
\usepackage{xfrac}
\usepackage{my_macros} %my macros
\usepackage[normalem]{ulem}
\usepackage{mdwtab}
\usepackage[font=footnotesize]{subfig}

\usepackage[ruled]{algorithm2e} % For algorithms

\SetAlFnt{\small}
\SetAlCapFnt{\small}
\SetAlCapNameFnt{\small}
\SetAlCapHSkip{0pt}

\setcopyright{none}

\makeatletter
\def\runningfoot{\def\@runningfoot{}}
\def\firstfoot{\def\@firstfoot{}}
\makeatother

\begin{document}

% Title portion
\title{A Monte Carlo Framework for Rendering Speckle Statistics in Scattering Media}

% DO NOT ENTER AUTHOR INFORMATION FOR ANONYMOUS TECHNICAL PAPER SUBMISSIONS TO SIGGRAPH 2019!
\author{Chen Bar}
\orcid{}
\affiliation{%
  \institution{Technion}
  \country{Israel}}
\email{}
\author{Marina Alterman}
\affiliation{%
  \institution{Technion}
  \country{Israel}
}
\email{}
\author{Ioannis Gkioulekas}
\affiliation{%
	\institution{Carnegie Mellon University}
	\country{USA}
}
\email{}
\author{Anat Levin}
\affiliation{%
 \institution{Technion}
 \country{Israel}}
\email{}

\makeatletter
\let\@authorsaddresses\@empty
\makeatother
%\renewcommand\shortauthors{Zhou, G. et al}

%\comment{

%\abstract{Scientific Abstract}
%\noindent{\large {\bf ~~~~~~~~~Coherent rendering and inverse rendering of scattering materials }}

\begin{abstract}
We present a Monte Carlo rendering framework for the physically-accurate simulation of speckle patterns arising from volumetric scattering of coherent waves. These noise-like patterns are characterized by strong statistical properties, such as the so-called memory effect, which are at the core of imaging techniques for applications as diverse as tissue imaging, motion tracking, and non-line-of-sight imaging. Our framework allows for these properties to be replicated computationally, in a way that is orders of magnitude more efficient than alternatives based on directly solving the wave equations. At the core of our framework is a path-space formulation for the covariance of speckle patterns arising from a scattering volume, which we derive from first principles. We use this formulation to develop two Monte Carlo rendering algorithms, for computing speckle covariance as well as directly speckle fields. While  approaches based on wave equation solvers require knowing the microscopic position of wavelength-sized scatterers, our approach takes as input only bulk parameters describing the statistical distribution of these scatterers inside a volume. We validate the accuracy of our framework by comparing against speckle patterns simulated using wave equation solvers, use it to simulate memory effect observations that were previously only possible through lab measurements, and demonstrate its applicability for computational imaging tasks.  
\end{abstract}





\begin{teaserfigure}
\begin{center}\begin{tabular}{ c c c }
        \begin{tabular}{c}
                \includegraphics[width= 0.275\textwidth]{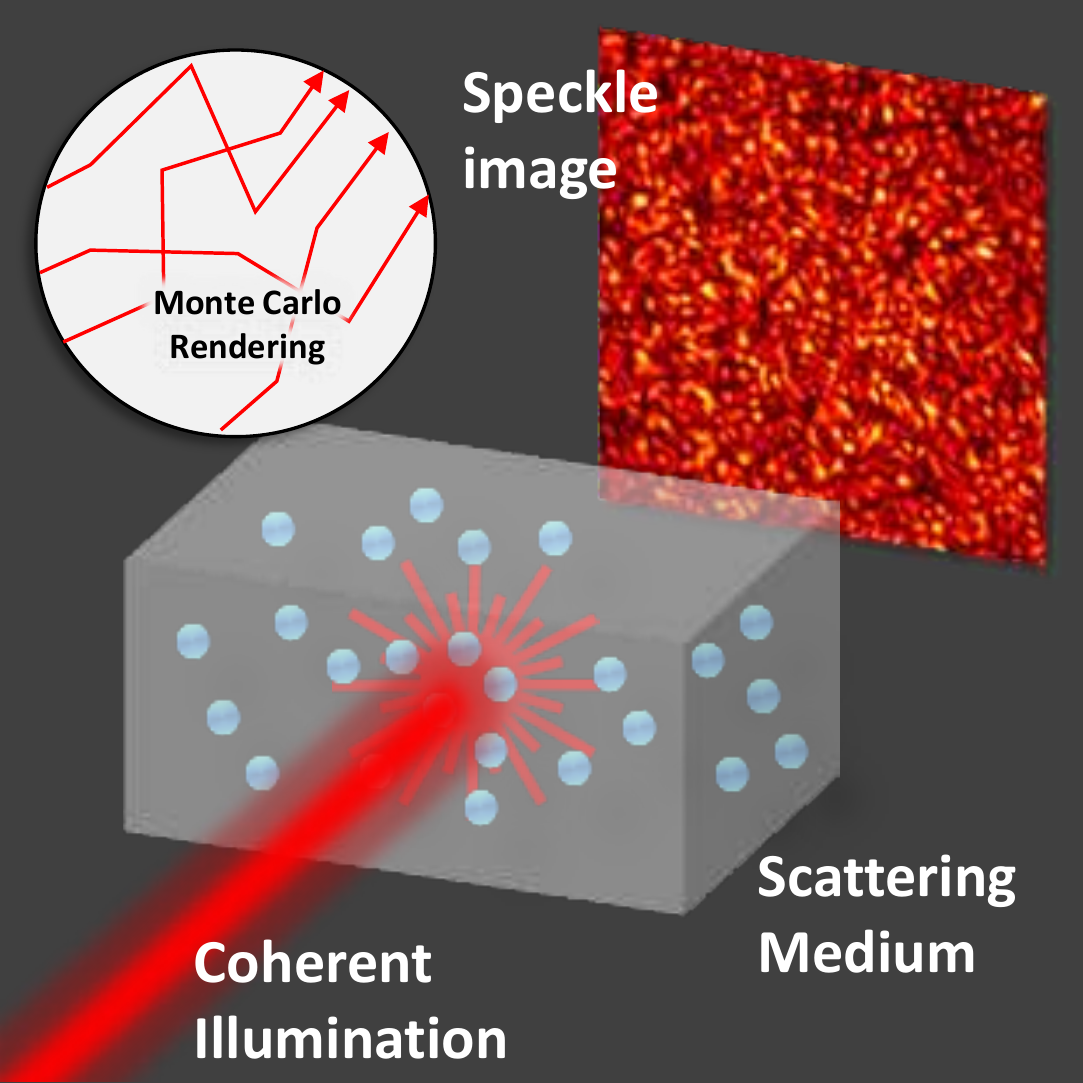}
                \vspace{-0.5cm}
        \end{tabular} &
    \hspace{-0.5cm}
        \begin{tabular}{c c c c}
                &$0\degree$&$0.0025\degree$&$0.0075\degree$\\
                    \raisebox{1cm}{\rotatebox[origin=t]{90}{$\bf g=0$}}&
                \includegraphics[width= 0.2\textwidth]{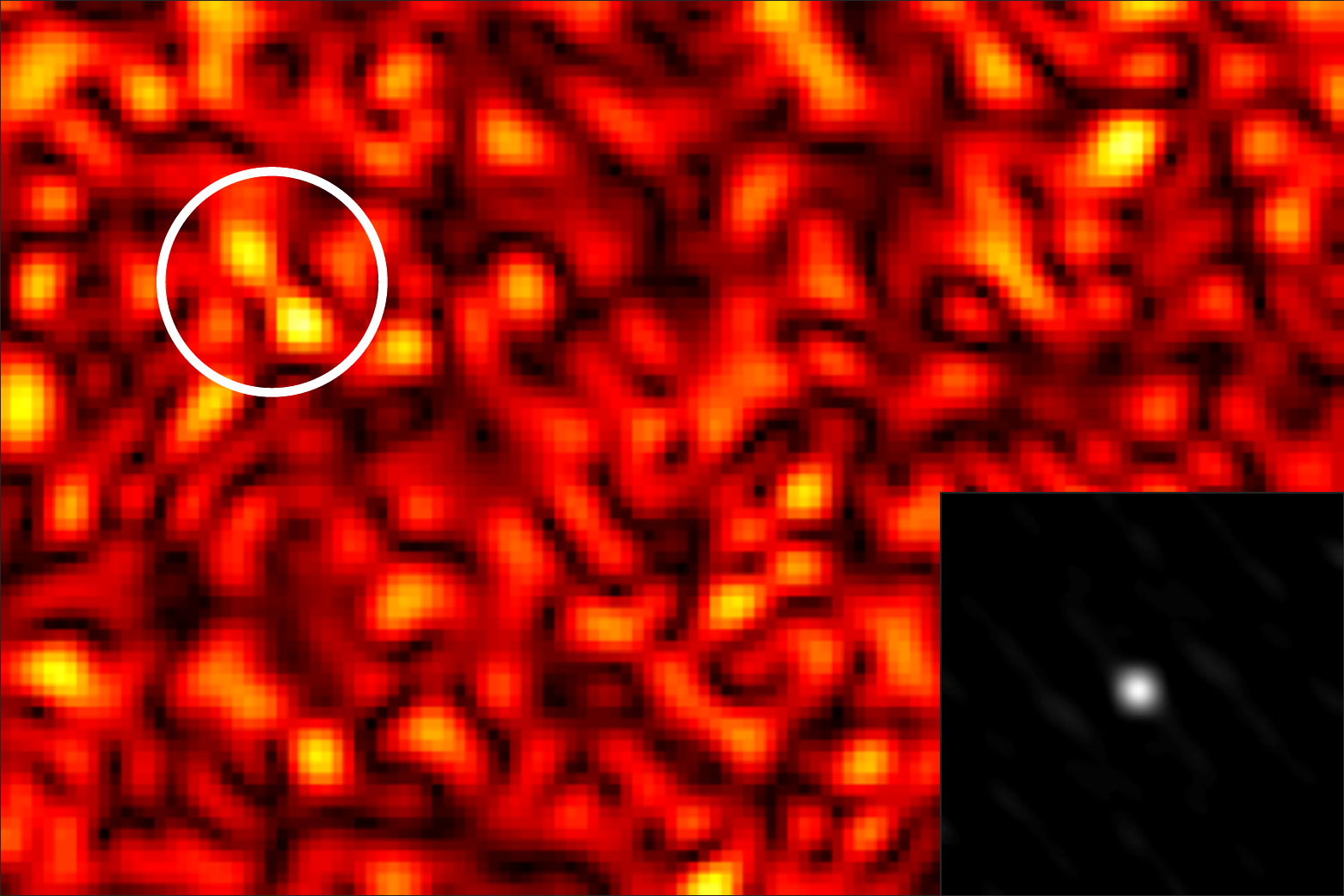} &
                \includegraphics[width= 0.2\textwidth]{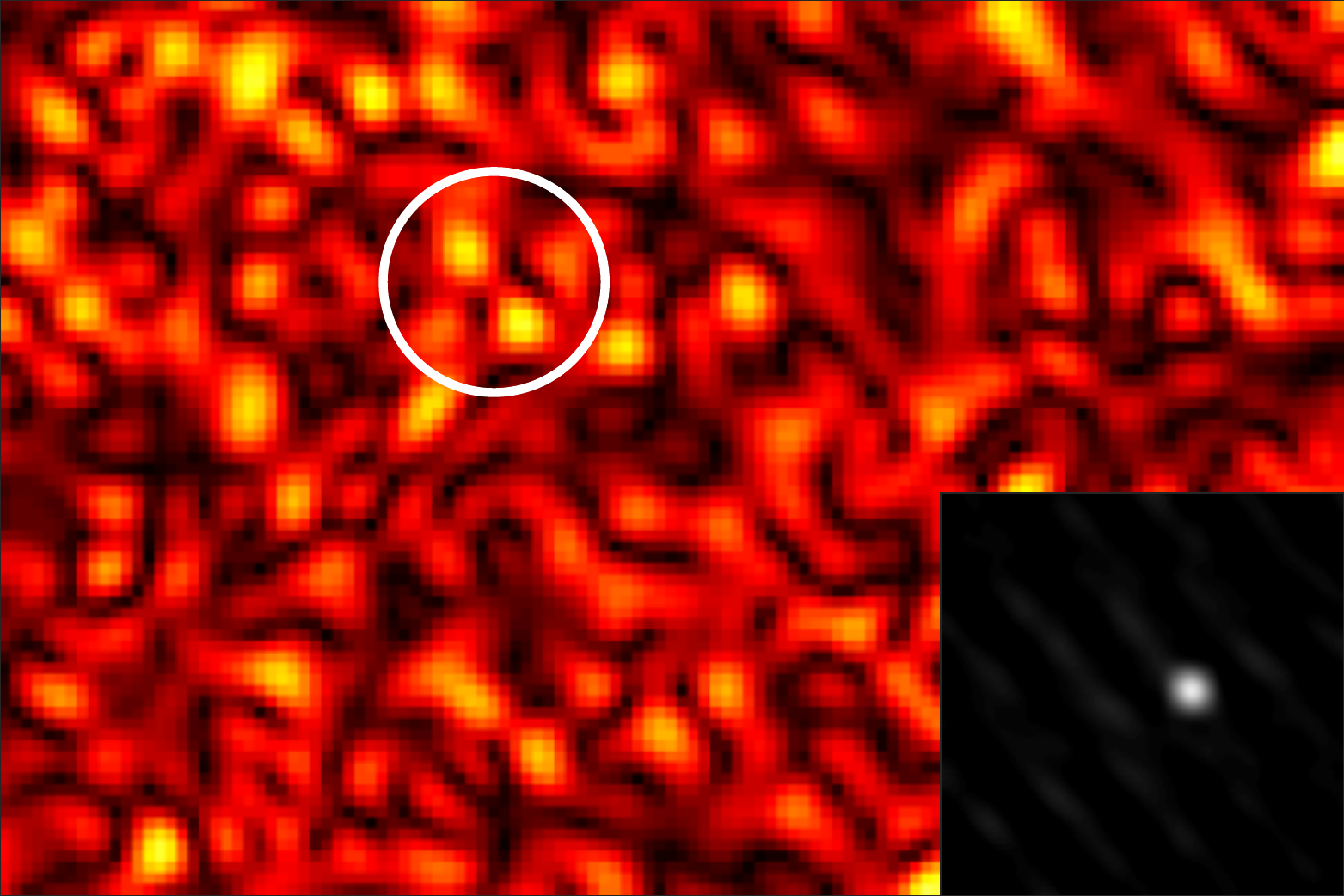} &
                \includegraphics[width= 0.2\textwidth]{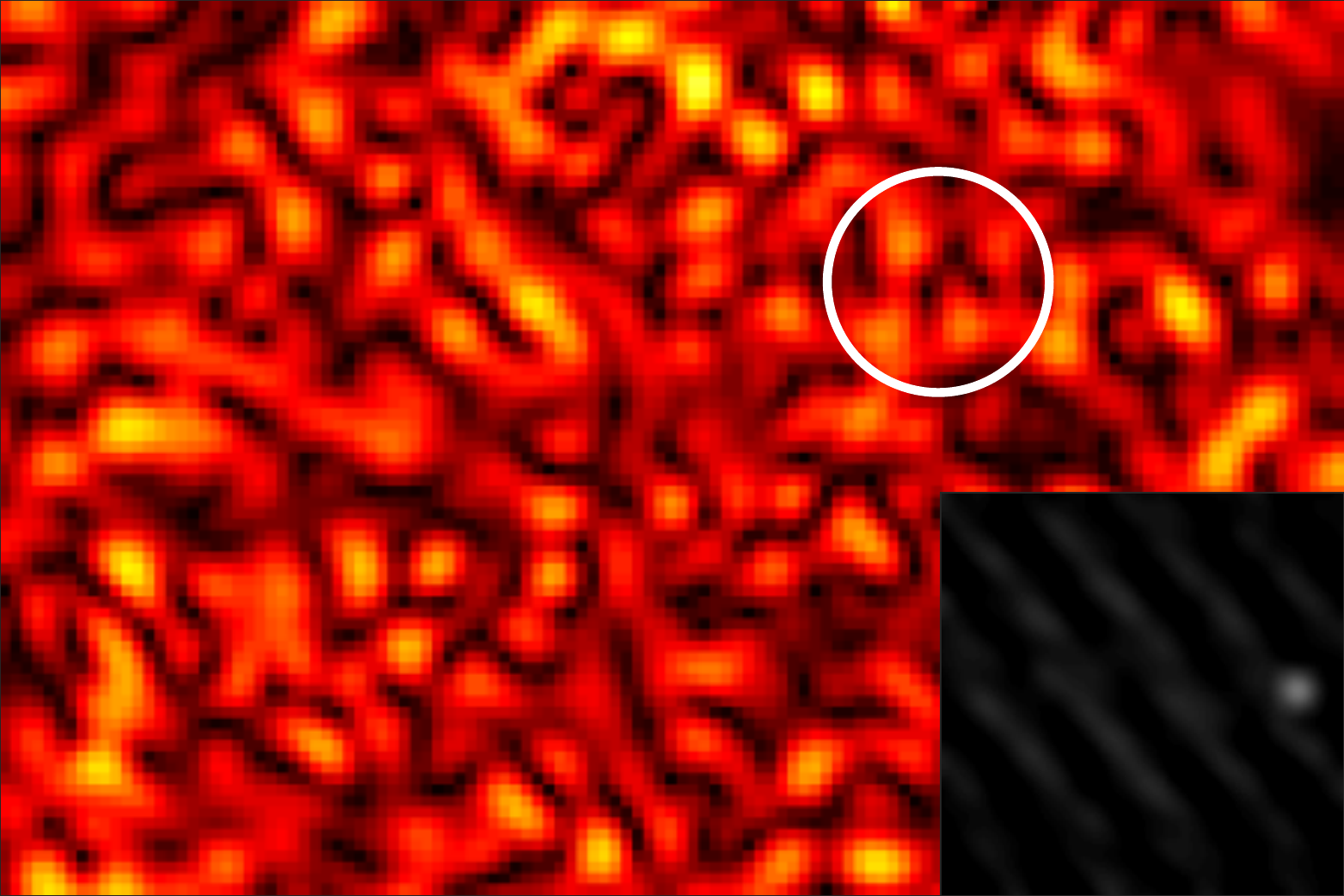} \\
                \raisebox{1cm}{\rotatebox[origin=t]{90}{$\bf g=0.9$}}&
                \includegraphics[width= 0.2\textwidth]{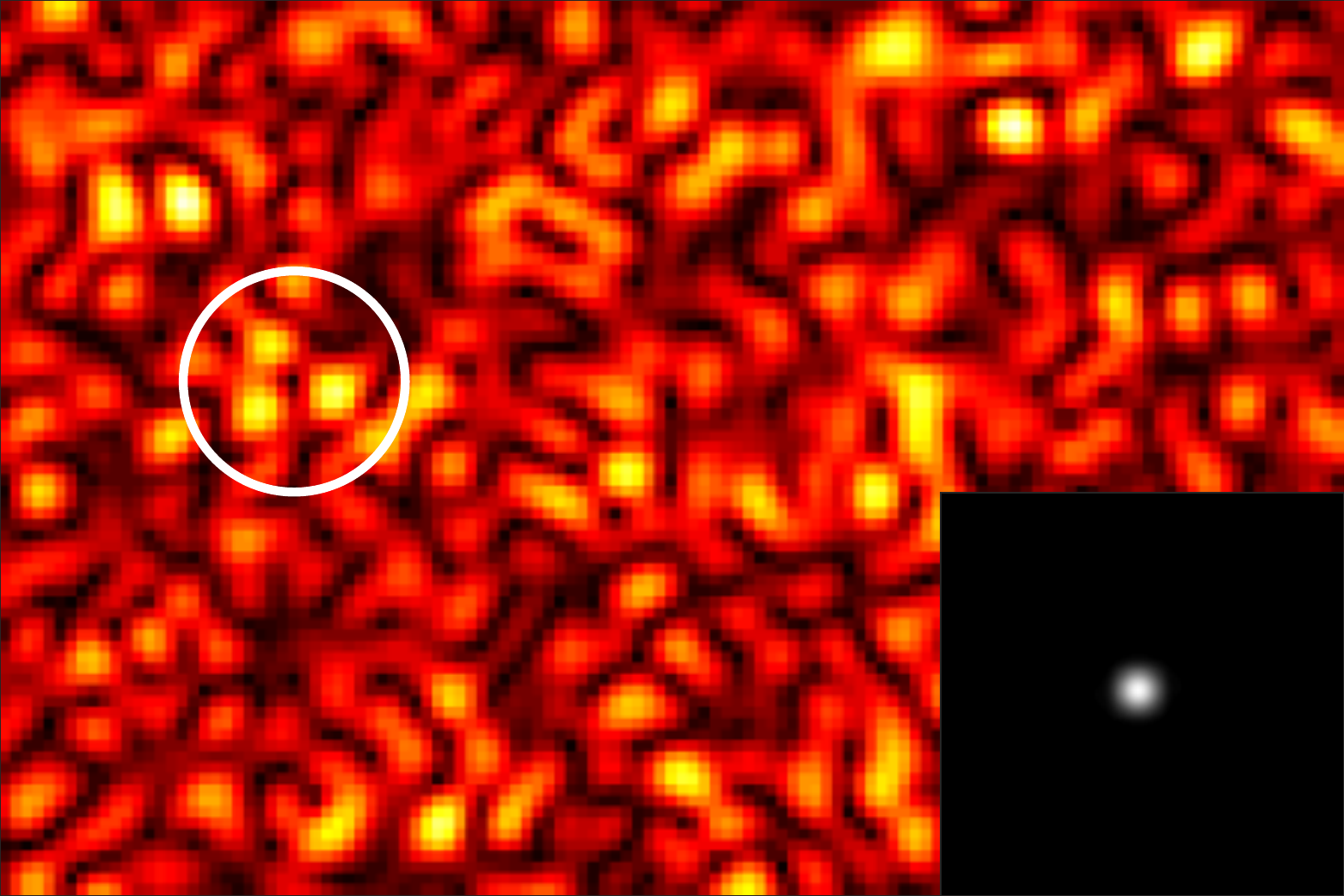} &
                \includegraphics[width= 0.2\textwidth]{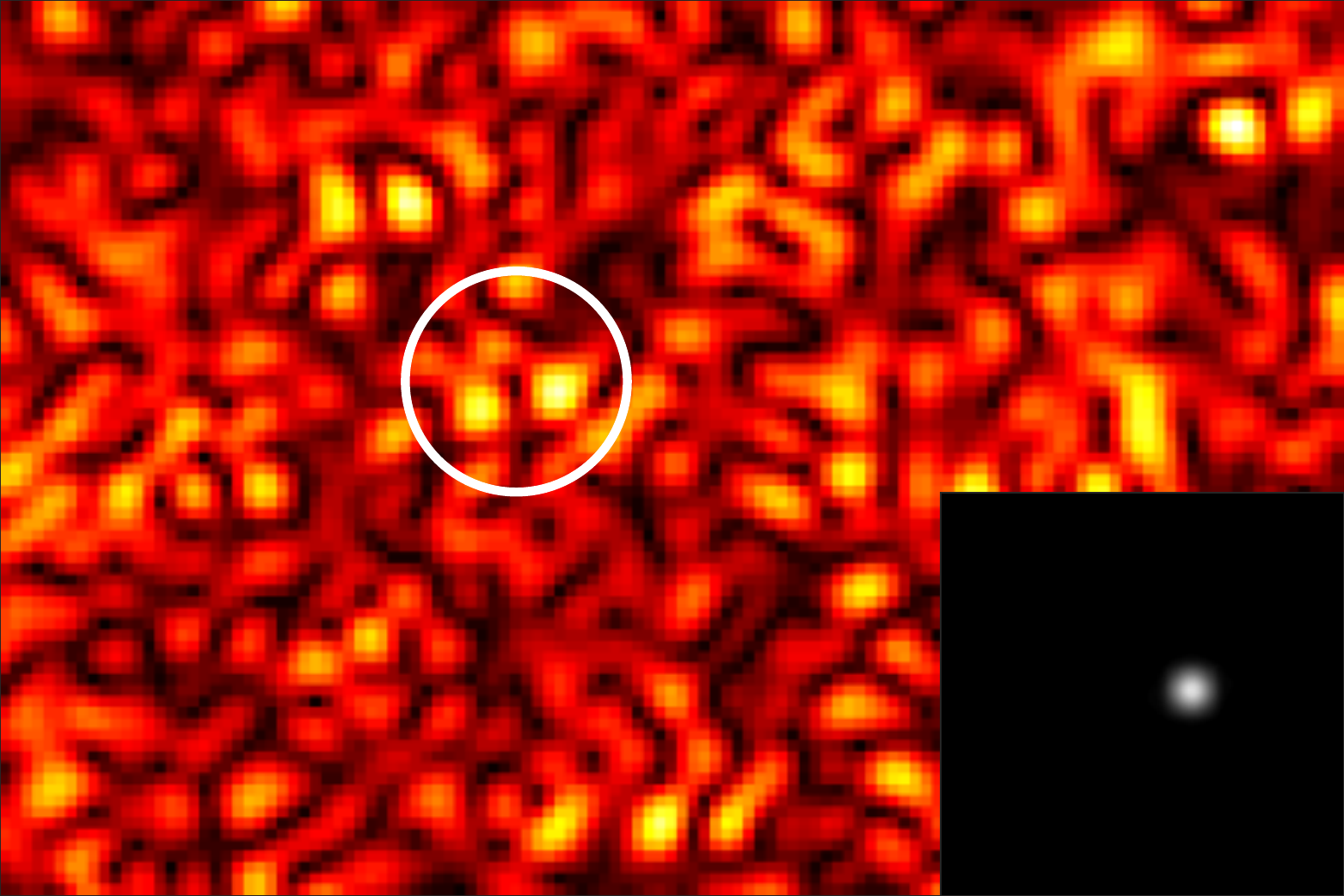} &
                \includegraphics[width= 0.2\textwidth]{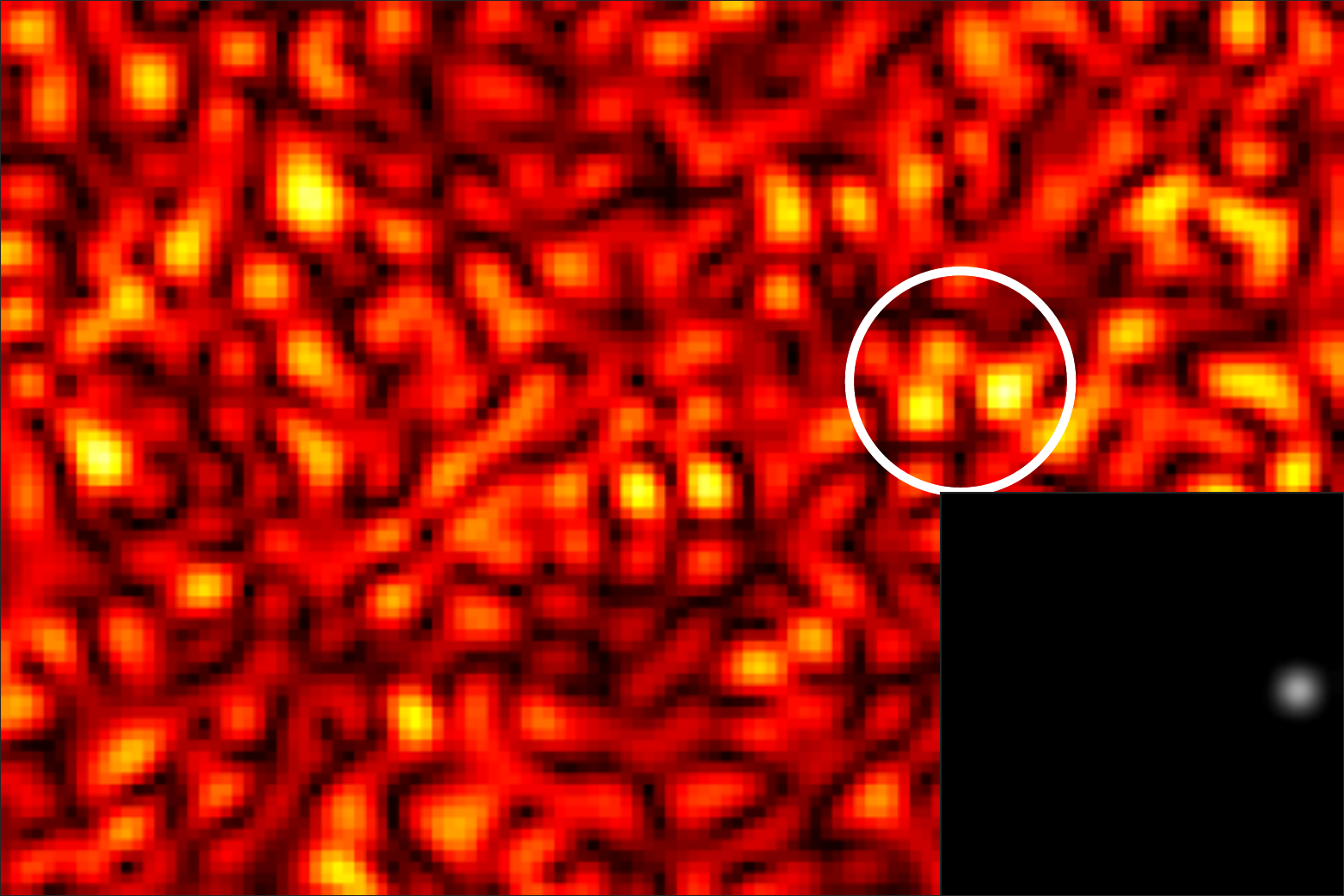} \\
        \end{tabular}
\end{tabular}
\caption{
Coherent images of translucent materials typically involve high fluctuations speckle structure. Despite their semi-random structure, speckles have strong statistical properties, in particular the memory effect stating that as one tilts the illumination direction the resulting speckles shift. This remarkable property was widely exploited in multiple computational imaging applications. The memory effect is usually valid over  a limited angular range that heavily depend  on material properties. Lacking analytical formulas, and given the wide practice applicability, memory effect properties of material of interest are often measured empirically in the lab. We present  a MC approach that can render physically consistent speckle images as well as their statistics as a function of material parameters. Here we show speckle images rendered by our algorithm for a few illumination directions, as well as their auto-correlation (black insets). The speckle shift is evident, but as the angle difference increases, correlation decays and the decay rate is different for different material parameters. This figure demonstrates two   Henyey-Greenstein (HG) phase functions with different anisotropy parameters $g$. We verify the accuracy of our algorithm against an exact yet computationally heavy  wave solver as well as against analytical formulas derived in limited  settings. }  
\label{fig:teaser}
\end{center}        
\end{teaserfigure}

\comment{
Coherent images of translucent materials typically involve high fluctuations speckle structure as in (b). Despite their semi-random structure, speckle have strong statistical properties, that have been exploited in many computational imaging tasks. For example, we replicate  a seeing through scattering layer application of~\cite{Katz2014}. A set of illuminators with the arrangement at the top of (c) generates a semi-random speckle image (b), yet due to speckle statistics, the auto-correlation of the speckle image is equivalent to the auto-correlation of the original illuminators and hence the illuminators can be recovered from the speckle image using phase retrieval algorithms.  However, the success of this algorithm heavily depends on adapting imaging configuration to material parameters. Lacking analytical models, these parameters are often adjusted empirically in the lab.   We present for the first time a MC approach that can render physically consistent speckle images as well as their statistics as a function of material parameters.  In (e,d) we show the auto-correlation and the corresponding illuminators reconstruction for different  material parameters, such as two   Henyey-Greenstein (HG) phase functions with different anisotropy parameters $g$, simulated with our speckle renderer.  The success of the algorithm depends on the validity of the statistical correlations the algorithm exploits in a given angular range for each type of material, and despite the seemingly subtle difference it succeeds with $g=0.9$ and fails with $g=0.85$.  This demonstrates the importance of numerically predicting speckle correlations in computational imaging.\\  {\Anat{Caption option 2 if we change the teaser}}}

\maketitle

\thispagestyle{empty}

\section{Introduction}

Scattering refers to the propagation of radiation (for instance, light or sound) in non-uniform media, composed of small discrete scatterers, usually particles of varying refractive properties: As an incident wave propagates through the medium, it will interact with scatterers multiple times, and each such interaction will change the wave's shape. Scattering is commonly encountered when visible light interacts with a large variety of materials, for instance biological tissues, minerals, the atmosphere and clouds, cosmetics, and many industrial chemicals. As a result of the ubiquity of scattering, its study has attracted numerous research efforts in computer graphics and vision, and much more broadly in medical imaging, remote sensing, seismic imaging, and almost any field of natural science.

The appearance of scattering materials is qualitatively very different, depending on whether they are imaged under \emph{incoherent}  or \emph{coherent} conditions. In the \emph{incoherent} case, scaterring results in images with smoothly-varying intensity distributions, often referred to as \emph{translucent appearance}. By contrast, under coherent imaging conditions, the appearance of scattering materials is characterized by \emph{speckles}, that is, pseudo-random high variations in the output waves and captured intensity images. Speckles have been the subject of multiple textbooks~\cite{GoodmanSpeckle,SpeckleMetrology,IntSpeckleLight,AdvSpeckleMetrology}, as despite their random structure, they have strong statistical properties that are characteristic of the underlying material. For example, a remarkable property of speckles is the \emph{memory effect}: speckle fields produced under small perturbations in imaging parameters (e.g., change in illumination direction) are highly correlated shifted versions of each other (see \figref{fig:teaser}). These speckle statistics have received strong attention since the invention of coherent laser illumination~\cite{PhysRevE.49.4530,BERKOVITS1994135,feng1988correlations,PhysRevLett.61.2328}, and are at the core of a large array of imaging techniques, with applications as diverse as motion tracking, estimating blood flow, looking around the corner, and seeing through scattering layers.

Unfortunately, and in stark contrast with the incoherent case, our ability to accurately simulate scattering in the coherent case is severely limited. Available algorithms generally fall into two categories. The first category consists of algorithms that compute output waves by numerically solving Maxwell's equations~\cite{mudiff,FDTDYee66,kWaves}. These algorithms are physically accurate, but require as input the \emph{microscopic} structure of the scattering medium, that is, knowledge of the exact (at sub-wavelength accuracy) locations of all scatterers in the medium. Even when such a microscopic characterization is available (e.g., specific samples examined with a microscope, or volumes with hallucinated scatterer locations), the high computational complexity of wave equation solvers makes them inapplicable for volumes larger than a few hundred cubic wavelengths, or containing more than a few hundred scatterers. The second category consists of approximate Monte Carlo rendering algorithms~\cite{Xu:04,Sawicki:08}, which accumulate the complex throughput (amplitude and phase) of paths sampled using standard volumetric path tracing. These algorithms are efficient, but cannot reproduce statistical properties of real speckles such as the memory effect. The lack of speckle rendering algorithms that are both \emph{physically accurate} and \emph{computationally efficient} is a significant obstacle in the wide range of fields interested in coherent imaging of scattering volumes. Symptomatic of these shortcomings of existing rendering tools is the fact that the only reliable way for estimating the memory effect has been by conducting painstaking optical lab experiments~\cite{Schott:15}.

In this paper, we change this state of affairs by developing a Monte Carlo framework for rendering speckles in volumetric scattering. Our framework builds on the following insight: Due to the central limit theorem, speckles are instances of a multivariate Gaussian distribution~\cite{GoodmanSpeckle}. Therefore, it is sufficient to model their (scene and material-dependent) mean and covariance. To achieve this, we draw inspiration from Monte Carlo volume rendering algorithms for the incoherent case: These algorithms treat the scattering medium as a {continuous volume}, inside which light can scatter randomly at any location. Given bulk parameters characterizing the \emph{statistical distribution} of scatterers in the medium, Monte Carlo algorithms synthesize images corresponding to the average distribution of scattered light across all scatterer instantiations that can be generated from the bulk parameters~\cite{moon2007rendering}. This macroscopic view of the medium enables efficient rendering, without the need to know and simulate the medium's microscopic structure.

To extend this approach to the coherent case, we begin by deriving from first principles a new path-integral formulation~\cite{veach1997robust} for the propagation of coherent light inside a scattering medium, which accurately encapsulates the first-order and second-order statistics of resulting speckle patterns. From this formulation, we derive two Monte Carlo rendering algorithms. The first algorithm estimates speckle covariance, which, together with an estimate of speckle mean obtained using a closed-form expression, can be subsequently used to sample multiple speckle images. The second algorithm directly simulates a physically-accurate speckle image, and operates by having sampled paths contribute to multiple pixels in a way that produces accurate speckle statistics. Both algorithms take as input only bulk macroscopic scattering parameters, as in the incoherent case. We validate our theory and algorithms in a few ways: First, we show that our approach can closely match ``groundtruth'' speckle estimates, obtained by averaging solutions of the wave equation across multiple particle instantiations, while also being orders of magnitude faster. Second, we show that our approach agrees with analytical formulas for speckle correlations derived for  specific cases (e.g., diffusion). Finally, we show that our approach can accurately reproduce well-documented properties of speckles, such as the memory effect and coherent backscattering. We show example applications of our framework, including simulating speckle-based computational imaging techniques, and evaluating the extent of their applicability. 

\subsection{Why render speckle patterns?}

There exist several imaging techniques that directly leverage second-order speckle statistics. Example applications include motion tracking~\cite{Jacquot:79,Jakobsen2012,Smith:2017:Speckle}, looking around the corner~\cite{FREUND199049,Katz2012,Batarseh2018}, and seeing through~\cite{Katz2014,Bertolotti2012} or focusing through~\cite{Mosk2012,Nixon2013,GenOptMemory17,Vellekoop:10} tissue and other scattering layers. Most of these imaging techniques rely on the \emph{memory effect} of speckles, a fact that has motivated significant research on quantifying this effect for different materials. Existing computational approaches generally attempt to derive closed-form expressions for the memory effect~\cite{Akkermans07,Baydoun2016,DOUGHERTY94,GenOptMemory17,Freund92,BERKOVITS1994135,Fried:82,feng1988correlations}. Unfortunately, these expressions only hold under assumptions such as diffusion or the Fokker-Planck limits, restricting their applicability. As a result, it has generally been necessary to measure the memory effect empirically using involved optical setups~\cite{Schott:15,Yang:14,mesradi:hal-01316109}. Our algorithm allows quantifying  the memory effect for arbitrary scattering materials computationally, through accurate yet efficient simulations. This can significantly enhance our understanding of the applicability of memory effect techniques to different materials. Additionally, this new simulation capability can save considerable lab effort for tasks such as discovering optimal settings for computational imaging systems, and evaluating new imaging configurations.

The ability to efficiently render speckle patterns can facilitate the widespread adoption of data-driven approaches in fields where coherent imaging of scattering is common, such as tissue imaging and material science. Previously, the lack of physically-accurate simulation tools meant that training datasets had to be collected using lab measurements, an approach that is not scalable.

Finally, speckle statistics can be beneficial for \emph{inverse rendering}, that is, retrieving material parameters from image measurements. While previous approaches  use  intensity measurements~\cite{InverseRendSA13,InvRenderingECCV16,VadimICCP,AviadICCV}, measurements of speckle statistics may  capture additional information and allow inverse rendering techniques to be applied in finer scales, where it is not possible to image without coherent effects.

\section{Related Work}\label{sec:related}

Monte Carlo rendering of wave optics effects has recently attracted increased attention in computer graphics. A primary focus has been on rendering diffraction and speckle effects generated by \emph{surface} microgeometry~\cite{yan2018rendering,Stam:1999:DS,Cuypers:2012:RMD:2231816.2231820,Werner17,Yeh:2013:WCI:2508363.2508420,vmv.20161357}, without tackling volumetric scattering. Some approaches focusing on scattering and speckle effects can be found in the optics literature~\cite{pan1995low,schmitt1997model,lu2004monte}. For instance, Xu et al.~\shortcite{Xu:04,Sawicki:08} modify volumetric path tracing, by tracking complex phase as a path is traced through the volume. By aggregating complex contributions from paths on the sensor, this technique produces images that resemble speckle patterns. However, because every pixel is rendered independently, this approach cannot reproduce spatial correlations between pixels. Additionally, it is impossible to use these approaches to reproduce correlations that exist across multiple  illumination directions as in the memory effect.

 There have been  attempts to use Monte Carlo algorithms to evaluate various properties of coherence and partial coherence of light after propagating through a scattering tissue~\cite{Shen:17,Pierrat:05}. Often these are based on using the radiative transfer equation (RTE) and intensity-based Monte Carlo rendering, then applying a Fourier transform on its result. Such approaches can be justified as a special case of our algorithm.

An important result in the study of speckle statistics, which can be used to derive Monte Carlo rendering algorithms, is the \emph{correlation transfer equation} (CTE)~\cite{ishimaru1999wave,Twersky64,DOUGHERTY94}. This integral equation extends the RTE, by modeling correlation of fields at different space points. As we show in \secref{sec:RES_Validation}, there are physical phenomena that are not accounted for by the CTE, such as coherent backscattering. While there exist some Monte Carlo rendering algorithms that take this effect into account~\cite{Sawicki:08,Ilyushin2012}, they only simulate intensity and not general covariance. We revisit the derivation of the CTE and its underlying assumptions, aiming to derive a more general rendering framework that accurately models both covariance and coherent backscattering.

Our derivation is fundamentally based on supplanting the true scattering volume, consisting of multiple discrete scatterers at fixed locations, with a continuous volume where scattering can happen randomly at any location. This macroscopic treatment of scattering underlies all current Monte Carlo volume rendering algorithms, and has also been used to accelerate rendering of so-called \emph{discrete random media}, where the scatterers can be arbitrarily large or dense~\cite{moon2007rendering,muller2016efficient,meng2015multi}. More recently, a number of works have used this approach to derive generalized versions of the RTE and Monte Carlo rendering algorithms, for media where the distribution of scatterer locations has spatial correlations, so-called \emph{non-exponential media}~\cite{bitterli2018radiative,Jarabo2018radiative,d2018reciprocal,d2018reciprocalpt2}. Even though we focus exclusively on exponential media, our work provides the foundations for future investigations of Monte Carlo rendering of speckles in non-exponential media.

Finally, there is also research on temporal correlations in the presence of scatterer motion, e.g., in liquid dispersions~\cite{DOUGHERTY94}. Many established techniques use these \emph{temporal speckle correlations} to estimate flow (e.g., blood flow~\cite{Durduran2010}) and liquid composition parameters. Example techniques include diffusing wave spectroscopy~\cite{pine1988diffusing}, laser speckle contrast imaging~\cite{Boas:97}, and dynamic light scattering~\cite{DynamicLightScattering}. Here we focus on spatial speckle correlations leaving  these temporal effects for future work.

\section{Modeling speckle statistics}\label{sce:speckeles-def}
\begin{figure*}[t]
        \centering
        \begin{tabular}{@{}c@{}c@{}c@{}c@{}c@{}c@{}}
                \includegraphics[height=0.155\linewidth]{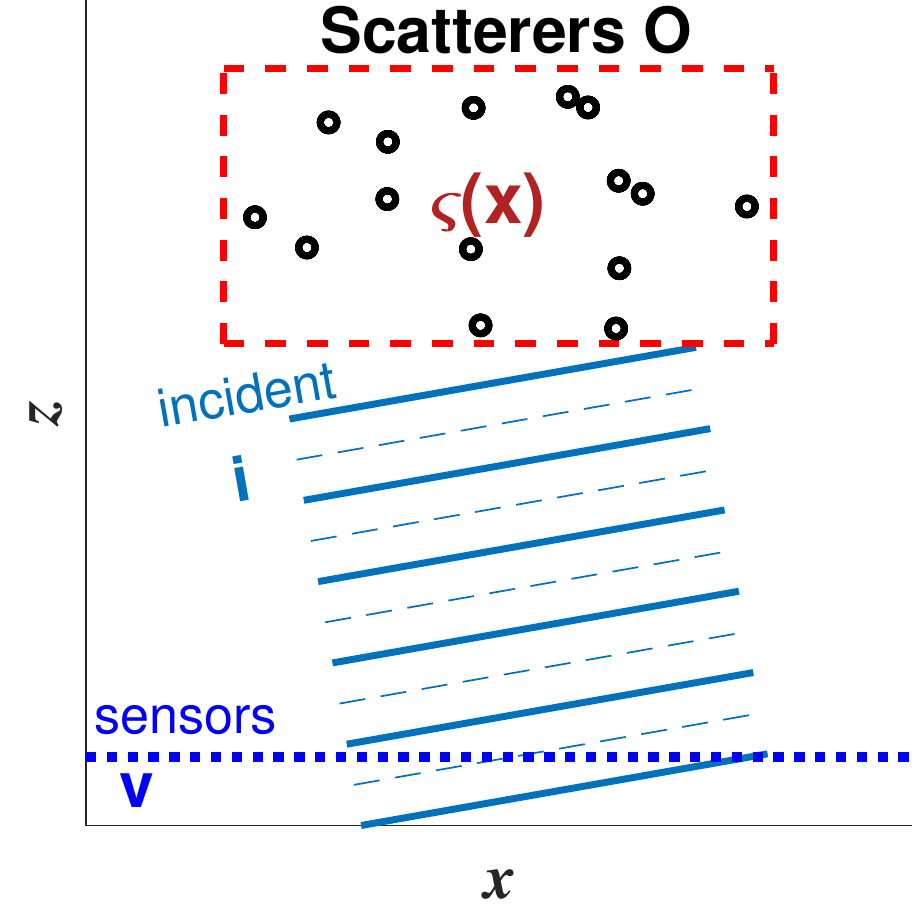}&
                \includegraphics[width=0.155\linewidth]{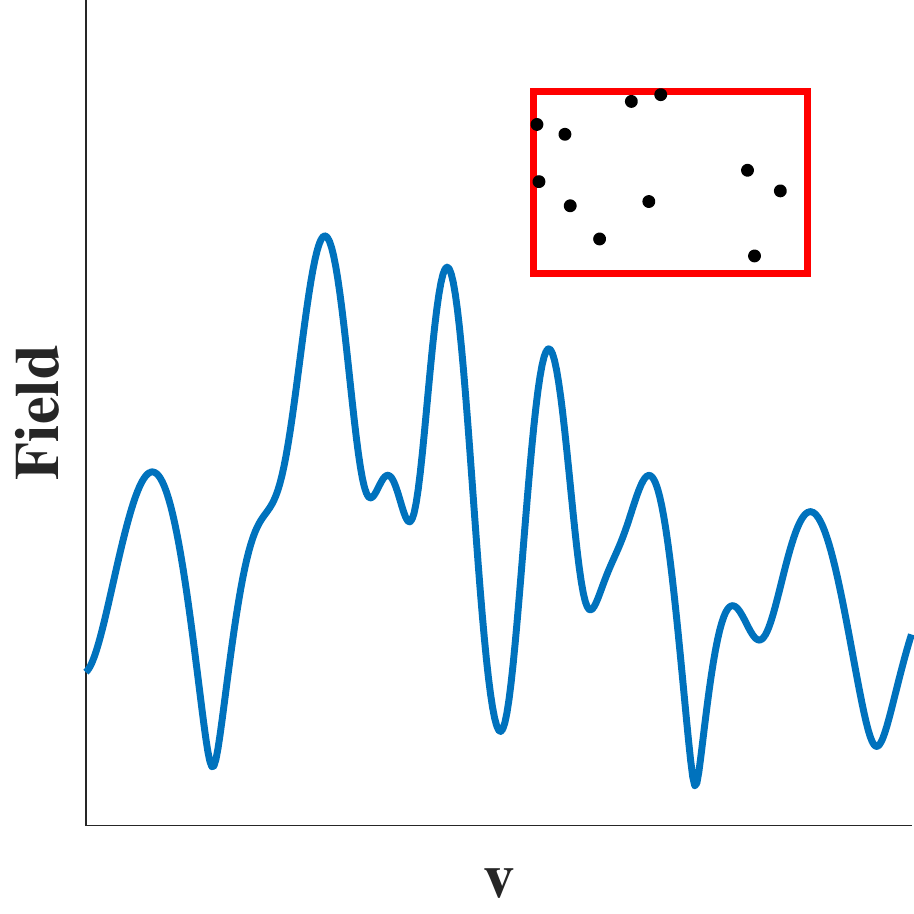}&
                \includegraphics[width=0.155\linewidth]{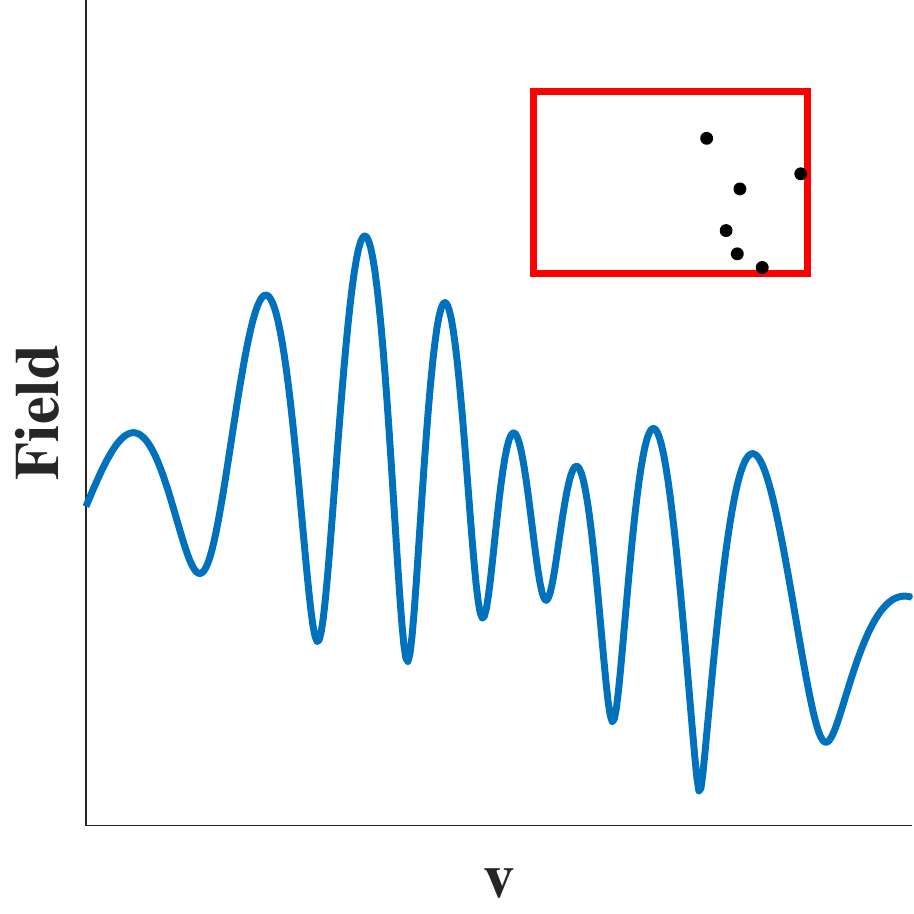}&
                \includegraphics[width=0.155\linewidth]{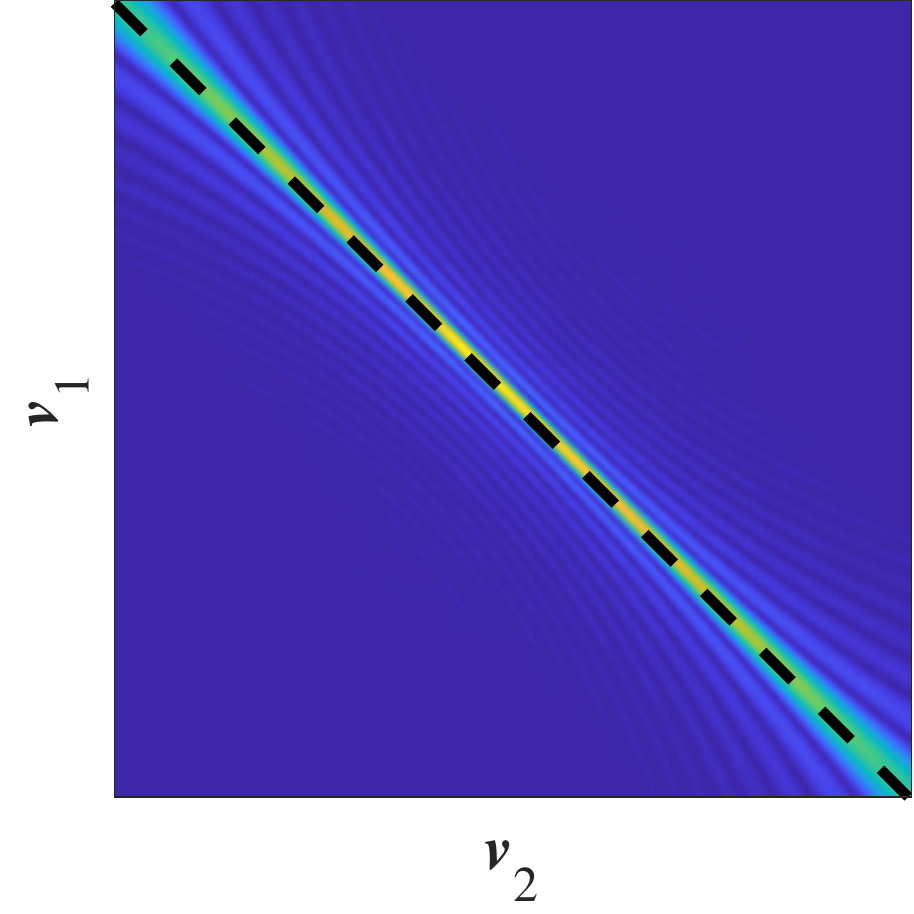}&
                \includegraphics[width=0.155\linewidth]{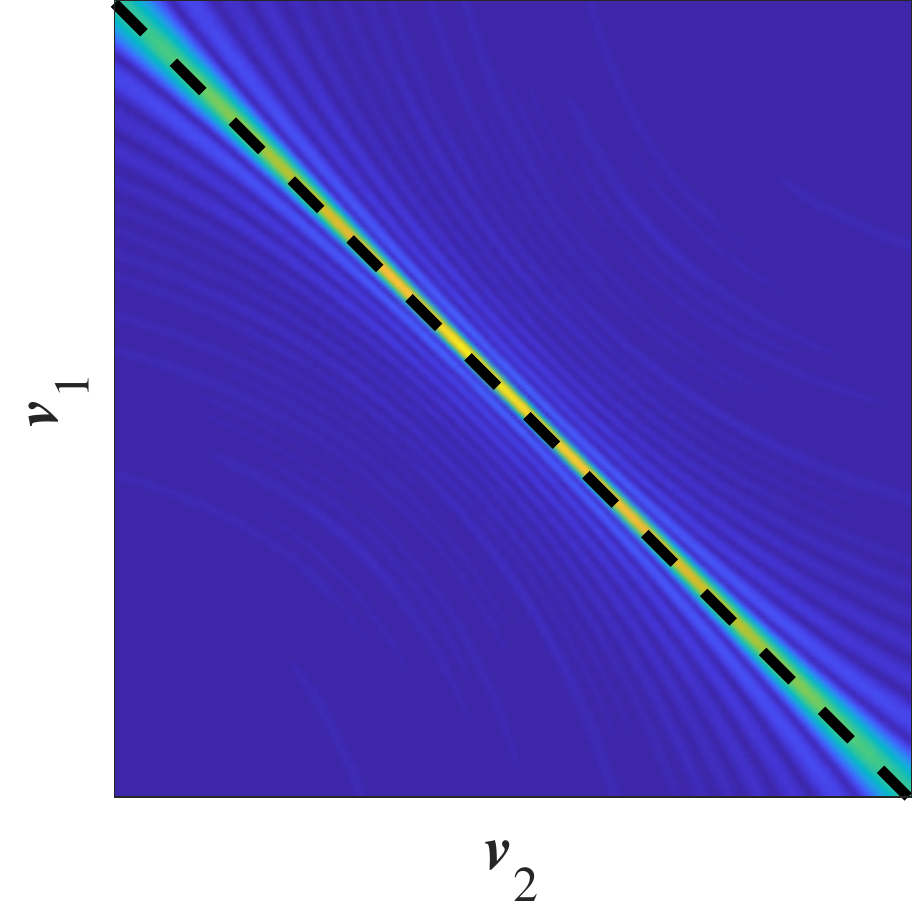}&
                \includegraphics[width=0.155\linewidth]{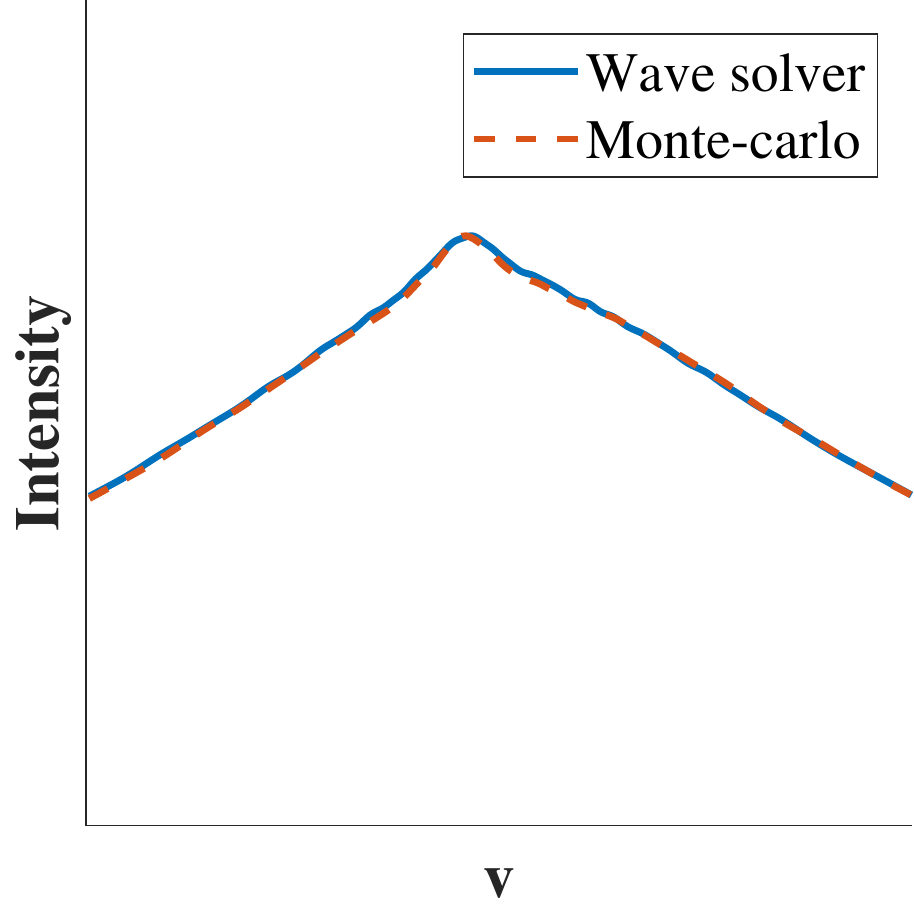}
                \\
                \begin{footnotesize}(a) Setup with one illumination\end{footnotesize}&\begin{footnotesize}(b) Sampled field 1\end{footnotesize}&\begin{footnotesize}(c) Sampled field 2\end{footnotesize}&\begin{footnotesize}(d) Wave solver cov\end{footnotesize}&\begin{footnotesize}(e) MC covariance\end{footnotesize}&\begin{footnotesize}(f) Diagonal plot from (d,e)\end{footnotesize}
                \\\\
                \includegraphics[height=0.155\linewidth]{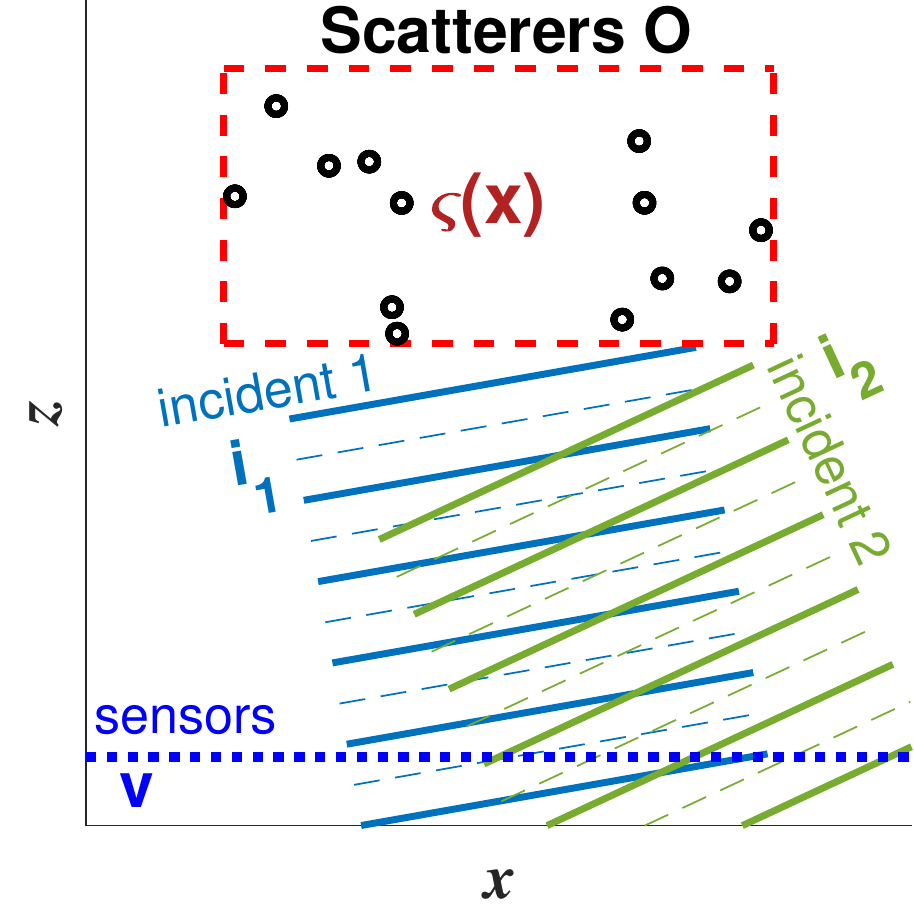}&
                \includegraphics[width=0.155\linewidth]{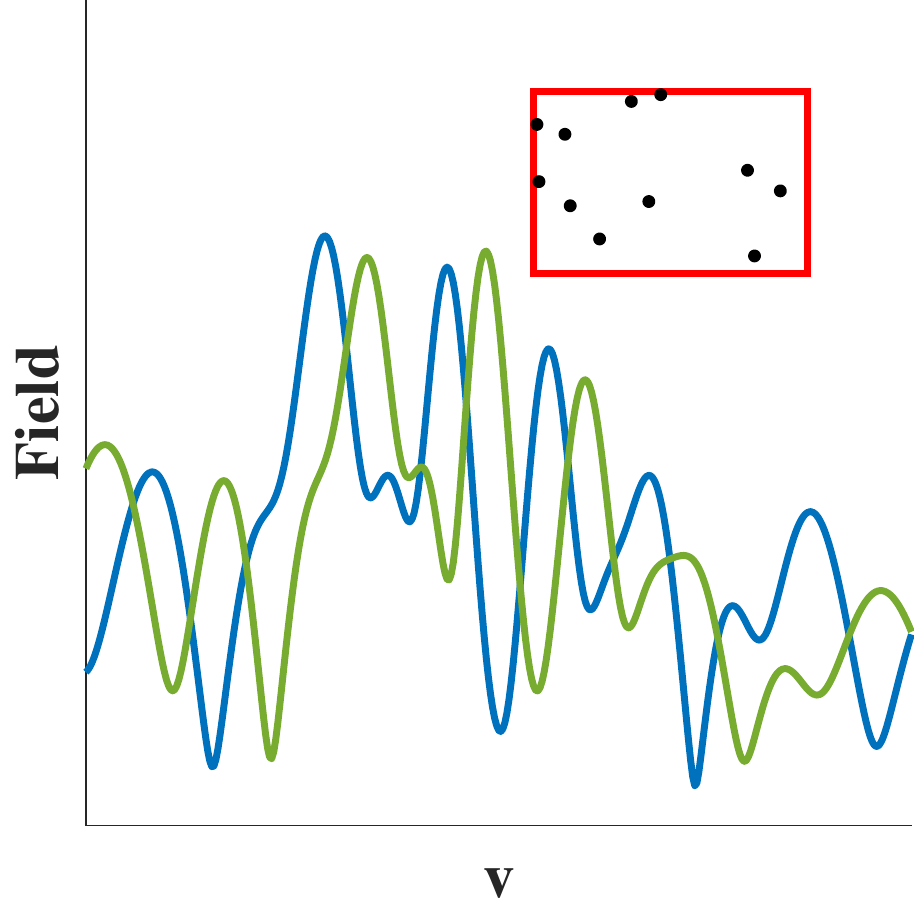}&
                \includegraphics[width=0.155\linewidth]{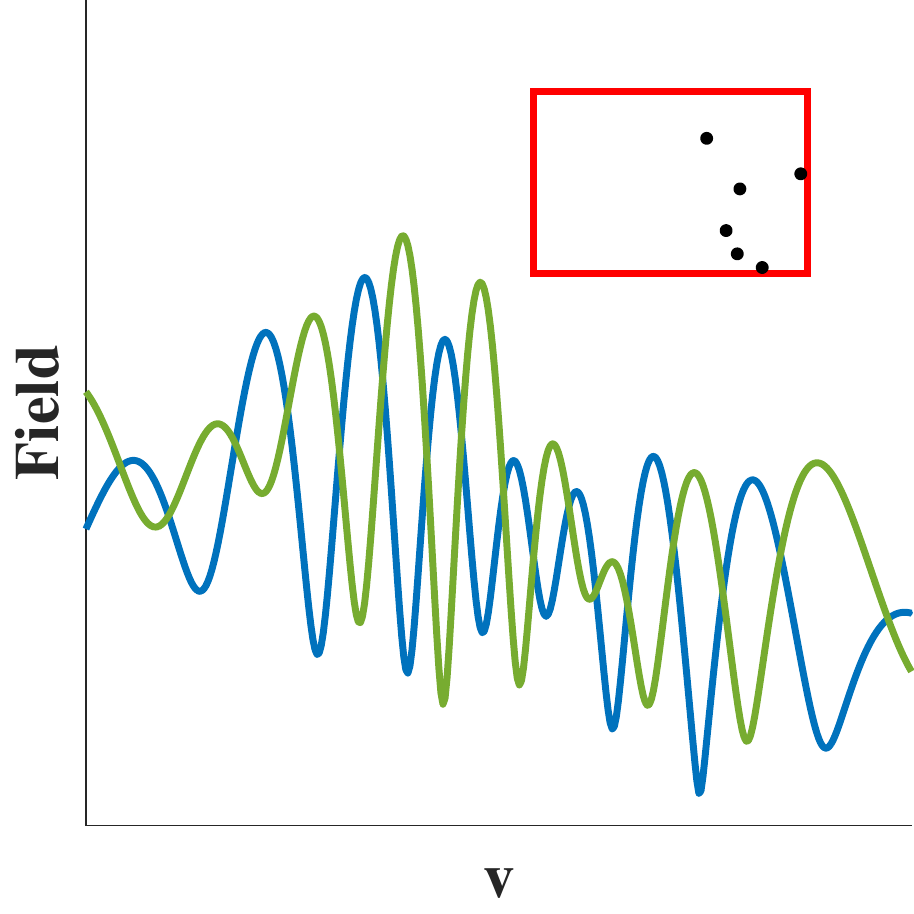}&
                \includegraphics[width=0.155\linewidth]{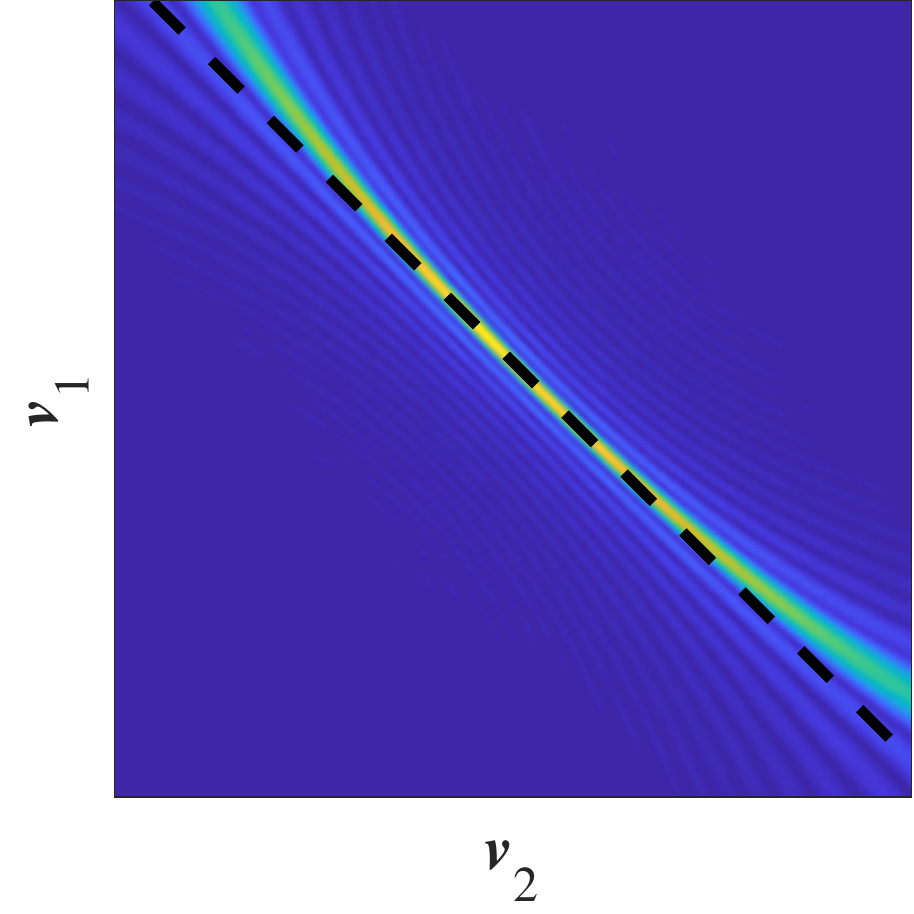}&
                \includegraphics[width=0.155\linewidth]{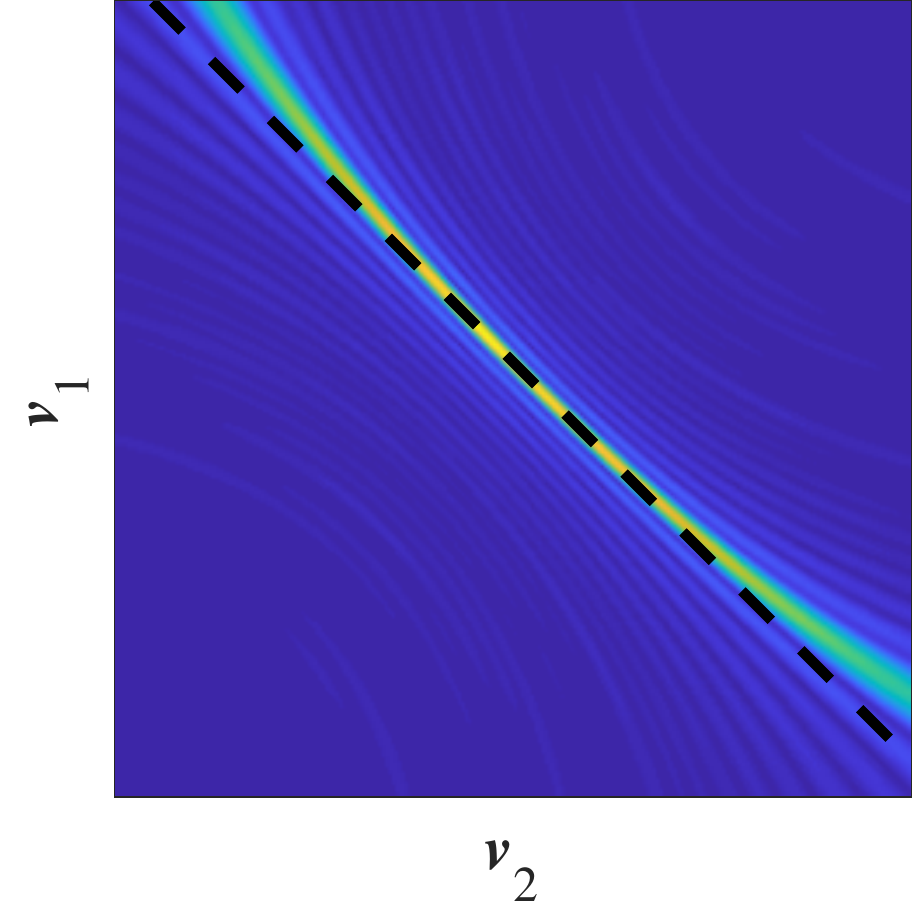}&
                \includegraphics[width=0.155\linewidth]{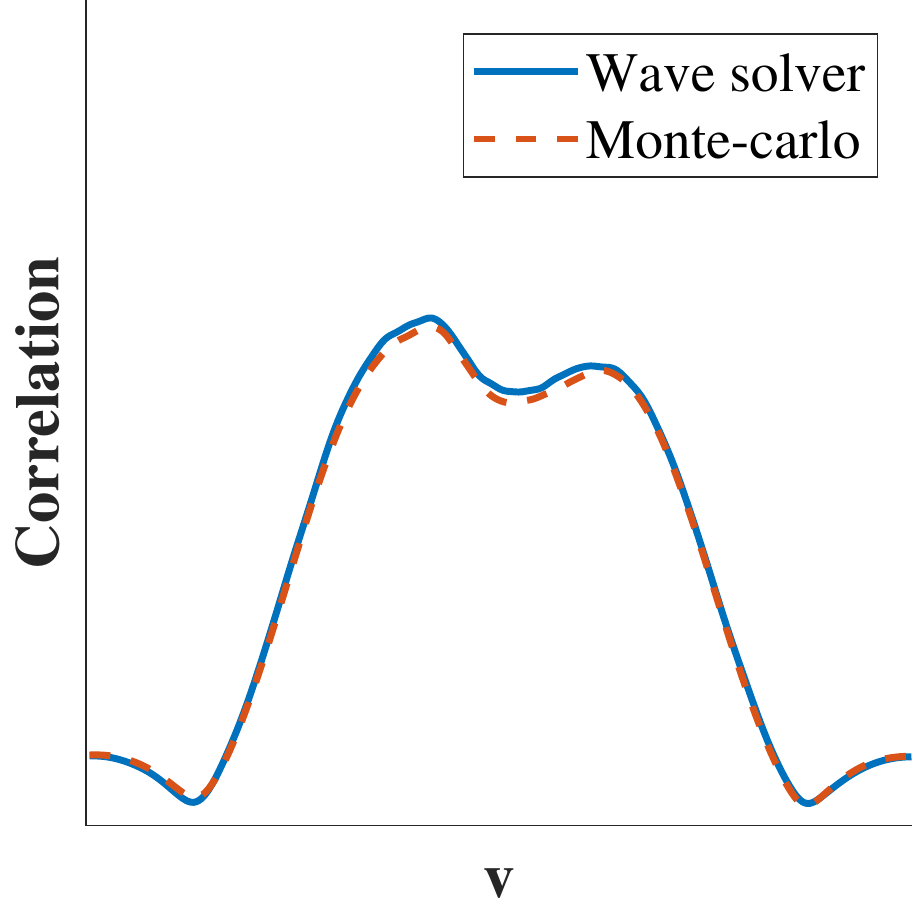}
                \\
                \begin{footnotesize}(g) Setup with two illuminations\end{footnotesize}&\begin{footnotesize}(h) Sampled field 3\end{footnotesize}&\begin{footnotesize}(i) Sampled field 4\end{footnotesize}&\begin{footnotesize}(j) Wave solver cov\end{footnotesize}&\begin{footnotesize}(k) MC covariance\end{footnotesize}&\begin{footnotesize}(l) Diagonal plot from (j,k)\end{footnotesize}
        \end{tabular}
        \caption{{\bf Simulating speckle and their statistics.} (a) Consider a rectangular scattering volume illuminated by a plane wave and a scattered field  sensed by collinear sensors. For each scatterer instantiation we solve the wave equation~using the package of \cite{mudiff} and compute the scattered field, shown in (b,c). Different scatterer positions lead to different high-fluctuation speckle fields. The empirical  covariance of multiple fields obtained with the wave solver is demonstrated in (d), and is closely matched by the covariance computed directly by our Monte Carlo algorithm (e). To demonstrate the good agreement we overlay a diagonal plot (f)% and a row plot (g)
        . The diagonal of the speckle covariance is equivalent to intensity images computed by standard incoherent intensity MC algorithms.  In the lower row (g) we consider a situation where the same scatterers instantiation is illuminated by two different incident directions highlighting that despite their semi-random structure speckles have strong statistical properties. In particular, the memory effect of speckles: when the same set of scatterers is illuminated by two incident directions the resulting speckle patterns are shifted versions of each other (h-i). This also implies that the covariance of the speckle fields (j) generated by two illumination directions has a shifted diagonal, where the diagonal offset corresponds to fields shift. Our Monte Carlo algorithm is physically correct and captures all such statistics, while having a computational complexity several orders of magnitude smaller than the wave equation solver.}  \label{fig:example-setup}    
\end{figure*}

\boldstart{Definitions and notation.} We use bold letters for three-dimensional vectors (e.g., points $\ptd,\inp,\snsp$), with a circumflex for unit vectors (e.g., directions $\omgv,\ind,\outd$). We also use $\dir{\ptd\ptdy}$ for the unit vector from $\ptd$ to $\ptdy$. Illumination and imaging can be at either the near or the far field: Near-field illumination is an isotropic source at point $\inp$, whereas far-field illumination is a directional plane wave source at direction $\ind$; and likewise for sensor points $\snsp$ and directions $\outd$. We often abuse the point notation $\inp,\snsp$ for both the far-field and near-field cases, except where context requires otherwise. We also restrict discussion to unpolarized illumination.

We consider scattering volumes $\V\in\R^3$ that satisfy four assumptions: First, they consist of scatterers with size comparable to the illumination wavelength, and which can therefore be considered infinitesimal. Second, the scatterers are far from each other, with an average pairwise distance (the \emph{mean free path}) an order of magnitude larger than the wavelength. Third, the locations of scatterers are statistically independent. Fourth, scatterers scatter incident waves in a way that is invariant to rotation. These assumptions underly classical radiative transfer~\cite{bitterli2018radiative}. To simplify notation, in the main paper we derive results assuming scatterers of a single type (same shape, size, and refractive index), and extend to the case of multiple types in Appendix~\ref{sec:bulkandmean}. We denote by $\materialDensity(\ptd),\,\ptd\in\V$ the, possibly spatially varying, density describing the distribution of scatterers in the medium. Finally, to simplify discussion, we do not model the interface of volume $\V$; interface events (reflection and refraction) can generally be incorporated in our resulting rendering algorithms same as in regular volume rendering. 

\boldstart{The scattered field.} An incident wave of wavelength $\lambda$ interacting with scatterers stimulates a \emph{scattered wave} $\wave$, which can be computed by solving the Helmholtz equation. When a single particle located at $\potd$ is illuminated from direction $\ind$, the scattered wave $\wave$ at distance $|\ptd-\potd| \gg \lambda$  is,
\begin{equation} \label{eq:farField}
u(\ptd) =\sqrt{\scatteringCrossSection} \cdot \ampf\left(\left<\ind\cdot\dir{\potd\ptd}\right>\right) \cdot \frac{\exp \left\{ ik|\ptd-\potd| \right\} }{|\ptd-\potd|},\,\,k=\frac{2\pi}{\lambda}.
\end{equation} 
The complex function $ \ampf(\cos\theta)$ is the \textit{scattering amplitude function}, describing  scattering at different angles. The scalar $\scatteringCrossSection$ is known as the \emph{scattering cross-section}, and accounts for the total energy scattered by the scatterer. Both quantities are a function of the wavelength and the scatterer's shape, size, interior and exterior refractive index; for spherical scatterers, they can be computed using Mie theory~\cite{BorenScattering,FCJ07}. We note that the scattering amplitude function is often defined as the product $\sqrt{\scatteringCrossSection} \cdot \ampf(\cos\theta)$. Here we separate the two terms and assume that $|\ampf(\cos\theta)|^2$ integrates to 1.

We now consider the geometry illustrated in \figref{fig:example-setup}a: Scatterers are placed at locations $O=\{\potd_1,\potd_2,\ldots\}$, each sampled \emph{independently} from the density $\materialDensity(\ptd )$. This configuration is illuminated from a source $\inp$, and imaged with a sensor $\snsp$. Knowing the exact scatterer locations, we can solve the wave equation to obtain the complex-valued scattered field $\wave^{\inp,O}_\snsp$, which typically contains large fluctuations with a semi-random noise structure known as speckles (see flatland  speckles in \figref{fig:example-setup}b,c).

\boldstart{Speckle statistics.} Images modeled using the radiative transfer equation correspond to the \emph{intensity} of the scattered field, averaged over all particle instantiations $O$ sampled from $\materialDensity(\ptd )   $, as in \figref{fig:example-setup}f:
\begin{equation}\label{eq:int-from-u} 
I^{\inp}_{\snsp}=E_{O}\left[\left|\wave^{\inp,O}_\snsp\right|^2\right].
\end{equation}
These incoherent intensity images are typically smooth, without speckles. This is because of the \emph{incoherent addition} in \equref{eq:int-from-u}: The expectation is formed by averaging intensities of waves, whereas speckles are the result of \emph{coherent addition} of complex valued waves. To capture speckle statistics, we can begin with the \emph{speckle mean},
\begin{equation}\label{eq:mean-from-u}
\meanspk_{\snsp}^{\inp}=E_{O}\left[\wave^{\inp,O}_{\snsp}\right].
\end{equation}
We can similarly define higher-order statistics of speckles. Of particular importance will be the \emph{speckle covariance},
\begin{equation}\label{eq:corr-from-u}
C^{\inp_1,\inp_2}_{\snsp_1,\snsp_2}=E_{O}\left[\wave^{\inp_1,O}_{\snsp_1}\cdot{\wave^{\inp_2,O}_{\snsp_2}}^*\right]-\meanspk_{\snsp_1}^{\inp_1}\cdot{\meanspk_{\snsp_2}^{\inp_2}}^*,
\end{equation}
where $(\cdot)^*$ denotes complex conjugation. In this case, $\wave^{\inp_1,O}_{\snsp_1}, \wave^{\inp_2,O}_{\snsp_2}$ are two speckle fields generated by the {\em same} scatterer configuration $O$, when illuminated by two incident waves from $\inp_1,\inp_2$, and measured at two sensors $\snsp_1,\snsp_2$. When $\inp_1=\inp_2=\inp,\snsp_1=\snsp_2=\snsp$ the term $C^{\inp,\inp}_{\snsp,\snsp}+|\meanspk_{\snsp}^{\inp}|^2$ from \equpref{eq:mean-from-u}{eq:corr-from-u} reduces to the intensity $I^{\inp}_{\snsp}$ of \equref{eq:int-from-u}.
As we discuss in \secref{sec:spk-mean}, the speckle mean can be computed using a closed-form expression; in fact, because the speckle mean is the aggregate of complex numbers of essentially randomly-varying phase, it is typically zero. Therefore, when characterizing speckle statistics, the most challenging part is computing the covariance.

\boldstart{Gaussianity of speckles.} Before we discuss ways to compute the speckle mean and covariance, one may wonder whether it is necessary to consider higher-order speckle statistics. The answer, in general, is negative: Classical results in optics~\cite{GoodmanSpeckle} state that the space of solutions $\wave^{\inp,O}_\snsp$ of the wave equation, for all particle configurations $O$ sampled from  $\materialDensity(\ptd )$, follows a multivariate Gaussian distribution with scene-dependent mean and covariance.
%, whose mean and covariance are a function of the illumination and sensing geometry, as well as the density and properties of the scatterers. 
The Gaussianity results from the central limit theorem, as the particle locations are independent random variables. 
%Even though most formal proofs of the Gaussianity property are stated for the cases of single or surface scattering, the same arguments extend to the case of multiple scattering. Additionally, the Gaussianity property has been verified empirically~\cite{gaussianityRefs}.\igkiou{Added these two sentences here, worth double-checking.} 
Consequently, the multivariate mean and covariance of \equpref{eq:mean-from-u}{eq:corr-from-u} provide sufficient statistics for speckle, and can be used to sample speckle patterns that are  indistinguisable from patterns generated by specifying exact particle positions and solving the wave equation.

\boldstart{Computing speckle statistics.} A straightforward approach for computing the speckle mean and covariance is to sample $N$ different scatterer configurations $O^1,\ldots O^N$, solve the wave equation for each configuration, and then compute the empirical moments: \begin{gather}\label{eq:mean-from-u-disc}
\meanspk_{\snsp}^{\inp}\approx\frac1N\sum_{n=1}^N \wave^{\inp,O^n}_{\snsp}, \\
C^{\inp_1,\inp_2}_{\snsp_1,\snsp_2}\approx\frac1N \sum_{n=1}^N \wave^{\inp_1,O^n}_{\snsp_1}\cdot{\wave^{\inp_2,O^n}_{\snsp_2}}^*-\meanspk_{\snsp_1}^{\inp_1}\cdot{\meanspk_{\snsp_2}^{\inp_2}}^*. \label{eq:corr-from-u-disc}
%=======
%C^{\inp_1,\inp_2}_{\snsp_1,\snsp_2}\approx\frac1N \sum_{n=1}^N \wave^{\inp_1,O^n}_{\snsp_1}\cdot{\wave^{\inp_2,O^n}_{\snsp_2}}^*-\meanspk_{\snsp_1}^{\inp_1}\cdot{\meanspk_{\snsp_2}^{\inp_2}}^*. \label{eq:corr-from-u-disc}
%>>>>>>> 676c8f04594cc8bf5075fa3a443c9dc22c835f4b
\end{gather}
\figref{fig:example-setup}(d,k) shows speckle covariances evaluated with this procedure. Solving the wave equation is only tractable for very small number of particles (a few thousands), and this computational cost is further exacerbated by the need to repeat this process multiple times. Our goal is to devise Monte Carlo algorithms that can compute speckle covariance directly and much more efficiently. 

\boldstart{Bulk parameters.} Unlike wave equation solvers, our algorithms are not tied to a specific position of scatterers. Instead, they rely only  on the \emph{distribution} of scatterers in the medium, as well as their size, shape, and refractive properties. As in the radiative transfer literature, we describe these using the scattering, absorption, and extinction coefficients, $\sigma_s, \sigma_a, \sigma_t$ respectively, defined as
\begin{gather}
\sctCoff(\ptd)= \Nomean (\ptd) \scatteringCrossSection,\;\; \atnCoff(\ptd)= \Nomean (\ptd) \atnCS+\atnCoff^{\text{med}},\;\; \Nomean(\ptd)=\frac{\materialDensity(\ptd)}{4/3\pi r^3},\\
\extCoff(\ptd)=\sctCoff(\ptd)+\atnCoff(\ptd),
\end{gather}
where $\scatteringCrossSection,\atnCS$ are the scattering and absorption cross-sections, denoting the energy scattered or absorbed upon interaction with one particle, $r$ is the radius of the particles, $\Nomean$ is the expected number of particles in a unit volume, and $\atnCoff^{\text{med}}$ is the absorption coefficient of the containing medium. We also use the \emph{phase function},   defined as $\rho(\cos\theta) = \left|  \ampf(\cos\theta) \right|^2$, explaining our earlier choice of normalization for the scattering amplitude function. The above definitions consider only particles of a single type, but it is not hard to extend them to multiple particle types, see Appendix~\ref{sec:bulkandmean}.

\section{Path-space view of speckle statistics}\label{sec:pathview}

In this section, we derive path-space expressions for the speckle mean and covariance. These expressions will form the basis for the Monte Carlo rendering algorithms of \secref{sec:rendering}. We note that, traditionally in computer graphics, path-space expressions are derived by recursively expanding integral equations such as the surface and volume rendering equations. Here, we start directly with a path-space view, and discuss the relationship with an integral equation known as the \emph{correlation transfer equation} (CTE) in Appendix~\ref{sec:CTE}.
\begin{figure}[t]
        \centering       
        \includegraphics[width=0.3\textwidth]{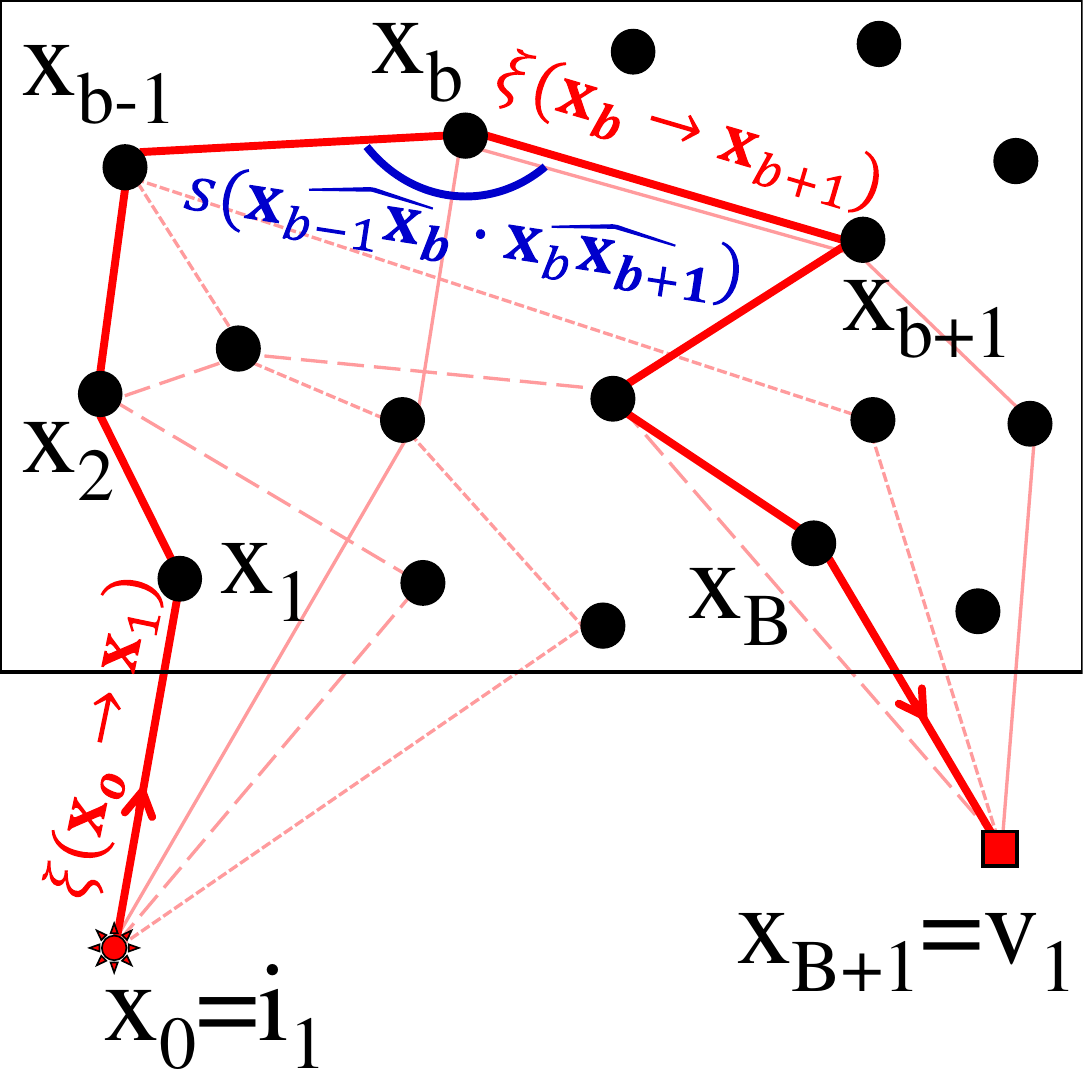}     
        \vspace{-0.4cm}
        \caption{The scattered field can be expressed as a sum of complex throughput $\sigsct(\pathseq)$ over all paths  in a scatterers set   $O$.}          \label{fig:field-path}
\end{figure}

\boldstart{Fields as path sums.} Our starting point is the classical theory of Twersky~\shortcite{Twersky64}: Given a configuration $O$ of particles, we can approximate the solution to the wave equation as the sum of contributions over all paths $\pathseq$ through $O$. That is, consider the (enumerable) set $\Path^{\inp,O}_{\snsp}$ of all ordered sequences:
\begin{equation}\label{eq:path-def} 
\!\!\pathseq=\ptd_0\!\rightarrow\!\dots\!\rightarrow\!\ptd_{B+1},\;\text{with}\;\ptd_0=\inp,\;\ptd_{B+1}=\snsp,\;\ptd_1,\dots,\ptd_{B}\in O,
\end{equation}
where $B=0,\dots,\infty$. Then, the scattered field can be expressed as
\begin{equation}\label{eq:field-paths} 
\!\!\wave^{\inp,O}_{\snsp}\!=\!\!\!\!\sum_{\pathseq\in\Path^{\inp,O}_{\snsp}}\!\!\sigsct(\pathseq)=\!\!\!\!\sum_{\pathseq\in\Path^{\inp,O}_{\snsp}}\!\!\sigsct(\ptd_0\!\to\!\ptd_1)\prod_{b=1}^{B}\sigsct(\ptd_{b-1}\!\rightarrow\!\ptd_b\!\rightarrow\!\ptd_{b+1}).
\end{equation}
These paths are illustrated in \figref{fig:field-path}. The \emph{complex throughput} terms $\sigsct(\cdot)$ correspond to the amplitude and phase change at each segment on the path, accounting for the scattering amplitude $\ampf$ and traveled length. They can be defined as
\begin{align}
\sigsct(\ptd_{b-1}\rightarrow\ptd_b\rightarrow\ptd_{b+1})&=\phase(\ptd_{b}\to{\ptd_{b+1}})\ampf(\dir{\ptd_{b-1}\ptd_{b}}\cdot\dir{\ptd_{b}\ptd_{b+1}}),\label{eq:sigsct-def}\\
\sigsct(\ptd_{0}\to{\ptd_{1}})&=\phase(\ptd_{0}\to{\ptd_{1}}).\label{eq:sigsct-0-def}
\end{align}
The \emph{complex transmission} terms $\phase(\cdot)$ account for phase change and radial decay between path vertices $\ptd_{b},\ptd_{b+1}$, defined for points at the near field and far field, respectively, as
\BE
\!\!\!\!\phase(\ptd_{1}\!\!\to\!\!{\ptd_2})\!=\! \frac{e^{ik|\ptd_{1}-{\ptd_2}|}}{|\ptd_{1}-\ptd_{2}|},\,\phase(\ind\!\!\to\!\!{\ptd})\!=\! e^{ik(\ind\cdot{\ptd})},\,\phase(({\ptd}\sto\outd)\!=\! e^{-ik(\outd\cdot{\ptd})}.
\EE

\boldstart{Speckle statistics as path integrals.} Using \equref{eq:field-paths}, we can now  express the mean and covariance by averaging over all particle configurations $O$ that can be sampled from the density $\materialDensity$: 
\begin{gather} \label{eq:mean-as-paths}
\meanspk^{\inp}_{\snsp}=E_{O}\left[\sum_{\pathseq\in\Path^{\inp,O}_{\snsp}}\sigsct(\pathseq)\right],\\
C^{\inp_1,\inp_2}_{\snsp_1,\snsp2_2}=E_{O}\left[\sum_{\pathseq^1\in\Path^{\inp_1,O}_{\snsp_1},\pathseq^2\in\Path^{\inp_2,O}_{\snsp_2}}\sigsct(\pathseq^1)\cdot{\sigsct(\pathseq^2)}^*\right]-\meanspk^{\inp_1}_{\snsp_1}\cdot{\meanspk^{\inp_2}_{\snsp_2}}^*.\label{eq:corr-as-paths}
\end{gather}
Note that, within the expectation, the summation is over paths $\pathseq^1,\pathseq^2$ through the \emph{same} particle instantiation $O$. By exchanging the order of expectation and summation in \equpref{eq:mean-as-paths}{eq:corr-as-paths}, we have: 
\begin{gather} \label{eq:meam-path-int}
\meanspk^{\inp}_{\snsp}=\int_{\Path^{\inp}_{\snsp}}p(\pathseq)\sigsct(\pathseq)\ud\pathseq,\\
C^{\inp_1,\inp_2}_{\snsp_1,\snsp_2}=\iint_{\Path^{\inp_1}_{\snsp_1},\Path^{\inp_2}_{\snsp_2}}p(\pathseq^1,\pathseq^2)\sigsct(\pathseq^1){\sigsct(\pathseq^2)}^*\ud\pathseq^1\ud\pathseq^2-\meanspk^{\inp_1}_{\snsp_1}{\meanspk^{\inp_2}_{\snsp_2}}^*,\label{eq:corr-path-int}
\end{gather}
where now the space $\Path^{\inp}_{\snsp}$ includes paths with vertices $\ptd_1,\dots,\ptd_B$ that can be \emph{anywhere} in the volume $\V$,  not only on fixed particle locations. Unlike $\Path^{\inp,O}_{\snsp}$, $\Path^{\inp}_{\snsp}$ is not an enumerable space, thus summation is replaced with integration. The term $p(\pathseq)$ is the probability that  the path $\pathseq$ is included in the enumerable path space $\Path^{\inp,O}_{\snsp}$ for some particle configuration $O$ sampled from $\materialDensity$; similarly $p(\pathseq^1,\pathseq^2)$  is the probability that all nodes on both $\pathseq^1,\pathseq^2$ are included in the {\emph {same}} particle configuration $O$.

In the following sections, we show that $\meanspk^{\inp}_{\snsp}$ can be computed in closed form, and we greatly simplify the path integral for $C^{\inp_1,\inp_2}_{\snsp_1,\snsp_2}$ by characterizing the pairs of paths that have non-zero contributions.

\subsection{The speckle mean}\label{sec:spk-mean}

Evaluating the speckle mean is addressed by standard textbooks on scattering. We present these results here, starting from a more general case, which subsumes the computation of speckle mean. The general case will also be useful for computing speckle covariance in the next section.

We consider a particle at $\ptd_1$, illuminated from a wave with incident direction $\omgv$. As this wave scatters, we want to evaluate the average contribution of all paths $\pathseq$ starting at $\ptd_1$ and arriving at a second point $\ptd_2$. The textbook result ~\cite{ishimaru1999wave,mishchenko2006multiple} states this average can be written as
%\begin{gather}\label{eq:int-paths-1}
%\int_{\Path_{\ptd_{1}}^{\ptd_{2}}} p(\pathseq)\sigsct(\pathseq)\ud\pathseq = \mulsct(\omgv\rightarrow\ptd_{1}\rightarrow\ptd_{2}),\\
%\text{with    }\;\;\mulsct(\omgv\rightarrow\ptd_{1}\rightarrow\ptd_{2})=
%\att(\ptd_{1},\ptd_{2})\cdot\sigsct(\omgv\rightarrow\ptd_{1}\rightarrow\ptd_{2}).\label{eq:mulsct-op}
%\end{gather}
\begin{equation}\label{eq:int-paths-1}
\int_{\Path_{\ptd_{1}}^{\ptd_{2}}} p(\pathseq)\sigsct(\pathseq)\ud\pathseq = \att(\ptd_{1},\ptd_{2})\cdot\sigsct(\omgv\sto\ptd_{1}\sto\ptd_{2}),
\end{equation}
where $\sigsct$ is defined as in \equref{eq:sigsct-def}. The term $\att$ is defined in the near and far fields as the probability of getting from $\ptd_1$ to $\ptd_2$ without meeting other particles, and is equal to
\BE\label{eq:att-def}
\!\!\att(\ptd_1,\ptd_2)\!=\!e^{\!-\frac12\int_{0}^{1}\!\!\extCoff(\alpha\ptd_1+(1-\alpha)\ptd_2)\ud\alpha},\,\att(\ind,\ptd)\!=\!e^{\!-\frac12\int_{0}^{\infty}\!\!\extCoff(\ptd_1-\alpha\ind)\ud\alpha}.
\EE
For a homogeneous medium $\att(\ptd_1,\ptd_2)=\exp(-\frac12\extCoff|\ptd_2-\ptd_1|).$ 
The factor $1/2$ in the exponent of \equref{eq:att-def} makes $\att$ the square root of the attenuation term used in standard radiative transfer. Intuitively, this is because we deal with the field rather than intensity. 

\begin{figure}[t]
        \centering       
        \subfloat[]{\label{fig:path-x1x2}\includegraphics[width=0.2\textwidth]
                {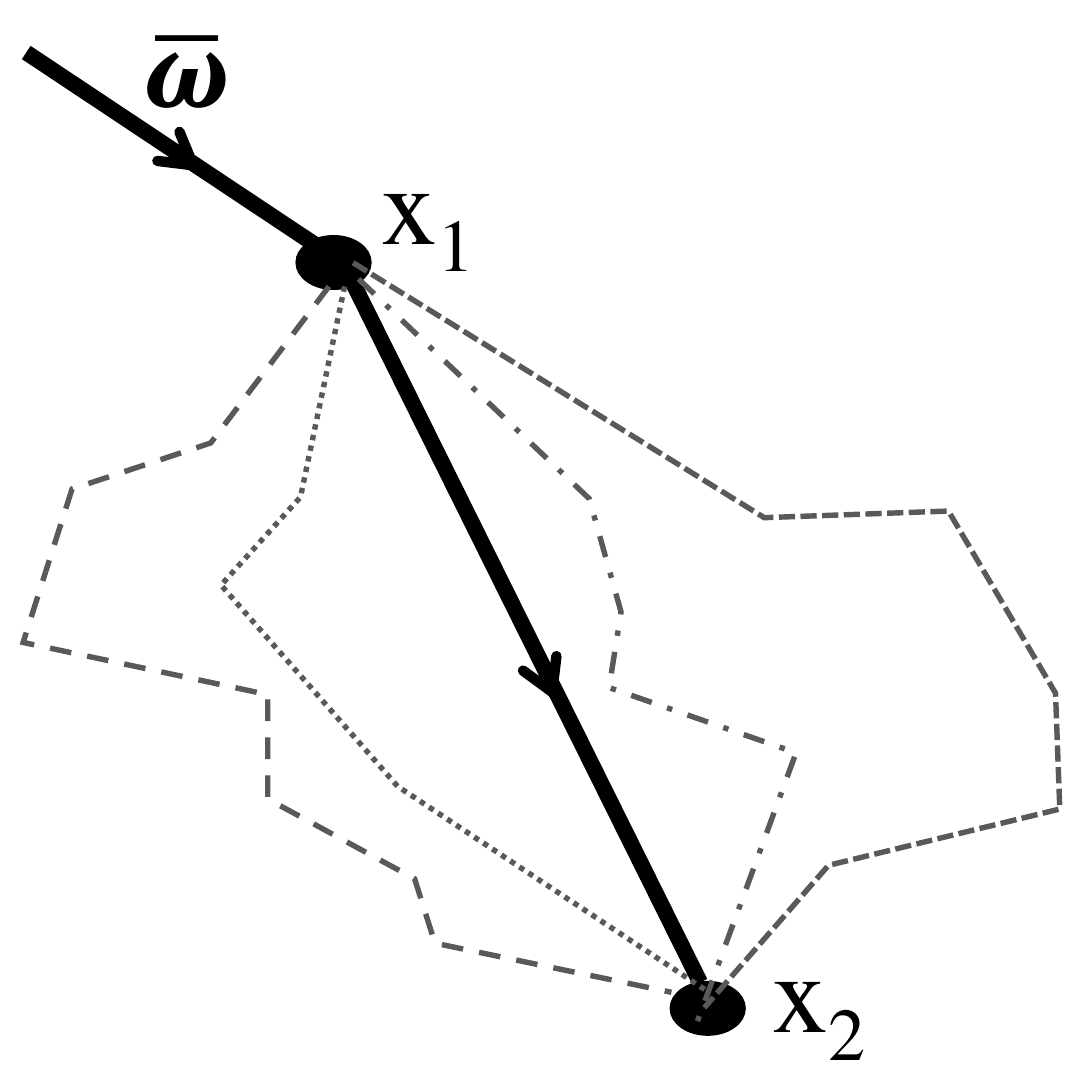}}\hfill
        \subfloat[]{\label{fig:point-source}\includegraphics[width=0.26\textwidth]
                {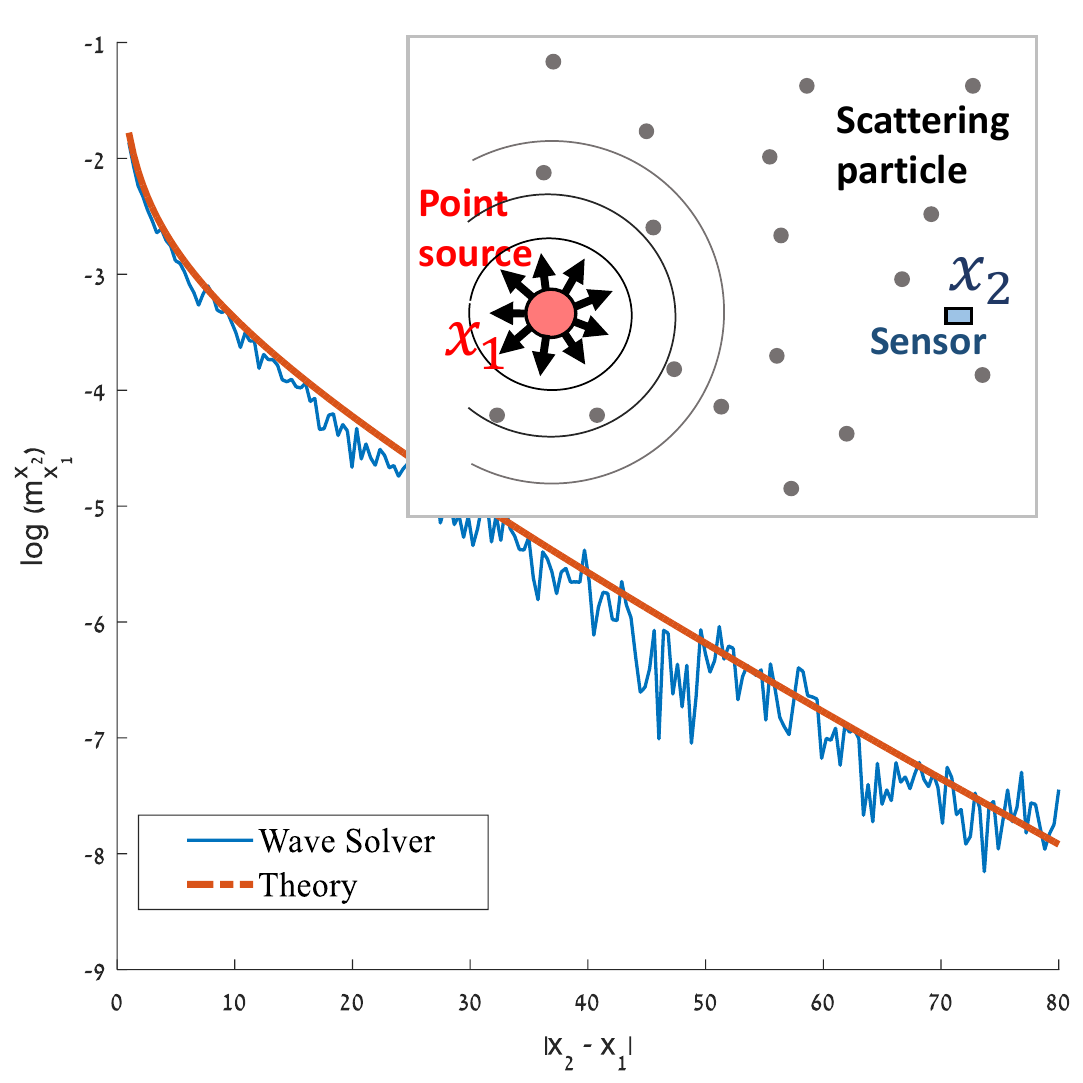}}
        \caption{(a) The average contribution of all paths connecting $x_1$ and $x_2$ (dashed lines) is the direct path (solid line). (b) Numerical simulation of speckle mean. Using the setup shown in the inset, we sampled multiple instantiations of particle positions and solved for the field scattered  from a point source at $x_1$ toward a sensor at $x_2$. Averaging scattered fields over multiple  instantiations of scattering particles provides a good agreement with the theory of \equref{eq:int-paths-1}.}  
        \label{fig:multiple-direct-x1x2}        
\end{figure}

The main intuition behind \equref{eq:int-paths-1} is that, as most paths contribute essentially random complex phases, they cancel each other out. Therefore, the total field from $\ptd_{1}$ to $\ptd_{2}$ equals the field that travels only along the \emph{direct path} between the two points, attenuated by the exponentially decaying probability $\att(\ptd_1,\ptd_2),$  see \figref{fig:multiple-direct-x1x2}(a). 
%}

\boldstart{Computing the speckle mean.} We can now adapt this result for the case of the speckle mean $\meanspk^{\inp}_{\snsp}$ of \equref{eq:meam-path-int}. If either the source or the sensor are at the near field, the speckle mean in \equref{eq:meam-path-int}   is
a special case of \equref{eq:int-paths-1}. Being the mean of paths from $\inp$ to $\snsp$ without conditioning on an incoming direction  $\omgv$, we
omit the 
$\ampf$ term representing scattering, and express
%\begin{equation}
%\meanspk^{\inp}_{\snsp}=\mulsct(\inp\leftrightarrow\snsp)=
%\att(\inp,\snsp)\cdot\sigsct(\inp\leftrightarrow\snsp).
%\end{equation}
\begin{equation}\label{eq:spk-mean-i-v}
\meanspk^{\inp}_{\snsp}=\int_{\Path^{\inp}_{\snsp}}p(\pathseq)\sigsct(\pathseq)\ud\pathseq= \att(\inp,\snsp)\cdot\sigsct(\inp\to\snsp),
\end{equation}
compare \equref{eq:sigsct-0-def} with \equref{eq:sigsct-def} for the definition of $\sigsct$. In Appendix 
\ref{sec:app-mean}, we show how to adjust this for the far field as well.
%=======
%If either the source or the sensor are at the near field (i.e. points rather than directions), the speckle mean in \equref{eq:meam-path-int} is a special case of \equref{eq:int-paths-1},  omitting the  $\ampf$ term representing the direction change,
%\BE \meanspk^{\inp}_{\snsp}=\mulsct(\inp\to\snsp)=
%\att(\inp,\snsp)\cdot\sigsct(\inp\to\snsp),
%\EE
%compare \equref{eq:sigsct-0-def} with \equref{eq:sigsct-def} for the definition of $\sigsct$. In Appendix  \ref{sec:app-mean}, we show how to adjust this for the far field as well.
%>>>>>>> 676c8f04594cc8bf5075fa3a443c9dc22c835f4b

The main consequence of this section is that computing the speckle mean becomes a \emph{direct illumination} problem, which can be solved analytically without the need for path integration. In \figref{fig:multiple-direct-x1x2}(b), we numerically evaluate the speckle mean by averaging multiple solutions of the wave equation as in \equref{eq:mean-from-u-disc},  showing a good agreement with the analytic formula of \equref{eq:spk-mean-i-v}. We note that, as the speckle mean decays exponentially with the distance, in most cases it is negligible, making the computation of covariance the main challenge in simulating speckle. We discuss this next.

\subsection{The speckle covariance}\label{cov-path-def}

We have shown in \equref{eq:corr-path-int} that the speckle covariance can be expressed as an integral over \emph{pairs} of paths $\pathseq^1$ from $\inp_1$ to $\snsp_1$ and $\pathseq^2$ from $\inp_2$ to $\snsp_2$. Unlike the mean, there is no closed-form expression for this integral. However, we can considerably simplify integration by characterizing the pairs of paths with non-zero integrand 
\BE \cll=p(\pathseq^1,\pathseq^2)\sigsct(\pathseq^1){\sigsct(\pathseq^2)}^*, \EE
and deriving a simple formula for $\cll$ for those pairs. Some of the arguments we use are also discussed in Mishchenko et al.~\shortcite{mishchenko2006multiple}. Here, we formalize these arguments and extend them to accurately account for both speckle covariance and, as we see below, coherent back-scattering. Our end-product is a new path-integral expression for covariance that lends itself to Monte Carlo integration.

%However, the fact that all paths connecting two nodes average to the direct path   will be reused below, and will allow us to prune many useless paths from the search space. 

%<<<<<<< HEAD
%\boldstart{Valid pairs of paths.} We first consider two paths $\pathseq^1,\pathseq^2$ that do not share any vertices. Intuitively, we expect these paths to not contribute to covariance. This intuition comes from the fact that $\cll$ is a complex number: As in the previous section, if we aggregate contributions $\cll$ from different pairs of paths with essentially random phase, they will average to zero. The exception to this argument is cases where $\cll$ is not complex; this can only happen when every segment $\ptd_{b}\rightarrow\ptd_{b+1}$ that appears in $\pathseq^1$ is also in $\pathseq^2$.
%=======
\boldstart{Valid pairs of paths.} Intuitively, as $\cll$ is a complex number, if we aggregate contributions $\cll$ from different pairs of paths with very different phases, they will likely average to zero. The exception to this argument is cases where $\cll$ is not complex; this happens when every segment $\ptd_{b}\rightarrow\ptd_{b+1}$ that appears in $\pathseq^1$  also appears in $\pathseq^2$.

\begin{figure*}[t]
        \centering       
        \subfloat[All paths]{\label{fig:paths-multiple}\includegraphics[width=0.22\textwidth]
                {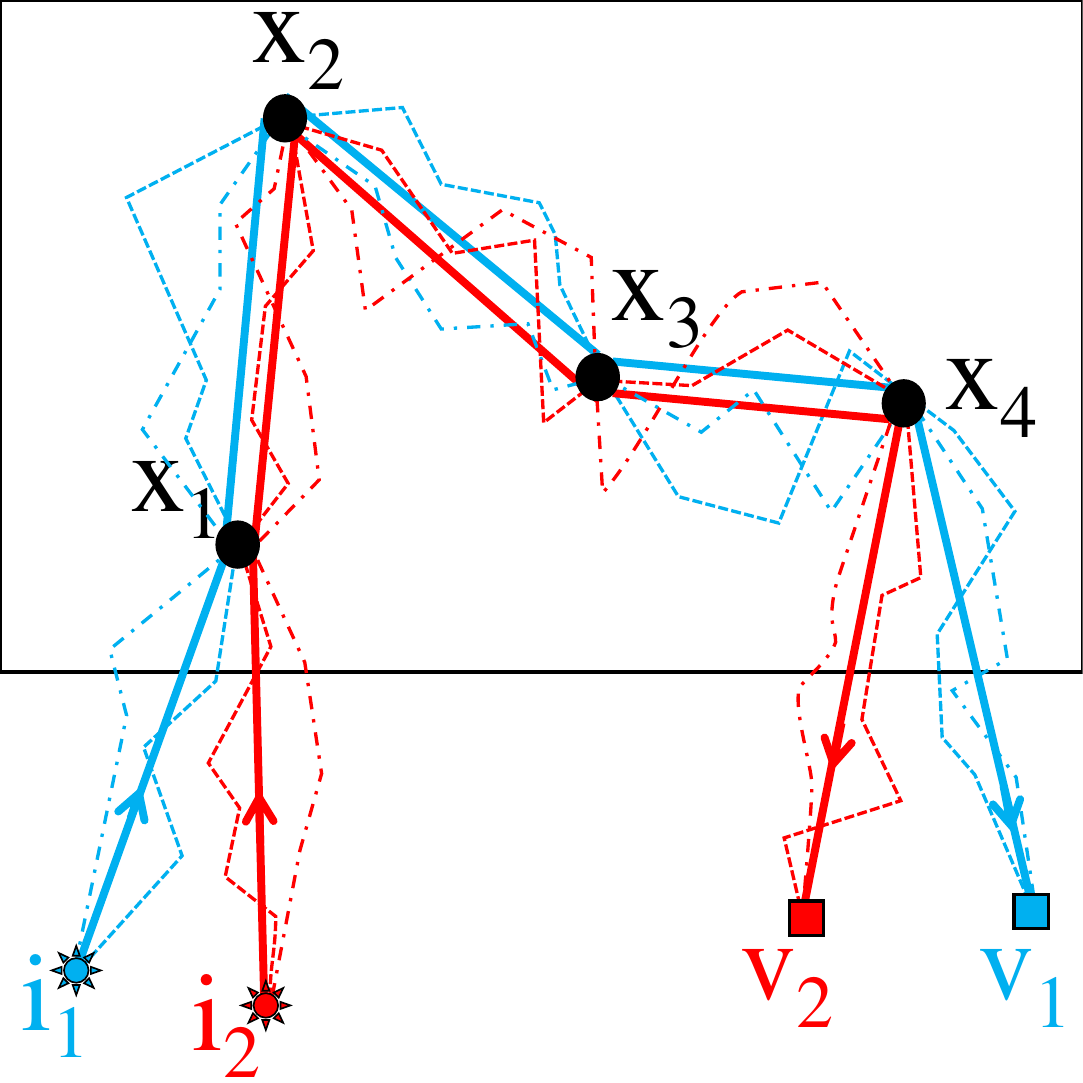}}\hspace{0.5cm}
        \subfloat[Permuted order of nodes]{\label{fig:paths-permute}\includegraphics[width=0.22\textwidth]
                {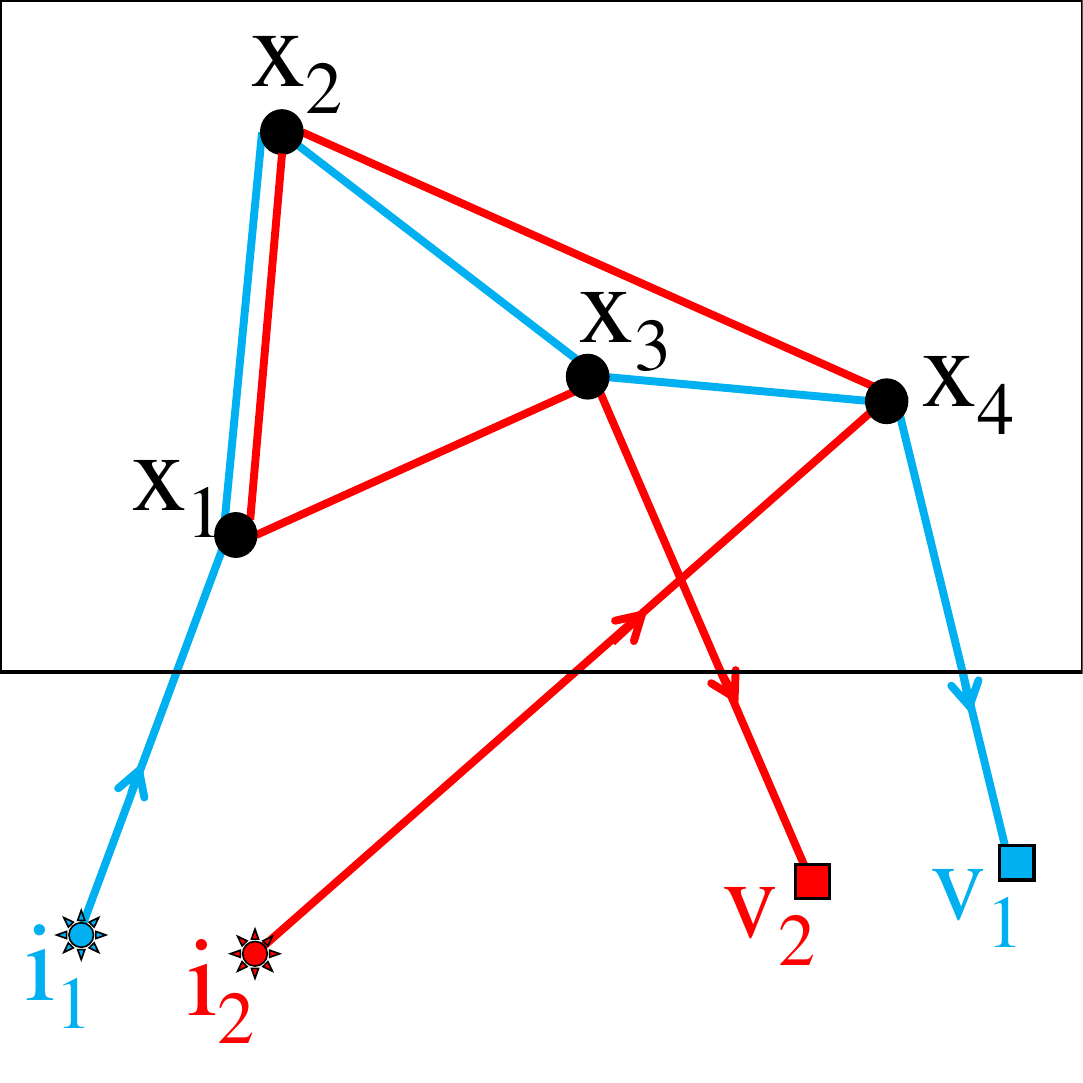}}  \hspace{0.5cm}  
        \subfloat[Same order of nodes]{\label{fig:paths-direct}\includegraphics[width=0.22\textwidth]
                {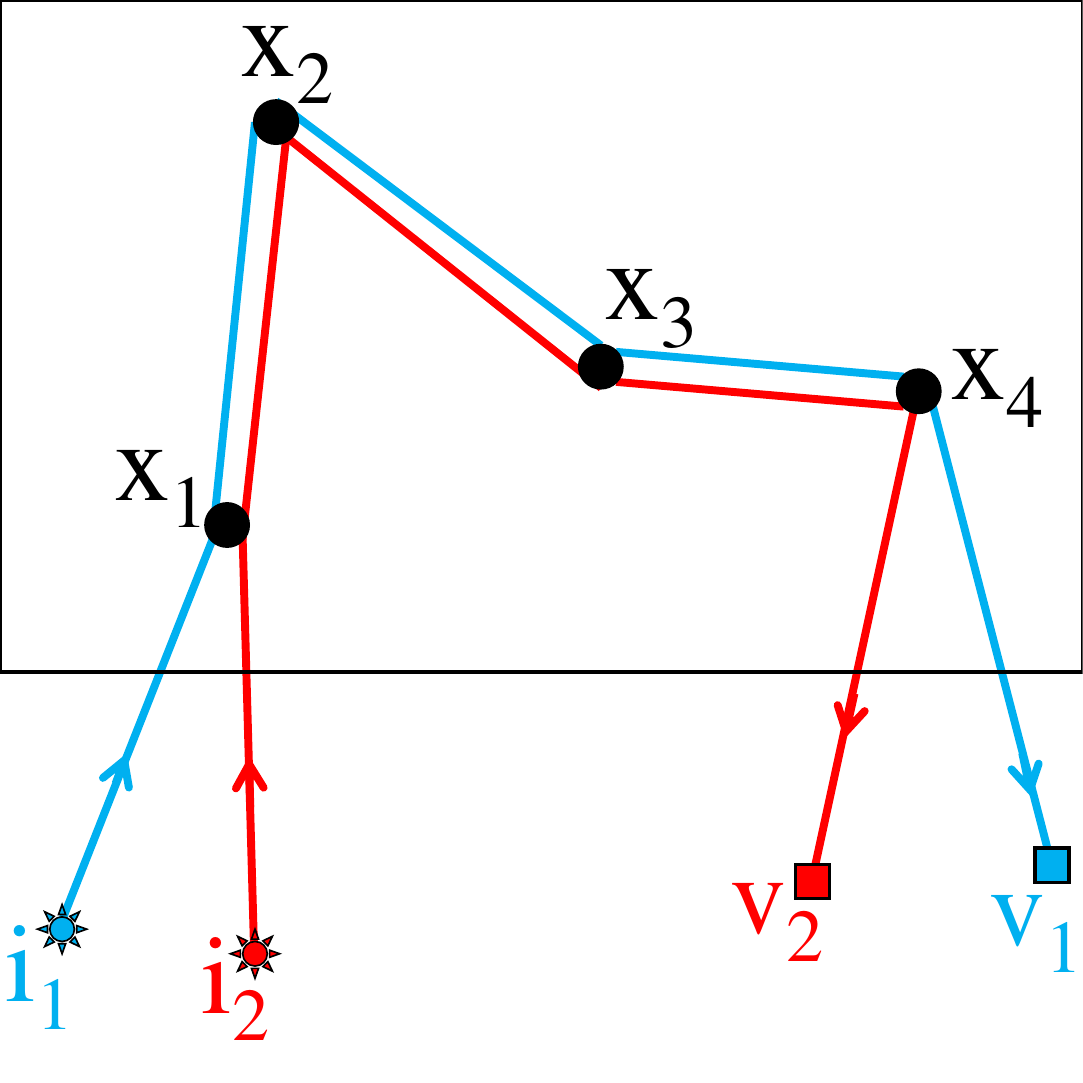}}\hspace{0.5cm}
        \subfloat[Reversed paths]{\label{fig:paths-reverse}\includegraphics[width=0.22\textwidth]
                {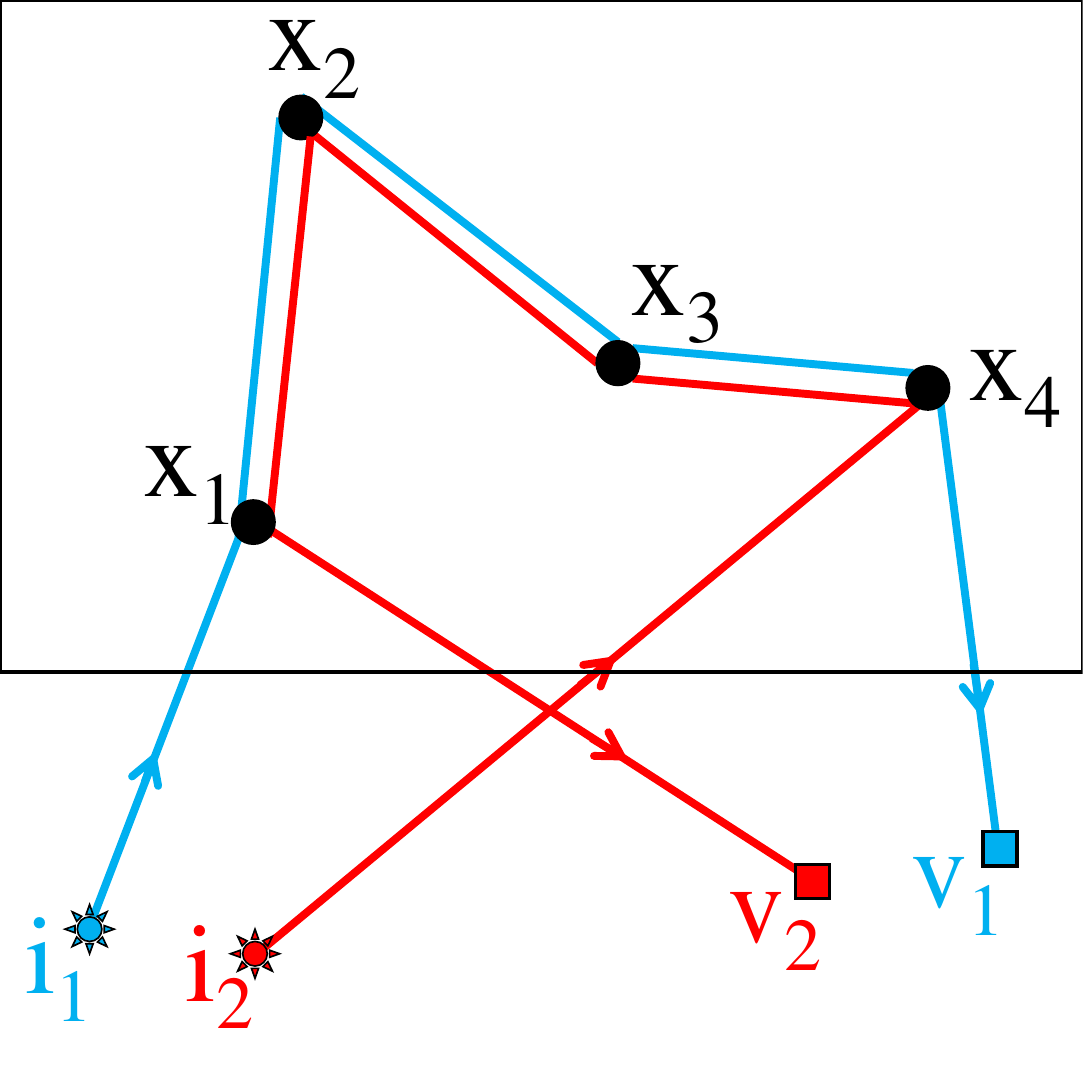}}
  \caption{Paths for covariance rendering. (a) Averaging path pairs reduces to the shared nodes of the two paths, as path segments from $\ptd_j$ to $\ptd_{j+1}$ with arbitrary length/phase cancel each other. (b) Averaging over path pairs sharing the same nodes at different orders are  also likely to cancel out due to length differences. Only paths that share the same ordered set of nodes $\ptd_1,...,\ptd_k$ (c) or its reverse set of nodes $\ptd_k,...,\ptd_1$, produce positive contribution to the average}  
   \label{fig:various-paths}        
\end{figure*}

Consider, as in \figref{fig:various-paths}(a), the set of path pairs $(\pathseq^1,\pathseq^2)$ that have an arbitrary number of vertices, but share only vertices $\ptd_1,\dots,\ptd_B$ (in any order). Then, as in \secref{sec:spk-mean}, we expect all the different path segments from $\ptd_b$ to $\ptd_{b+1}$ to average to the direct path between these points. In Appendix \ref{sec:path-integrals}, we prove that indeed all path pairs with disjoint nodes collapse to their joint nodes, and the average contribution of all pairs of paths sharing nodes $\ptd_1,\dots,\ptd_B$ is
\begin{gather}
\cll=\mulsct(\pathseq^1)\cdot{\mulsct(\pathseq^2)}^*\cdot\Pi_{b=1}^{B}\sctCoff(\ptd_b),\label{eq:path-cont-markov}\\
\text{where}\quad \mulsct(\pathseq) = \mulsct(\ptd_0\to\ptd_1)\Pi_{b=1}^B\mulsct(\ptd_{b-1}\rightarrow\ptd_b\rightarrow\ptd_{b+1}).\label{eq:mulsct-op}
\end{gather}
The \emph{complex volumetric throughput} terms $\mulsct(\cdot)$ combine the volumetric attenuation of \equref{eq:att-def} with the complex throughput of \equpref{eq:sigsct-def}{eq:sigsct-0-def}. They can be defined as
\begin{align}
\mulsct(\ptd_{b-1}\rightarrow\ptd_b\rightarrow\ptd_{b+1})&=\att(\ptd_b,\ptd_{b+1})\sigsct(\ptd_{b-1}\!\rightarrow\!\ptd_b\!\rightarrow\!\ptd_{b+1}),\label{eq:mulsct} \\
\mulsct(\ptd_{0}\to\ptd_1)&=\att(\ptd_0,\ptd_1)\sigsct(\ptd_{0}\to\ptd_{1}).\label{eq:mulsct-0}
\end{align}
To recap, the complex volumetric throughput is a direct term, the  product of three factors (i)  the attenuation $\att$, (ii) the complex transmission $\phase$, whose phase is the  segment length, (iii) the scattering amplitude function $\ampf$ of the direction turn (for paths of lengths $>1$). The different terms are summarized in \figref{fig:algo-notation}  .

We can therefore restrict the integration space of \equref{eq:corr-path-int} to only pairs of paths that share all vertices except, perhaps, their endpoints. The contribution of such pairs, given by \equref{eq:path-cont-markov}, is Markovian and can be computed analytically. Next, we further constrain the integration space, by examining when pairs of paths sharing the same vertices but in \emph{different order} have non-zero contribution.

\boldstart{Vertex permutations.} We now consider the contribution of a pair of paths sharing the same vertices $\ptd_1,\dots,\ptd_B$, but in different permutations. The phase of the segment $\ptd_b\to\ptd_{b+1}$ is proportional to the length of that segment. Permutations that do not trace the nodes in the same order have segments with different lengths (see \figref{fig:paths-permute}), and thus different phases. Intuitively, as in \secref{sec:spk-mean}, they are likely to average to zero. However, for each \emph{ordered} set of vertices $\ptd_1\rightarrow\dots\rightarrow\ptd_B$, there is one important permutation for which this argument does not apply, as the central segments have the same length: the \emph{reversed} permutation (\figref{fig:paths-direct} and \ref{fig:paths-reverse}). Therefore, we need to consider contributions from pairs involving four paths \cite{mishchenko2006multiple},
\begin{equation}
\begin{aligned}\label{eq:fb-paths}
%\BEA\label{eq:fb-paths}
\!\!\pathseq^1&=\inp_1\!\!\rightarrow\!\!\ptd_{1}\!\!\rightarrow\!\!\dots\!\!\rightarrow\!\!\ptd_{B}\!\!\rightarrow\!\!\snsp_1,\,&\pathseq^2=\inp_2\!\!\rightarrow\!\!\ptd_{1}\!\!\rightarrow\!\!\dots\!\!\rightarrow\!\!\ptd_{B}\!\!\rightarrow\!\!\snsp_2,\\ 
\!\!\pathseq^{1,r}&=\inp_1\!\!\rightarrow\!\!\ptd_{B}\!\!\rightarrow\!\!\dots\!\!\rightarrow\!\!\ptd_{1}\!\!\rightarrow\!\!\snsp_1,\,&\pathseq^{2,r}=\inp_2\!\!\rightarrow\!\!\ptd_{B}\!\!\rightarrow\!\!\dots\!\!\rightarrow\!\!\ptd_{1}\!\!\rightarrow\!\!\snsp_2.
%\EEA
\end{aligned}
\end{equation}
The reversed paths are the cause of the well-documented phenomenon of {\em coherent backscattering}, which happens when one measures backscattering from a dense scattering volume, with far-field coherent illumination and sensing. When the backscattering and illumination directions are exactly equal, the scattered intensity is increased compared to nearby directions. 

For intuition behind this effect, we first note that every particle instantiation $O$ that contains the path $\ptd_1\rightarrow\dots\rightarrow\ptd_B$, also contains the reversed path $\ptd_B\rightarrow\dots\rightarrow\ptd_1$; that is, {\em the forward and reversed paths are not independent events}. Consequently, their contribution in \equref{eq:corr-as-paths} is $(\sigsct(\pathseq^1)+\sigsct(\pathseq^{1,r}))\cdot(\sigsct(\pathseq^2)+\sigsct(\pathseq^{2,r}))^*$ rather than $\sigsct(\pathseq^1){\sigsct(\pathseq^2)}^*+\sigsct(\pathseq^{1,r}){\sigsct(\pathseq^{1,r})}^*$. We can compute the difference between these two terms, by considering the case $\ind_1=\ind_2=\ind$, $\outd_2=\outd_2=\outd$ and neglecting the scattering amplitude $\ampf$. Then, the contribution of the forward and reversed paths becomes,
\begin{align}
\Big|\sigsct(\pathseq)+\sigsct(\pathseq^r)\Big|^2=&\left|\phase(\ind\sto\ptd_{1})\phase(\ptd_B\sto\outd)+\phase(\ind\sto\ptd_B)\phase(\ptd_{1}\sto\outd)\right|^2\nonumber\\
\cdot&\left|\Pi_{b=1}^{B-1}\phase(\ptd_b\sto\ptd_{b+1})\right|^2.\label{eq:fr-cont} 
\end{align}
The shared intermediate segments have the same phase, therefore,
  \begin{align}\Big|\sigsct(\pathseq)+\sigsct(\pathseq^r)\Big|^2&=\Big|\phase(\ind\sto{\ptd_{1}})\phase({\ptd_B}\sto\outd)+\phase(\ind\sto{\ptd_B})\phase({\ptd_{1}}\sto\outd)\Big|^2\nonumber\\
&=\left|e^{ik\left(\ind^T \ptd_1-\outd^T\ptd_B \right)}+e^{ik\left(\ind^T \ptd_B-\outd^T\ptd_1\right)}\right|^2\nonumber\\
&=2+2Re\left(e^{ik\left(\ind+\outd\right)^T\left( \ptd_1-\ptd_B \right)} \right).\label{eq:frd-bkrd-paths-cont}
\end{align}
When $\ind+\outd $ is large, the average of the real term in \equref{eq:frd-bkrd-paths-cont} over all space points is low. However, when $\ind\sim -\outd$, as in coherent backscattering, the real term approaches unity, and therefore the total contribution is doubled. In other words, we get \emph{constructive interference} between the forward and reversed paths.
%\BE2Re\left(e^{\frac{2\pi i}{\lambda}\left(\ind+\outd\right)^T\left( \ptd^1-\ptd^k \right)}\right) \to2,\EE 

\boldstart{Covariance path integral.} We can now state concretely our path integral formulation for the speckle covariance: Consider the space $\Path$ of \emph{sub-paths} $\subpath=\ptd_1\rightarrow\dots\rightarrow\ptd_B$, where each vertex can be everywhere in the volume $\V$, and $B=0,\dots,\infty$. We  write:
\begin{equation}\label{eq:cov-integral-w-mean}
C^{\inp_1,\inp_2}_{\snsp_1,\snsp_2} = \int_\Path c(\subpath) \ud\subpath-\meanspk^{\inp_1}_{\snsp_1}\cdot{\meanspk^{\inp_2}_{\snsp_2}}^*.
\end{equation}
To define the integrand $c(\subpath)$, we first form the four complete paths of \equref{eq:fb-paths}, by connecting the forward and reversed versions of $\subpath$ to the illumination and sensing conditions $\inp_1, \snsp_1$ and $\inp_2, \snsp_2$. Then,
%By considering all four combinations of forward and reversed paths, we can write the total covariance contribution as
\begin{equation}
c(\subpath) = c_{\pathseq^1,\pathseq^2}+c_{\pathseq^1,\pathseq^{2,r}}+c_{\pathseq^{1,r},\pathseq^2}+c_{\pathseq^{1,r},\pathseq^{2,r}},
\end{equation}
where the summands are defined in \equref{eq:path-cont-markov}. By expanding the equations, and considering that now the pairs of paths have identical intermediate segments, we can rewrite this sum as
\begin{align}
c(\subpath) = f(\subpath)\cdot&\Big(\mulsct(\ptd_2\!\rightarrow\!\ptd_1\!\rightarrow\!\inp_1)\mulsct(\ptd_{B-1}\!\rightarrow\!\ptd_B\!\rightarrow\!\snsp_1)\nonumber\\
&\quad+\mulsct(\ptd_{B-1}\!\rightarrow\!\ptd_B\!\rightarrow\!\inp_1)\mulsct(\ptd_{2}\!\rightarrow\!\ptd_1\!\rightarrow\!\snsp_1)\Big)\nonumber\\
\cdot&\Big(\mulsct(\ptd_2\!\rightarrow\!\ptd_1\!\rightarrow\!\inp_2)\mulsct(\ptd_{B-1}\!\rightarrow\!\ptd_B\!\rightarrow\!\snsp_2)\nonumber\\
&\quad+\mulsct(\ptd_{B-1}\!\rightarrow\!\ptd_B\!\rightarrow\!\inp_2)\mulsct(\ptd_{2}\!\rightarrow\!\ptd_1\!\rightarrow\!\snsp_2)\Big)^*,\label{eq:covariance-contrib}
\end{align}
where $f(\subpath)$ is the standard \emph{radiometric throughput} of $\subpath$, augmented by the scattering coefficients at the first and last vertices,
\BEA\label{eq:radiometric}
\!\!\!\!f(\subpath)\!\!&\!\!\!\!=\!\!\!\!&\!|\mulsct(\ptd_1\sto\ptd_2)|^{2}\prod_{b=2}^{B-1}|\mulsct(\ptd_{b-1}\sto\ptd_b\sto\ptd_{b+1})|^2\prod_{b=1}^{B}\!\sctCoff(\ptd_b)\\&\!\!\!\!=\!\!\!\!&\!\!\sctCoff(\ptd_1)\sctCoff(\ptd_B)\att^2(\ptd_{1},\ptd_{2})\!\nonumber\!\prod_{b=2}^{B-1}\!\!\rho(\dir{\ptd_{b-1}\ptd_{b}}\!\cdot\!\dir{\ptd_{b}\ptd_{b+1}}) \att^2(\ptd_{b},\ptd_{b+1})\sctCoff(\ptd_b)\nonumber.
\EEA
where the phase function was defined as $\rho(\cos\theta)=|\ampf(\cos\theta)|^2$. The 4 volumetric throughput connections $\mulsct$ of \equref{eq:covariance-contrib} are illustrated
in \figref{fig:algo-notation}, while  $f(\subpath)$ is the volumetric throughput of the central segments (gray path in \figref{fig:algo-notation}).
As we see in the next section, the radiometric throughput term in \equref{eq:radiometric} allows us to reuse path sampling algorithms from intensity rendering also for covariance rendering.

As the mean products $\meanspk^{\inp_1}_{\snsp_1}\cdot{\meanspk^{\inp_2}_{\snsp_2}}^*$ are essentially the throughput contribution of paths from $\inp_1$ to $\snsp_1$ and from $\inp_2$ to $\snsp_2$ without shared nodes, we can drop this term from \equref{eq:cov-integral-w-mean} by restricting the subpath space $\Path$ to paths of length $B\geq1$, and define\vspace{-0.2cm}
\begin{equation}\label{eq:cov-integral}
C^{\inp_1,\inp_2}_{\snsp_1,\snsp_2} = \int_\Path c(\subpath) \ud\subpath.
\end{equation}
We make two notes about the path integral formulation of \equref{eq:cov-integral}. First, if one does not consider the reverse paths, then the resulting path-integral formulation is equivalent to what can be obtained from the \emph{correlation transfer equation} (CTE). We discuss this in Appendix \ref{sec:CTE}, and we also discuss how the Monte Carlo algorithms we derive in the next section compare to Monte Carlo algorithms derived from the CTE. In the evaluation of \secref{sec:RES_Validation} we show that considering only forward paths can provide a good approximation in many cases; however, in cases where the sensor is close to collocated with the source, we should consider reversed paths as well.

Second, at the start of this section, we argued informally that pairs of paths with different permutations of $\ptd_1,\ldots,\ptd_B$ do not contribute to covariance. In Appendix \ref{sec:permutations}, we discuss this in more detail, and additionally show empirical evidence for ignoring these pairs. Likewise, the results in \secref{sec:RES_Validation} show that accounting for only the forward and reversed path is accurate enough.
% Yet, in Appendix \ref{sec:permutations} we suggest a Monte Carlo algorithm %that takes into account all permutations, at the cost of additional computational %complexity.

\vspace{-0.1cm}
\section{Monte Carlo rendering algorithms}\label{sec:rendering}
\vspace{-0.05cm}
We use the results of the previous section, to derive two Monte Carlo rendering algorithms. The first algorithm directly computes the speckle covariance, which we can use, together with an estimate of the speckle mean, to sample multiple speckle patterns. The second algorithm directly renders a speckle pattern, so that the empirical mean and covariance of multiple renderings is accurate.
\vspace{-0.1cm}
\subsection{Rendering speckle covariance}\label{sec:MCalg}

%To define a MC algorithm for covariance evaluation,  we define a path sampling %strategy and approximate the correlation integral of \equref{eq:corr-path-int} %using importance sampling.

%Following the above discussion, we consider  two versions of a MC algorithm, a simpler version that considers only forward paths and a more accurate version that consider both forward and reversed paths.
To approximate the covariance integral of \equref{eq:cov-integral}, we define a strategy that samples \emph{sub-paths} $\subpathn$ from a distribution $q(\subpathn)$ that will be defined below. We
use them to form a Monte Carlo estimate of the covariance as\vspace{-0.0cm}
\begin{equation}
C^{\ind_1,\ind_2}_{\snsp_1,\snsp_2}\approx\frac1N\sum_{n=1}^N\frac{c(\subpathn)}{q(\subpathn)+q(\rsubpathn)}\label{eq:cov-app-fb}
\end{equation}
The denominator of \equref{eq:cov-app-fb} is the sampling probability. As it is possible to independently sample both the forward and reserved version of a subpath, the total probability is $q(\subpath)+q(\rsubpath)$.

\begin{figure*}[t]
        \centering    
        \begin{tabular}{cc}
                \begin{tabular}{c}$$ \begin{array}{lcl}
\text{Complex transmission:}&&\phase(\ptd_{b}\!\!\to\!\!{\ptd_{b+1}})\!=\! \frac{e^{ik|\ptd_{b}-{\ptd_{b+1}}|}}{|\ptd_{b}-\ptd_{b+1}|}\\
\text{Scattering  amplitude function:}&&\ampf(\dir{\ptd_{b-1}\ptd_{b}}\cdot\dir{\ptd_{b}\ptd_{b+1}})
\\
\text{Complex throughput:}&&\sigsct(\ptd_{b-1}\sto\ptd_b\sto\ptd_{b+1})=\quad\phase(\ptd_{b}\sto{\ptd_{b+1}})\ampf(\dir{\ptd_{b-1}\ptd_{b}}\cdot\dir{\ptd_{b}\ptd_{b+1}})\\
\text{Attenuation:}&&\att(\ptd_b,\ptd_{b+1})=e^{-\frac12\extCoff|\ptd_b-\ptd_{b+1}|}\\
\text{Complex volumetric throughput:}&&\mulsct(\ptd_{b-1}\sto\ptd_b\sto\ptd_{b+1})=\att(\ptd_b,\ptd_{b+1})\sigsct(\ptd_{b-1}\sto\ptd_b\sto\ptd_{b+1})\\ 
\text{Radiometric throughput:}&&f(\ptd_{b-1}\sto\ptd_b\sto\ptd_{b+1})=\sctCoff(\ptd_b)|\mulsct(\ptd_{b-1}\sto\ptd_b\sto\ptd_{b+1})|^2
        \end{array}$$
        \end{tabular}&\begin{tabular}{c}   
        \includegraphics[width=0.25\textwidth]{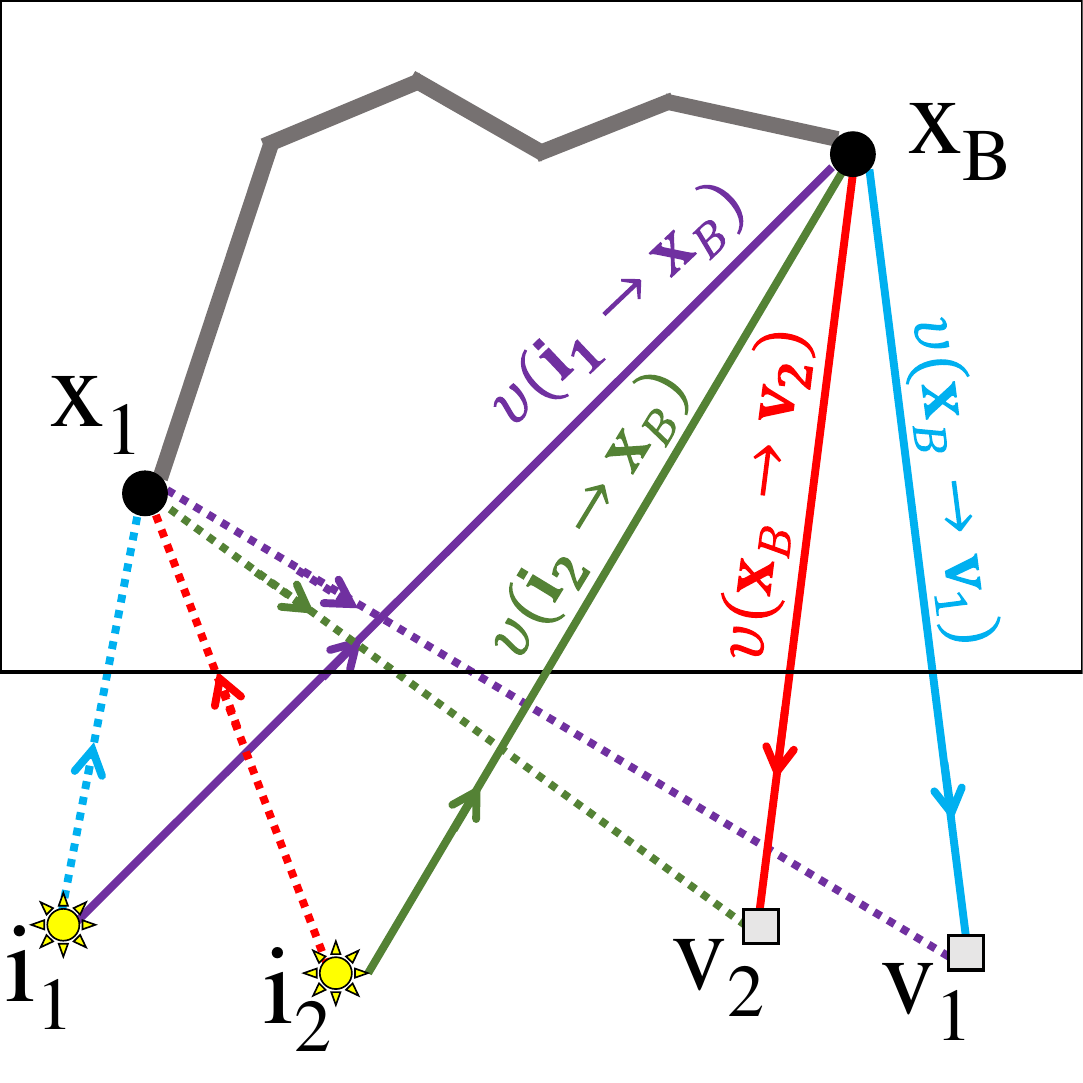}\end{tabular}\end{tabular}
        \caption{Summary of notation and relationships between different throughput terms used in our Monte Carlo algorithms.}  
        \label{fig:algo-notation} \vspace{0.0cm}       
\end{figure*}

\begin{algorithm}[!t]
\mbox{}\\
        \mbox{}\hfill {\stcom\it{Initialize covariance estimate.}}\\
        Set $C=0$. \\
        \For{$\text{iteration}=1:N$}{ 
                \mbox{}\hfill {\stcom\it{Sample first vertex of subpath.}}\\
                Sample point $\ptd_1\sim q_{o}(\ptd_1)$ .\\ 
                Sample uniformly direction $\omgv_1$.\\
                \mbox{}\hfill {\stcom\it{Update covariance with single scattering path.}}\\
                Update $C\pluseq V\cdot\mulsct(\inp_1\sto \ptd_1)\mulsct(\inp_1\sto\ptd_1\sto\snsp_1){\mulsct(\inp_2\sto\ptd_1)}^*{\mulsct(\inp_2\sto\ptd_1\sto\snsp_2)}^*$.\\
                \mbox{}\hfill {\stcom\it{Continue tracing the subpath.}}\\
                \mbox{}\hfill {\stcom\it{Sample second vertex of subpath.}}\\
                Sample distance $d\sim \sctCoff(\ptd_1)|\att(\ptd_1,\ptd_{1}+d\omgv_{1})|^2$.\\ 
                Set point $\ptd_2=\ptd_{1}+d\omgv_{1}$.\\
                Set $b=2$.\\
                \While{ $\ptd_b$ inside medium}{
                        \mbox{}\hfill {\stcom\it{Update covariance with  next-event estimation.}}\\
                        Update $C\pluseq\frac{V}{2}\Big(\mulsct(\ptd_2\sto\ptd_1\sto\inp_1)\mulsct(\ptd_{b-1}\sto\ptd_b\sto\snsp_1)$\\ \quad\quad\quad\quad\quad\quad\quad\quad$+\mulsct(\ptd_{b-1}\sto\ptd_b\sto\inp_1)\mulsct(\ptd_2\sto\ptd_1\sto\snsp_1)\Big)$\\\quad\quad\quad\quad\quad\quad\quad$\cdot\Big(\mulsct(\ptd_2\sto\ptd_1\sto\inp_2)\mulsct(\ptd_{b-1}\sto\ptd_b\sto\snsp_2)$\\\quad\quad\quad\quad\quad\quad\quad\quad$+\mulsct(\ptd_{b-1}\sto\ptd_b\sto\inp_2)\mulsct(\ptd_2\sto\ptd_1\sto\snsp_2)\Big)^*$.\\
                        \mbox{}\hfill {\stcom\it{Sample next vertex of subpath.}}\\
                        Sample direction $\omgv_b\sim \rho(\omgv_{b-1},\omgv_b)$.\\
                        Sample distance $d\sim\sctCoff(\ptd_b) |\att(\ptd_{b},\ptd_{b}+d\omgv_{b})|^2$.\\
                        Set point $\ptd_{b+1}=\ptd_{b}+d\omgv_{b}$.\\
                        \mbox{}\hfill {\stcom\it{Account for absorption.}}\\ 
                        Sample scalar $a\sim\text{Unif}[0,1]$.\\
                        \If {$a>\sctCoff(\ptd_{b+1})/\extCoff(\ptd_{b+1})$}{
                                \mbox{}\hfill {\stcom\it{Terminate subpath at absorption event.}}\\ 
                                break\\ 
                        }
                        Set $b=b+1$.
                }
        }
        \mbox{}\hfill {\stcom\it{Produce final covariance estimate.}}\\ 
        Update $C = \frac{1}{N}C$.\\
        \Return{$C$.}\\\mbox{}\\\mbox{}
        \caption{Monte Carlo rendering of covariance $C^{\inp_1,\inp_2}_{\snsp_1,\snsp_2}$.}
        \label{alg:MCpair}\vspace{0.0cm}
 \end{algorithm}

 \begin{algorithm}[!t]\mbox{}\\
        \mbox{}\hfill {\stcom\it{Initialize field estimate.}}\\
        Set $\wavevec=\zerovec$. \\
        \For{$\text{iteration}=1:N$}{ 
%               \mbox{}\hfill {\stcom\it{Sample term for mean correction.}}\\
                Sample random phase $\zeta\sim\text{Unif}[0,1]$.\\
                Set $z=e^{2\pi i \zeta}$.\\
                \mbox{}\hfill {\stcom\it{Sample first vertex of subpath.}}\\
                 Sample point $\ptd_1\sim q_o(\ptd_1)$.\\ 
                \mbox{}\hfill {\stcom\it{Update field with single scattering path}}\\
                Update $\forall j,~ \wavevec_j\pluseq z\cdot\sqrt{\frac{V}{2}}\cdot\mulsct(\inp_j\sto\ptd_1)\mulsct(\inp_j\sto\ptd_1\sto\snsp_j)$. \\ 
                \mbox{}\hfill {\stcom\it{Continue tracing the subpath.}}\\
                \mbox{}\hfill {\stcom\it{Sample second vertex of subpath.}}\\
                Sample uniformly direction  $\omgv_1$.\\ 
                Sample $d\sim \sctCoff(\ptd_1)|\att(\ptd_1,\ptd_{1}+d\omgv_{1})|^2$\\ Set $\ptd_2=\ptd_{1}+d\omgv_{1}$.\\
                Set $b=2$.\\
                \While{ $\ptd_b$ inside medium}{
%                       \mbox{}\hfill {\stcom\it{Sample term for mean correction.}}\\
                        Sample random phase $\zeta\sim\text{Unif}[0,1]$.\\
                        Set $z=e^{2\pi i \zeta}$.\\
                        \mbox{}\hfill {\stcom\it{Update field with next-event estimation.}}\\
                        Update $\forall j,~\wavevec_j \pluseq z\cdot\sqrt{\frac{V}{2}}\cdot\Big(\mulsct(\ptd_2\sto\ptd_1\sto\inp_j)\mulsct(\ptd_{b-1}\sto\ptd_b\sto\snsp_j)$\\\quad\quad\quad\quad\quad\quad\quad\quad\quad$+\mulsct(\ptd_{2}\sto\ptd_1\sto\snsp_j)\mulsct(\ptd_{b-1}\sto\ptd_b\sto\inp_j)\Big)$.\\ 
                        \mbox{}\hfill {\stcom\it{Sample next vertex of subpath.}}\\
                        Sample direction $\omgv_k\sim \rho(\omgv_{b-1},\omgv_b)$.\\
                        Sample distance $d\sim\sctCoff(\ptd_{b}) |\att(\ptd_{b},\ptd_{b}+d\omgv_{b})|^2$.\\ 
                        Set point $\ptd_{b+1}=\ptd_{b}+d\omgv_{b}$.\\
                        \mbox{}\hfill {\stcom\it{Account for absorption.}}\\ 
                        Sample scalar $a\sim\text{Unif}[0,1]$.\\
                        \If {$a>\sctCoff(\ptd_{b+1})/\extCoff(\ptd_{b+1})$}{
                                \mbox{}\hfill {\stcom\it{Terminate subpath at absorption event.}}\\ 
                                break\\ 
                        }
                        Update $b=b+1$.
                }
        }
        \mbox{}\hfill {\stcom\it{Produce final field with correct mean.}}\\ 
        Update $\forall j,~\wavevec_j=\meanspk^{\inp_j}_{\snsp_j}+\sqrt{\frac{1}{{N}}}\wavevec_j$.\\
        \Return{$\wavevec$.\mbox{}\\\mbox{}}
        \caption{Monte Carlo rendering of $J\!\times \!1$ field $\wavevec$ for $\{(\inp,\!\snsp)_j\!\}_{j=1}^J$.}
        \label{alg:MCfield}
 \end{algorithm}

The variance of the estimator in \equref{eq:cov-app-fb} reduces when $q(\subpath)$ is a good approximation to $c(\subpath)$. As $c(\subpath)$ in \equref{eq:covariance-contrib} is Markovian, that is, expressed as a product of the contributions of individual segments, it lends itself to local sampling procedures. The sampling algorithm we use operates as follows: We sample the first vertex $\ptd_1$ according to the volume density, using the probability distribution $q_o$ defined as:
\BE q_o(\ptd)=\frac{\sctCoff(\ptd)}{V} \;\;\text{with}\;\;\;V={\int \sctCoff(\ptd) d\ptd}.\EE 
For a homogeneous volume, $q_o$ reduces to the uniform density. 
%We note that, when using only forward paths, it is possible to define a better importance sampling procedure for the first vertex (see Appendix~\ref{sec:cte}). 
Then, taking advantage of the fact that $c(\subpath)$ includes the radiometric throughput $f(\subpath)$, we sample all other vertices of $\subpath$ using volume path tracing~\cite{veach1997robust,AGI}. Finally, as we trace $\subpath$, we perform \emph{next event estimation}, connecting each vertex to the endpoints of the forward and reverse paths of \equref{eq:fb-paths}, as illustrated in \figref{fig:algo-notation}.  The MC process is summarized in Algorithm~\ref{alg:MCpair}, which also details how to handle single-scattering subpaths consisting of only one node.

The probability of a sub-path $\subpath$ sampled as above, and its contribution in \equref{eq:cov-app-fb}, become:\vspace{-0.0cm}
%\begin{align}
%q(\subpath)&=q_o(\ptd_1)\mulsct(\ptd_1\leftrightarrow\ptd_2)\prod_{b=2}^{B-1}\mulsct(\ptd_{b-1}\rightarrow\ptd_{b}\rightarrow\ptd_{b+1})\prod_{b=2}^{B}\sctCoff(\ptd_b)\nonumber\\
%&=\frac1V\mulsct(\ptd_1\leftrightarrow\ptd_2)\prod_{b=2}^{B-1}\mulsct(\ptd_{b-1}\rightarrow\ptd_{b}\rightarrow\ptd_{b+1})\prod_{b=1}^{B}\sctCoff(\ptd_b).\label{eq:q-path-samp}
%\end{align}
\begin{equation}\label{eq:update-fb-paths}
q(\subpath)=\frac1V f(\subpath),\quad\text{and}\quad\frac{c(\subpath)}{q(\subpath)+q(\rsubpath)}=\frac{V}{2}\frac{c(\subpath)}{f(\subpath)}.
\end{equation}\vspace{0cm}
After term cancellations, we end up having to compute only the terms involving $\mulsct(\cdot)$ in \equref{eq:covariance-contrib}.
This 4 next event estimation connections are illustrated in \figref{fig:algo-notation}. %\clearpage
\subsection{Rendering speckle fields}\label{sec:renderingimage}

As discussed in \secref{sce:speckeles-def}, the space of speckle images follows a multi-variate Gaussian distribution. Thus the mean and covariance provide sufficient statistics, which we can use to sample physically-correct speckle images, statistically indistinguishable from ones generated through an exact solution to the wave equation. However, with this approach, sampling an image of $J$ pixels requires that we first render an $J\times J$ covariance matrix. While this is significantly more efficient than solving the wave equation, for large $J$ values this can still be costly. To address this, we present a second rendering algorithm that can synthesize speckle images directly.

Our starting point is the following observation: Let $\covmat$ be the $J\times J$ covariance matrix corresponding to all pairwise combinations of $J$ illumination and sensing conditions $\{(\inp, \snsp)_j\}_{j=1}^J$. Then, from \equpref{eq:cov-integral}{eq:covariance-contrib}, we can write $\covmat$ as an integral of rank-1 matrices,
%\BE\label{eq:covariance_rank1_sum}
%C(j_1,j_2)=\int_{\Path} p(\subpath) \avec_{j_1}(\subpath)\cdot{\avec_{j_2}(\subpath)}^* \ud \subpath
%\EE
\begin{equation}\label{eq:covariance_rank1_sum}
\covmat=\int_{\Path} f(\subpath) \cdot \avec(\subpath)\cdot\avec^*(\subpath) \ud \subpath,
\end{equation}
where: $f(\subpath)$ is defined in \equref{eq:radiometric}, and $\avec(\subpath)$ is a $J\times 1$ vector with $j$-th entry equal to the $\mulsct(\cdot)$ terms in \equref{eq:covariance-contrib} applied to $\inp_j$ and $\snsp_j$,
\begin{align}
\avec_j(\subpath)=&\Big(\mulsct(\ptd_2\rightarrow\ptd_1\rightarrow\inp_j)\mulsct(\ptd_{B-1}\rightarrow\ptd_B\rightarrow\snsp_j)\nonumber\\
&+\mulsct(\ptd_{B-1}\rightarrow\ptd_B\rightarrow\inp_j)\mulsct(\ptd_2\rightarrow\ptd_1\rightarrow\snsp_j)\Big).\label{eq:a-mc-update}
\end{align}

Sampling a $J\times 1$ field $\wavevec$ from a multivariate Gaussian with a covariance as in \equref{eq:covariance_rank1_sum} can be done by first initializing $\wavevec$ to the zero vector,
% of the mean speckle  $u=\meanspk^{\inp_j}_{\snsp_j}$ defined in \equref{}, 
then repeating the following:
% from  $p(\ptd_1,\ptd_2,\omgv_1,\omgv_2)$ 
(i) Sample a subpath $\subpath$ as in Algorithm \ref{alg:MCpair}. (ii)  Sample a complex number $z$ of unit magnitude and random phase. (iii) Increment $\wavevec$ by $\sqrt{\frac{1}{{N}}}\cdot\sqrt{\frac{V}{2}}\cdot z\cdot\avec(\subpath)$ (where $\sqrt{V/2}$ is the square root of the scale in \equref{eq:update-fb-paths}). This is summarized in Algorithm \ref{alg:MCfield}, which also shows how to handle single-scattering subpaths consisting of length $B=1$.

We elaborate on two details of the above procedure: First, a single sample drawn according to this algorithm has the right covariance, but may not follow a Gaussian distribution. By averaging multiple samples, the central limit theorem implies that their average will converge to a Gaussian distribution. To keep the total variance of the average independent of the number of samples $N$, we scale each sample by $\sqrt{1/N}$. Second, we draw the random variable $z$ to ensure that the mean of the samples is zero; we subsequently add the desired mean (computed analytically as described in Section~\ref{sec:spk-mean}) to the final estimate. 

\boldstart{Relationship to path tracing algorithms.} This algorithm is similar to volumetric path tracing for rendering intensity: The weight $\mulsct$ is a \emph{complex} square root of the next-event-estimation weight used in intensity estimation. We can see this from \equref{eq:mulsct}, where $\mulsct$ is defined as the product of three terms: (i) the amplitude function $\ampf$, which is the complex square root of the phase function $\rho$; (ii) the attenuation term $\att$, which is the square root of the attenuation term of intensity estimation; and (iii) the unit-magnitude phase term $\phase$.  

We note, however, an important difference: Every sampled subpath is used to update {\em all sources and sensors}. This is the key for generating speckle images with accurate second-order statistics, and is the fundamental difference with previously proposed speckle rendering algorithms~\cite{Xu:04,Sawicki:08}: As those algorithms update different pixels {\em independently}, they cannot reproduce correlations between pixels or across different illumination conditions. We demonstrate this in \secref{sec:resultsrendering}.

%An alternative approach is to replace the random phase of $z$ with the length of the path from $\ptd_1$ to $\ptd_2$. This approach works well when $\meanspk_\snsp^\inp=0$ but otherwise produce a non zero mean that scales with the number of samples $N$.

%\input{figure-phantom-size-diag-width}

\section{Experiments and Applications}\label{sec:results}

We perform two sets of experiments. First, we seek to validate the accuracy of our algorithm, by comparing with a wave equation solver. Second, we use our algorithm to quantify the memory effect of speckles, and replicate computational imaging techniques based on that effect.

\subsection{Validation against a wave-solver} \label{sec:RES_Validation}
To validate the correctness of our rendering algorithms, we compare their outputs with ``groundtruth'' computed as in \equref{eq:corr-from-u-disc}, by solving the wave equation for multiple particle instantiations.

\begin{figure}[t]
        \centering
        \subfloat[Far Field]{\label{fig:fig1_farField}\includegraphics[width=0.22\textwidth,height=0.25\textwidth]
                {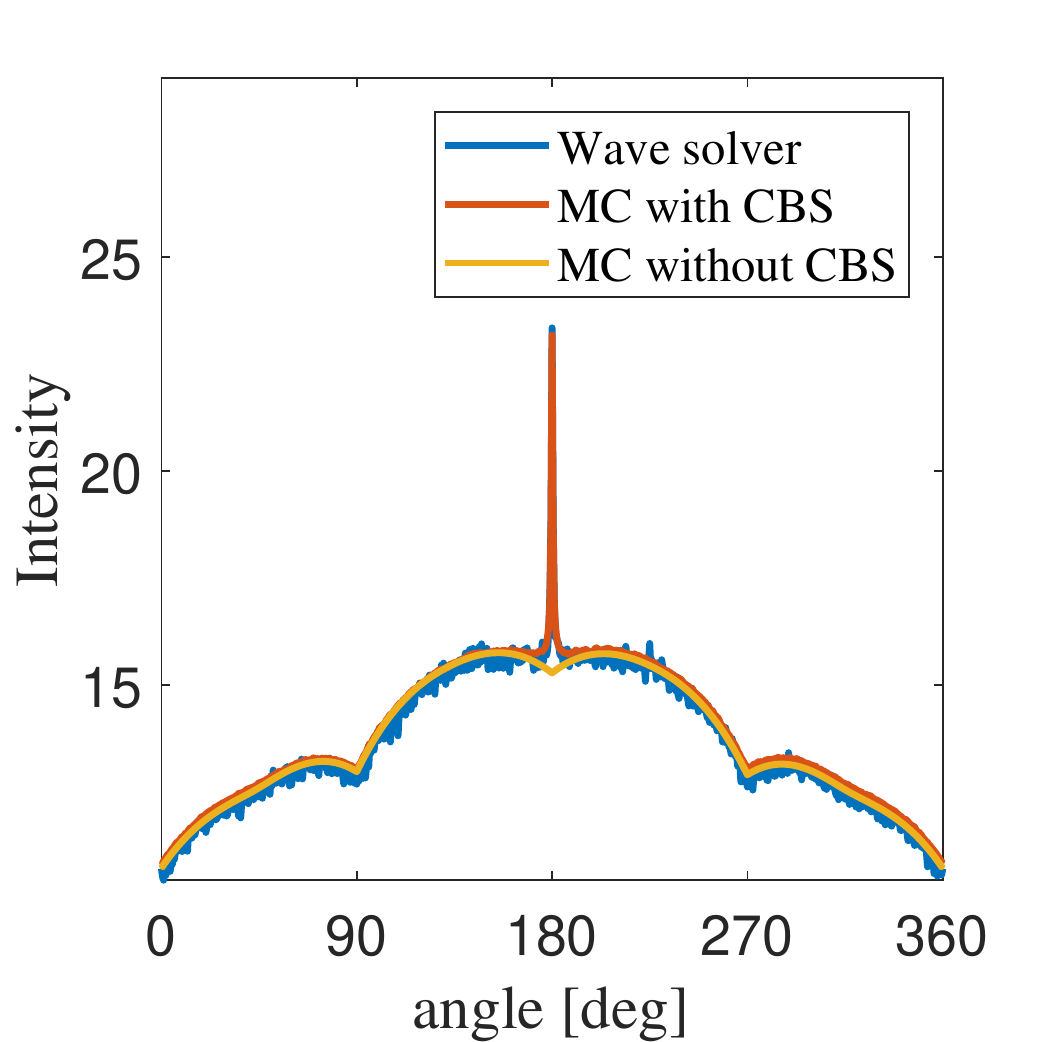}}
        \subfloat[Near Field]{\label{fig:fig1_nearField}\includegraphics[width=0.24\textwidth,height=0.25\textwidth]
                {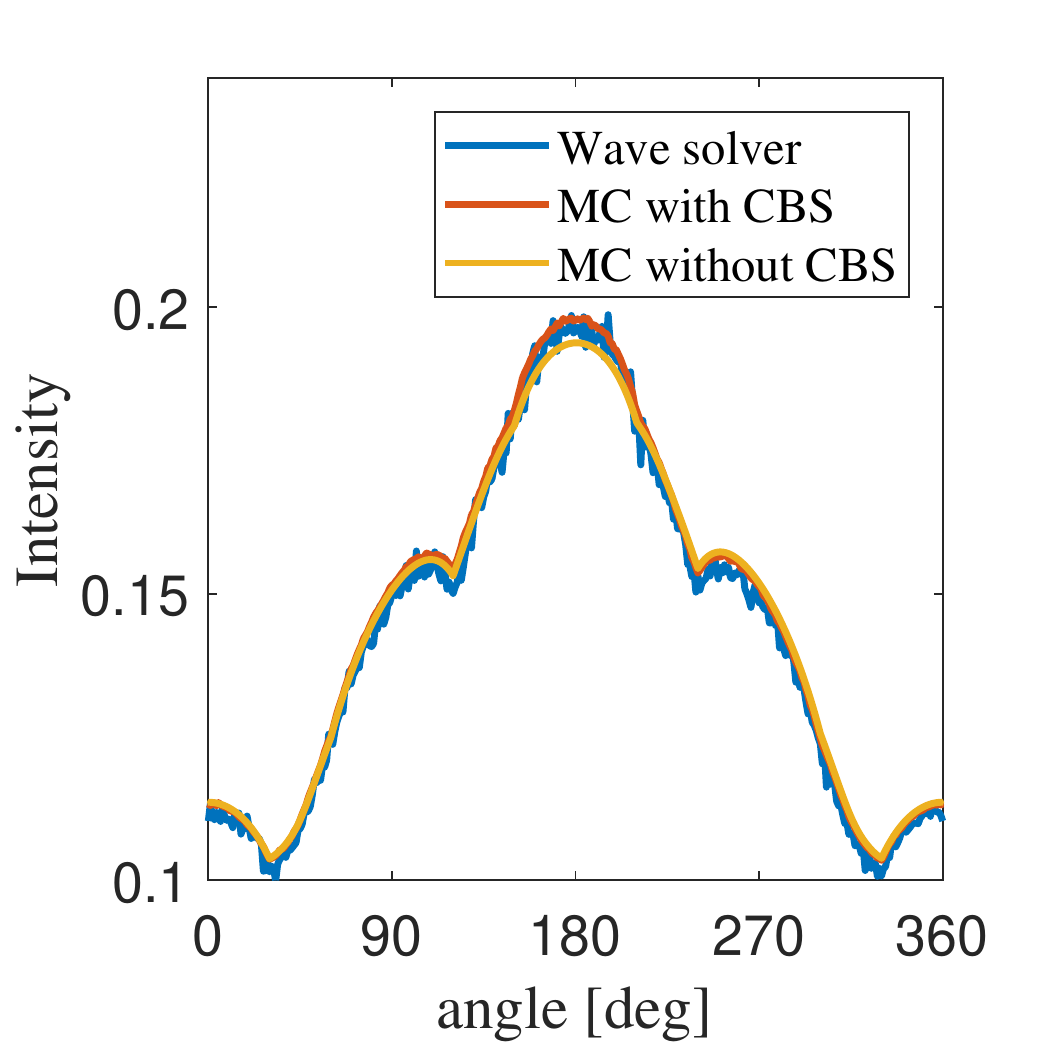}}
  \caption{Coherent Back-Scattering (CBS). We used a $100\lambda\times 100\lambda$ target with an O.D. of 2 to validate CBS. (a) Intensity as a function of sensor angle in the far field. Notice that our MC algorithm, including both forward and reversed paths (red), matches the intensity of the wave solver (blue). On the contrary, neglecting reversed paths (orange), mismatches for the exact back-scattering direction. (b) Intensity as a function of sensor angle in the near field. Here, due to the absence of CBS, both versions of our MC algorithm agree with the wave solver.}  
   \label{fig:cbs-f-nf}        
\end{figure}

\begin{figure*}[t]
        \centering
        \subfloat{\makebox[0.07\textwidth][c]{   }}
        \subfloat{\makebox[0.22\textwidth][c]{$(0^\circ,0^\circ)$}}
        \subfloat{\makebox[0.22\textwidth][c]{$(0^\circ,1^\circ)$}}
        \subfloat{\makebox[0.22\textwidth][c]{$(0^\circ,4^\circ)$}}
        \subfloat{\makebox[0.27\textwidth][c]{$(0^\circ,20^\circ)$}}
        \\
     \subfloat{\setcounter{subfigure}{0}\raisebox{2cm}{\rotatebox[origin=t]{90}{\bf MC, O.D.=2}}}\hfill
     \subfloat{\raisebox{2cm}{\rotatebox[origin=t]{90}{angle [deg]}}}\hfill
     \subfloat{\label{fig:fig2_MFP10_mc_direction_0}\includegraphics[width=0.22\textwidth]
                {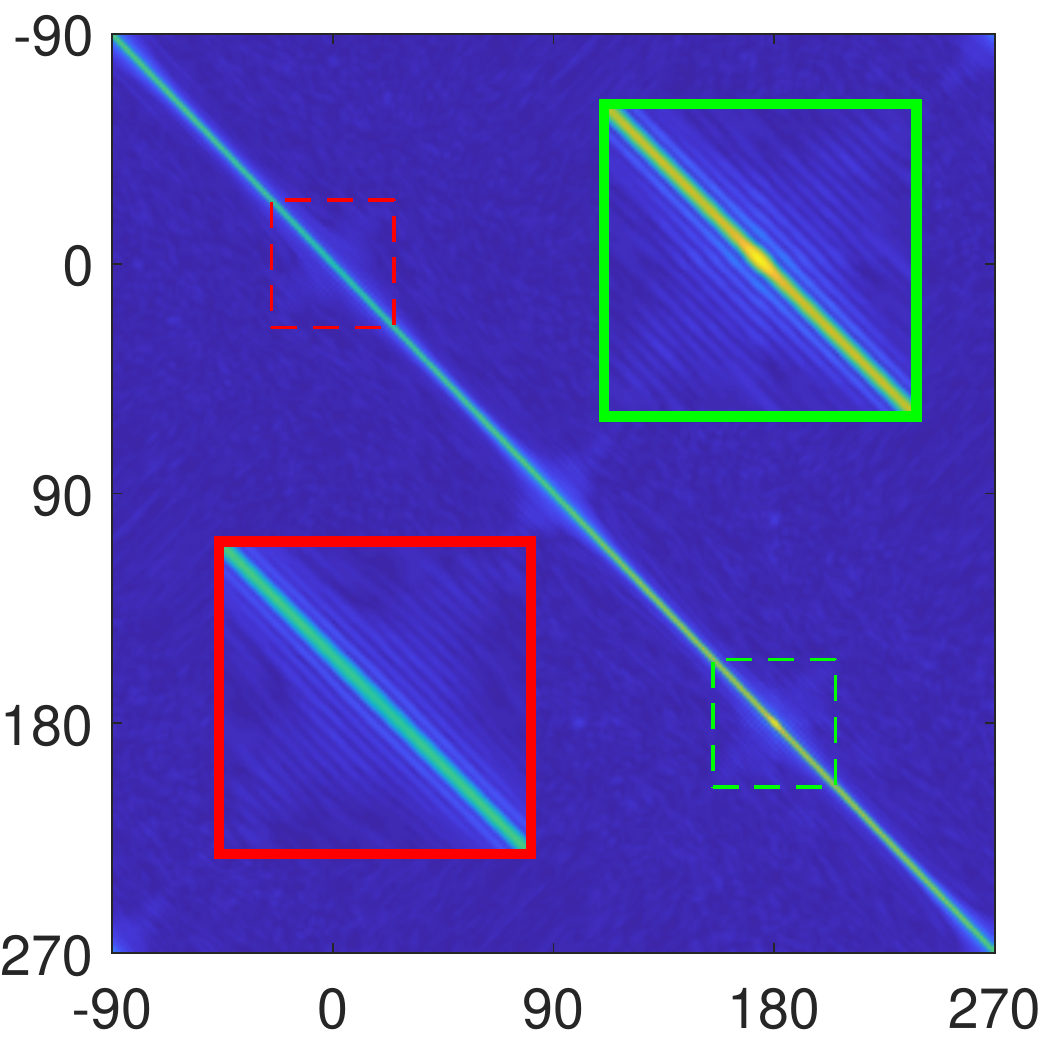}}
     \subfloat{\label{fig:fig2_MFP10_mc_direction_1}\includegraphics[width=0.22\textwidth]
				{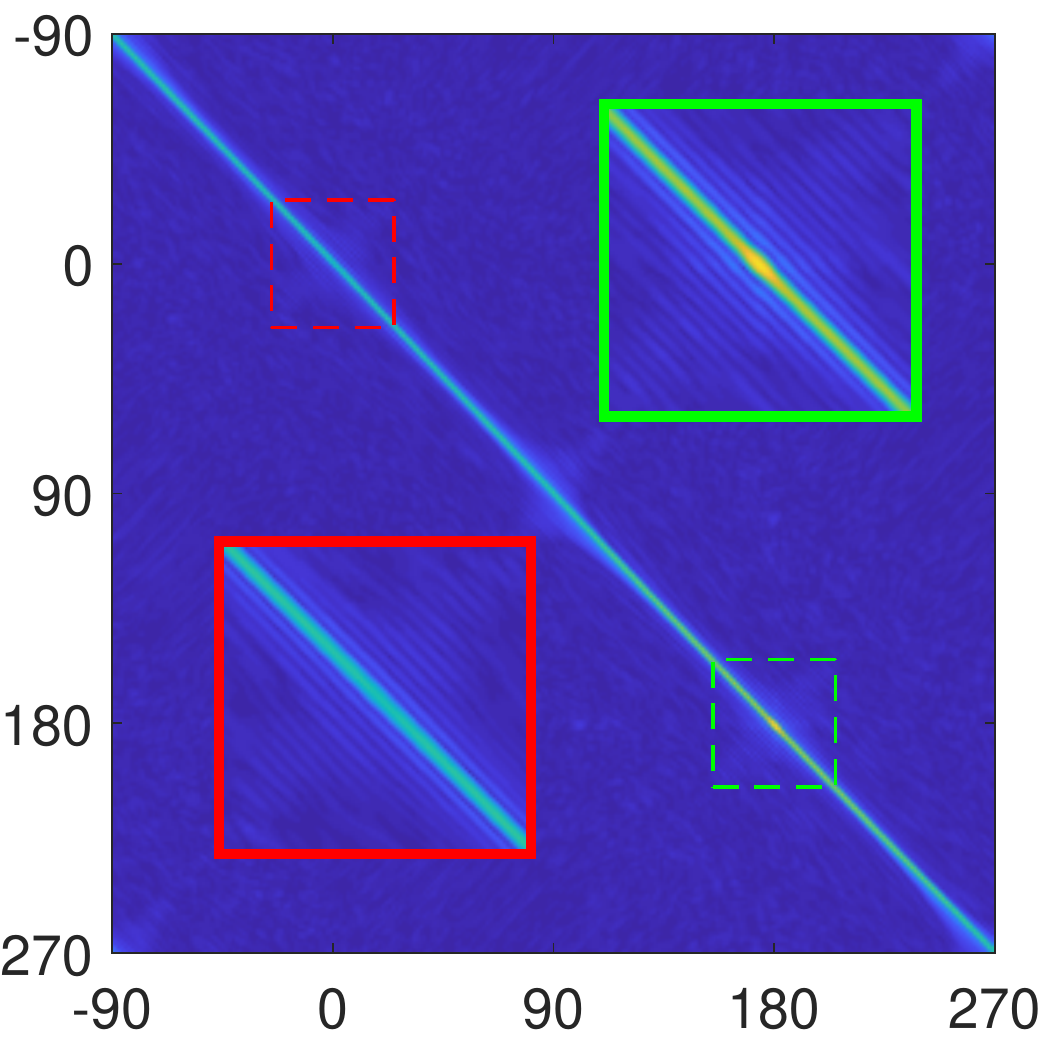}}
     \subfloat{\label{fig:fig2_MFP10_mc_direction_4}\includegraphics[width=0.22\textwidth]
				{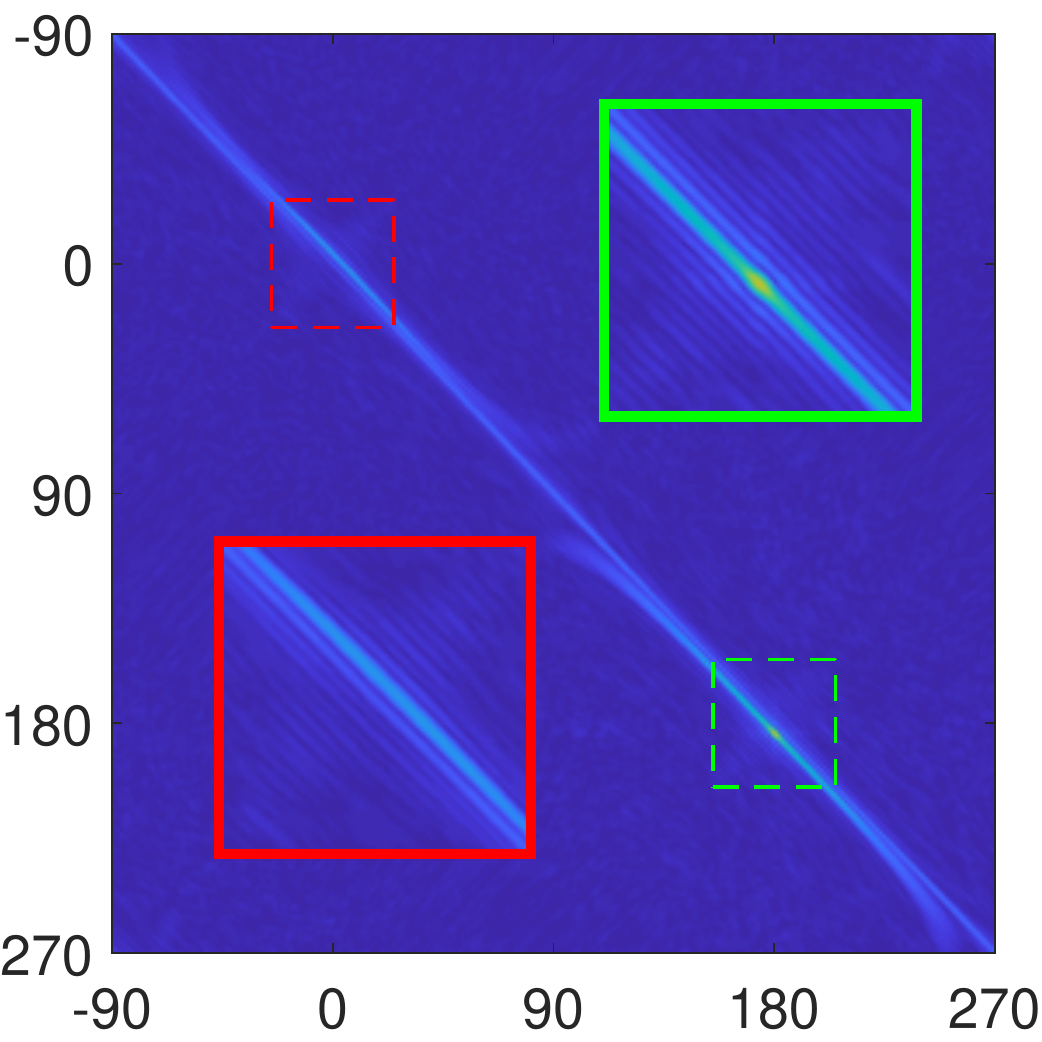}}
     \subfloat{\label{fig:fig2_MFP10_mc_direction_20}\includegraphics[width=0.27\textwidth]
				{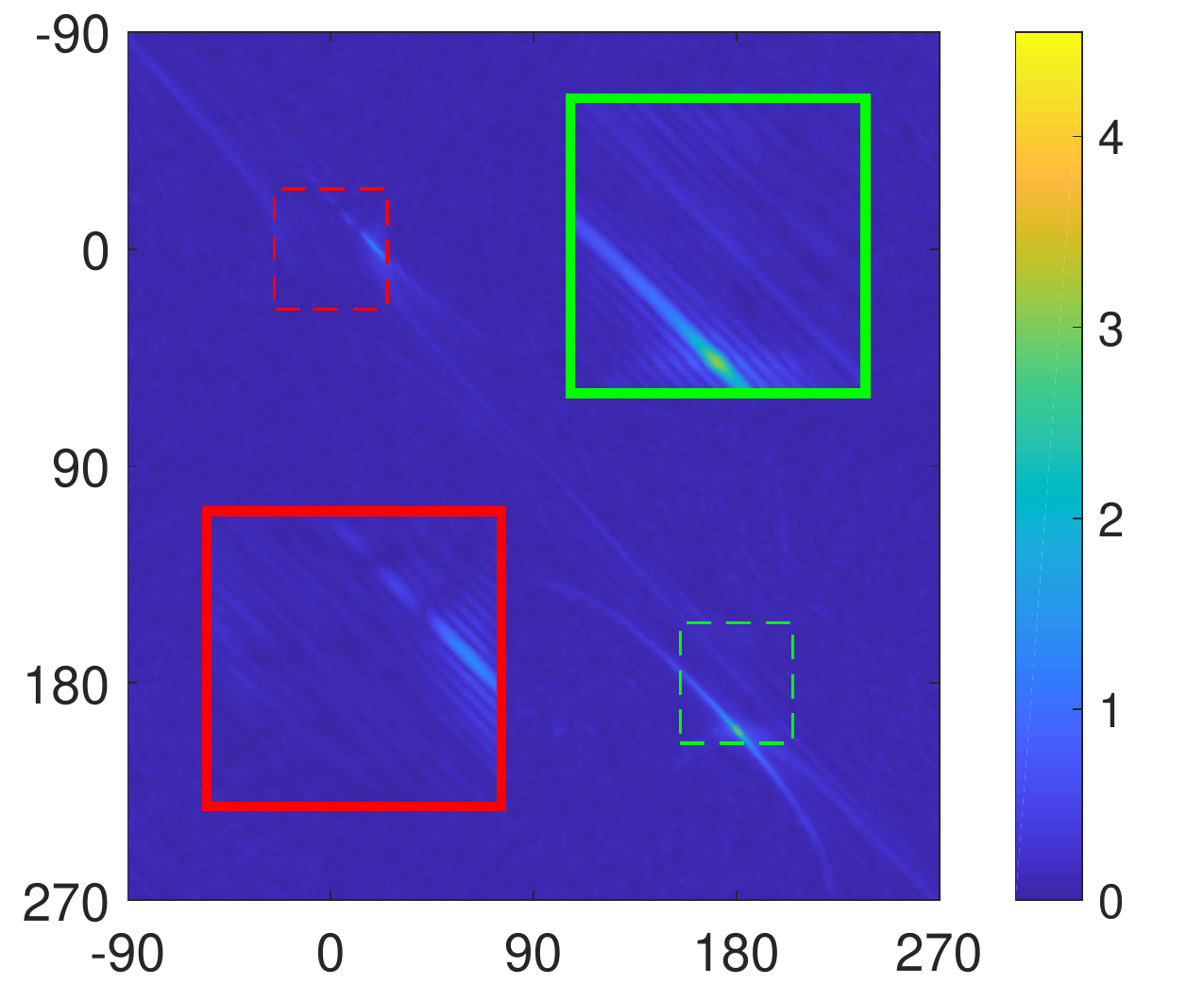}}
\\\vspace{-0.3cm}
        \centering
	\subfloat{\setcounter{subfigure}{4}\raisebox{2cm}{\rotatebox[origin=t]{90}{\bf Wave Solver, O.D.=2}}}\hfill
	     \subfloat{\raisebox{2cm}{\rotatebox[origin=t]{90}{angle [deg]}}}\hfill
	\subfloat{\label{fig:fig2_MFP10_mudiff_direction_0}\includegraphics[width=0.22\textwidth]
				{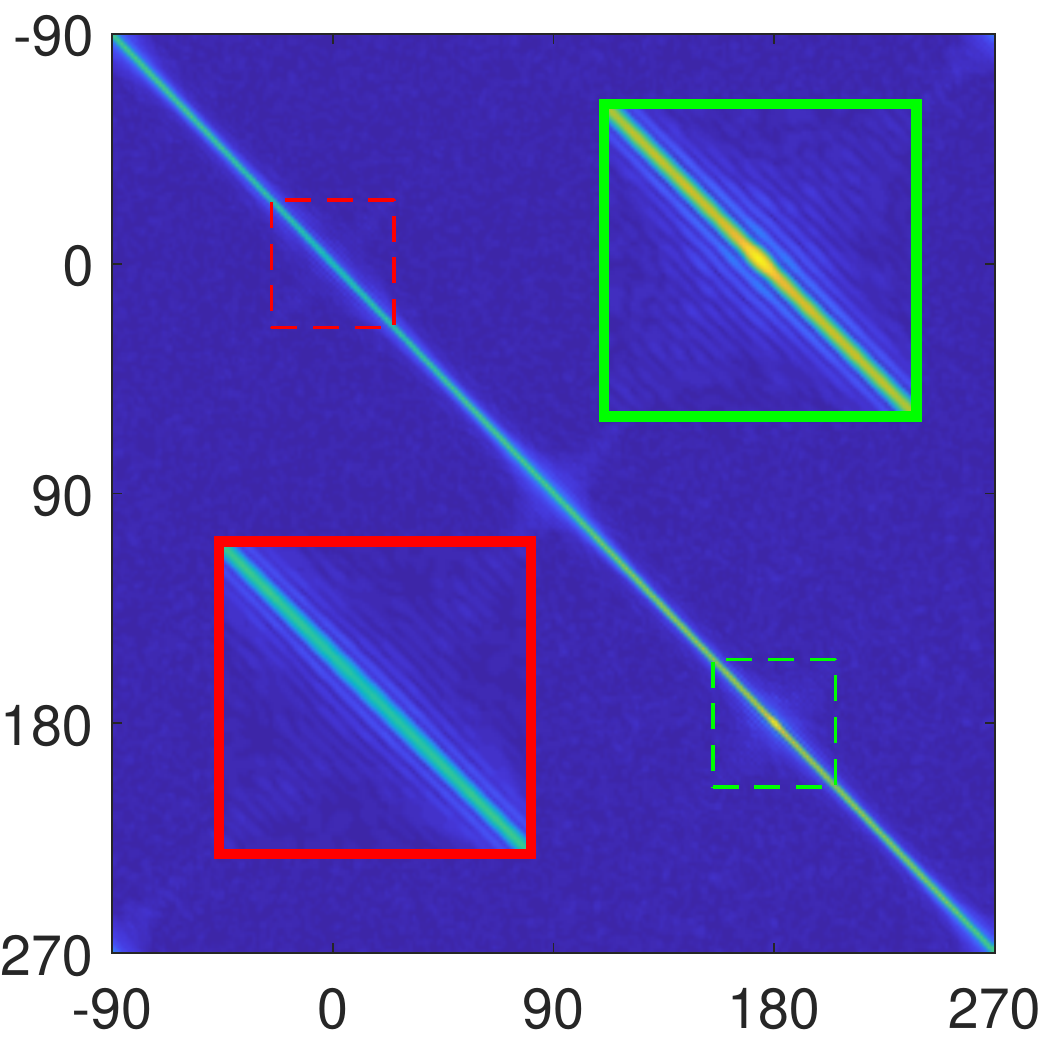}}
	\subfloat{\label{fig:fig2_MFP10_mudiff_direction_1}\includegraphics[width=0.22\textwidth]
				{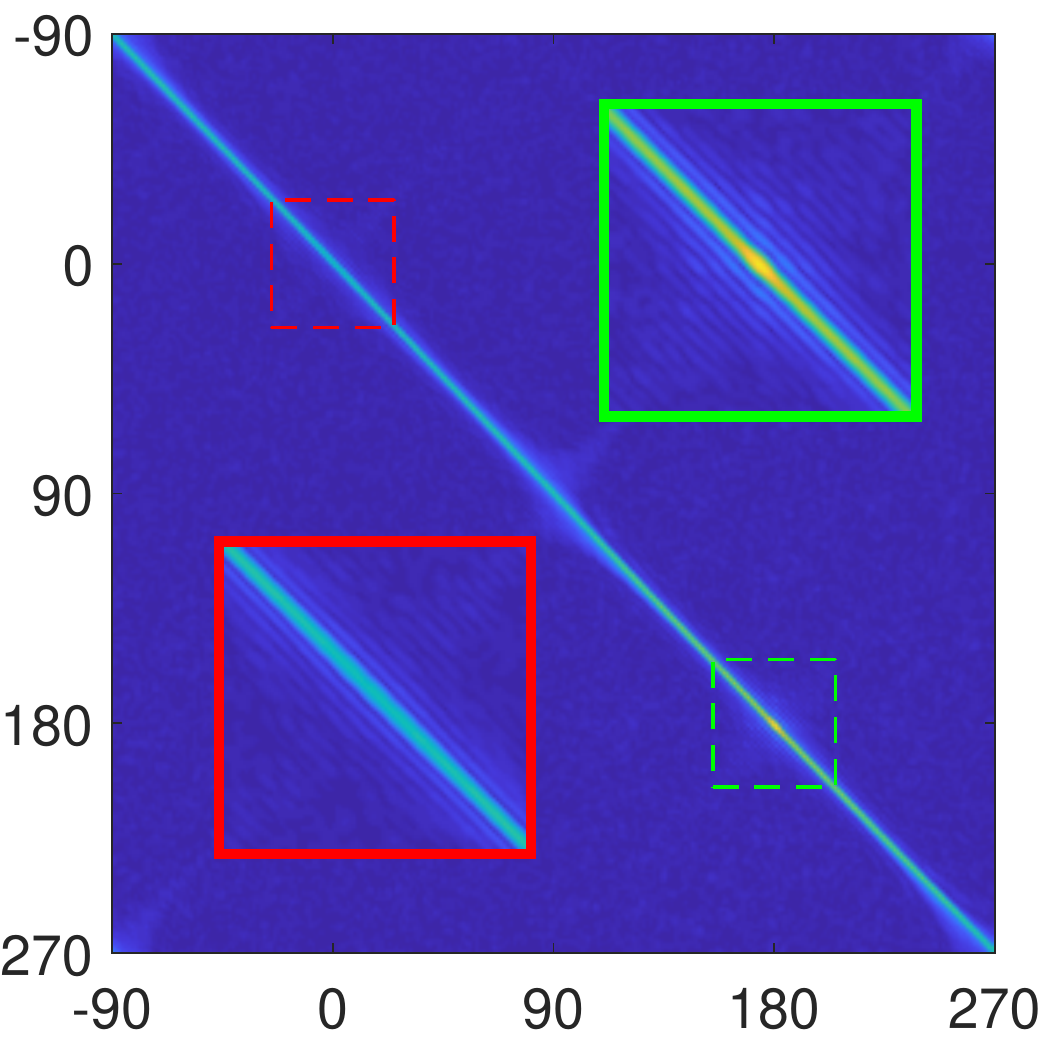}}
	\subfloat{\label{fig:fig2_MFP10_mudiff_direction_4}\includegraphics[width=0.22\textwidth]
				{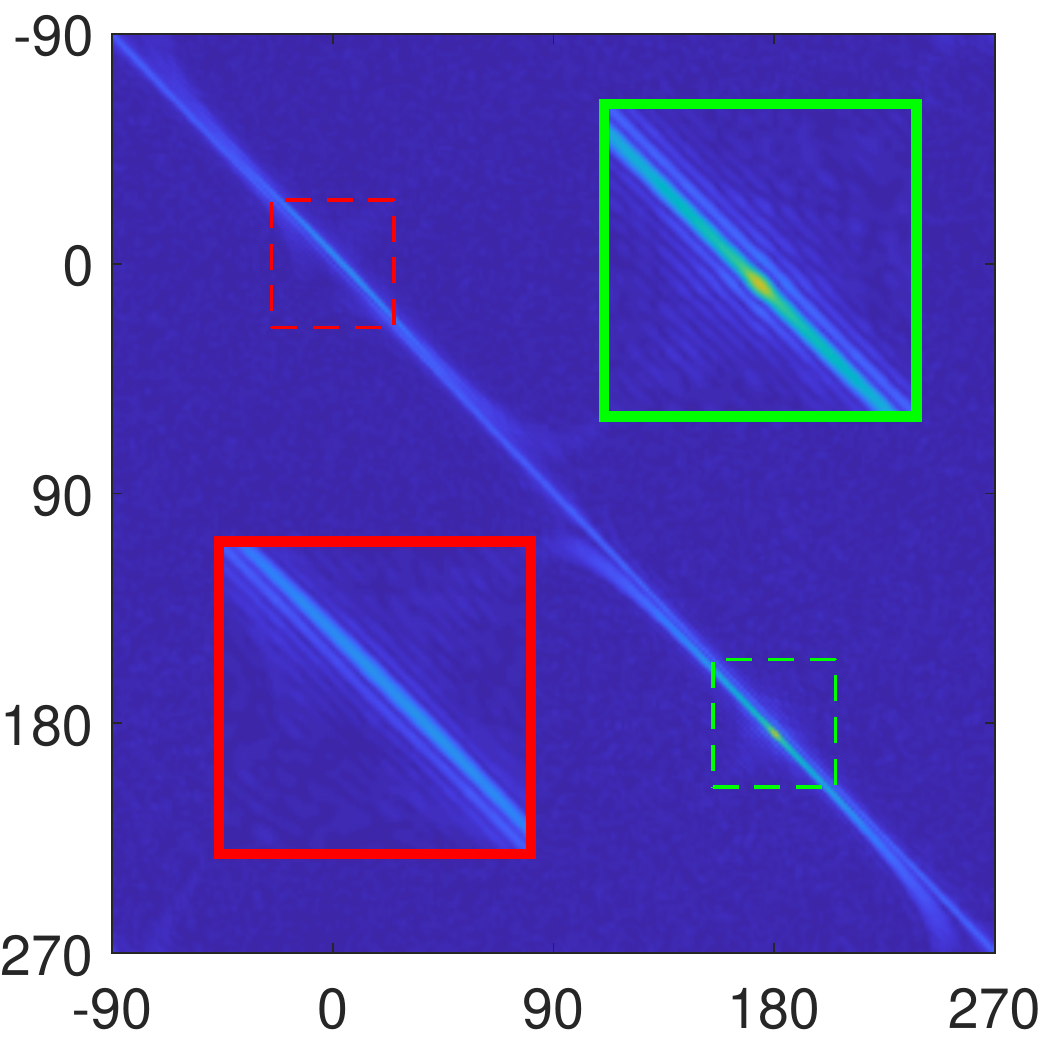}}
	\subfloat{\label{fig:fig2_MFP10_mudiff_direction_20}\includegraphics[width=0.27\textwidth]
				{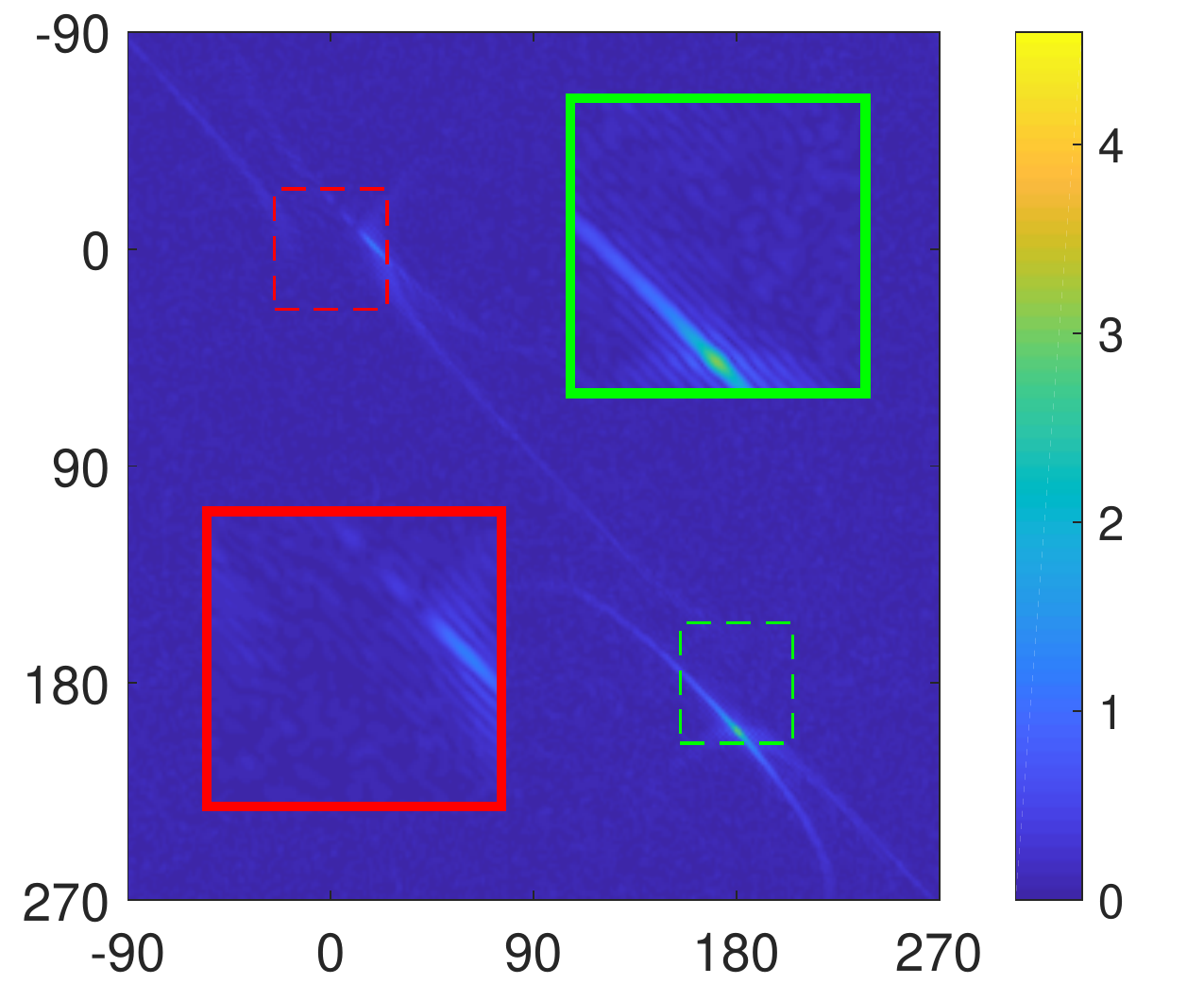}}
\\\vspace{-0.3cm}
	\subfloat{\setcounter{subfigure}{8}\raisebox{2cm}{\rotatebox[origin=t]{90}{\bf MC, O.D.=0.5}}}\hfill
	     \subfloat{\raisebox{2cm}{\rotatebox[origin=t]{90}{angle [deg]}}}\hfill
	\subfloat{\label{fig:fig2_MFP40_mc_direction_0}\includegraphics[width=0.22\textwidth]
				{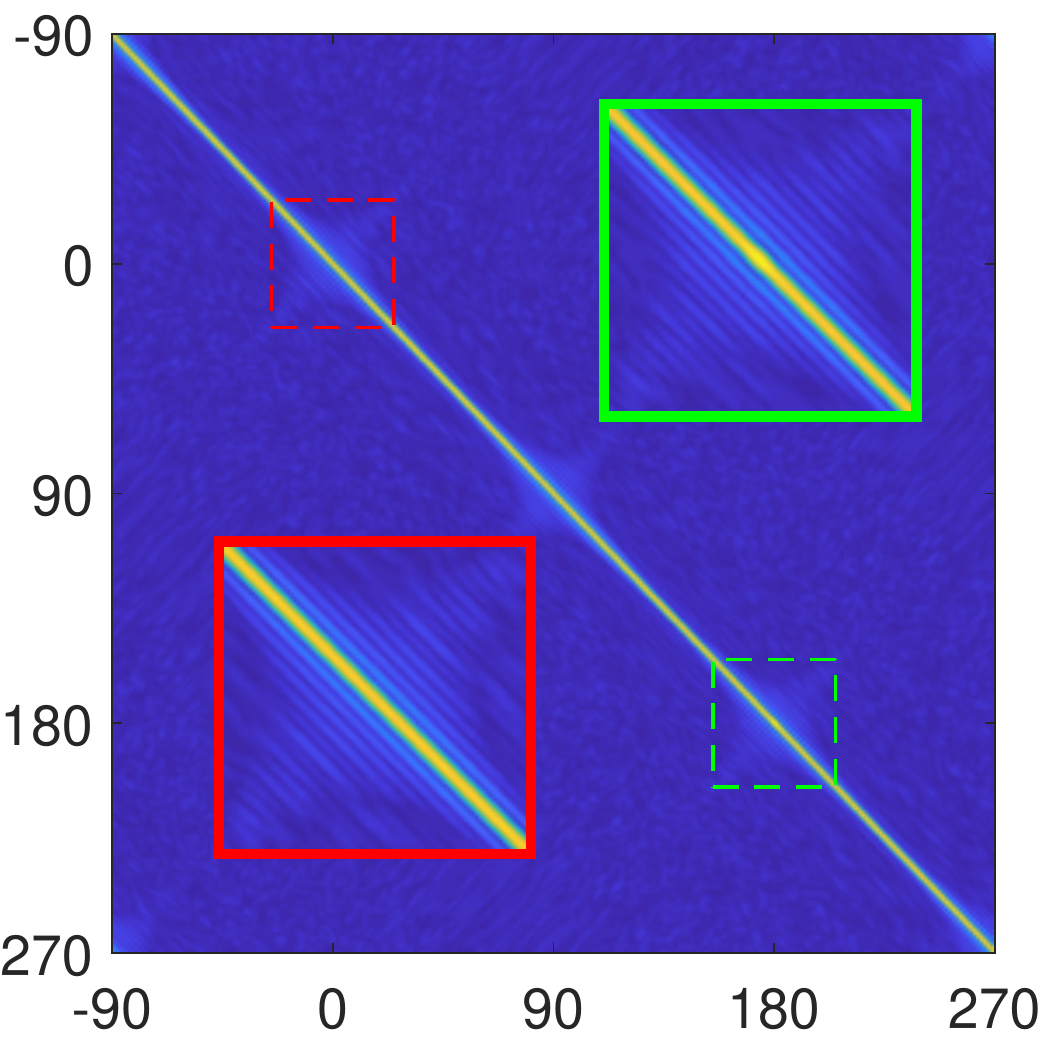}}
	\subfloat{\label{fig:fig2_MFP40_mc_direction_1}\includegraphics[width=0.22\textwidth]
				{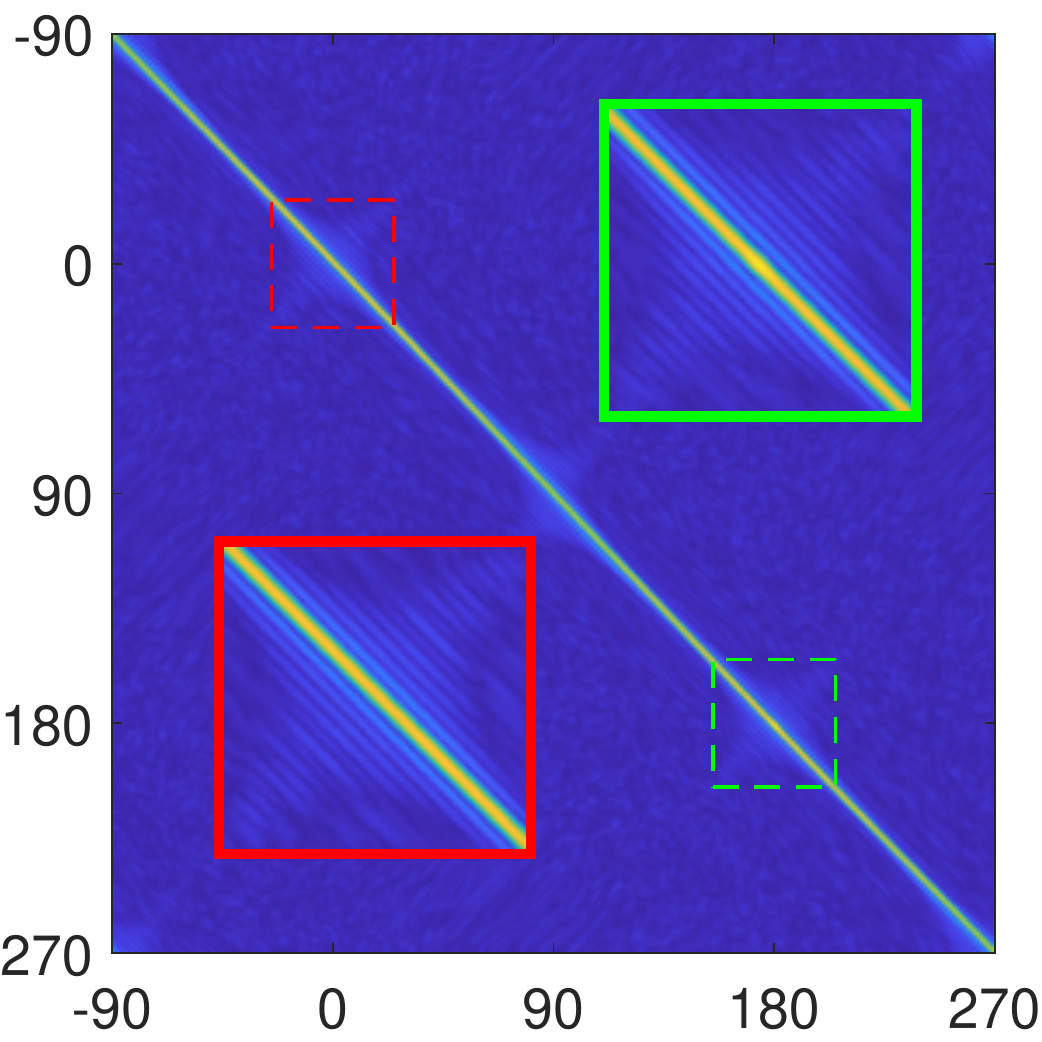}}
	\subfloat{\label{fig:fig2_MFP40_mc_direction_4}\includegraphics[width=0.22\textwidth]
				{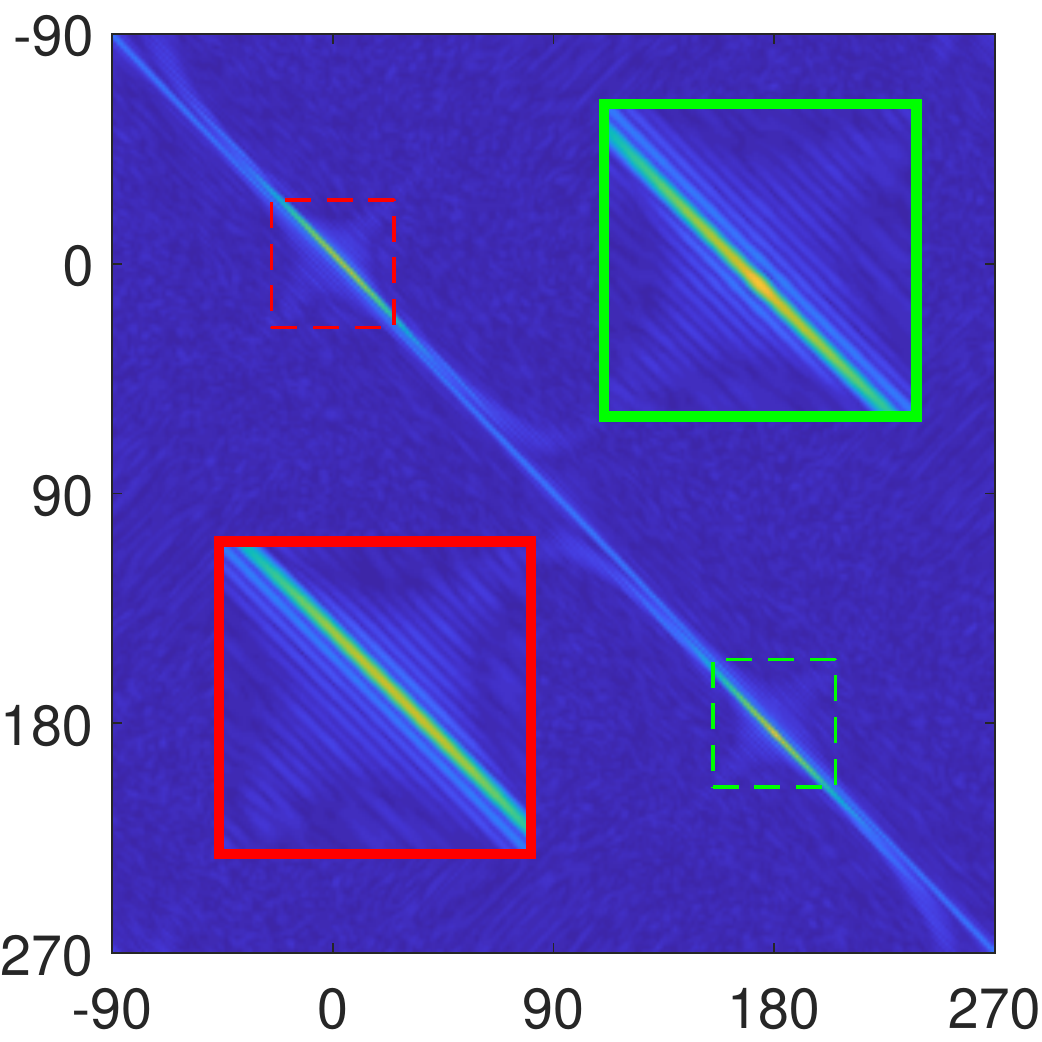}}
	\subfloat{\label{fig:fig2_MFP40_mc_direction_20}\includegraphics[width=0.27\textwidth]
				{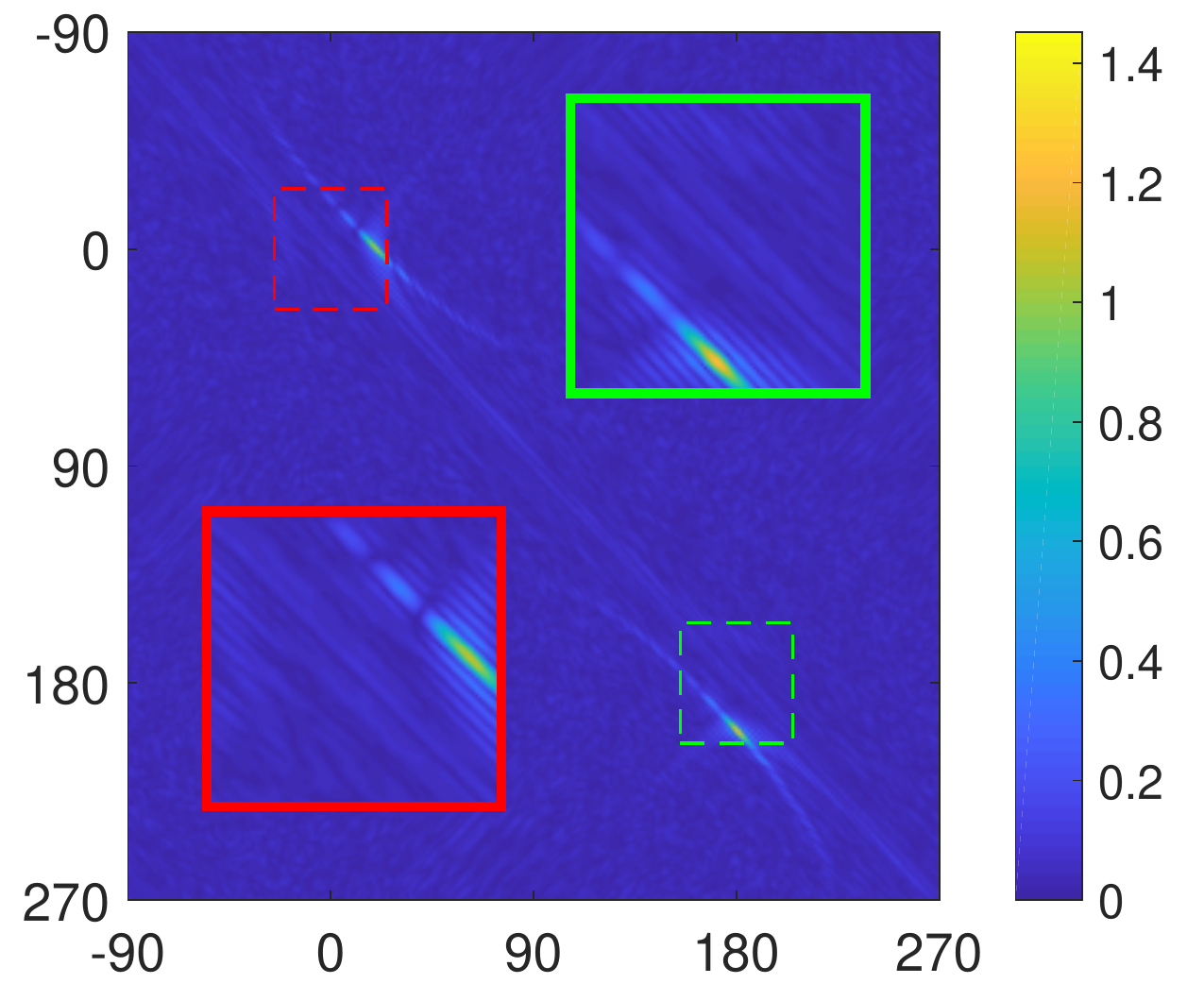}}
\\\vspace{-0.3cm}
			\centering
	\subfloat{\setcounter{subfigure}{12}\raisebox{2cm}{\rotatebox[origin=t]{90}{\bf Wave Solver, O.D.=0.5}}}\hfill
	     \subfloat{\raisebox{2cm}{\rotatebox[origin=t]{90}{angle [deg]}}}\hfill
	\subfloat{\label{fig:fig2_MFP40_mudiff_direction_0}\includegraphics[width=0.22\textwidth]
				{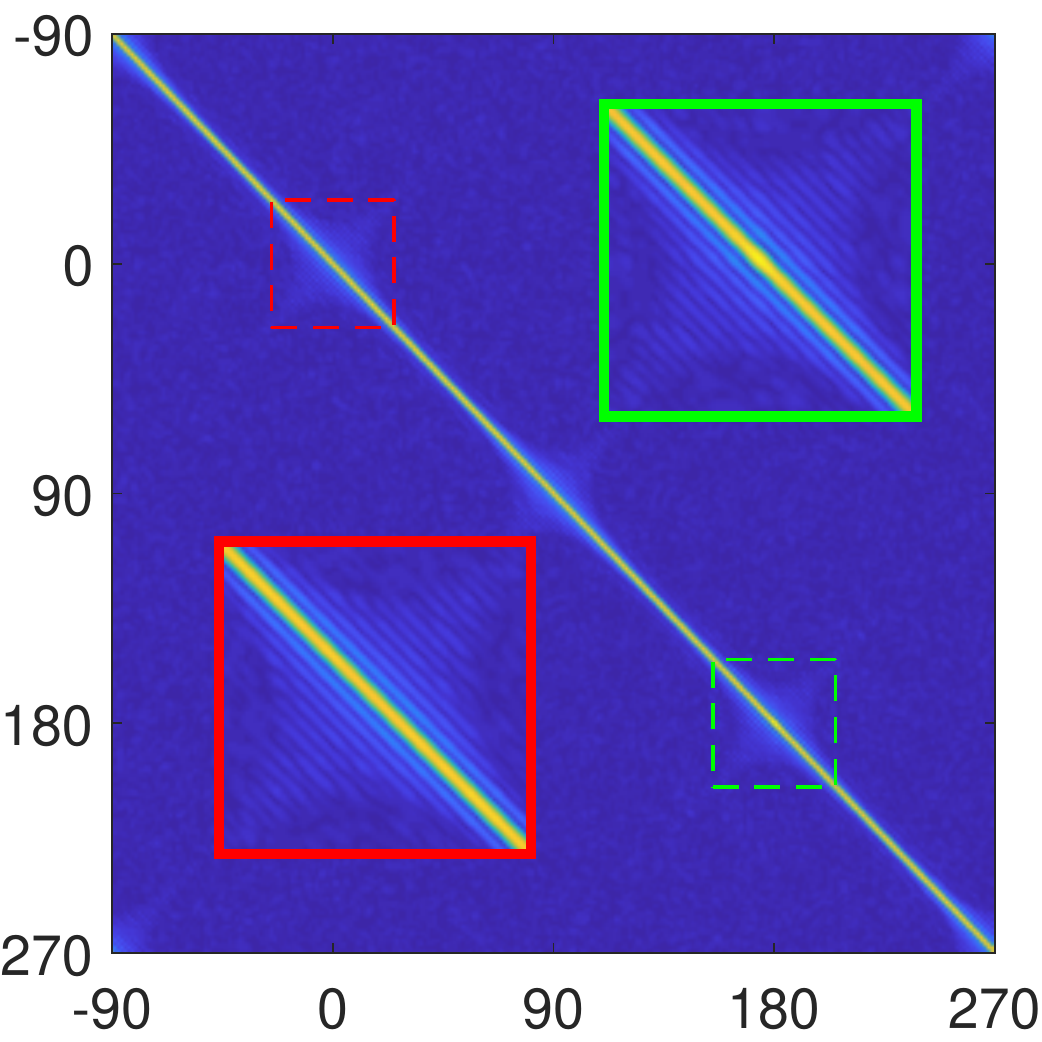}}
	\subfloat{\label{fig:fig2_MFP40_mudiff_direction_1}\includegraphics[width=0.22\textwidth]
				{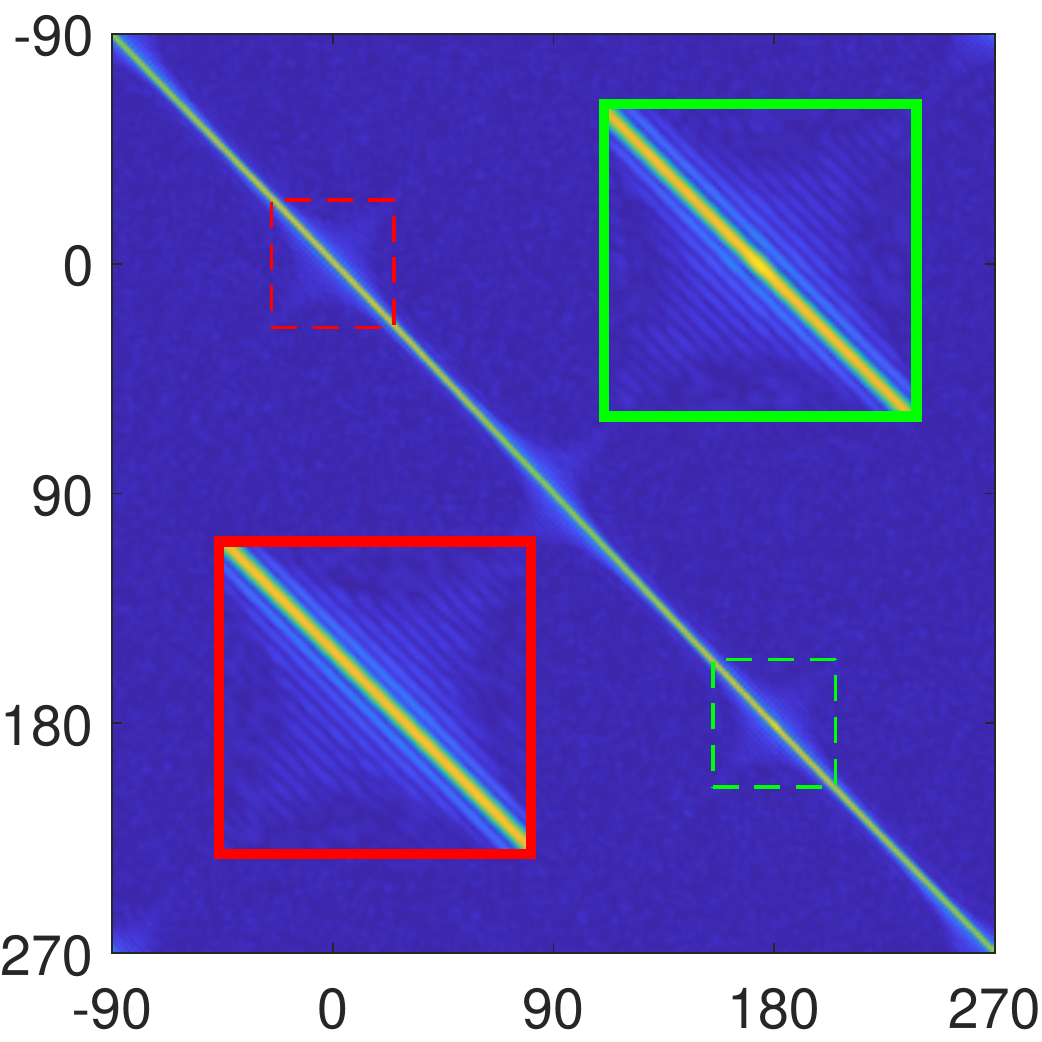}}
	\subfloat{\label{fig:fig2_MFP40_mudiff_direction_4}\includegraphics[width=0.22\textwidth]
				{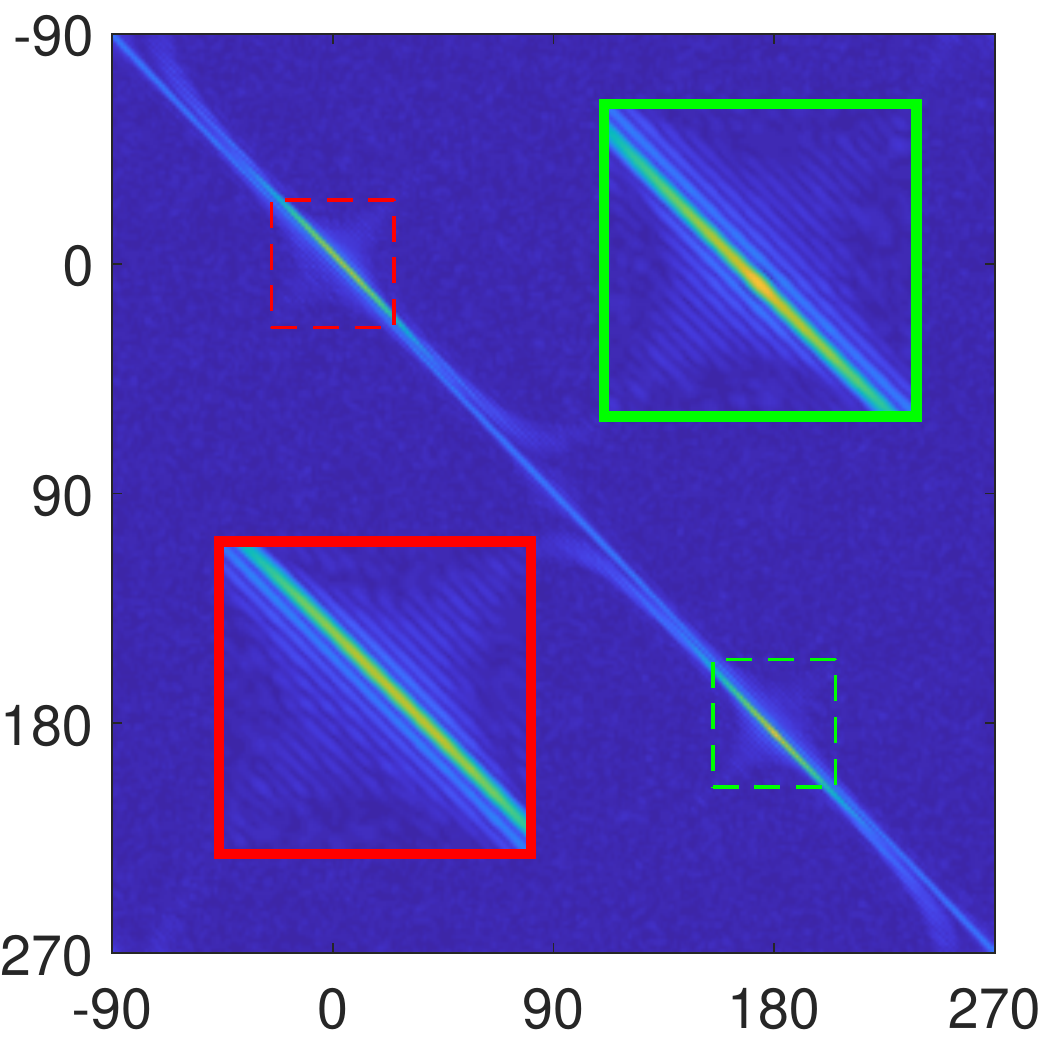}}
	\subfloat{\label{fig:fig2_MFP40_mudiff_direction_20}\includegraphics[width=0.27\textwidth]
				{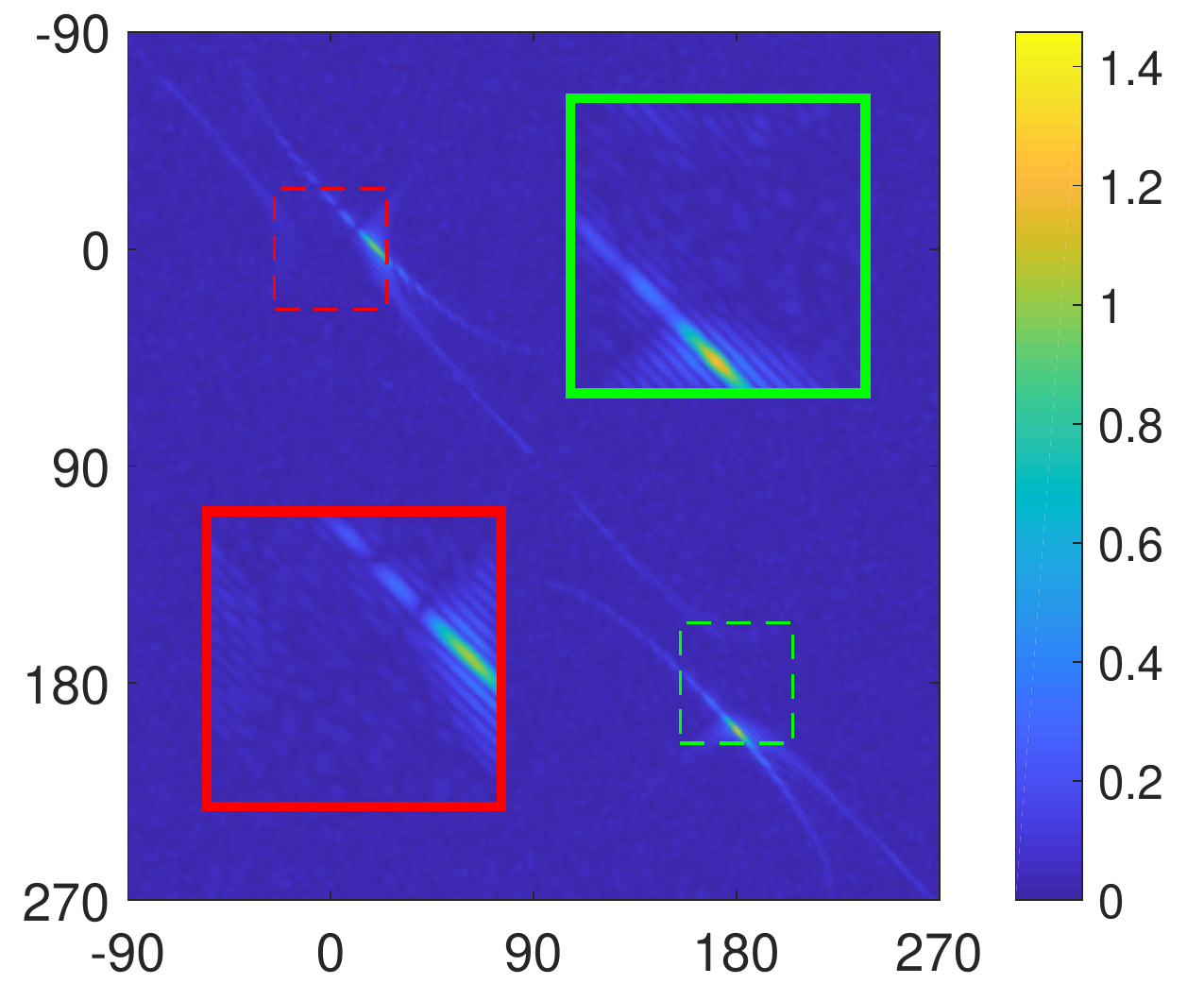}}
			\\\vspace{-0.3cm}
			        \subfloat{\makebox[0.03\textwidth][c]{    }}
			\subfloat{\makebox[0.2275\textwidth][c]{angle [deg]}}
			\subfloat{\makebox[0.2275\textwidth][c]{angle [deg]}}
			\subfloat{\makebox[0.2275\textwidth][c]{angle [deg]}}
			\subfloat{\makebox[0.2795\textwidth][c]{angle [deg]}}
			
  \caption{Memory Effect Validation. Covariance matrices for illumination with two mutually coherent plane waves, evaluated at the far field over $360\degree$ viewing directions. We compare our MC algorithm (rows 1 and 3) with a wave solver (rows 2 and 4). We demonstrate the memory effect for four different pairs of illumination angles and two target optical depths.}  
   \label{fig:beam-cov}        
\end{figure*}

{\boldstart{Wave solvers:}} We have experimented with two popular approaches for solving the wave equation. The classical approach uses finite-difference time-domain (FDTD)
methods~\cite{FDTDYee66,kWaves},  relying on a sub-wavelength discretization of the target. As a result, the approach has high memory and CPU consumption and  does not scale to a 2D cube that is more than a few dozen wavelengths wide. For the specific case of spherical particles, toolboxes such as $\mu$-diff \cite{mudiff}  use the integral version of Helmholtz equation. This is significantly more efficient than FDTD approaches, but the complexity is still cubic in the number of scatterers. Therefore, even these approaches become impractical for volumes with more than a few thousand particles. For the purposes of validating our algorithms, we use the $\mu$-diff solver.  As it is restricted to 2D configurations, we run our comparison in 2D. 

\begin{figure*}[!t]
\begin{center}\begin{tabular}{lcc@{}c@{}c@{}c@{}c}
&&(a) Cov. Rendering&\;\;\;\;\;(b) Field samples&(c) Independent&(d) Diagonal plot\\
\raisebox{2cm}{\rotatebox[origin=t]{90}{$(0\degree,0\degree)$}}&\raisebox{2cm}{\rotatebox[origin=t]{90}{angle [deg]}}&
\includegraphics[width= 0.208\textwidth]{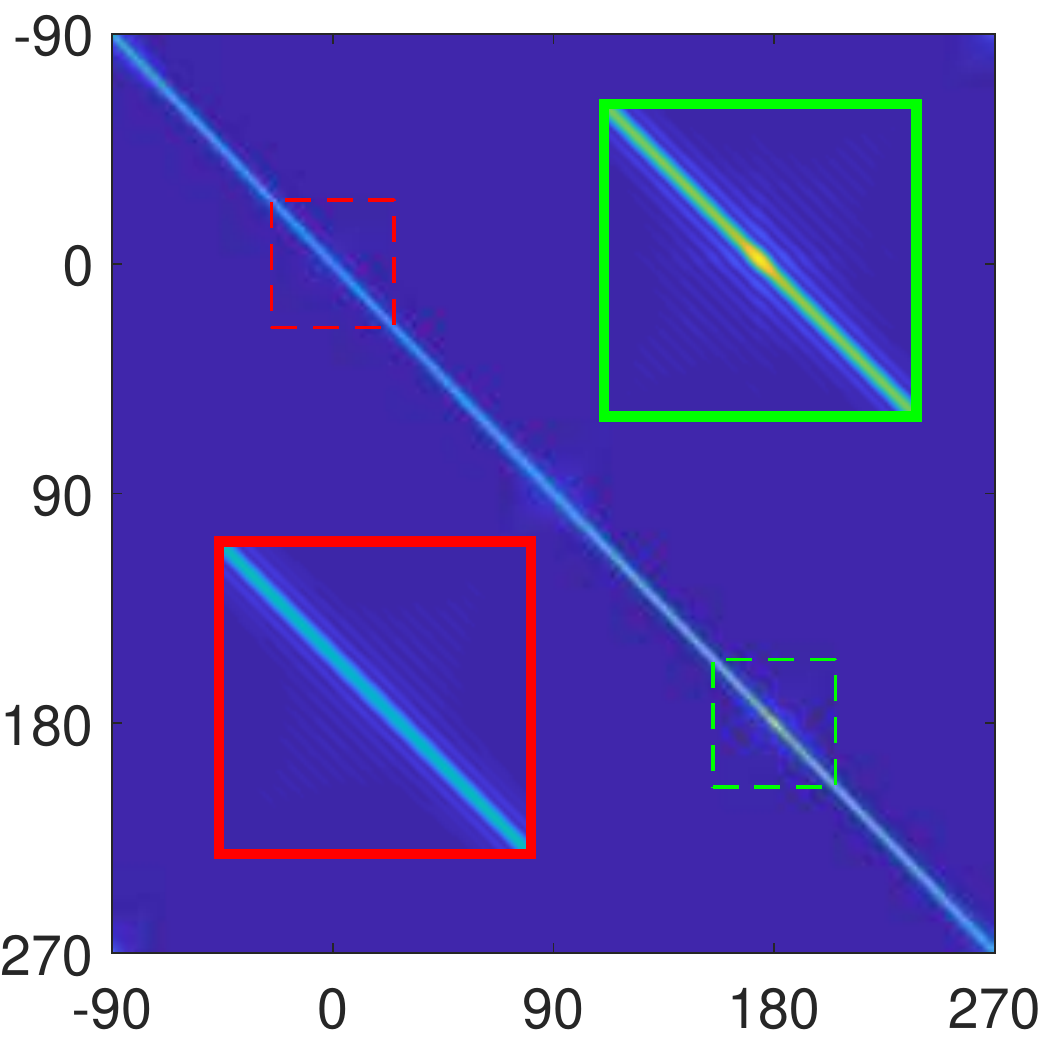}&
\includegraphics[width= 0.208\textwidth]{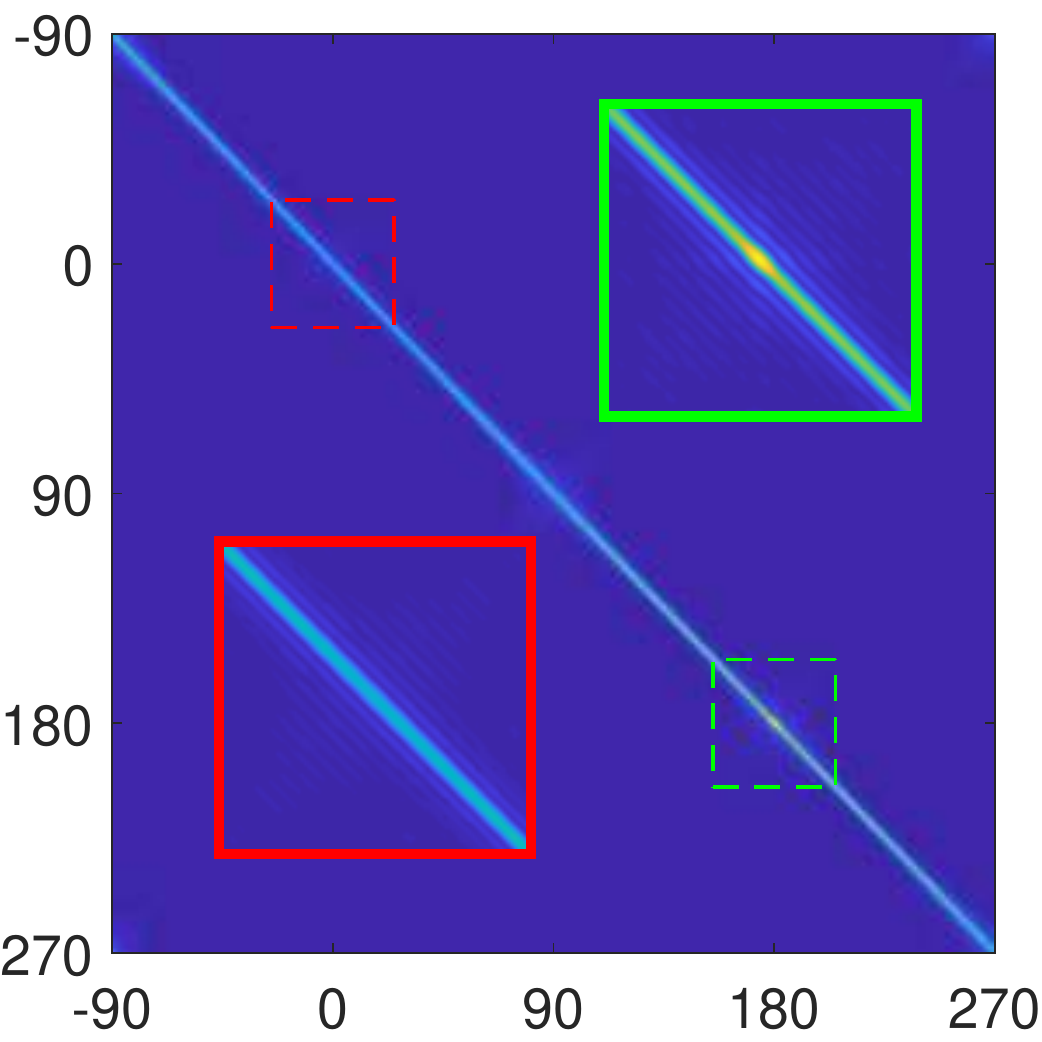}&
\includegraphics[width= 0.256\textwidth]{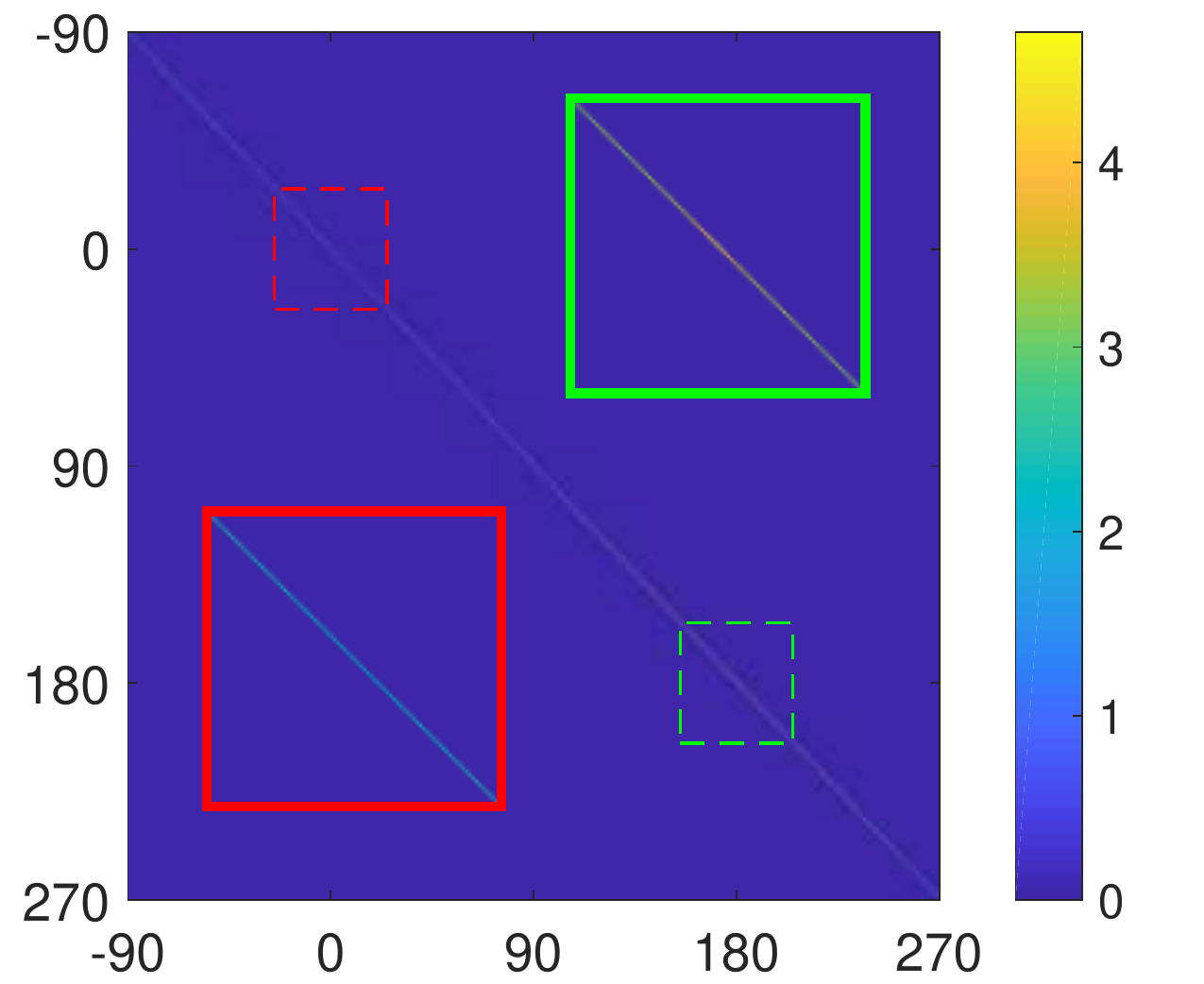}&
\includegraphics[width= 0.208\textwidth]{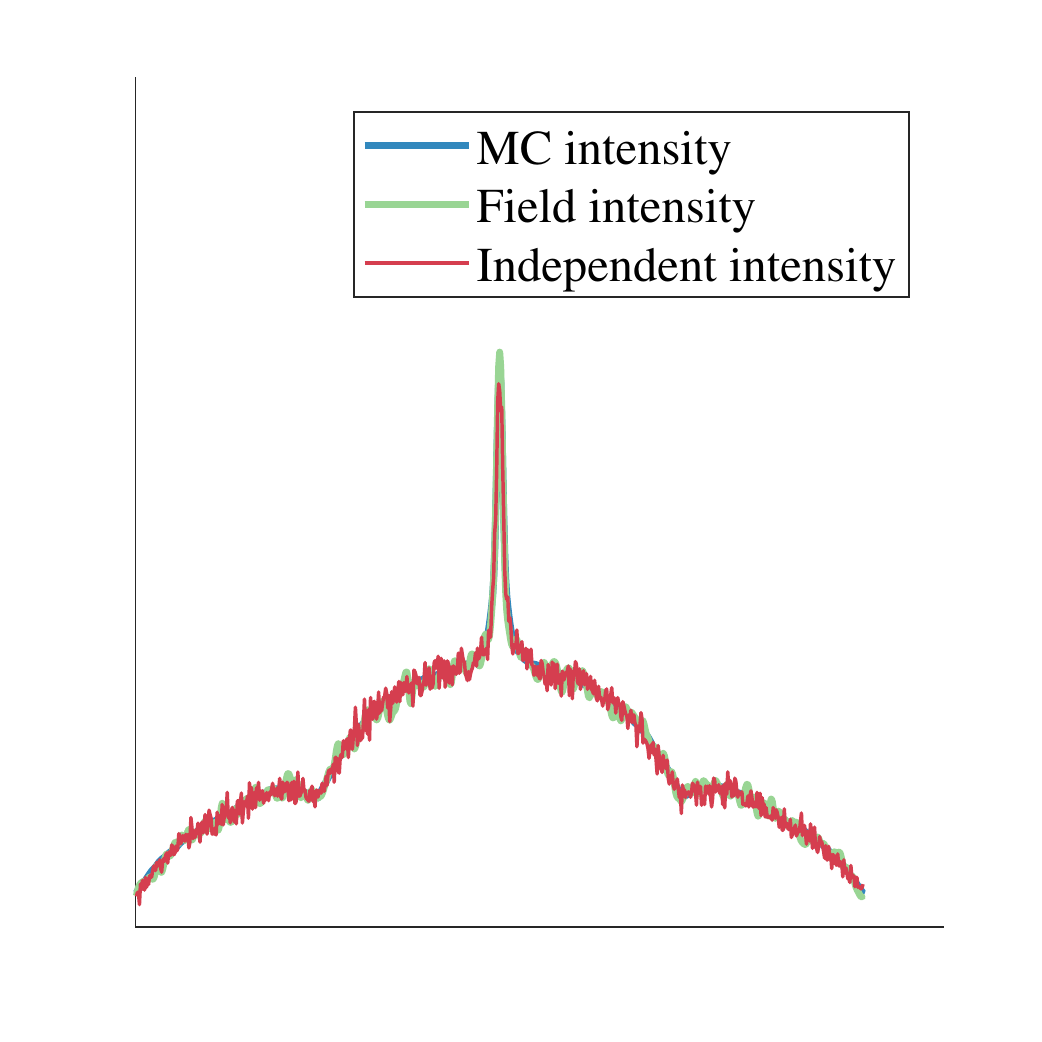}\\
\raisebox{2cm}{\rotatebox[origin=t]{90}{$(0\degree,4\degree)$}}&\raisebox{2cm}{\rotatebox[origin=t]{90}{angle [deg]}}&
\includegraphics[width= 0.208\textwidth]{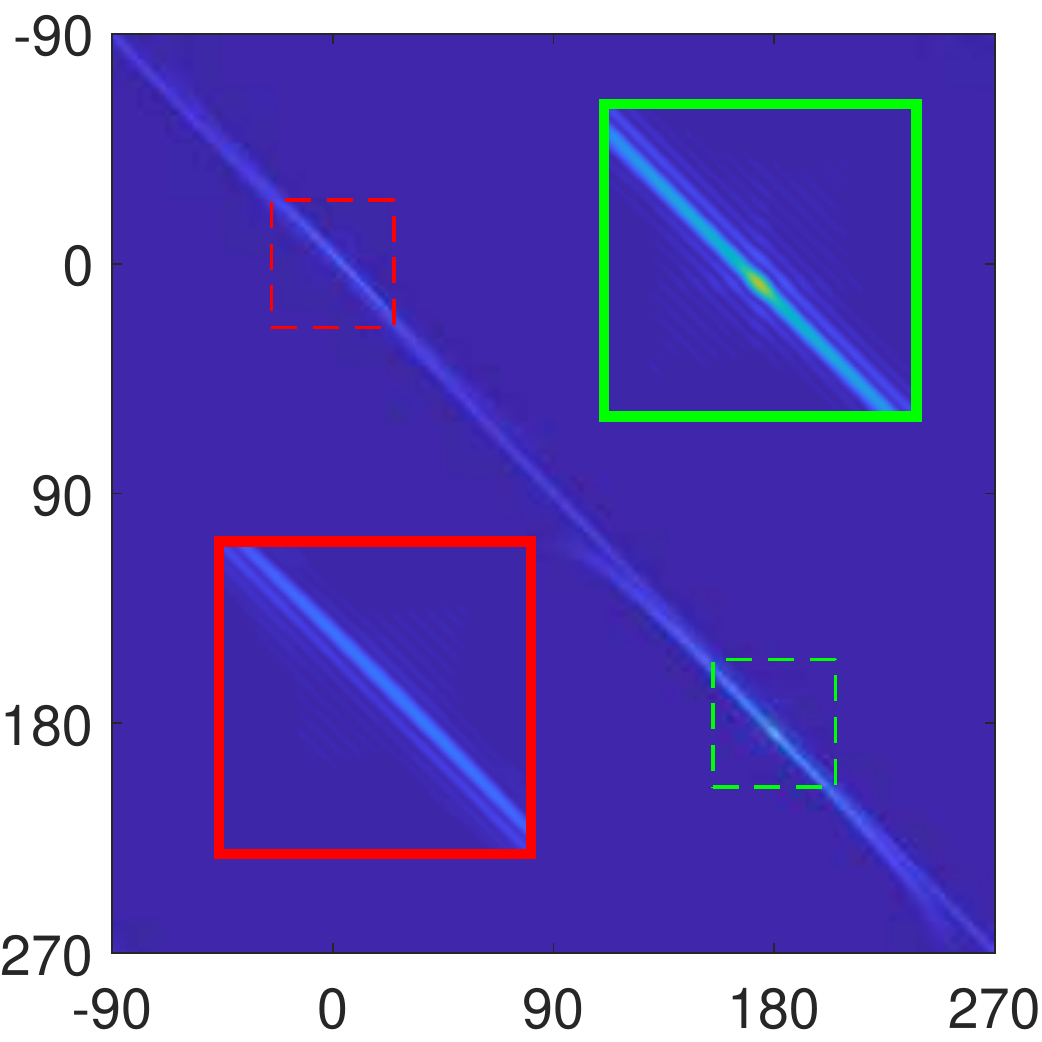}&
\includegraphics[width= 0.208\textwidth]{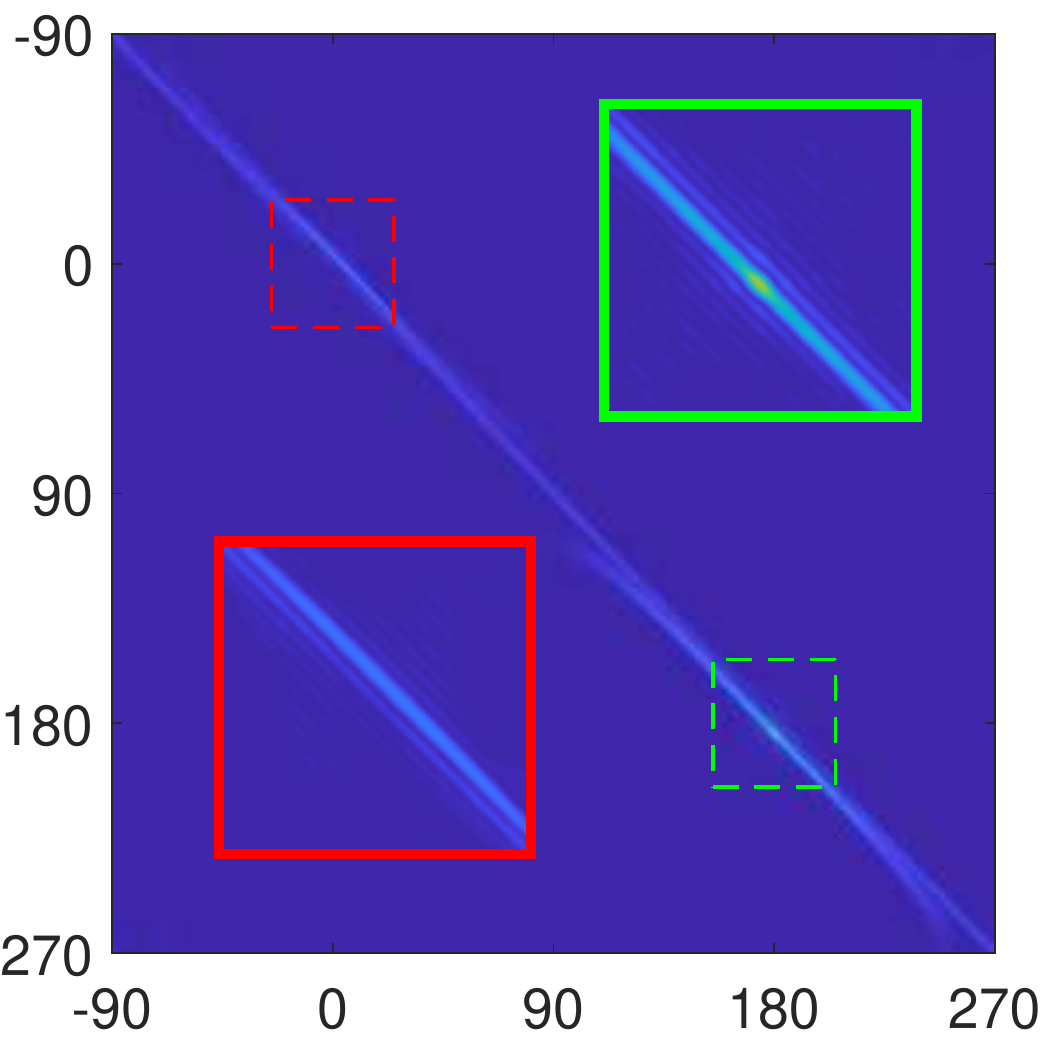}&
\includegraphics[width= 0.256\textwidth]{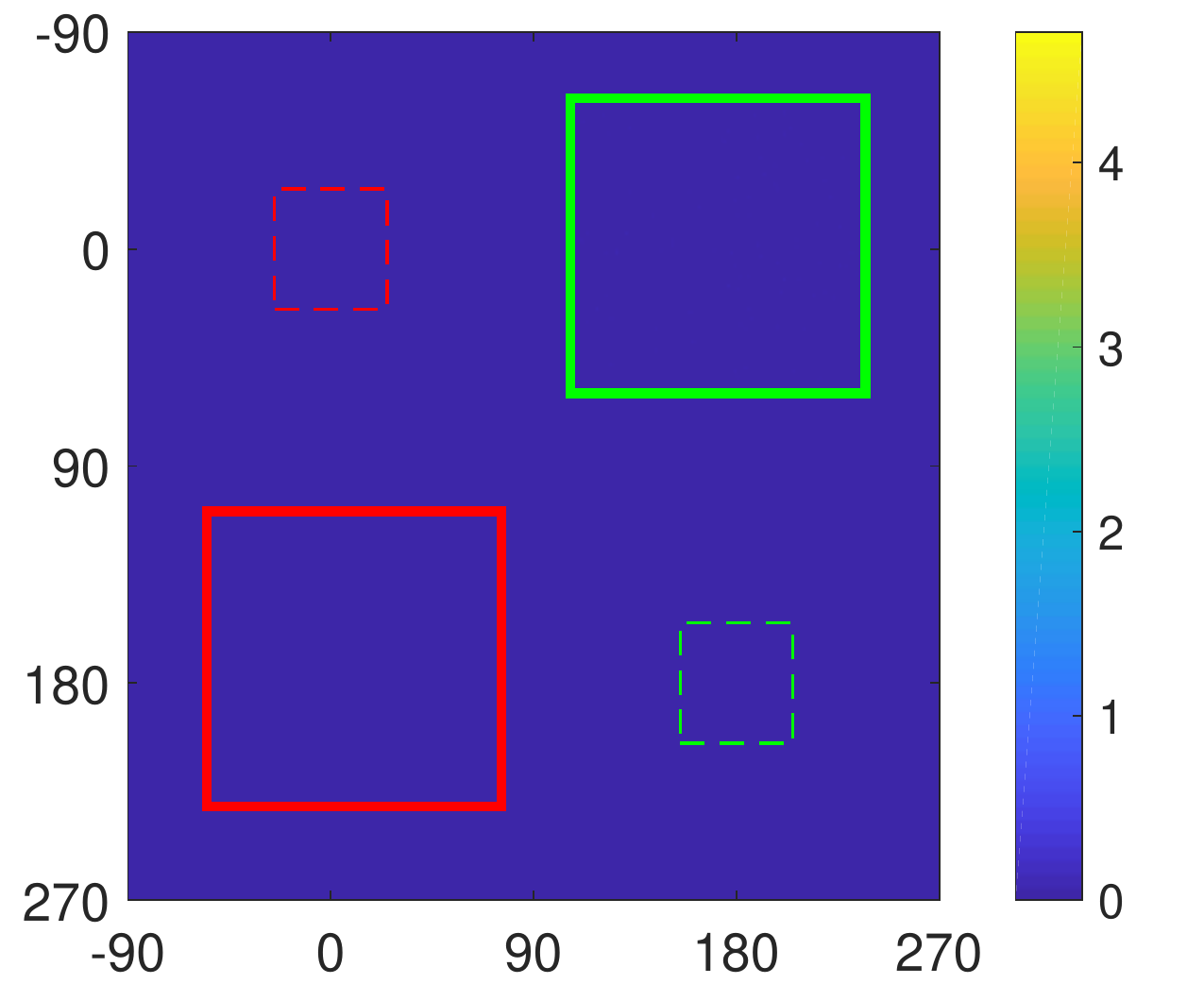}&
\includegraphics[width= 0.208\textwidth]{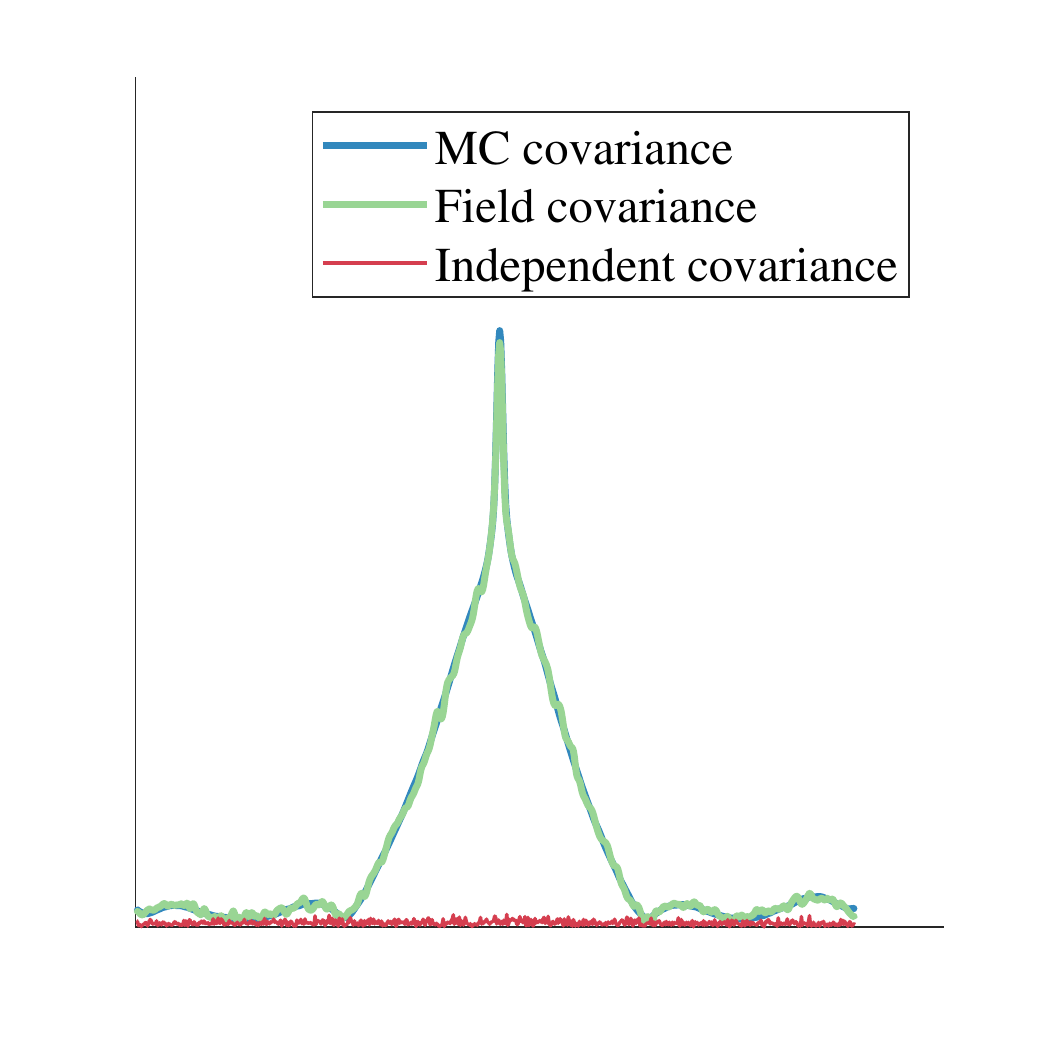}\\
&&angle [deg]&angle [deg]&angle [deg]&angle [deg]
\end{tabular}
\caption{Evaluating speckle fields:   The covariance of  speckle fields sampled by Alg. \ref{alg:MCfield} (b), is equivalent to the covariance computed directly using Alg \ref{alg:MCpair} in (a). In contrast, a simpler speckle rendering approach~\cite{Sawicki:08} that samples paths independently produces images with no spatial correlation, yet it can produce the correct intensity as evident by the diagonal plot in (d). Top row shows correlation between illuminations $\ind_1=\ind_2=0\degree$ whose diagonal is intensity, while the lower row shows correlation between different illumination directions $\ind_1=0\degree, \ind_2=4\degree$.   }  
\label{fig:mc-field-cov-indep}
\end{center}        
\end{figure*}

{\boldstart{Coherent backscattering.}} \figref{fig:cbs-f-nf} demonstrates coherent backscattering intensity, rendered using our algorithm with $\inp_1=\inp_2,\snsp_1=\snsp_2$. We use a target of size $100\lambda\times 100\lambda$, with a mean free path of $50\lambda$, leading to  an optical depth (that is, average number of scattering events) O.D.= 2. We simulate far-field sensors through all $360\degree$ around the target, and near-field sensors located on a $360\degree$ circle of diameter $200\lambda$ around the target. We compare the mean speckle intensity obtained from the electromagnetic solver with our Monte Carlo algorithm, considering forward and reversed paths, and with a simpler algorithm considering only forward paths derived in Appendix \ref{sec:CTE}. For far-field sensors, we see that when the viewing direction approaches the inverse of the illumination direction, a narrow peak in brightness occurs, which is the manifestation of coherent backscattering. This peak is not predicted when using forward-only paths, but is indeed explained when using both forward and reversed paths. For near-field sensors, coherent backscattering is less pronounced and the two Monte Carlo algorithms are closer to each other.

{\boldstart{Memory effect.}} In \figref{fig:beam-cov}, we show simulated covariance matrices for a target of size $20\lambda\times 20\lambda$ at O.D. $2$ and $0.5$. We visualize covariance matrices of a target illuminated by two plane waves, measured at the far-field over $360\degree$ viewing directions. In the covariance matrices, the memory effect is evident by the fact that, for small angles, the strongest correlation is obtained at a diagonal that is offset from the main diagonal, and the offset increases with the illumination angle difference. When the angle difference is large, the classical version of the memory effect no longer holds and the covariance is no longer a shifted diagonal. However, one can still observe some correlation along a curved set of viewing directions. To our best knowledge such correlations have not yet been explored, and provide an exciting direction of future research. In particular they may allow expanding the angular range of existing computational imaging techniques relying on the memory effect. We note also that, while the shape of the correlation curve is consistent, its exact value is a function of density, as seen from the two optical depths simulated in \figref{fig:beam-cov}.

{\boldstart{Runtime comparison.}} Overall, \figpref{fig:cbs-f-nf}{fig:beam-cov} illustrate that our Monte Carlo simulations provide useful and accurate predictions of speckle correlations, which is orders of magnitudes  more efficiently than the approach based on the wave solver. To quantify the performance difference, in the small example of \figref{fig:beam-cov}, simulating the covariance with the wave solver approach took 6 hours on a 50-core cluster, using the $\mu$-diff solver~\cite{mudiff}. By contrast, our Monte Carlo algorithm produced the same estimate in 45 minutes on a single core, using an unoptimized Matlab implementation. The difference in performance becomes even more drastic as the number of scatterers increases. 
\begin{figure*}[t]
        \centering
        \subfloat{\setcounter{subfigure}{0}\raisebox{1.5cm}{\rotatebox[origin=t]{90}{$C(\theta)$}}}
        \subfloat[]{\label{fig:mem-MFP}\includegraphics[width=0.24\textwidth]
                {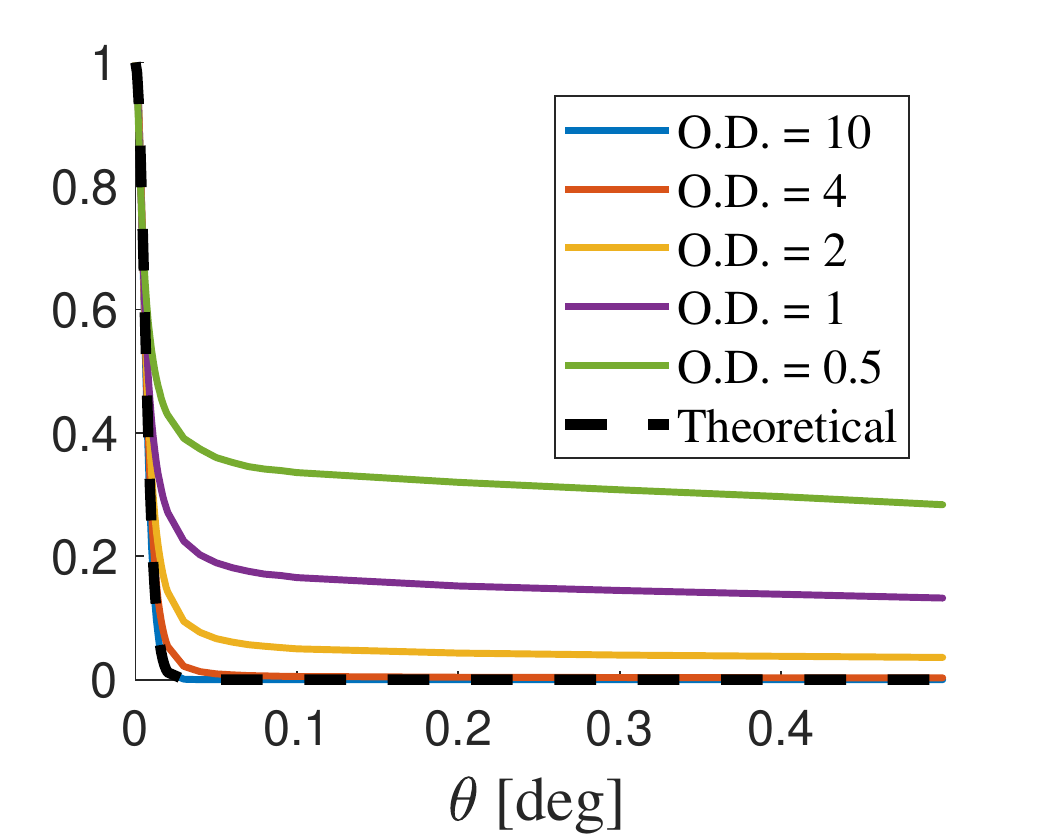}}
        \subfloat[]{\label{fig:mem-albedo}\includegraphics[width=0.24\textwidth]
                {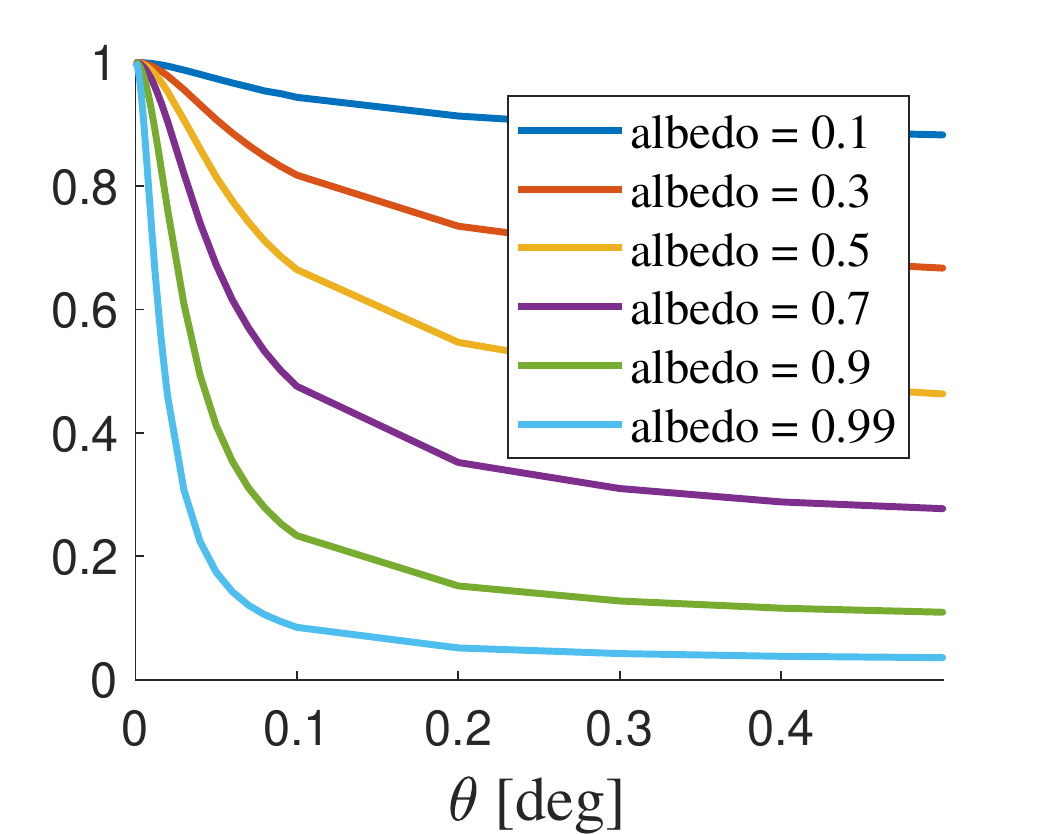}}
                 \subfloat[]{\label{fig:mem-g}\includegraphics[width=0.24\textwidth]
                {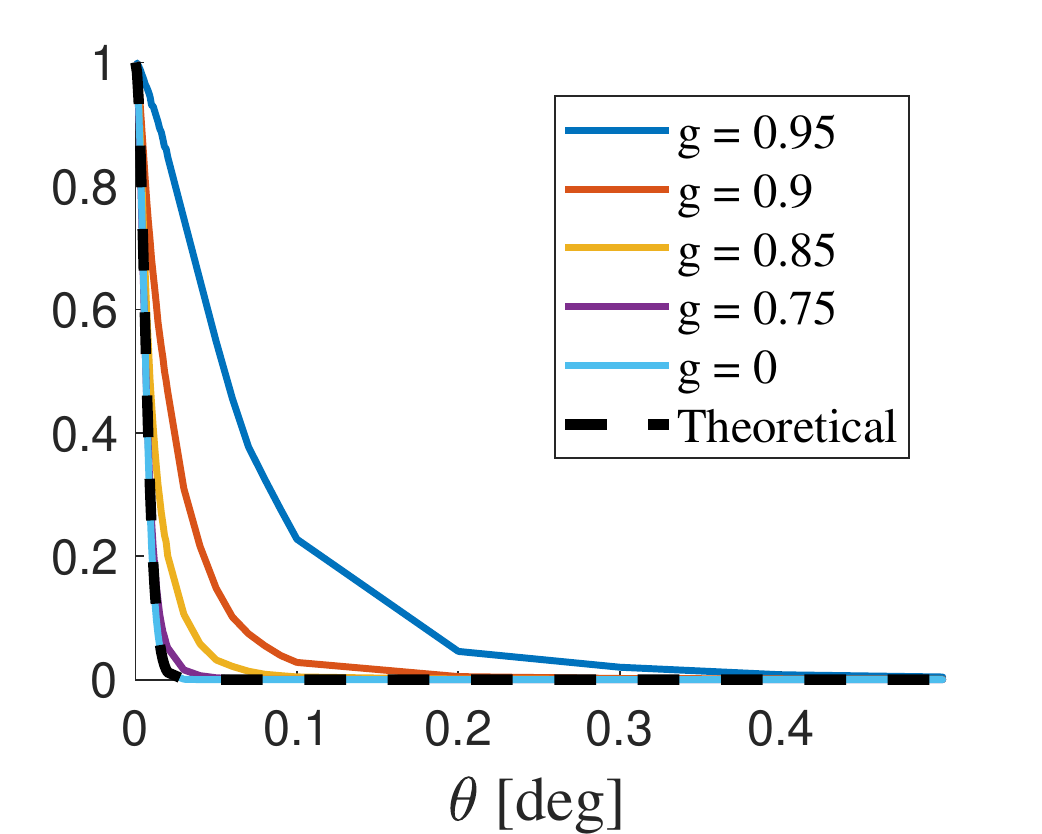}}
                 \subfloat[]{\label{fig:sim-g}\includegraphics[width=0.24\textwidth]
                {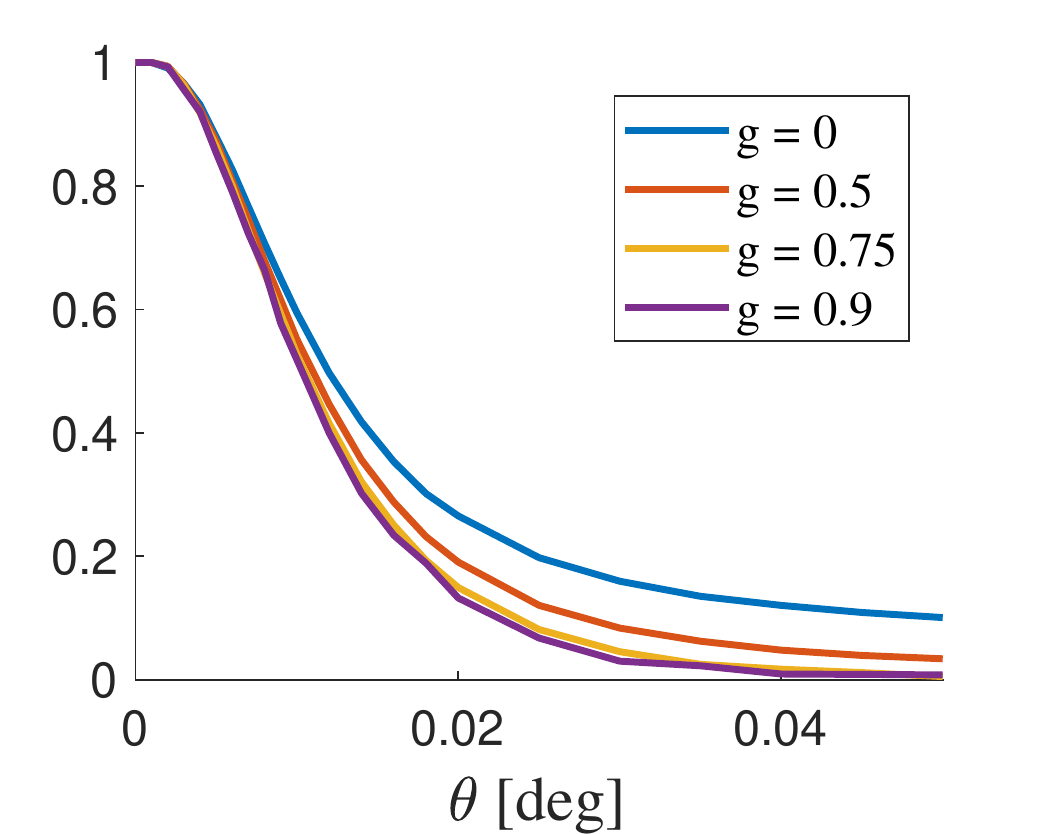}}
  \caption{Theoretical (dash) and numerical calculations (solid) of the correlation as a function of angle $\theta$. (a) Varying the optical depth (O.D.) for forward scattering configuration and an isotropic phase function $g=0$. For a high O.D. the computed correlation agrees with theory. As the O.D. decreases (mean free path increases) the range of the memory effect increases.  (b) Varying albedo for a backscattering configuration with a fixed $g$ and O.D. Observe the longer memory effect for highly absorbing materials. (c) Varying the anisotropy parameter of the phase function for a fixed O.D. and a forward scattering setup. Notice that the memory effect range increases with $g$. (d) Validating similarity theory for $g$ $MFP$ pairs  keeping the  ratio $(1-g)/MFP$ constant, configurations that should be equivalent by similarity theory leads to similar, yet non identical correlation curves.}  
   \label{fig:memory-effect-plots}        
\end{figure*}

{\boldstart{Field samples.}} We use Alg \ref{alg:MCfield} to sample multiple speckle fields, as visualized in \figref{fig:teaser}.  \figref{fig:mc-field-cov-indep} demonstrates their empirical covariance, showing good agreement with the covariance rendered directly by Alg. \ref{alg:MCpair}. We also compare with the simpler ``electric field Monte Carlo''  speckle rendering algorithm~\cite{Xu:04,Sawicki:08}. This approach extends MC algorithms rendering intensities by using the length of the path as the phase. The main difference is that each sampled path is used to update only one sensor point, and therefore different illumination and viewing directions are updated {\em independently}. As a consequence, while this approach can accurately render intensity and even simulate coherent backscattering, it cannot reproduce spatial correlation. The target size and densities in \figref{fig:mc-field-cov-indep} are equivalent to the setup of \figref{fig:beam-cov} at O.D. 2. It should be noted that~\cite{Xu:04,Sawicki:08} focus on modeling polarization correctly, while polarization is not addressed in our work.

\subsection{Quantifying the memory effect of speckles}\label{sec:memoryeff}

The memory effect of speckles is an active research area in optics, and has been a central property allowing researchers to develop techniques for seeing through scattering layers and around corners.  \comment{ For thin scattering media,  a light-emitting object hidden behind a scattering layer is numerically reconstructed from its transmitted speckle pattern~\cite{Katz2014,Bertolotti2012}, using speckle  autocorrelation. As a result of the memory effect, speckle images produced by different light positions are a shifted version of each other, hence the speckle auto-correlation is equivalent to the  auto-correlation of source positions.  Another approach that is valid when the transmission matrix is available   (e.g., by placing a point source or detector behind the scattering
layer)
 is to actively modulate  the incident wavefront and focus inside turbid media~\cite{Mosk2012,Nixon2013}.  The memory effect plays again an important task as if one measured the modulation pattern at one position (e.g. by recording a flurecent source), a shift/tilt of the same pattern can be used at nearby points~\cite{GenOptMemory17,Vellekoop:10}. 

As all these approach are relying on the memory effect,} Given its wide applicability, it is crucial to understand the range of illumination and viewing configurations for which we can indeed expect a high correlation between speckles.

There have been multiple attempts~\cite{GenOptMemory17,BERKOVITS1994135,Fried:82} to derive closed-form formulas for speckle correlations. The complexity of multiple scattering means that this is only possible under various assumptions, which limit the approximation accuracy and the applicability of such formulas. The most commonly used result~\cite{feng1988correlations,Akkermans07} is a formula derived under diffusion (i.e., high-order scattering) assumption,
 \BE\label{eq:corr-diffusion}
 C(\theta)\approx\frac{(k \theta L)^2 }{sinh^2(k \theta L)}
 \EE
where $\theta$ is the angle between illumination and viewing directions, $L$ is the material thickness, and $ C(\theta)$ is the correlation between intensity images (rather than complex fields) $I^\ind_\outd$ and $I^{\ind+\theta}_{\outd+\theta}$. The correlation of \equref{eq:corr-diffusion} decays to zero exponentially fast as soon as $k\theta L>1$, hence the angular range at which the memory effect is valid is proportional to $1/(kL)$. In \secref{sec:understandingME}, we show that the Monte Carlo formulation can help understand this result.

The diffusion assumption used to derive \equref{eq:corr-diffusion} means that the formula applies only when the average number of scattering events on a path is large. However, empirical observations  suggest that, in practice, the memory effect is valid through a much wider range. A few scenarios that have been  observed to increase this range are (i) an average number of scattering events that is lower than the diffusive regime, (ii) absorption, (iii) forward scattering phase functions \cite{Schott:15}. Forward scattering is particularly important in practice, as tissue is known to be highly forward scattering and is usually described by an Henyey-Greenstein (HG) phase function with anisotropy parameter  $g\in [0.85-0.95]$. Given the lack of analytic formulas and the practical importance of the problem, there have been multiple attempts to empirically  measure the range of the memory effect of materials of interest in the lab~\cite{Schott:15,Yang:14,mesradi:hal-01316109}.
  \begin{figure*}[t]
        \centering
        \subfloat{\makebox[0.17\textwidth][c]{Predicted correlation        }}
        \subfloat{\makebox[0.13\textwidth][c]{$0\Delta$}}
        \subfloat{\makebox[0.13\textwidth][c]{$\Delta$}}
        \subfloat{\makebox[0.13\textwidth][c]{$2\Delta$}}
        \subfloat{\makebox[0.13\textwidth][c]{$3\Delta$}}
        \subfloat{\makebox[0.13\textwidth][c]{$4\Delta$}}
        \subfloat{\makebox[0.13\textwidth][c]{$10\Delta$}}
            \\\vspace{-0.2cm}
        \subfloat{\raisebox{1cm}{\rotatebox[origin=t]{90}{$g=0$}}}
                 \subfloat{\label{fig:g_0}\includegraphics[width=0.17\textwidth]{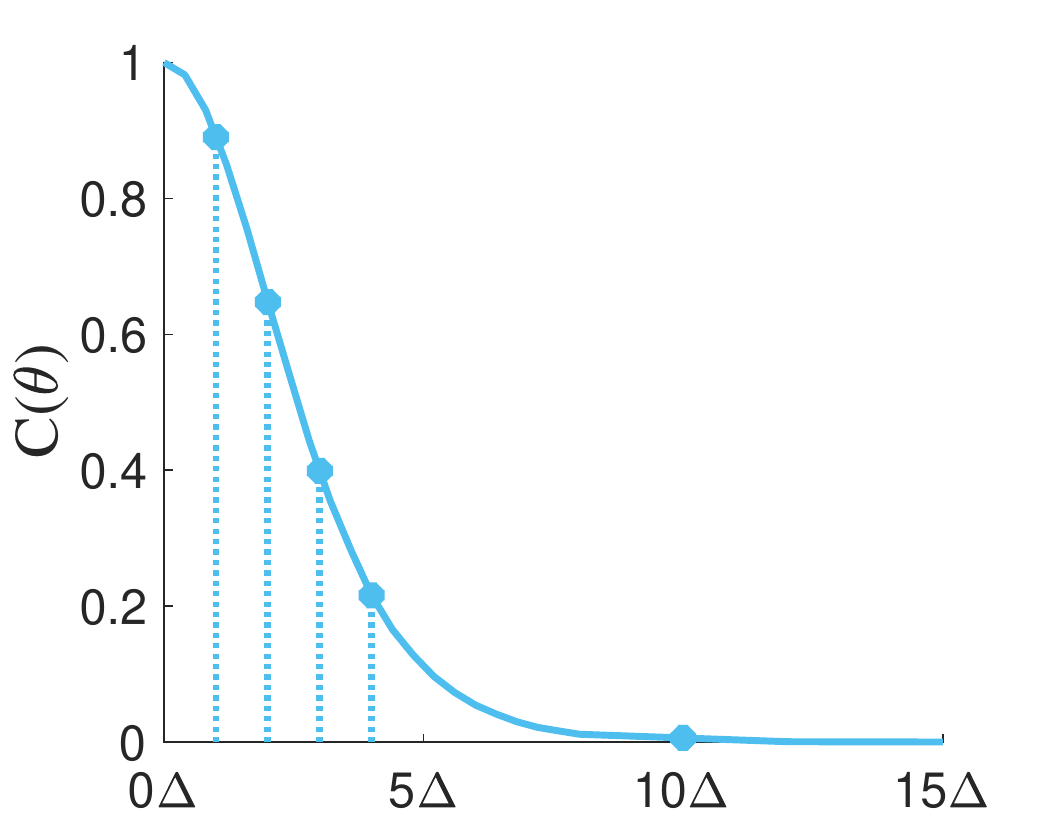}}\hfill
        \subfloat{\label{fig:g_0_1}\includegraphics[width=0.13\textwidth]
                {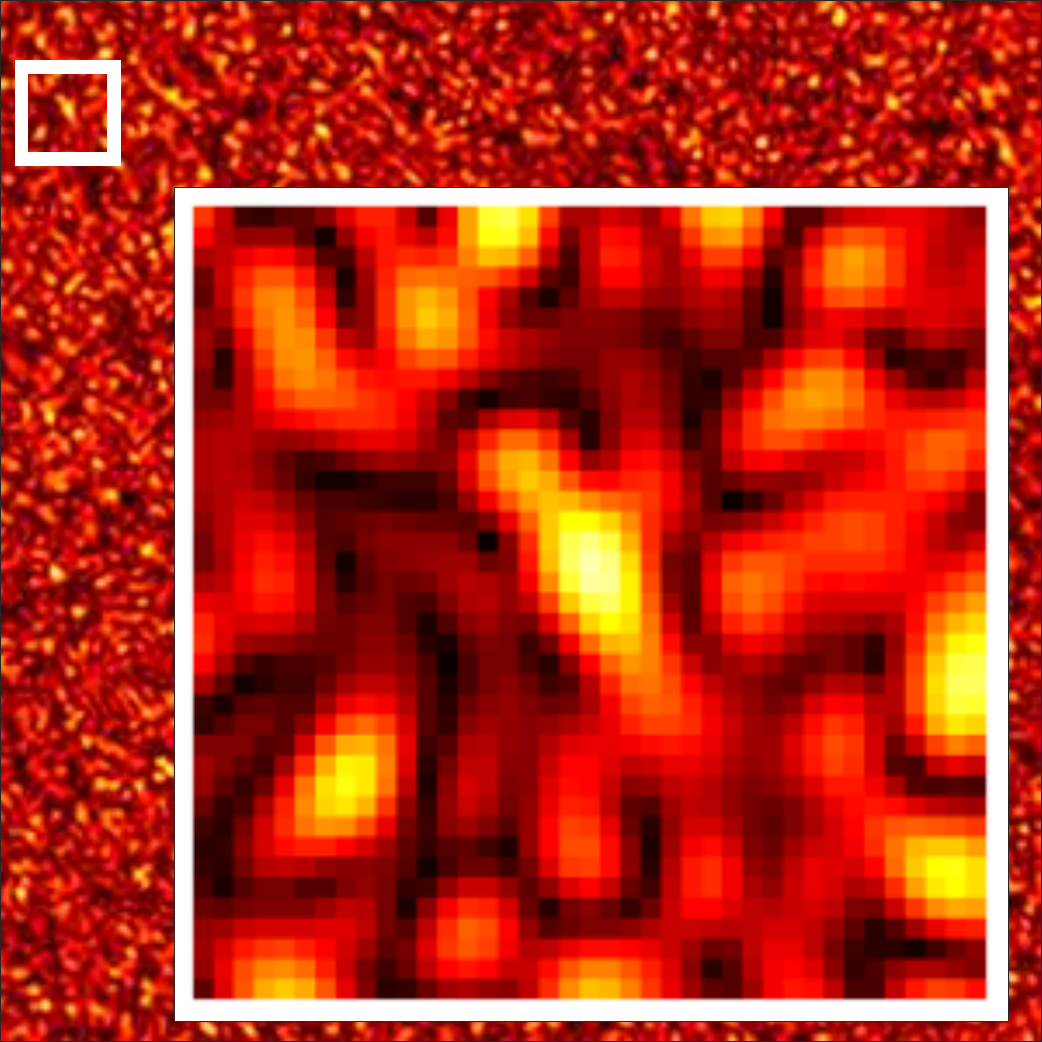}}\hfill
        \subfloat{\label{fig:g_0_2}\includegraphics[width=0.13\textwidth]
                {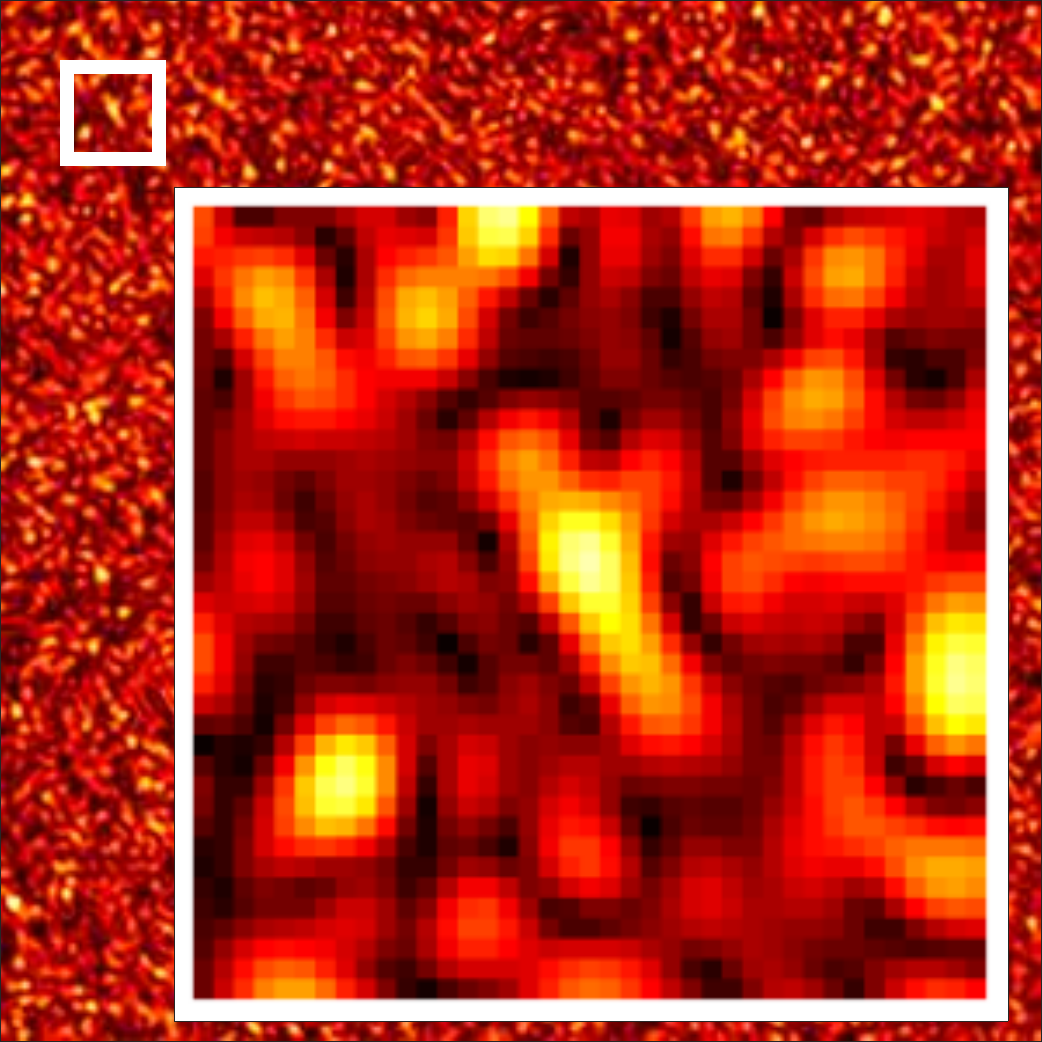}}\hfill
       \subfloat{\label{fig:g_0_3}\includegraphics[width=0.13\textwidth]
        {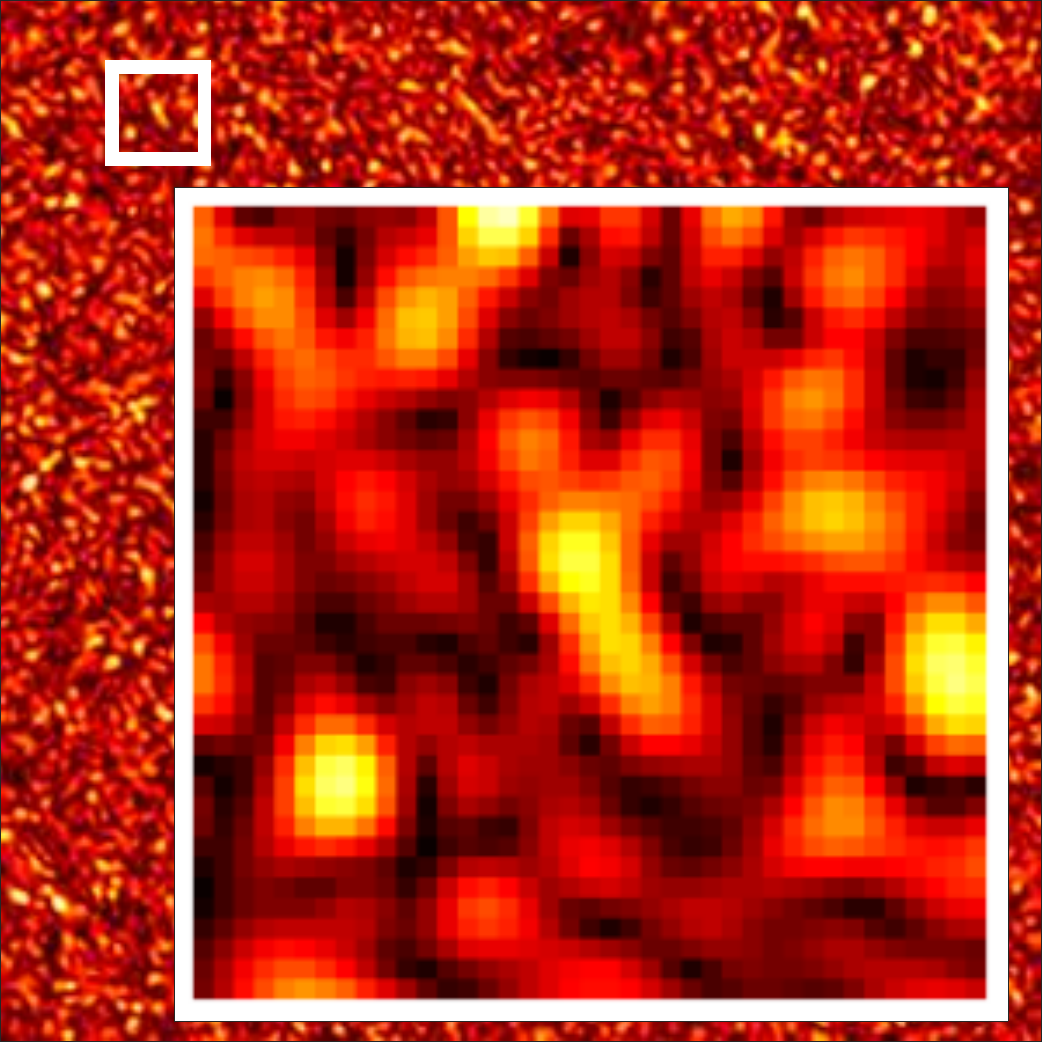}}\hfill
        \subfloat{\label{fig:g_0_4}\includegraphics[width=0.13\textwidth]
                {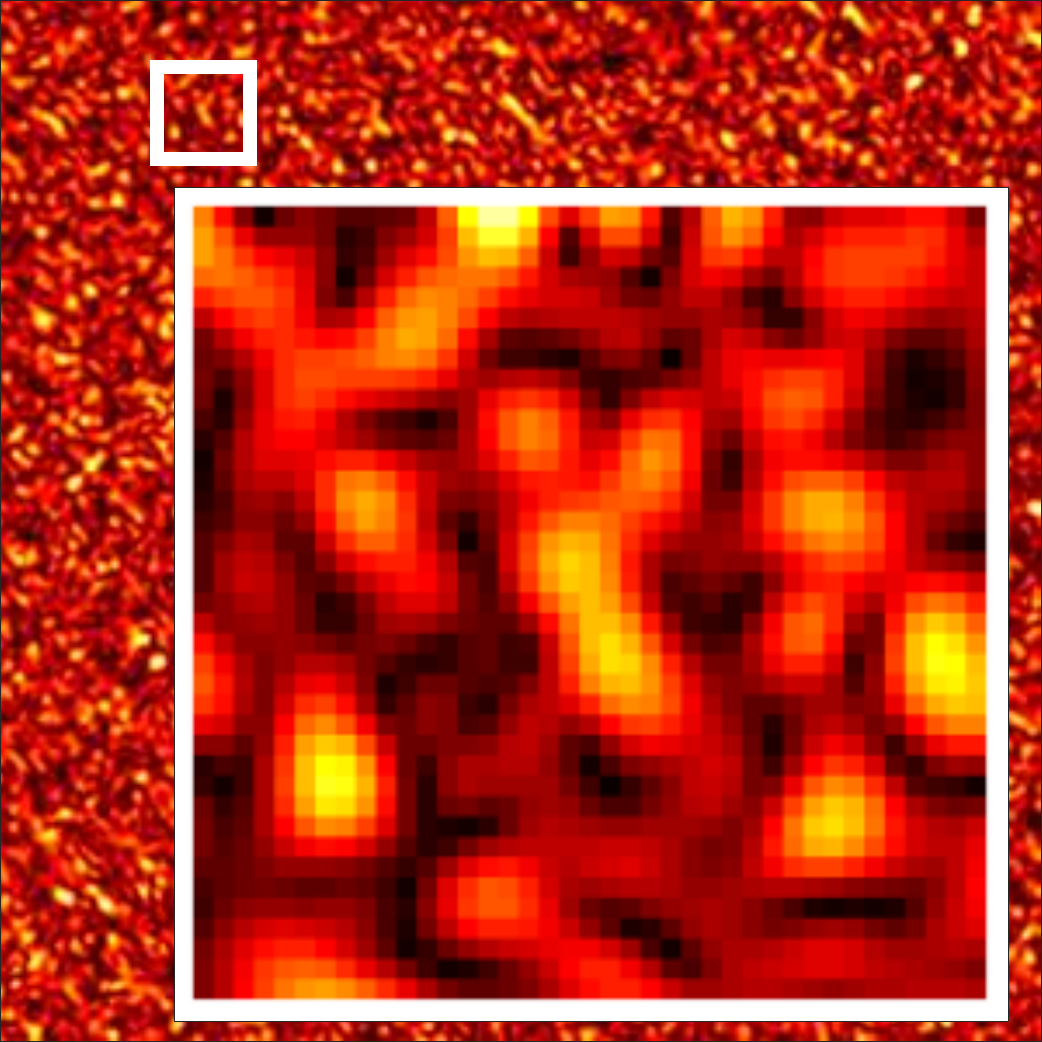}}\hfill
        \subfloat{\label{fig:g_0_5}\includegraphics[width=0.13\textwidth]
                {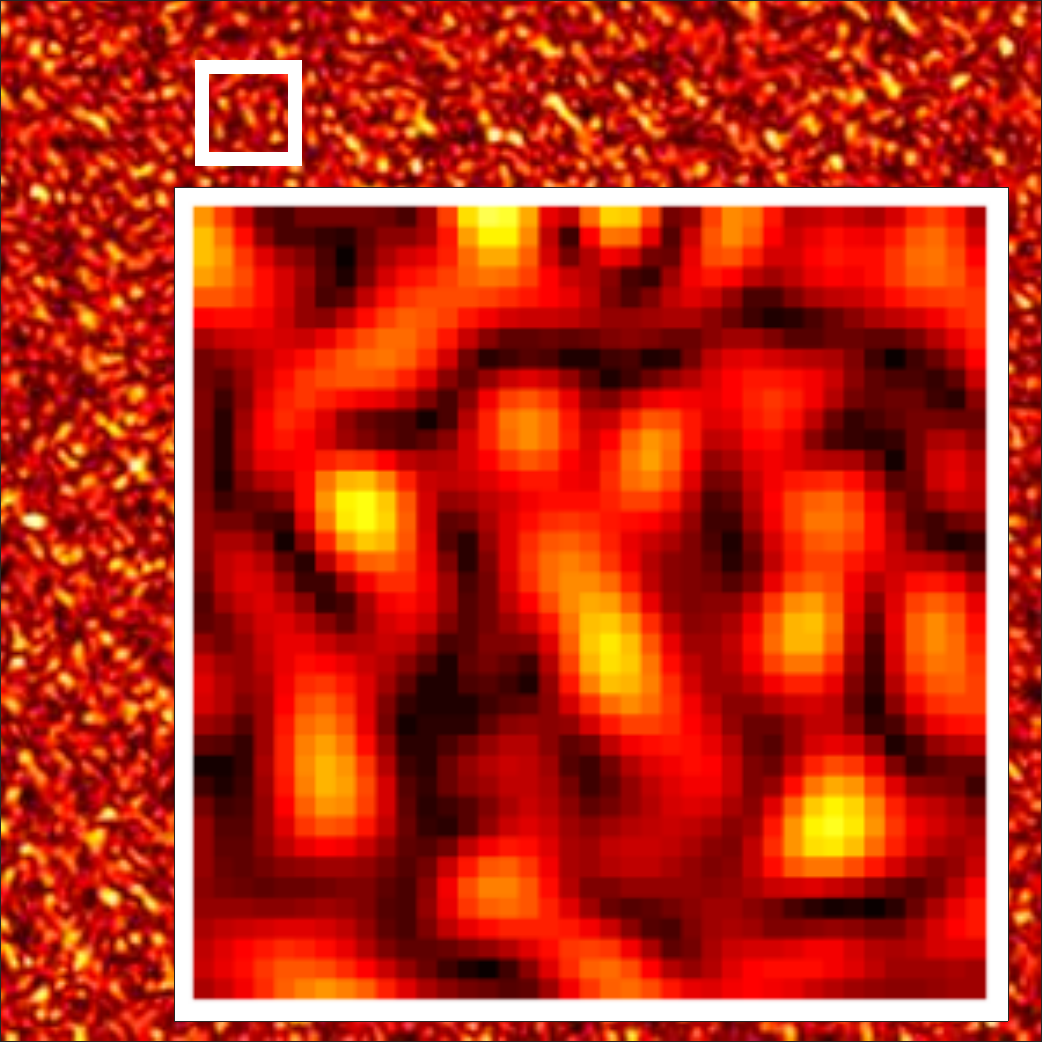}}\hfill
        \subfloat{\raisebox{+1cm}{ ...}}\hfill
        \subfloat{\label{fig:g_0_6}\includegraphics[width=0.13\textwidth]
                {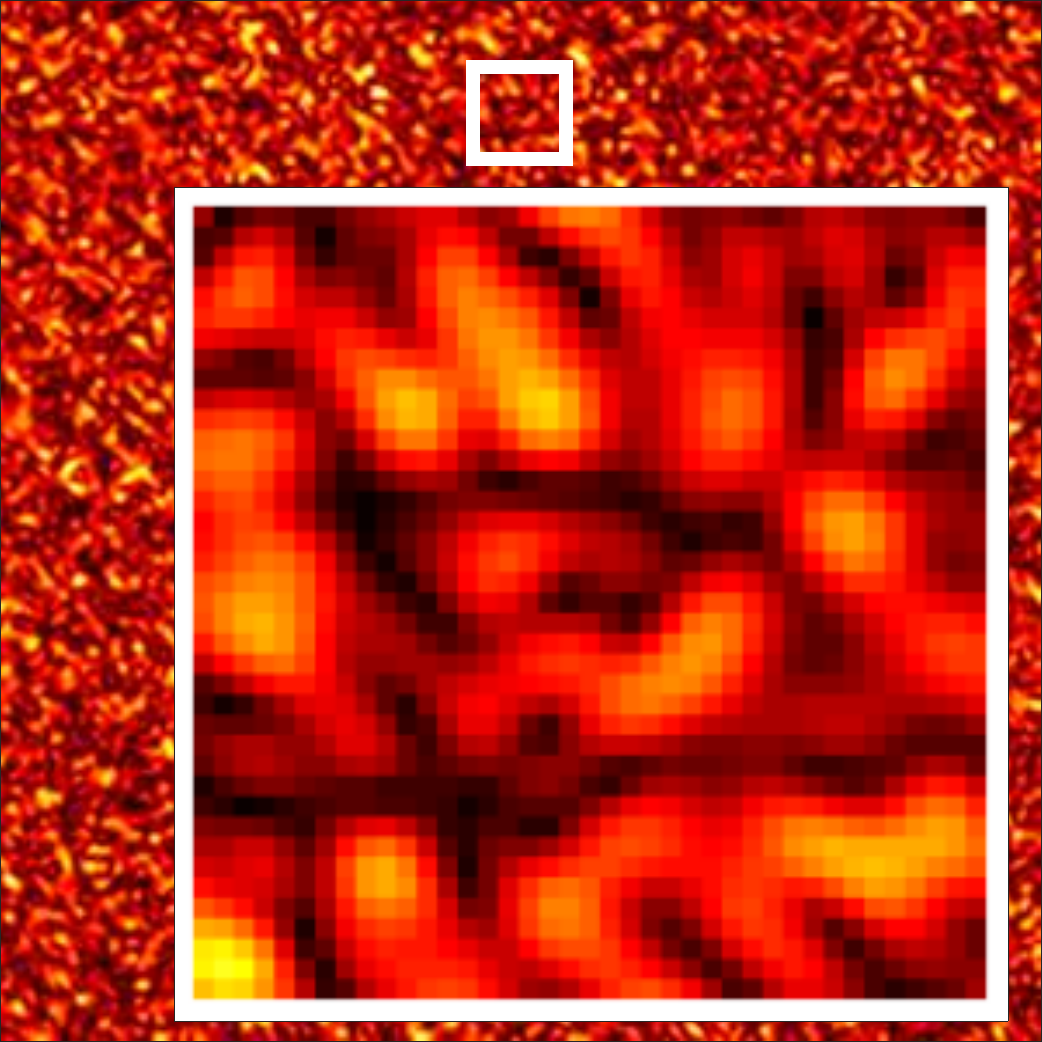}}\hfill
        \\\vspace{-0.2cm}
        \subfloat{\raisebox{1cm}{\rotatebox[origin=t]{90}{$g=0.85$}}}
        \subfloat{\label{fig:g_085}\includegraphics[width=0.17\textwidth]{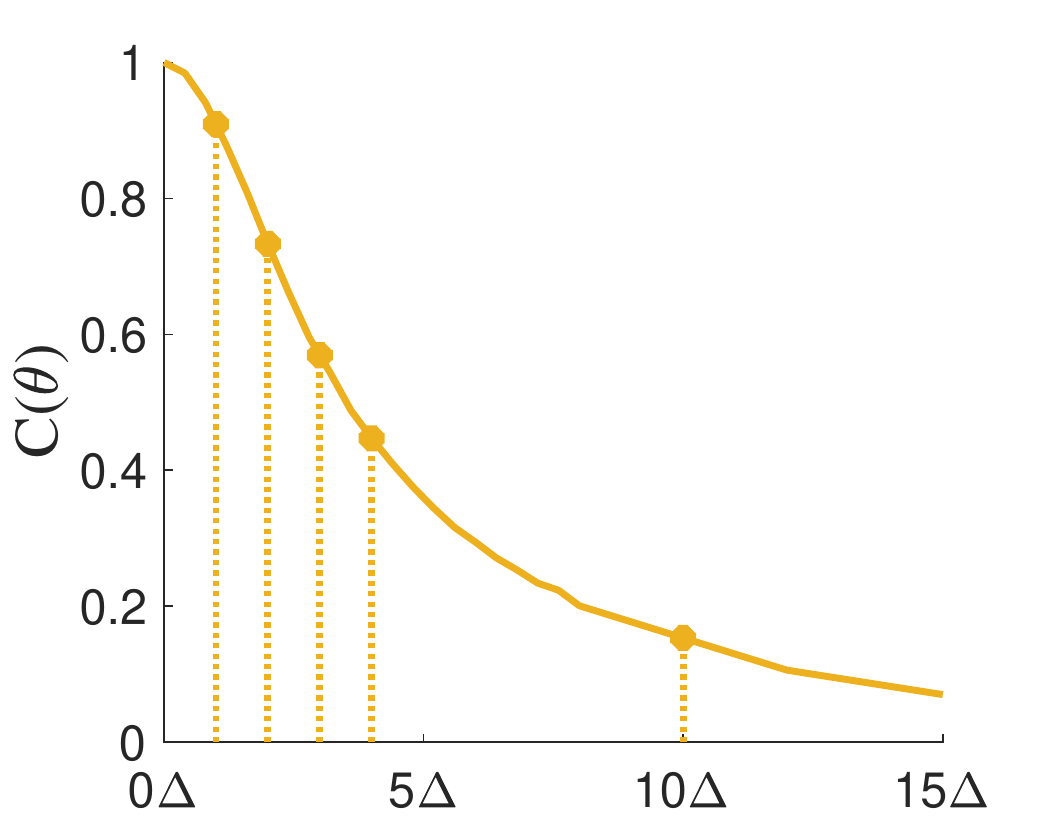}}\hfill
        \subfloat{\label{fig:g_075_1}\includegraphics[width=0.13\textwidth]
                {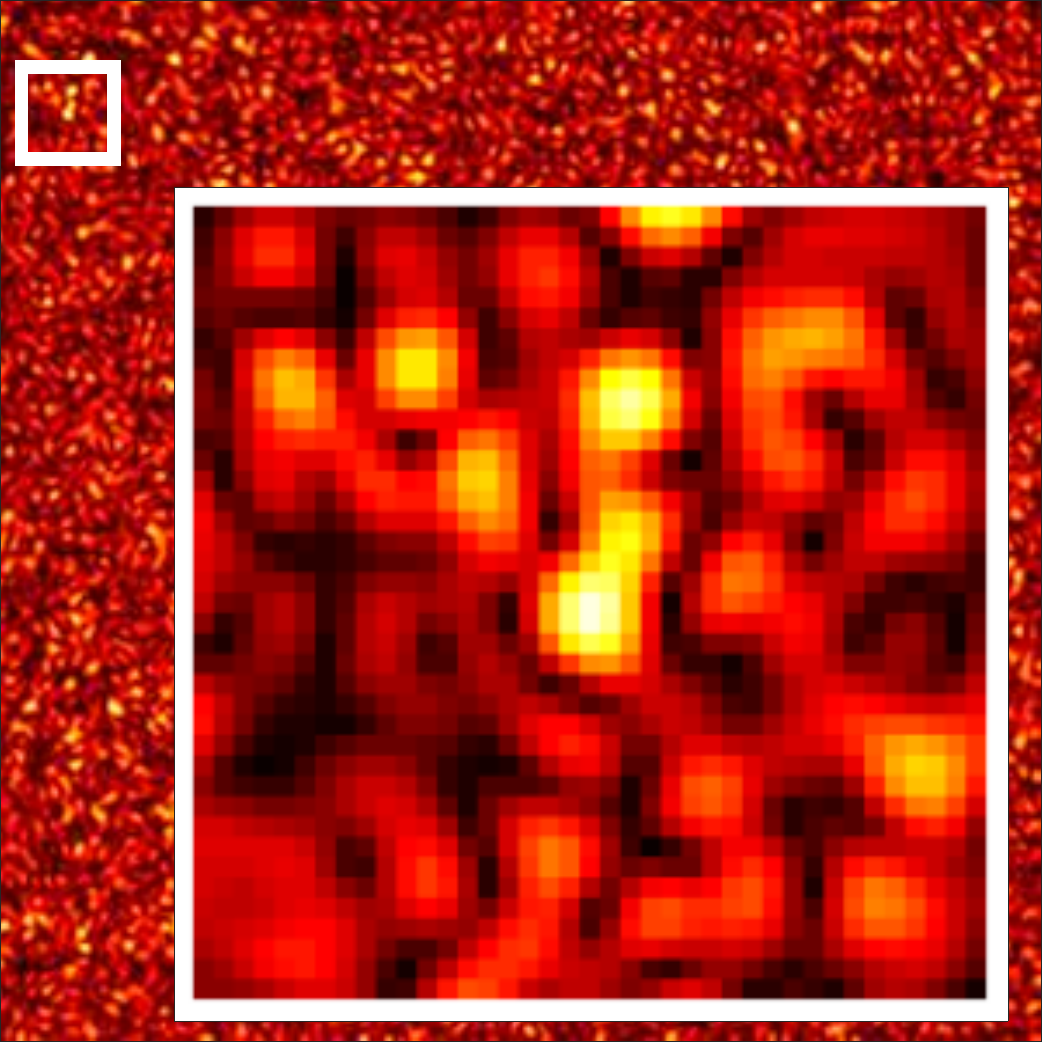}}\hfill
        \subfloat{\label{fig:g_075_2}\includegraphics[width=0.13\textwidth]
                {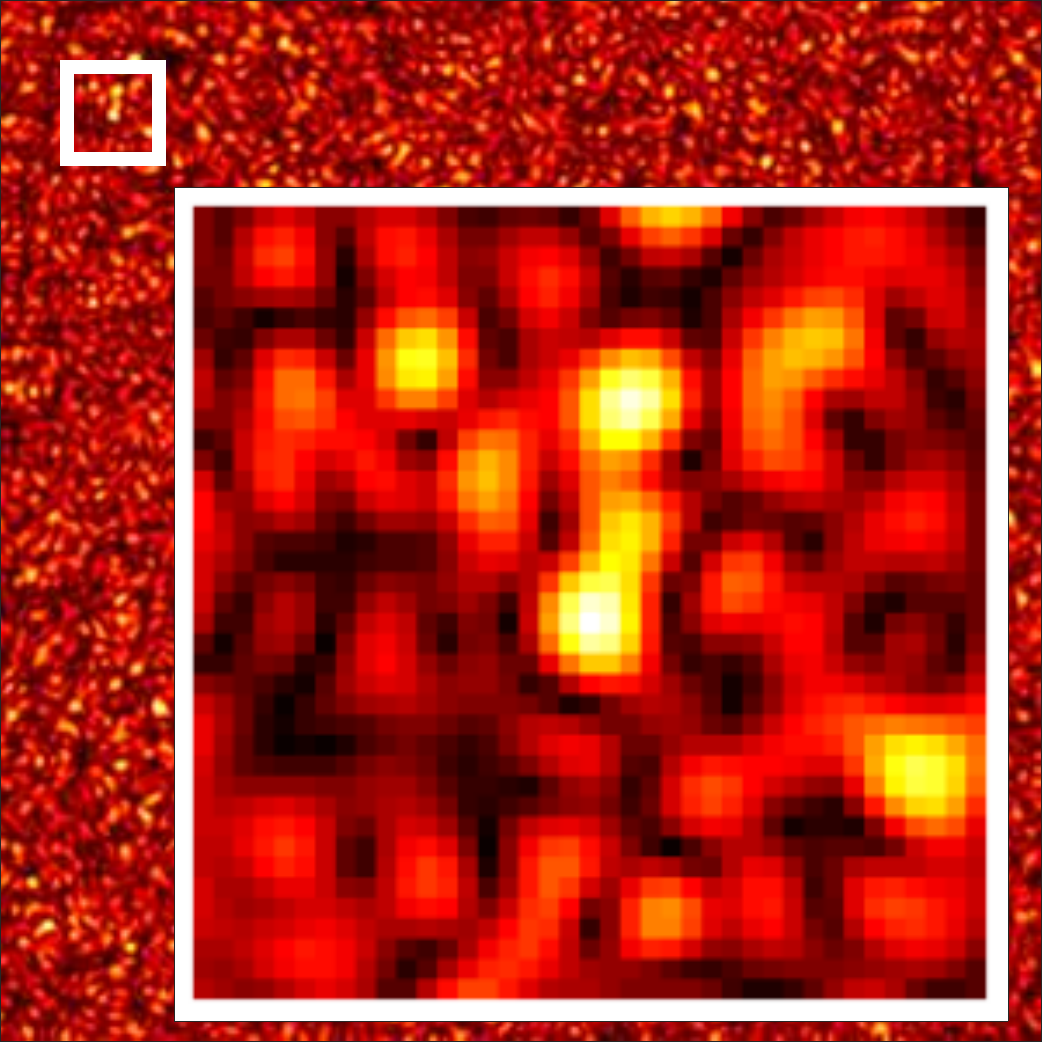}}\hfill
        \subfloat{\label{fig:g_075_3}\includegraphics[width=0.13\textwidth]
                {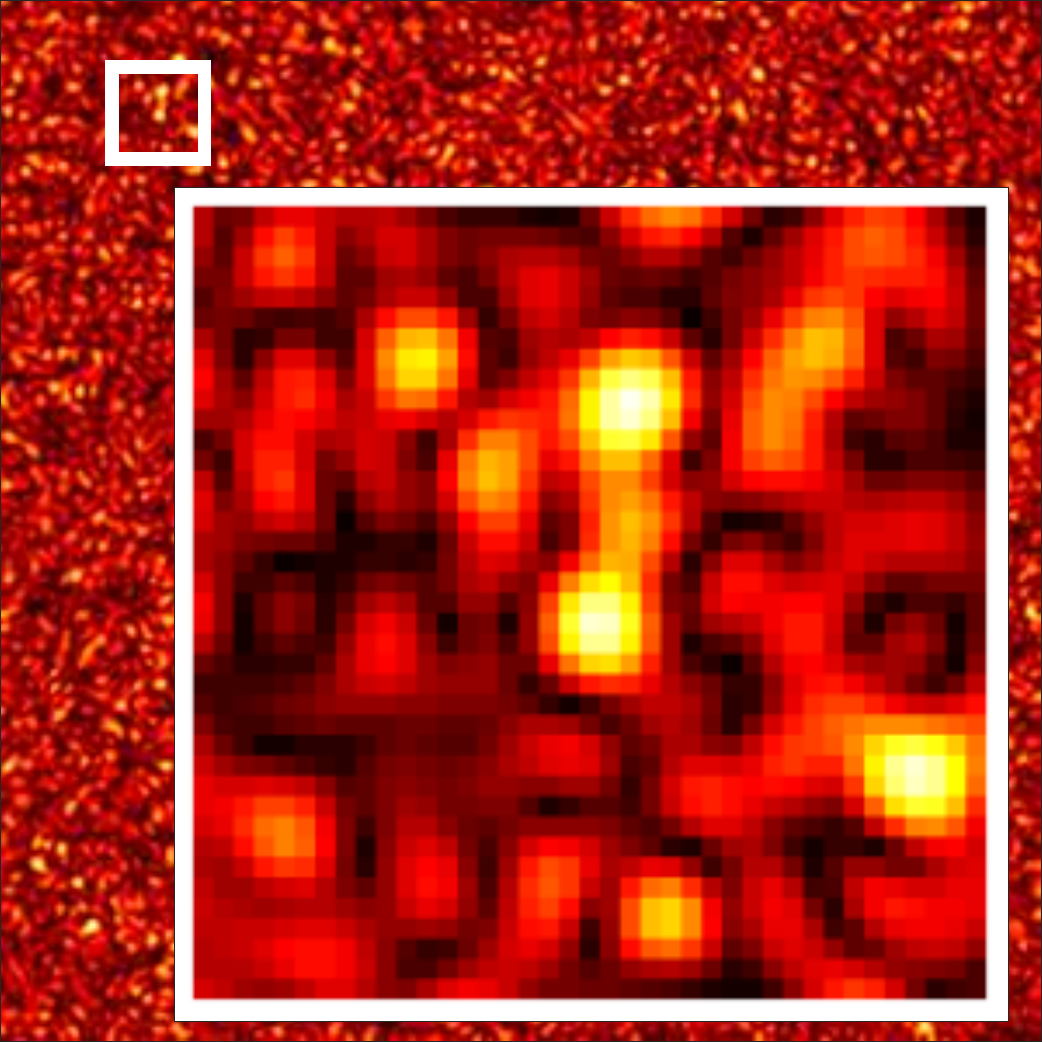}}\hfill
        \subfloat{\label{fig:g_075_4}\includegraphics[width=0.13\textwidth]
                {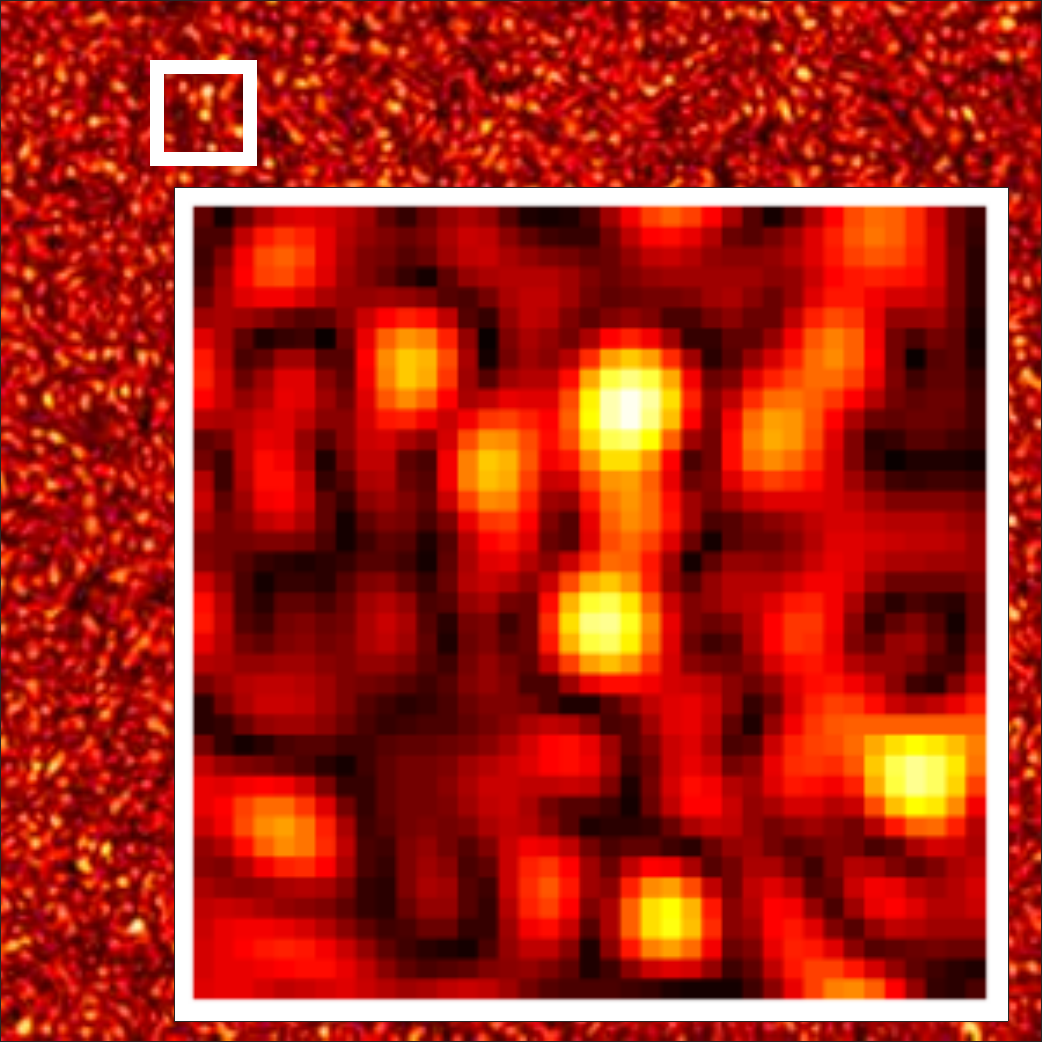}}\hfill
        \subfloat{\label{fig:g_075_5}\includegraphics[width=0.13\textwidth]
                {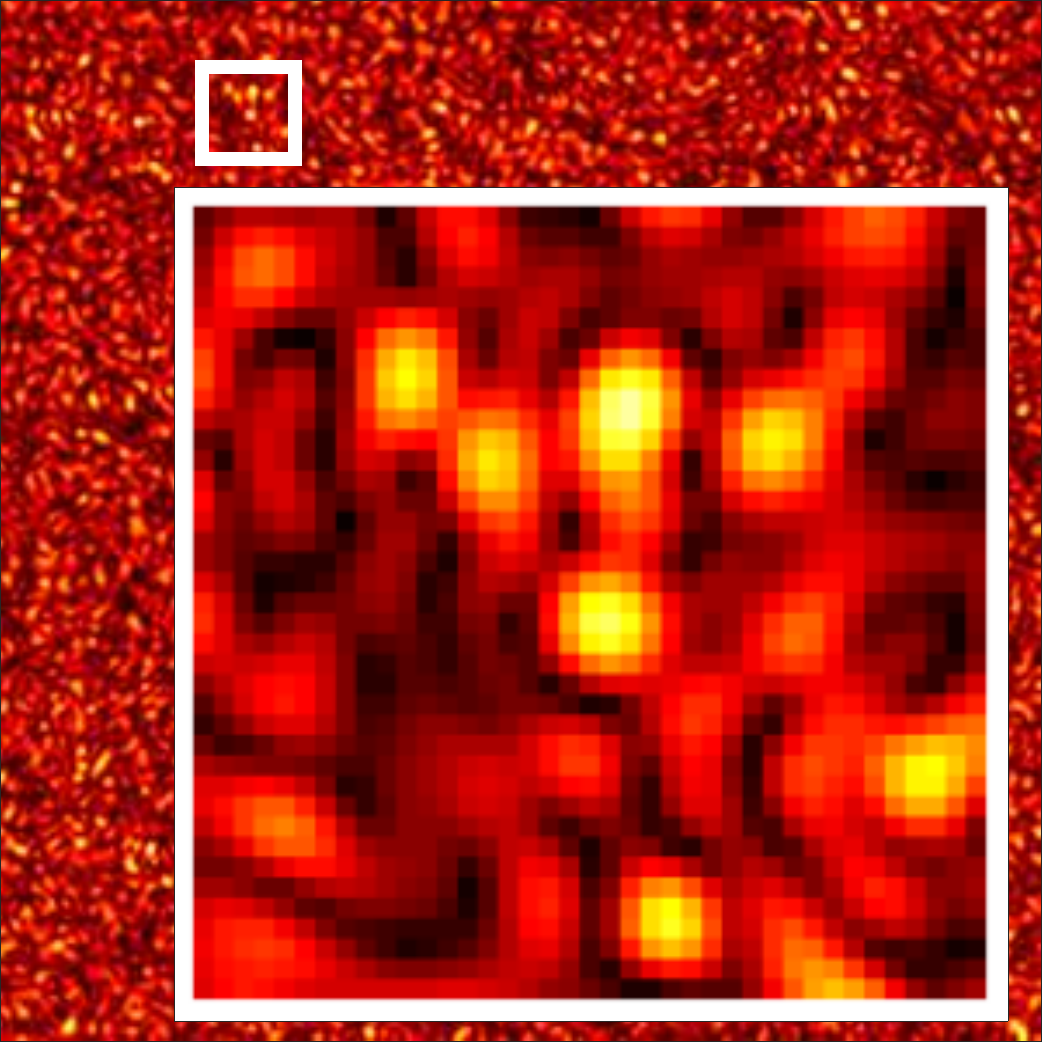}}\hfill
                \subfloat{\raisebox{+1cm}{...}}
        \subfloat{\label{fig:g_075_6}\includegraphics[width=0.13\textwidth]
                {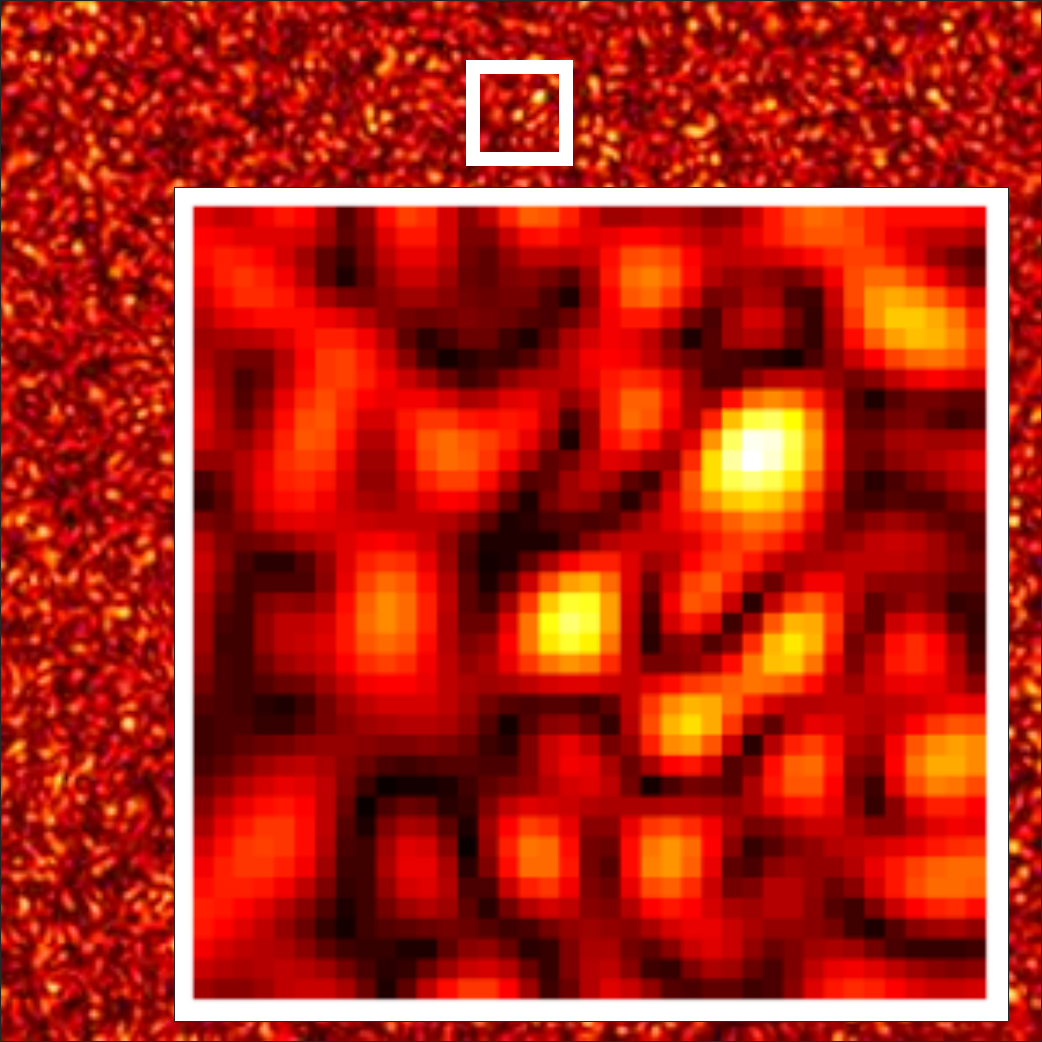}}\hfill
        \\\vspace{-0.2cm}
        \subfloat{\raisebox{1cm}{\rotatebox[origin=t]{90}{$g=0.9$}}}
        \subfloat{\label{fig:g_0}\includegraphics[width=0.17\textwidth]{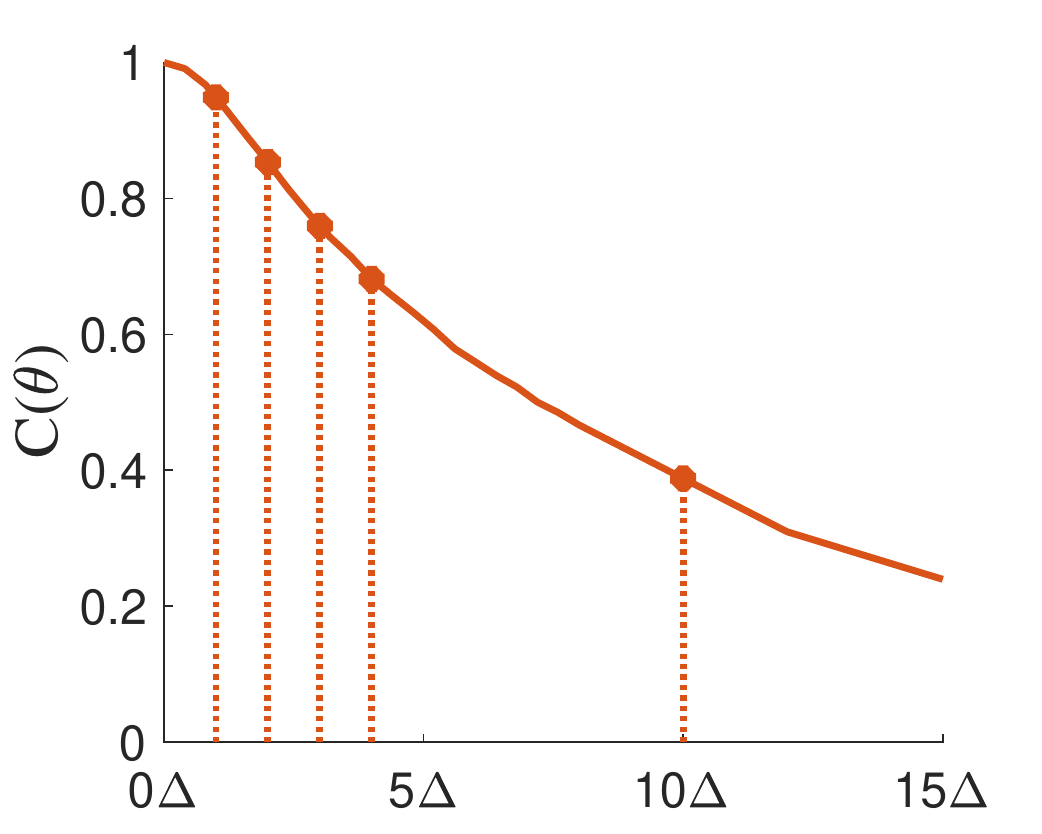}}\hfill
         \subfloat{\label{fig:g_09_1}\includegraphics[width=0.13\textwidth]
                {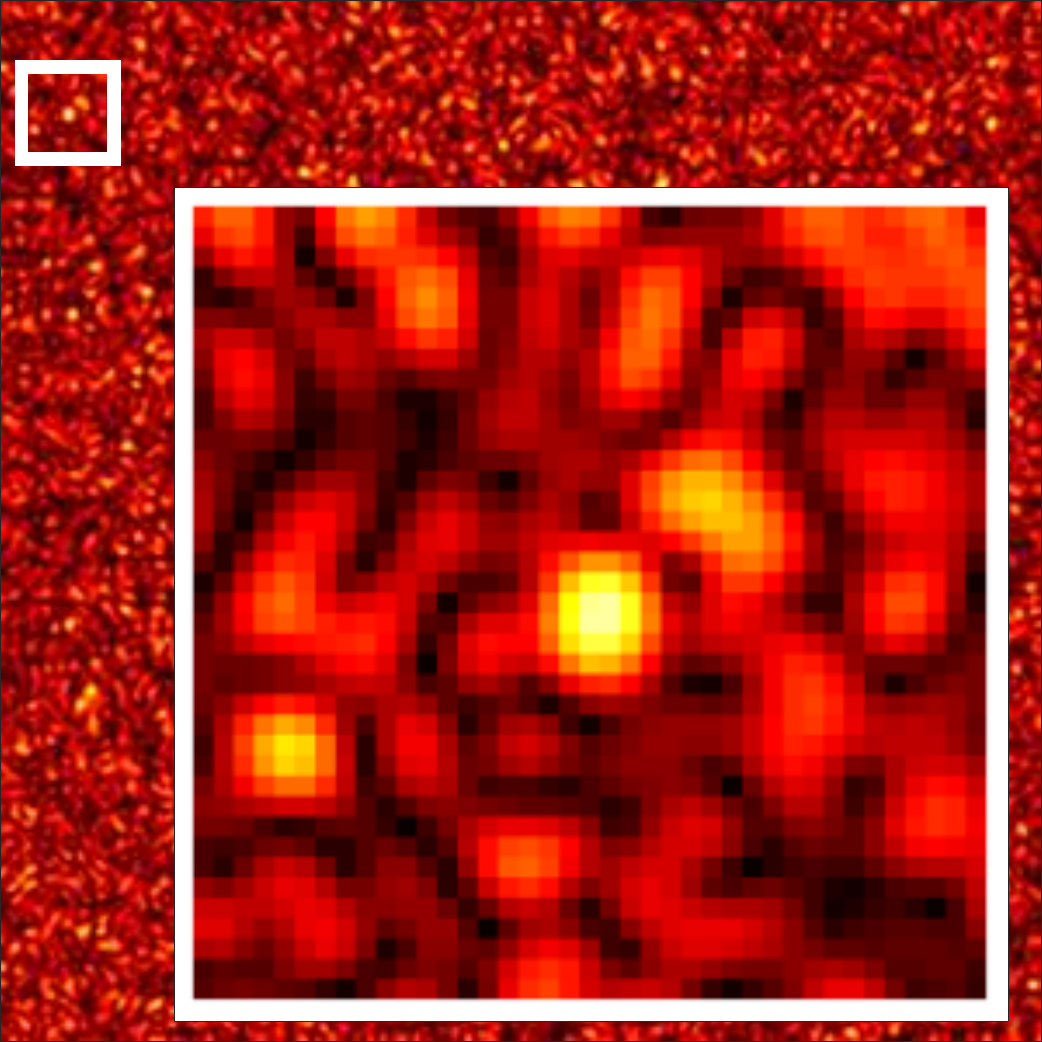}}\hfill
         \subfloat{\label{fig:g_09_2}\includegraphics[width=0.13\textwidth]
                {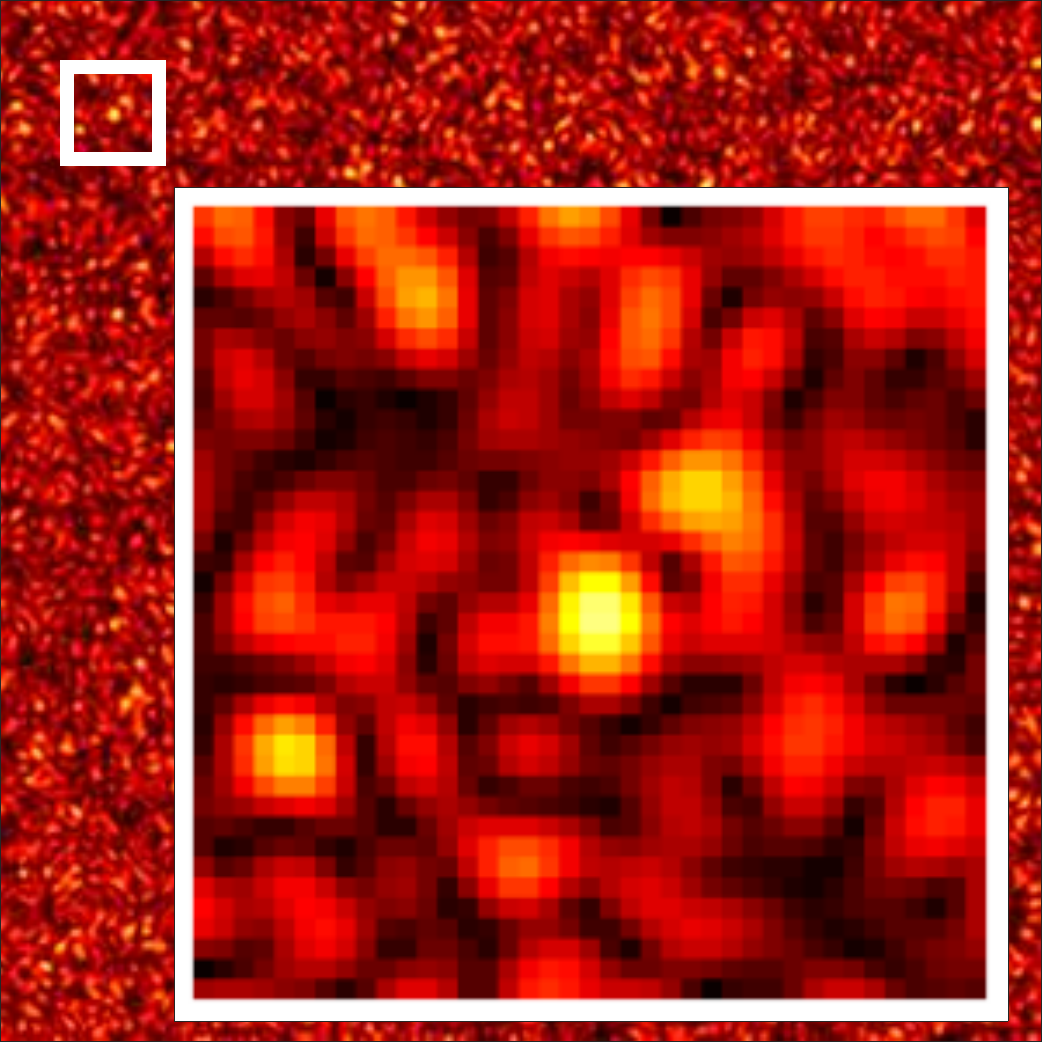}}\hfill
         \subfloat{\label{fig:g_09_3}\includegraphics[width=0.13\textwidth]
                {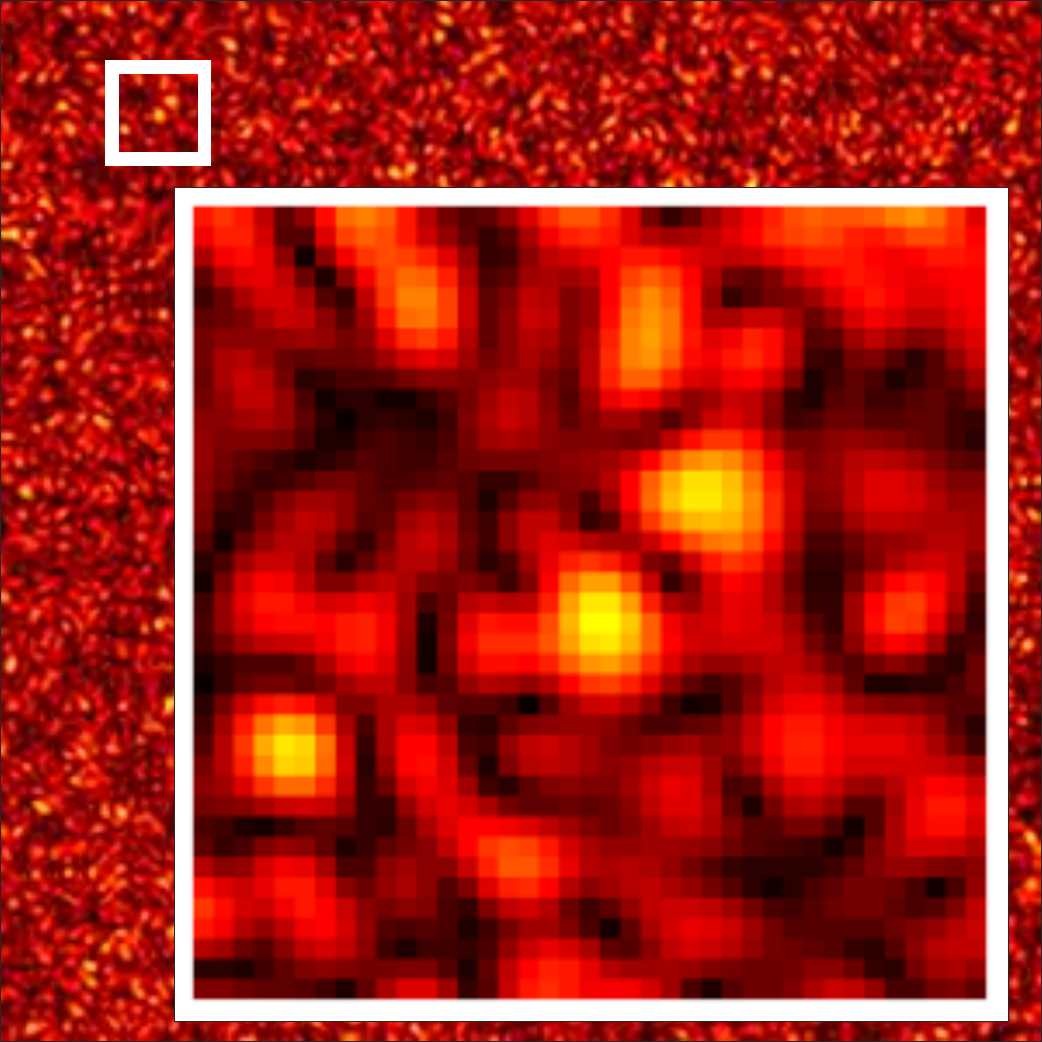}}\hfill
                \subfloat{\label{fig:g_09_4}\includegraphics[width=0.13\textwidth]
                {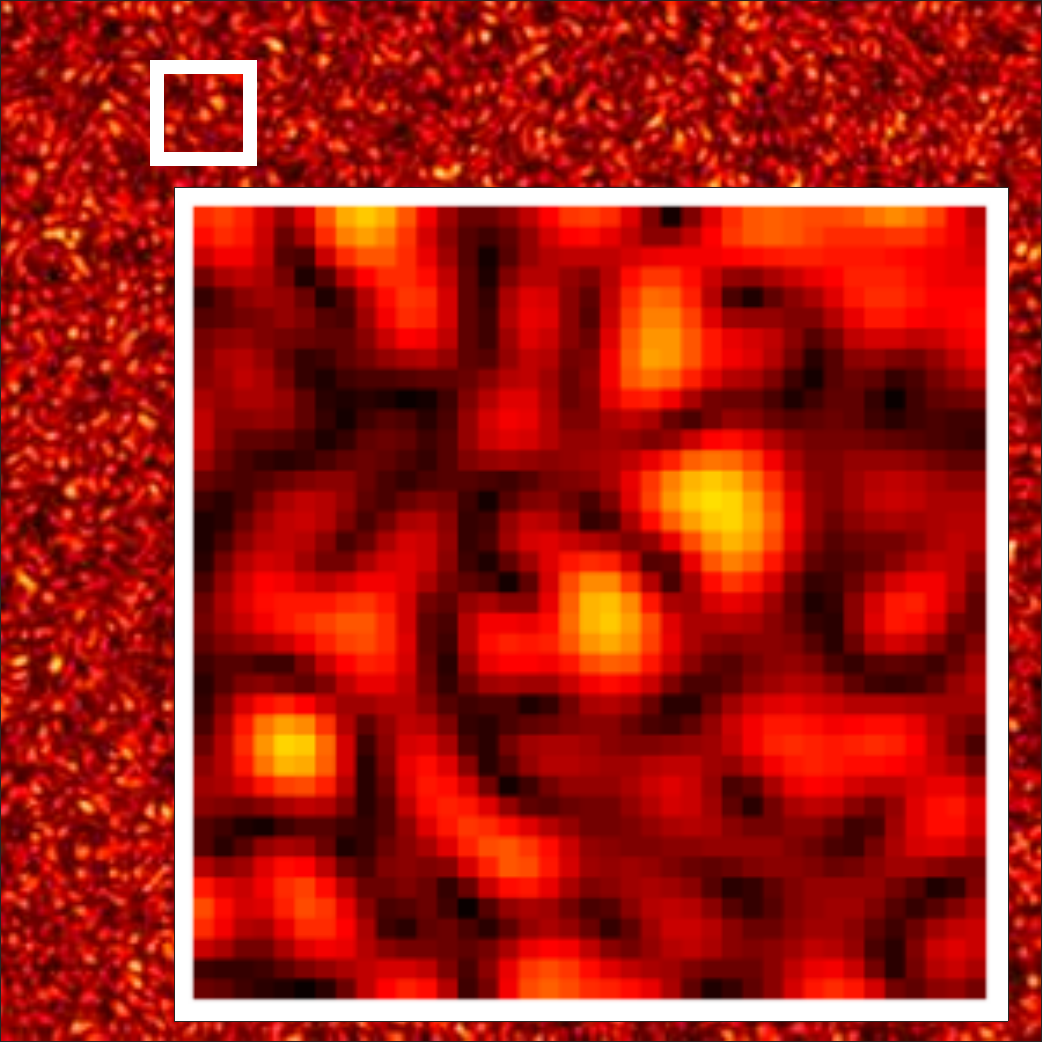}}\hfill
         \subfloat{\label{fig:g_09_5}\includegraphics[width=0.13\textwidth]
                {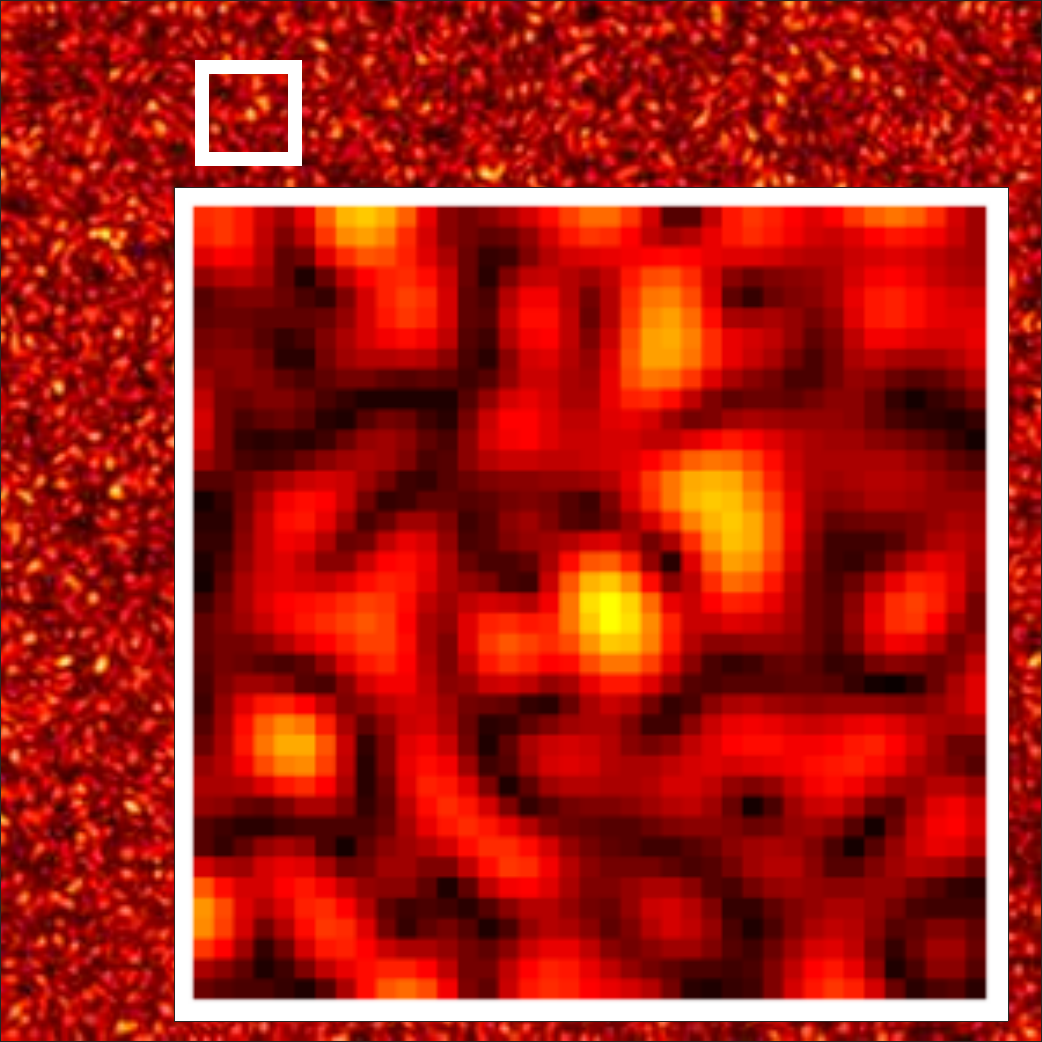}}\hfill
         \subfloat{\raisebox{+1cm}{...}}\hfill
         \subfloat{\label{fig:g_09_6}\includegraphics[width=0.13\textwidth]
                {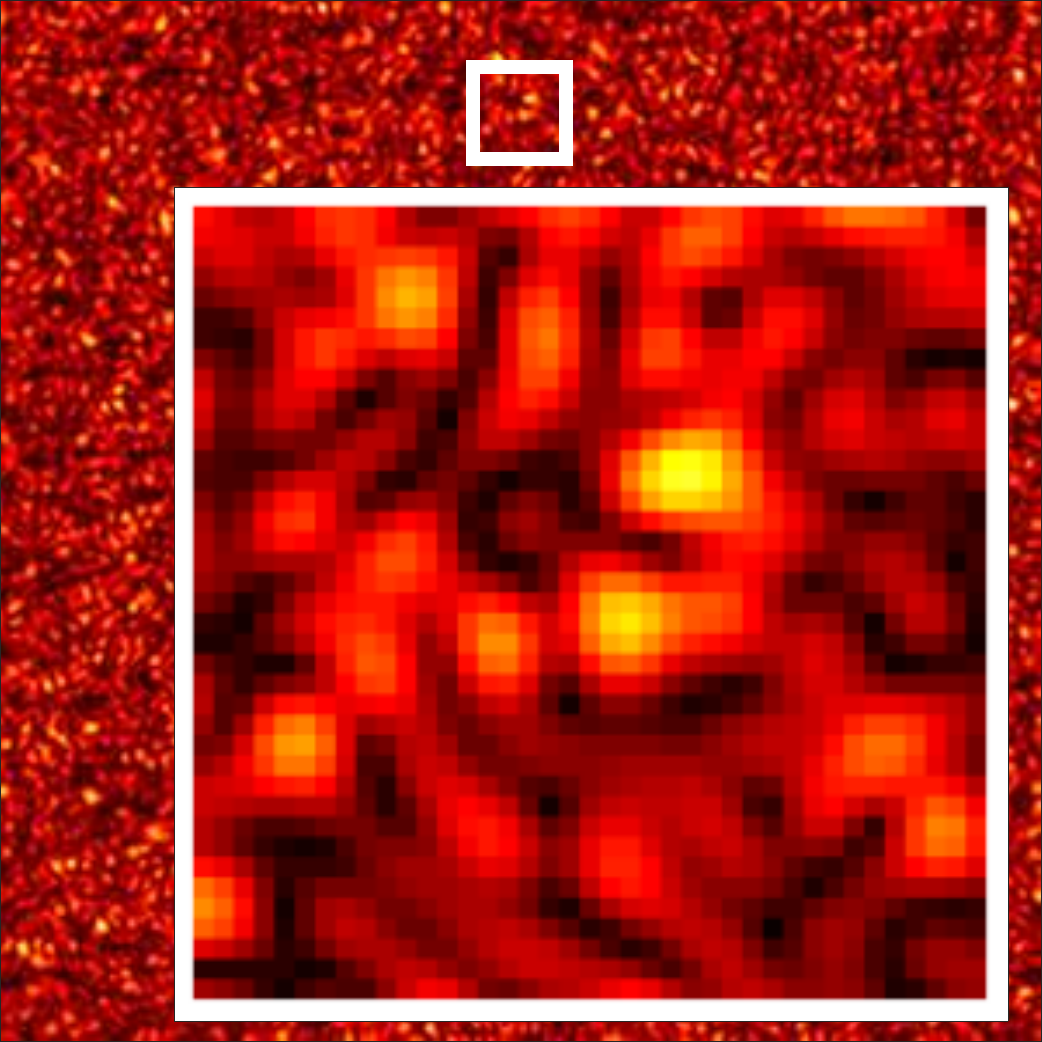}}\hfill
         \\\vspace{-0.2cm}        
                 \subfloat{\raisebox{1cm}{\rotatebox[origin=t]{90}{Indp. MC}}}\hfill
             \subfloat{\makebox[0.17\textwidth][c]{        }}\hfill
         \subfloat{\label{fig:g_09_d1}\includegraphics[width=0.13\textwidth]
                {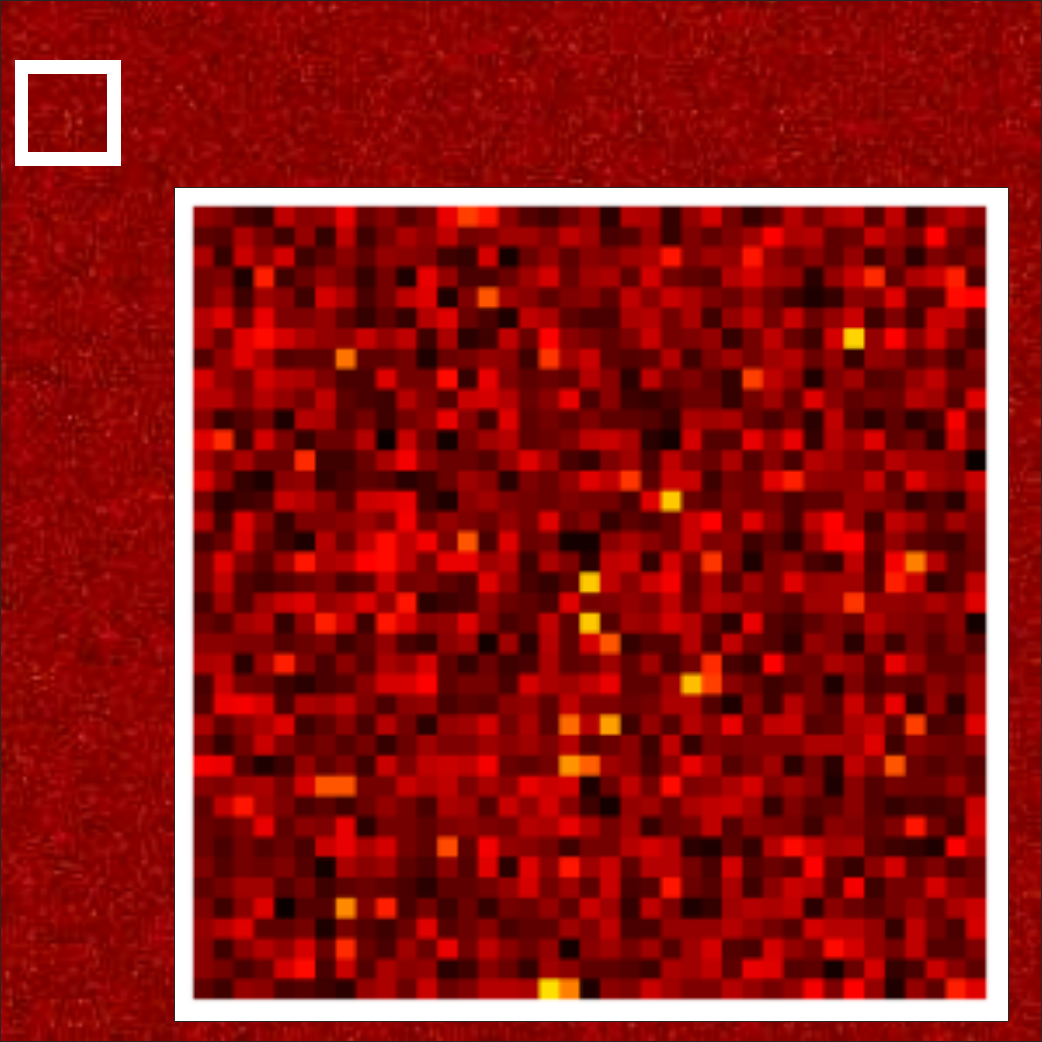}}\hfill
         \subfloat{\label{fig:g_09_d2}\includegraphics[width=0.13\textwidth]
                {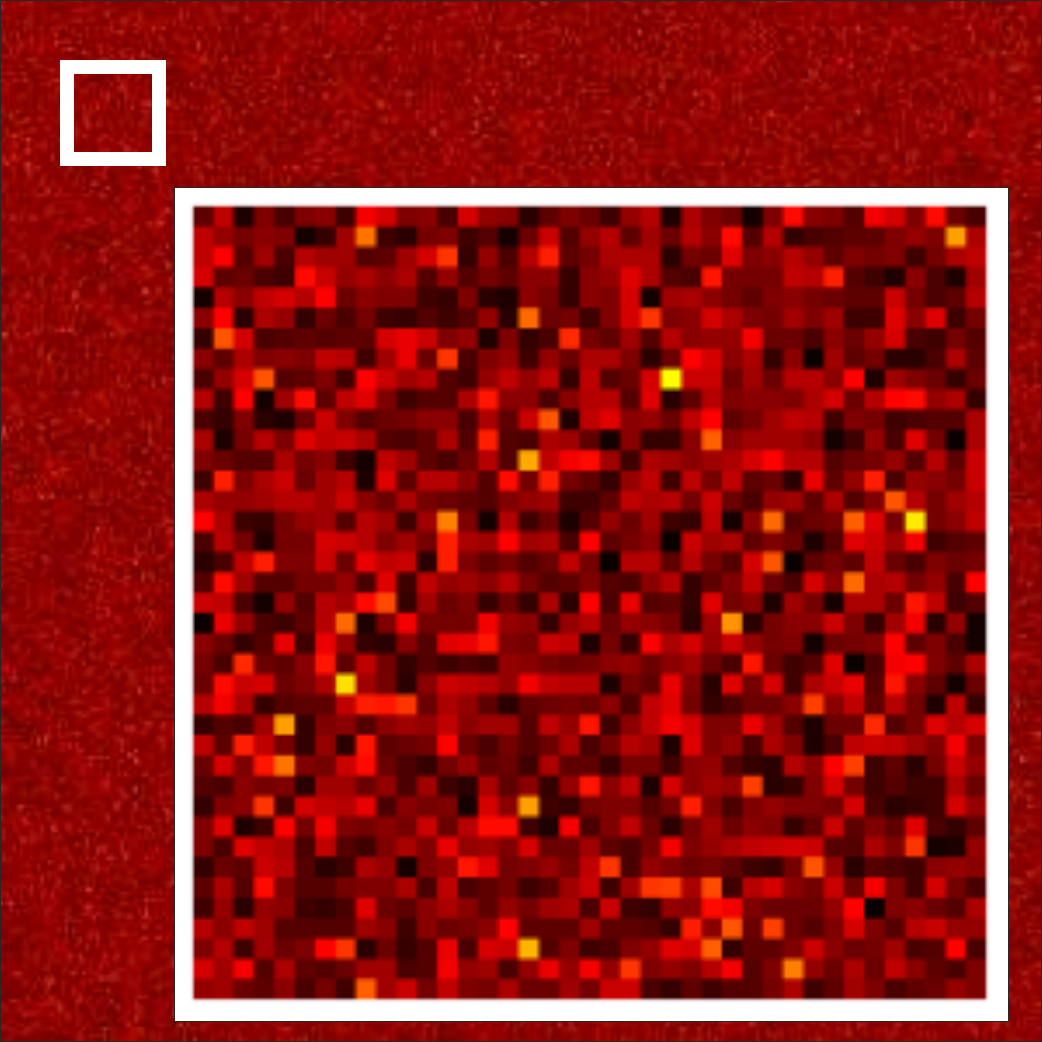}}\hfill
         \subfloat{\label{fig:g_09_d3}\includegraphics[width=0.13\textwidth]
                {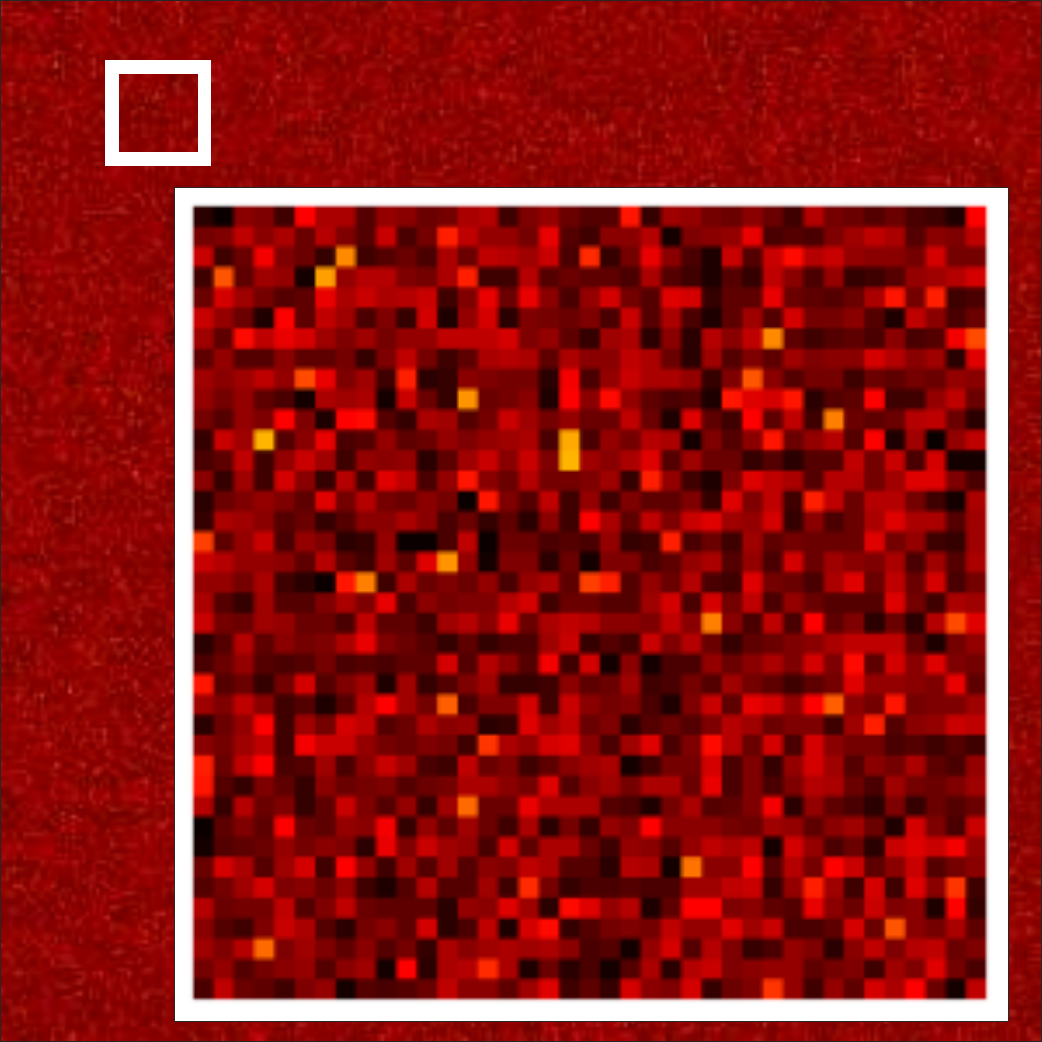}}\hfill
         \subfloat{\label{fig:g_09_d4}\includegraphics[width=0.13\textwidth]
                {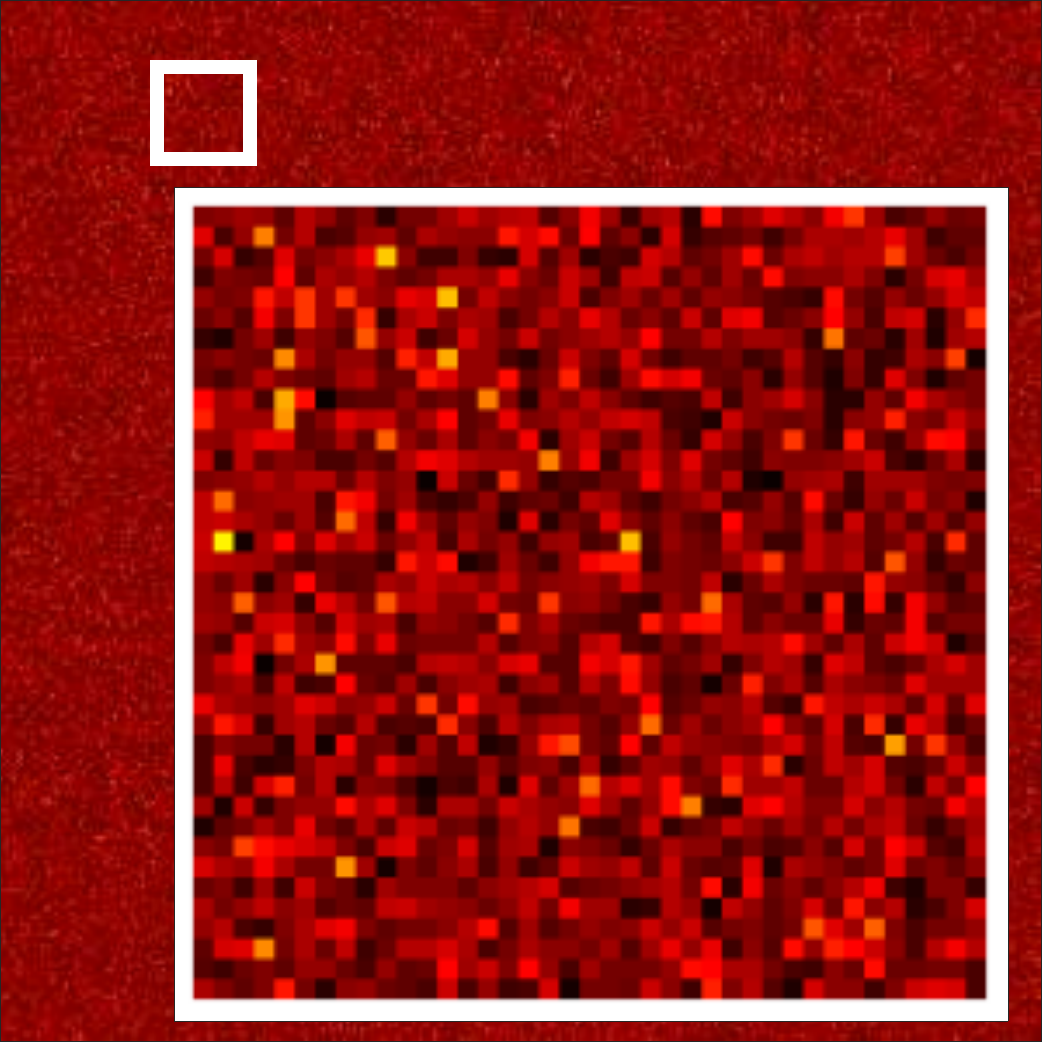}}\hfill
         \subfloat{\label{fig:g_09_d5}\includegraphics[width=0.13\textwidth]
                {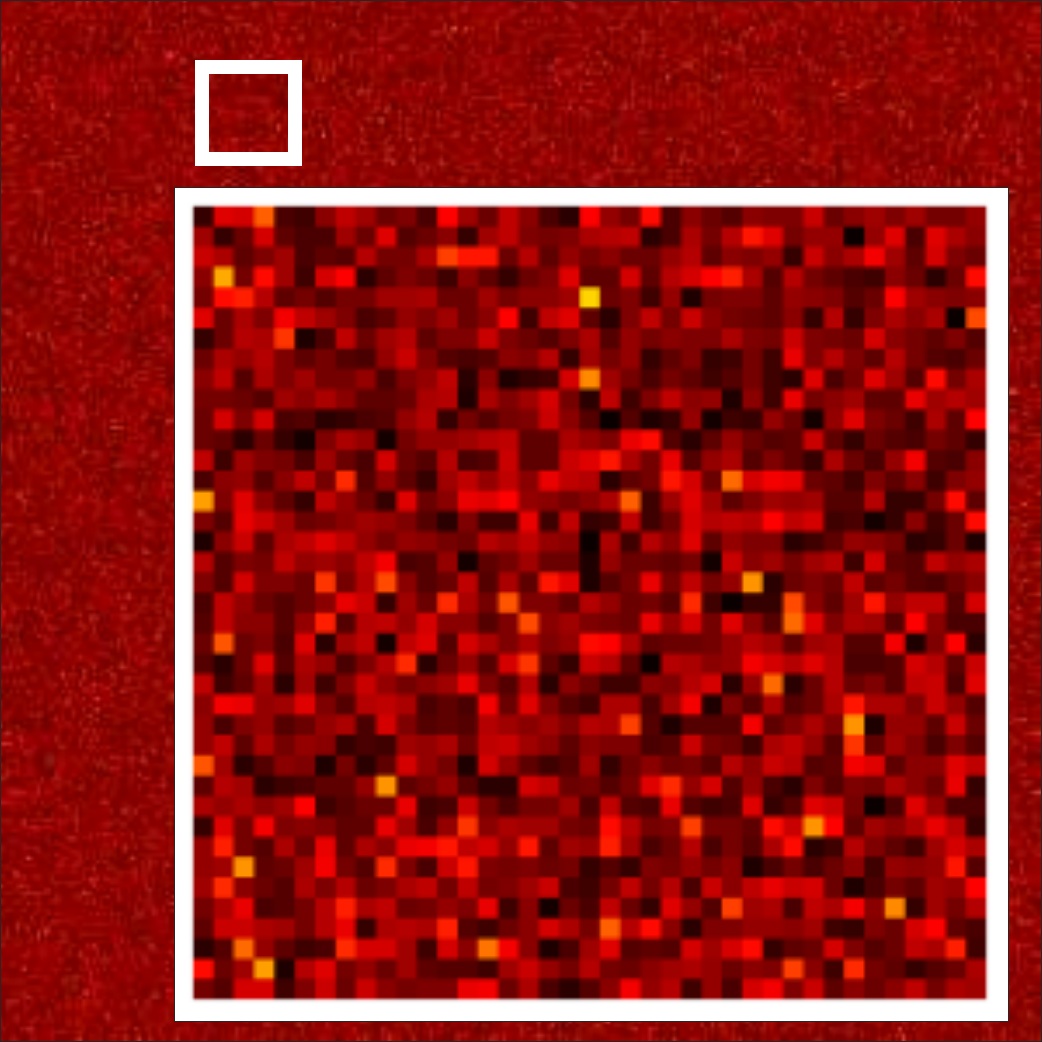}}\hfill
         \subfloat{\raisebox{+1cm}{\large ...}}\hfill
         \subfloat{\label{fig:g_09_d6}\includegraphics[width=0.13\textwidth]
                {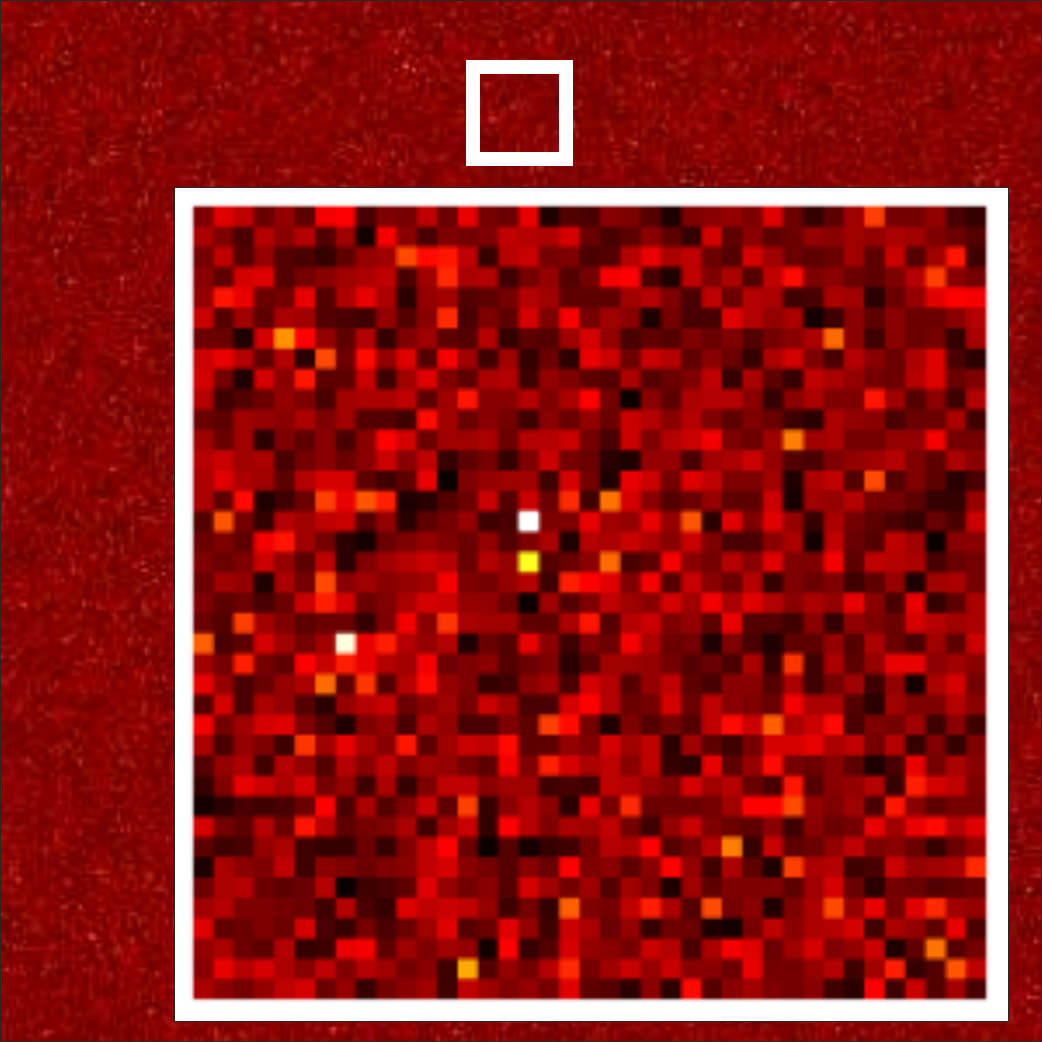}}\hfill
  \caption{Sampling speckle images as a function of illumination shift angle with ($\Delta=0.0025\degree$) and anisotropy parameter $g$. The memory effect can be clearly observed  in the first three rows. Notice that for $g=0$ the correlation is lost already at a shift of $3\Delta$, while for $g=0.9$, the speckle survives even after an illumination shift of $10\Delta$. In the lower row, we show sample images obtained by a naive MC approach as in \cite{Xu:04}, where each view is sampled independently.}  
   \label{fig:memory-effect}        
\end{figure*}

Our Monte Carlo algorithm can compute the expected correlations directly, without the need for approximations or lab measurements.  Correlations are computed as a function of simple material parameters such as thickness, $\extCoff$, $\sctCoff$ and phase function. In \figref{fig:memory-effect-plots}, we show numerical calculations of the expected correlation as a function of angle $\theta$. In \figref{fig:mem-MFP} we use a forward scattering configuration, a sample of thickness $L=1mm$ at illumination wavelength $\lambda=500nm$, $\atnCoff=0$, isotropic phase function $g=0$, and varying mean free path ($MFP$) values. For a high optical depth, the correlation computed by our algorithm agrees with the theoretical formula of \equref{eq:corr-diffusion}, and as the optical depth decreases (mean free path increases) the range of the memory effect increases. In \figref{fig:mem-albedo} we simulate a backscattering configuration for fixed $g=0$,  $MFP=0.1mm$, $\extCoff=1/MFP$, and varying albedo $\sctCoff/\extCoff$. As expected, the memory effect is stronger as absorption increases (albedo decreases).

 In \figref{fig:mem-g} we keep the thickness and mean free path fixed to $L=1mm$, $MFP=0.1mm$  and vary the anisotropy parameter $g$ of the phase function. As previous empirical observations report~\cite{Schott:15}, increasing $g$ increases the \emph{transport} mean free path, and thus the memory effect range expands. Finally, in \figref{fig:sim-g} we investigate another common analytical approximation, the similarity theory~\cite{wyman1989similarity,zhao2014high}, stating that scattering coefficient and phase functions satisfying $\sctCoff^*(1-g^*)=\sctCoff(1-g)$ should produce indistinguishable measurements. Using $L=1mm$, $\atnCoff=0$, we set at $g=0$ a mean free path of $250\mu m$ (leading to $O.D.=4$),  and then vary $g$ and $\sctCoff=\extCoff=1/MFP$ while maintaining the similarity relation. The graphs in \figref{fig:sim-g} show that similarity theory is  reasonably accurate, though low $g$ values have a somewhat heavier tail.  \comment{In \figref{fig:mem-chickenBreast} we considered the specific case of chicken breast tissue that was imaged  by~\cite{Schott:15}. Using the mean free path and $g=0.96$ values they report we simulated the exact same material thickness reported in this work and predicted the range of the memory effect, comparing it with the reported empirical measurements.\Anat{Need to add their values}}

Our MC algorithm computes correlations of complex fields while \equref{eq:corr-diffusion} evaluates intensity correlations. Field covariances can be easily converted to intensity covariances using $2|C^{\inp_1,\inp_2}_{\snsp_1,\snsp_2}|^2$.

\begin{figure}
\begin{center}
%\begin{tabular}{l@{~~}c@{~}c@{~~~~~}c@{~~~~~}c@{~}}
\begin{tabular}{cccc}
&(a) Source &(b) $g=0.9$&(c) $g=0.85$ \\
\raisebox{0.85cm}{\rotatebox[origin=t]{90}{Illuminators}}\hfill&
\includegraphics[width=0.13\textwidth]{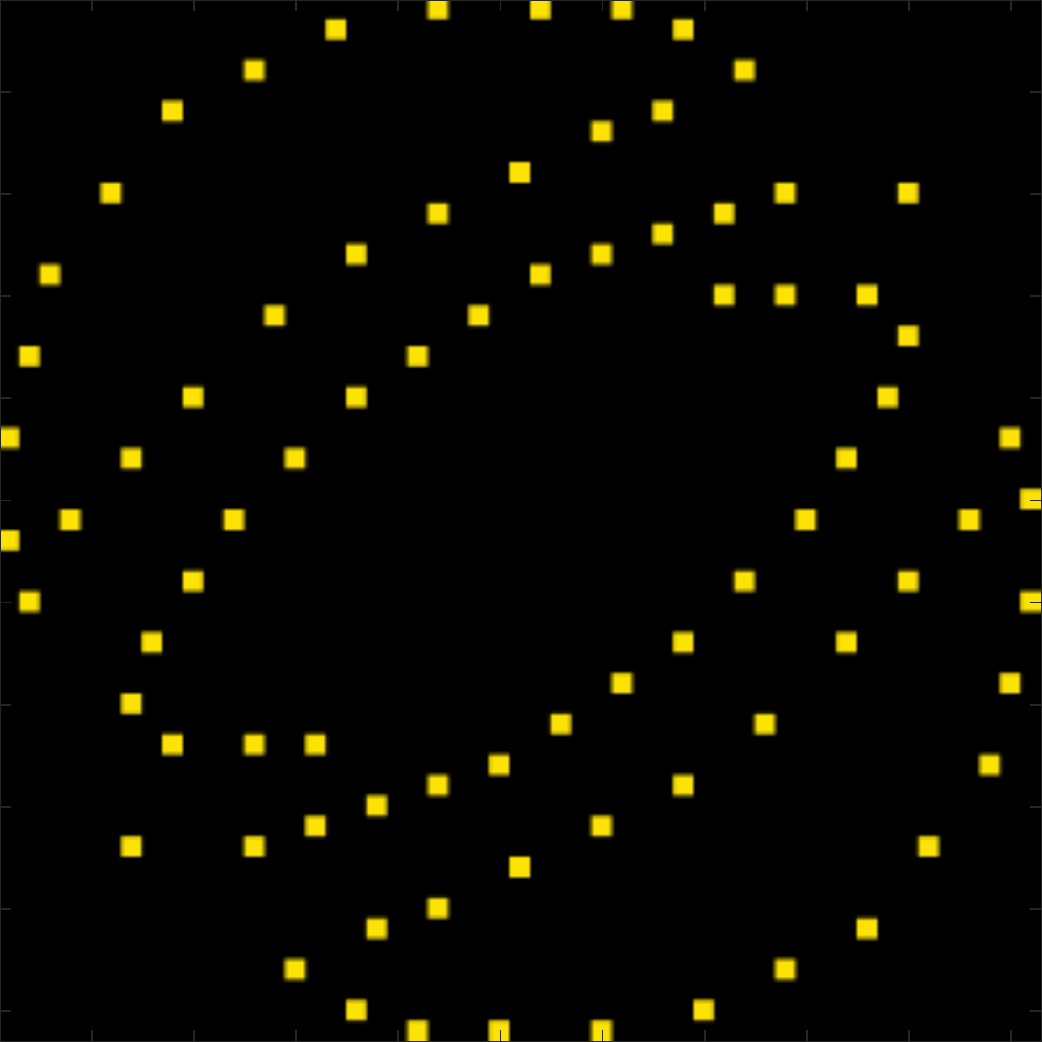}&
     \includegraphics[width= 0.13\textwidth]{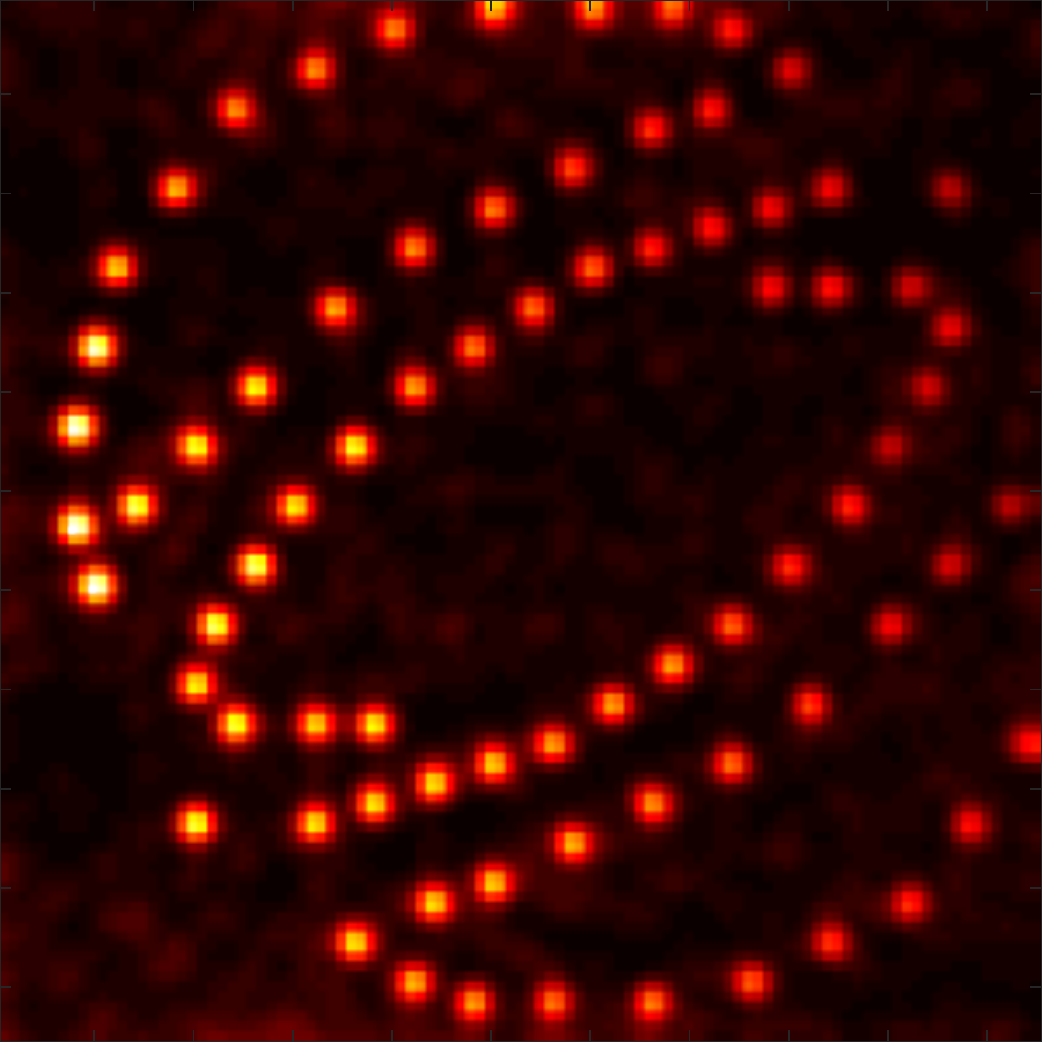}&
     \includegraphics[width= 0.13\textwidth]{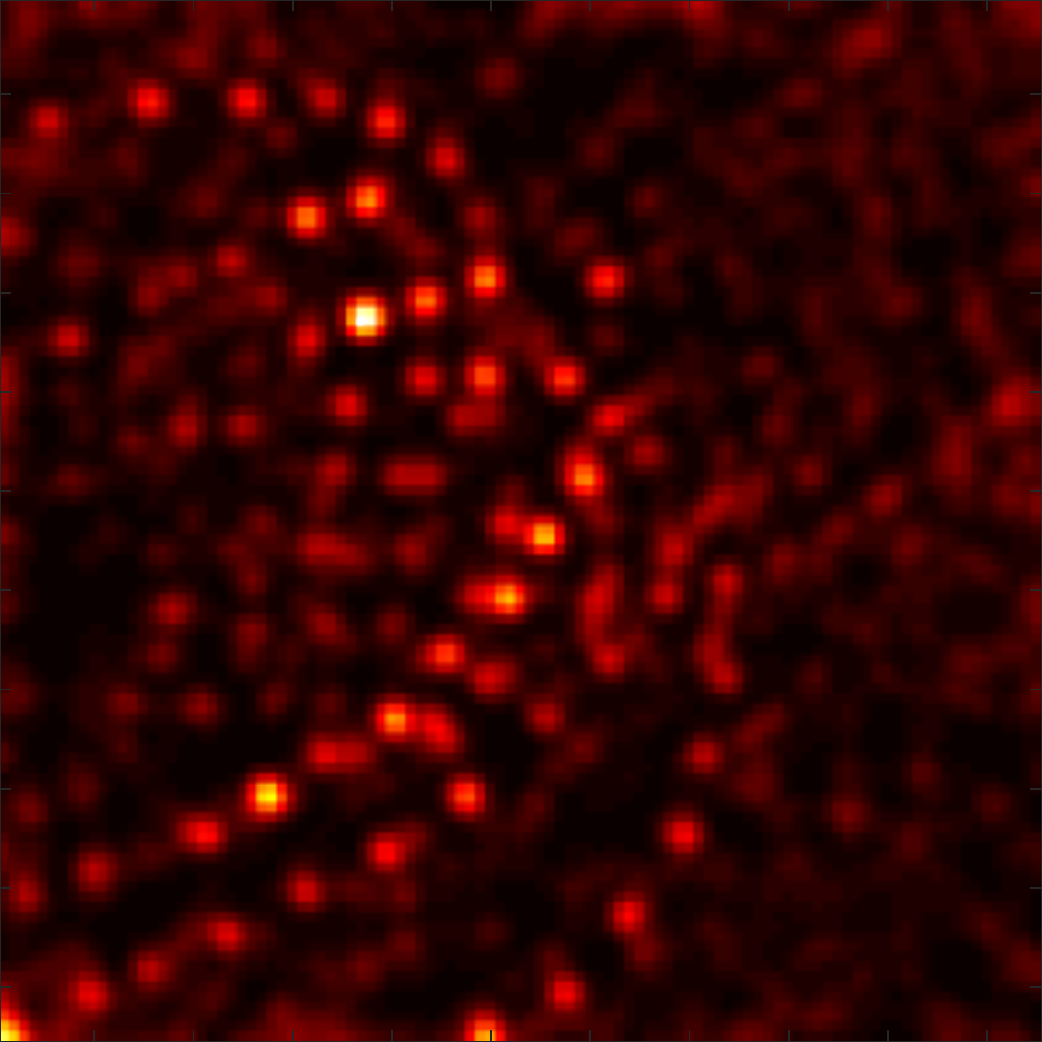}
\\
\raisebox{0.85cm}{\rotatebox[origin=t]{90}{Auto-corr}}\hfill&
                \includegraphics[width= 0.13\textwidth]{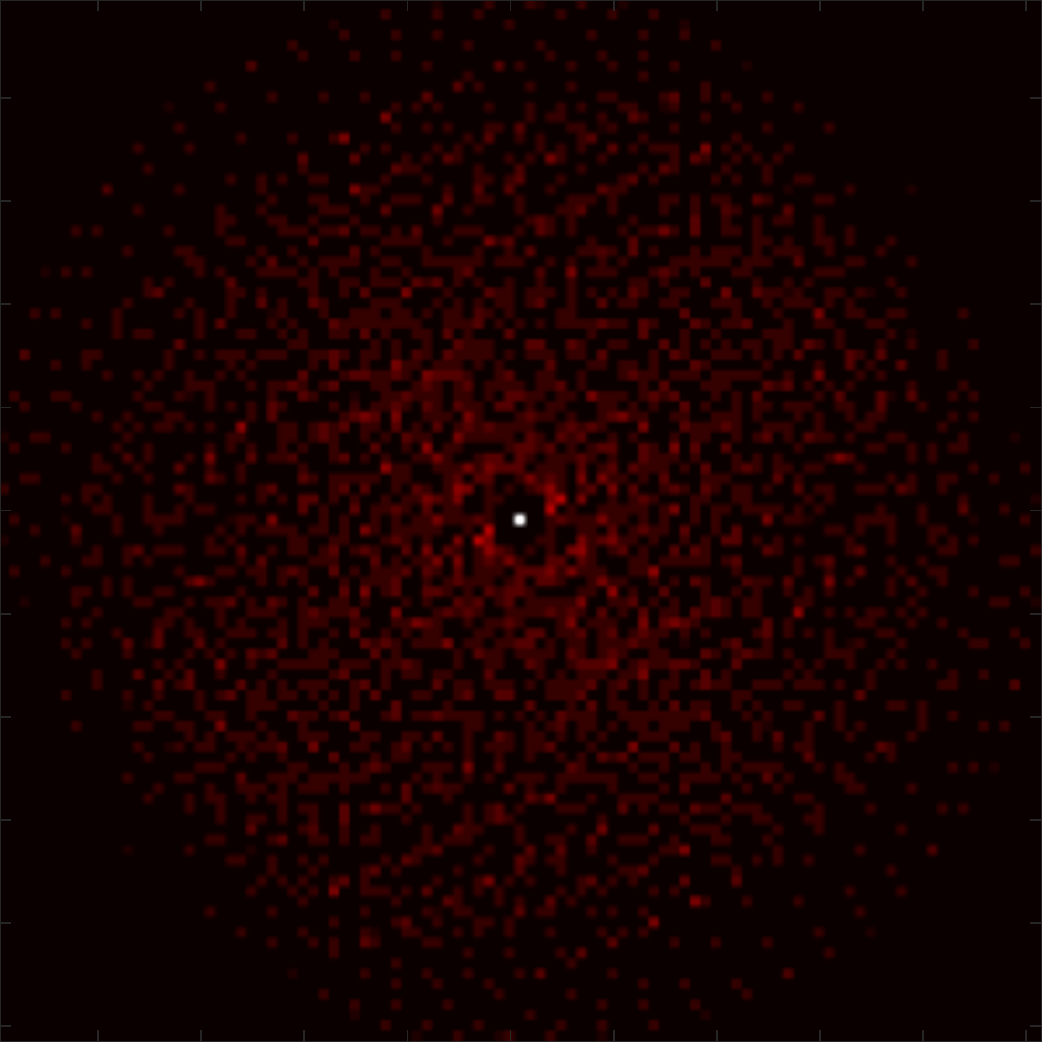} &
                \includegraphics[width= 0.13\textwidth]{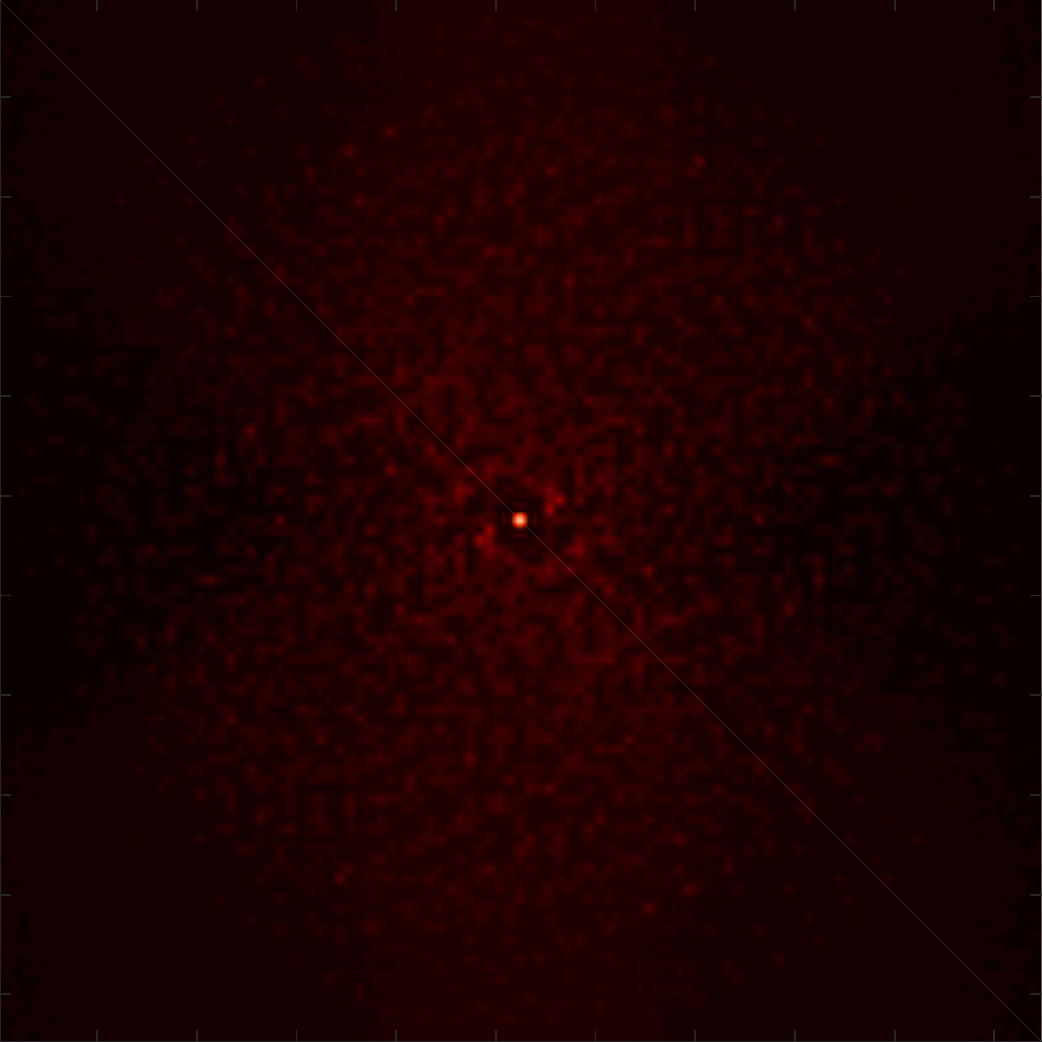} &
                \includegraphics[width= 0.13\textwidth]{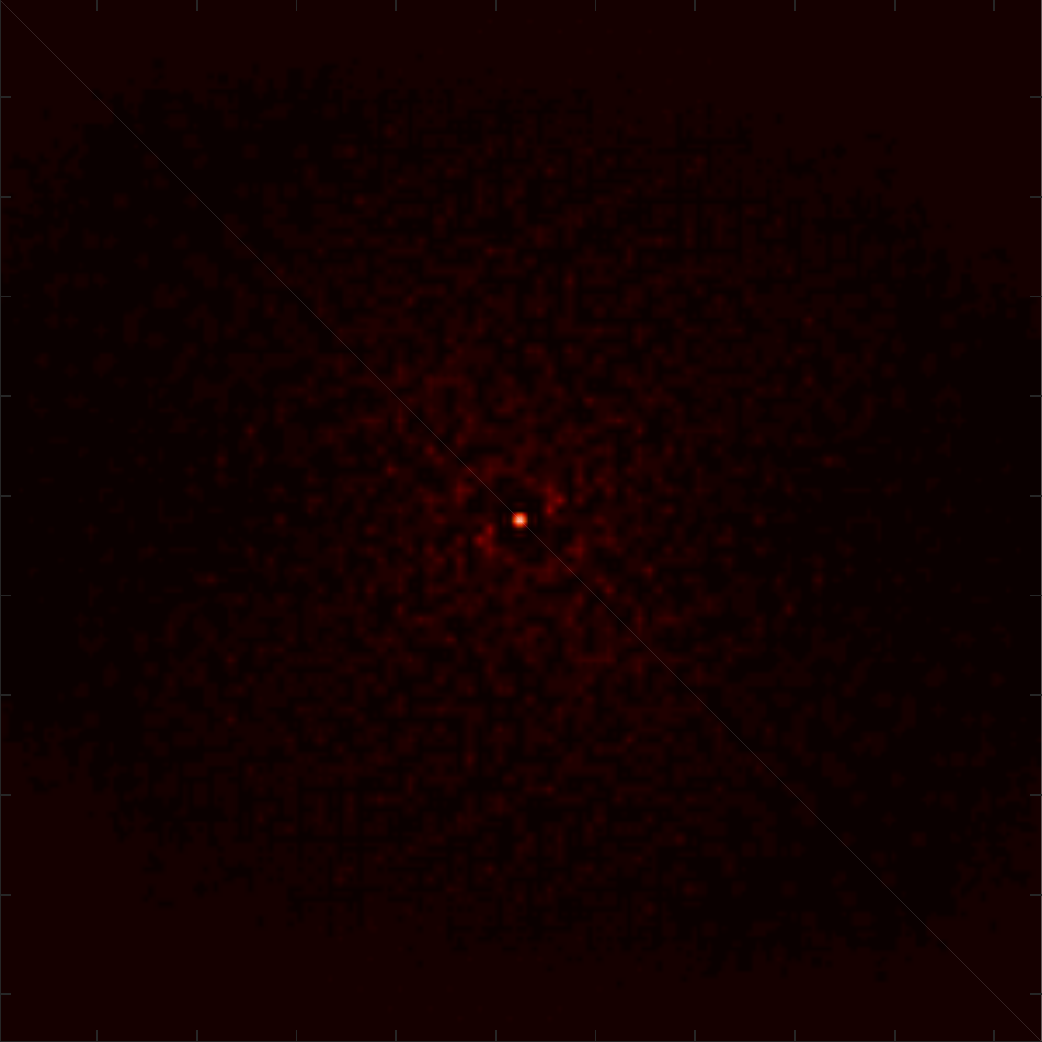} 
\end{tabular}
\caption{Replication of a seeing through scattering layer application of~\cite{Katz2014}. A set of illuminators with the arrangement at the top of (a) generates a semi-random speckle image, yet the auto-correlation of the speckle image is similar to the auto-correlation of the original illuminators and hence the illuminators can be recovered from the speckle image using phase retrieval algorithms.  In (b,c) we show the auto-correlation and the corresponding reconstruction for different material parameters simulated with our speckle renderer.  The success of the algorithm depends on the validity of the memory effect in this angular range for each type of material. }  
\label{fig:mc-phaseRetrival}
\end{center}        
\end{figure}

\subsubsection{Sampling speckle images}\label{sec:resultsrendering}
 In \figref{fig:memory-effect} we use the sampling algorithm of \secref{sec:renderingimage} to sample speckle images as seen from a sensor at infinity over a viewing range of $0.1\degree$, when the illumination direction is shifting (from $0\degree$ to $0.025\degree$, at $\Delta=0.0025\degree$ intervals). As can be seen, these images reproduce the memory effect: For small changes in illumination angle the speckles appear as shifted versions of each other. When the illumination angle difference increases, the correlation decays. We show this simulation for a few anisotropy parameters $g$  and as illustrated in \figref{fig:mem-g},  when the anisotropy increases  the memory effect can be observed over a wider angular range. In the last row of \figref{fig:memory-effect} we show simulations using the electric field Monte Carlo approach~\cite{Xu:04,Sawicki:08}, which updates different viewing and illumination directions independently. We observe that no joint speckle statistics are produced and the resulting images appear as independent noise.

\subsubsection{Example application}

To demonstrate an application of speckle correlations, we reproduced the algorithm of Katz et al.~\shortcite{Katz2014}. This algorithm attempts to recover a set of incoherent light sources located behind a scattering layer. Remarkably, due to the memory effect, the auto-correlation of the  speckle image should be  {\em equivalent} to the auto-correlation of light sources positions. Thus, given the seemingly random speckle image, one can recover the position of light sources behind it by applying an iterative phase retrieval algorithm~\cite{Fienup:82}.
In \figref{fig:mc-phaseRetrival} we show the result of this reconstruction applied on speckle images rendered with Algorithm \ref{alg:MCfield}. We use two of the materials in \figref{fig:mem-g}, with anisotropy parameters $g=0.85,g=0.9$. The hidden source is placed over an angular range of $0.0125\degree=5\Delta$. As evaluated in \figpref{fig:mem-g}{fig:memory-effect}, for this angular range the correlation for $g=0.9$ is high,  but for $g=0.85$ we are already outside the memory effect range. Indeed the $g=0.9$ speckle auto-correlation at the bottom of \figref{fig:mc-phaseRetrival}b is almost equivalent to the source auto-correlation (\figref{fig:mc-phaseRetrival}a[bottom]), while the auto-correlation of speckles rendered with $g=0.85$ is darker due to the lower correlation (\figref{fig:mc-phaseRetrival}c[bottom]). As a result, phase retrieval with the $g=0.9$ speckles provides a good reconstruction of the original illuminator arrangement (\figref{fig:mc-phaseRetrival}b[top]). For $g=0.85$ (\figref{fig:mc-phaseRetrival}c[top])  only a cropped version of the illuminator pattern is recovered (along with background noise), as within this subset of illuminators  the  angular differences are smaller and the correlation is stronger. Experiments of this kind can be used to evaluate the applicability of the imaging technique of Katz et al.~\shortcite{Katz2014} under different conditions, and to select optimal values for various parameters involved in an optical implementation of the technique.
  
  \subsubsection{Understanding the memory effect bounds}\label{sec:understandingME}
Before concluding this section, it is worth mentioning that the MC path integral formulation can provide an intuitive way to understand the memory effect range derived in \equref{eq:corr-diffusion}. Consider two pairs of illumination and viewing directions $\ind_1,\ind_2,\outd_1,\outd_2$ s.t. $\ind_1-\ind_2=\outd_1-\outd_2=\omgv$, and consider a path starting at $\ptd_1$ and ending at $\ptd_B$. Dropping attenuation,  the phase contributed by this path to the correlation is  \BE e^{ik\left((\ind_1-\ind_2)\ptd_1 - (\outd_1-\outd_2)\ptd_B\right)}= e^{ik\omgv(\ptd_1-\ptd_B)} \EE
If tis complex number can have highly varying phases, than summing over multiple random paths averages to zero. The different paths interfere constructively only if the phase difference is negligible, roughly when $k|\omgv||\ptd_1-\ptd_n|<1$. Intuitively, the average distance between an entrance point and an exit point on the target scales with the target depth, and it is reasonable to expect that $E[|\ptd_1-\ptd_n|]$ is proportional to $ L$. This implies that the memory effect holds when    $k|\omgv| L<1$, in agreement with \equref{eq:corr-diffusion}.

\section{Single-scattering approximation for covariance rendering}\label{sec:bounces}
Before we conclude, we report an interesting property of speckle covariance, which can be used to accelerate its estimation under certain illumination and imaging conditions.

\begin{figure*}[!t]       
        \begin{center}\begin{tabular}{cccc}
                        
                        \includegraphics[width= 0.22  \textwidth]{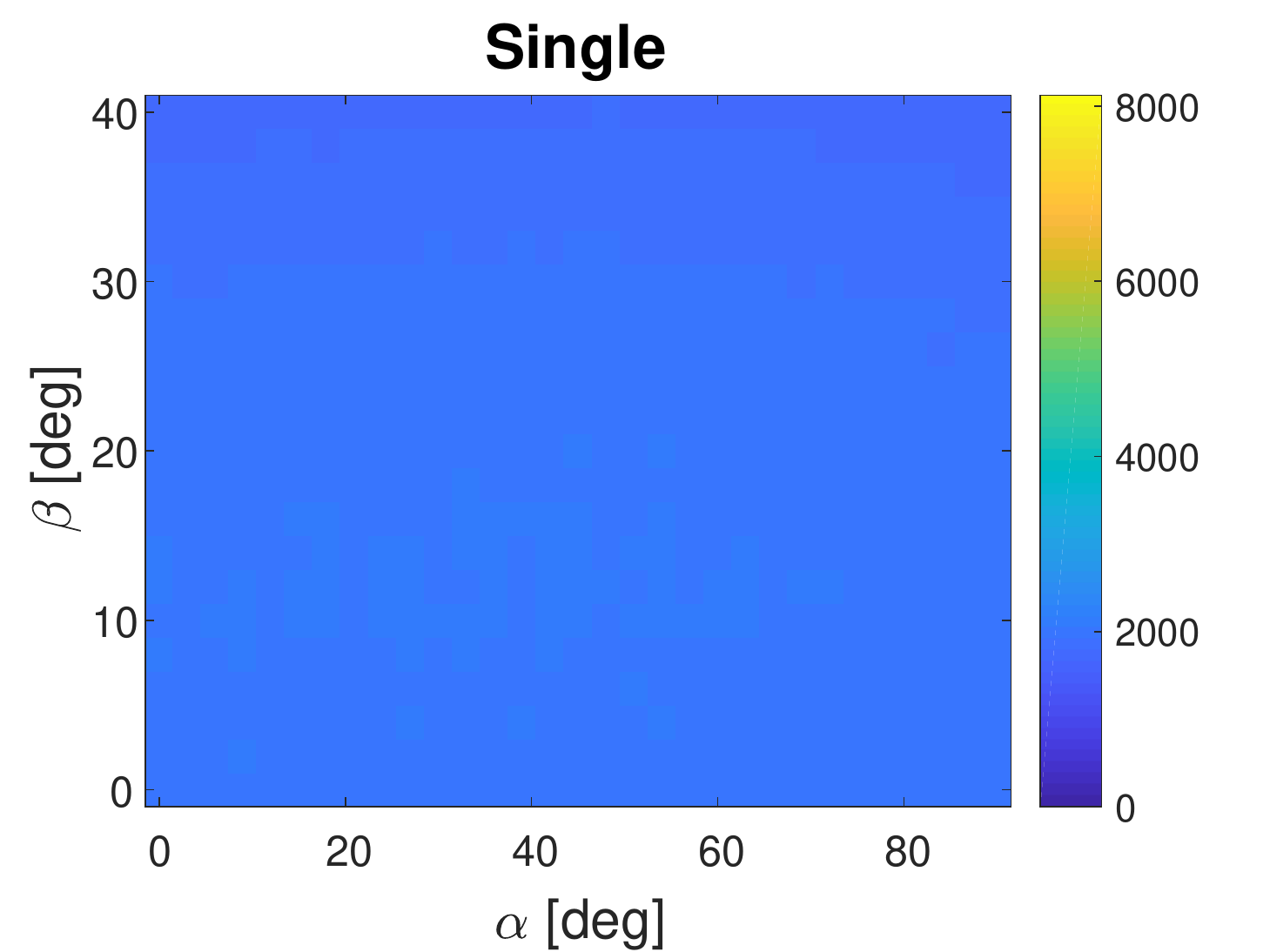}&
                        \includegraphics[width= 0.22 \textwidth]{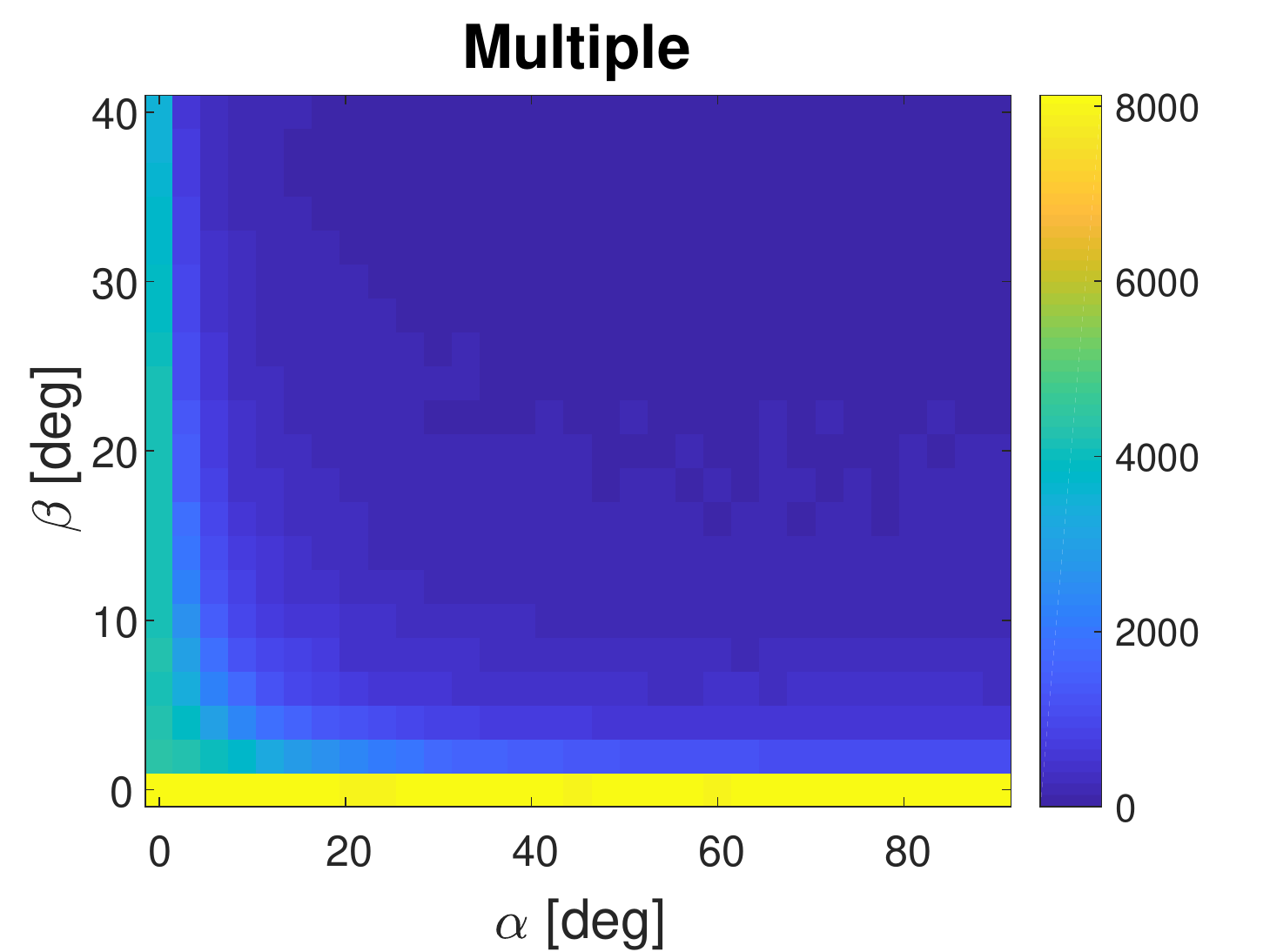}&
                        \includegraphics[width= 0.22  \textwidth]{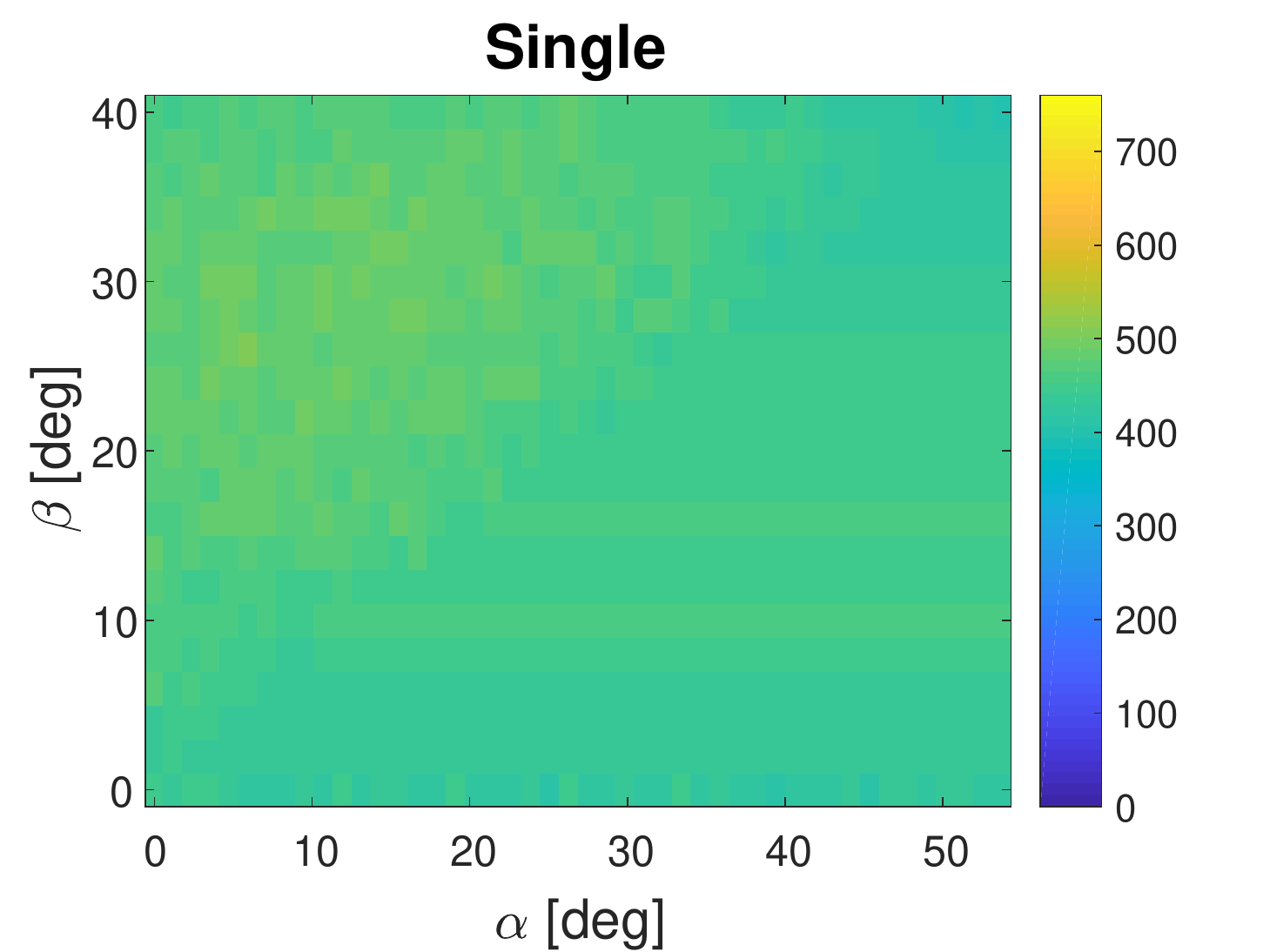}&
                        \includegraphics[width= 0.22  \textwidth]{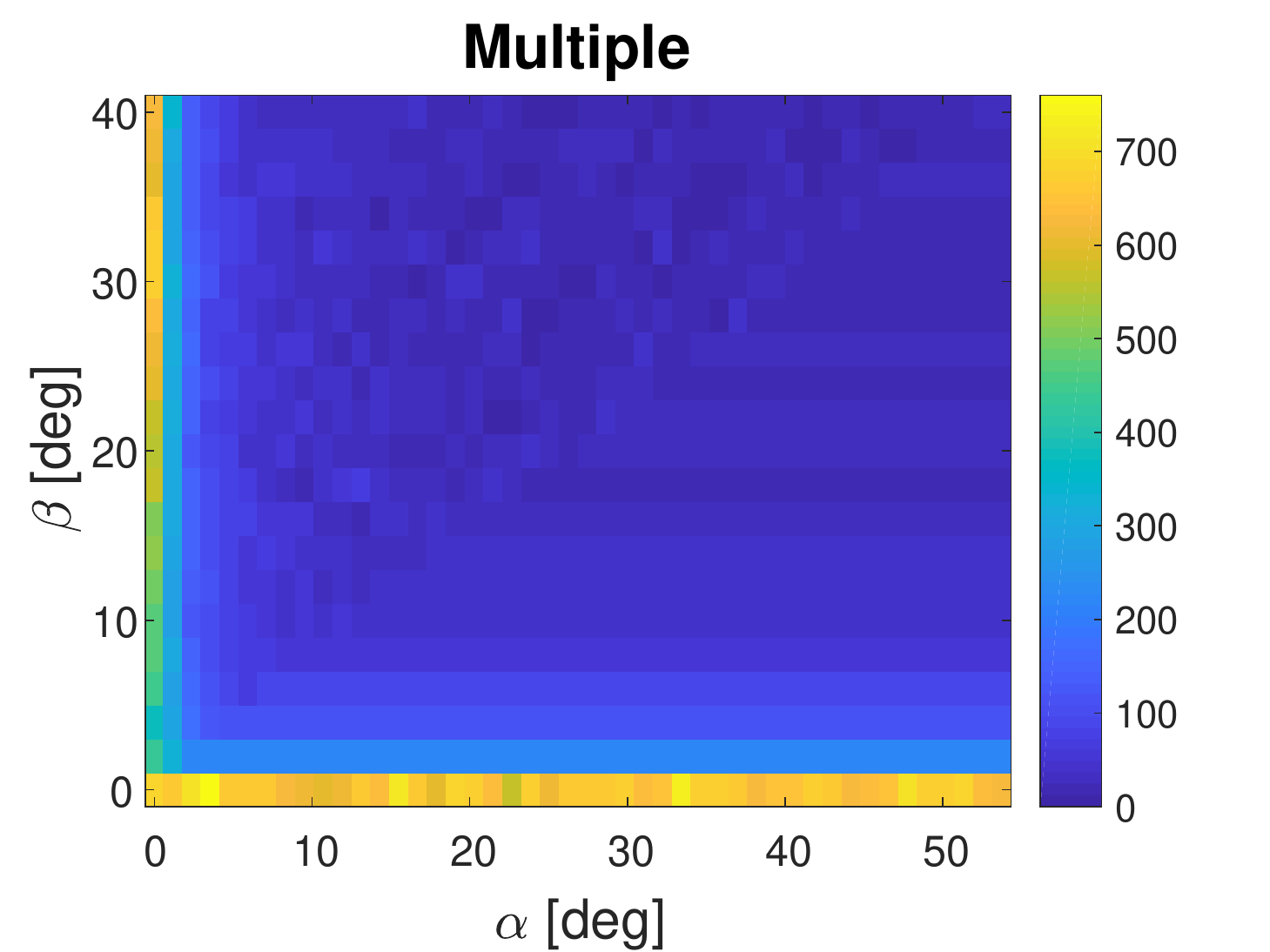}\\
                        \multicolumn{2}{c}{(a) $\bomg=\bzero$}&\multicolumn{2}{c}{(b) $\bomg=(0.007\frac{2\pi}{\lambda},0,0)$}
                        %\begin{footnotesize}Single\end{footnotesize}&\begin{footnotesize}Multiple\end{footnotesize} %&\begin{footnotesize}Single \end{footnotesize}
                        %&\begin{footnotesize}Multiple  \end{footnotesize}
                \end{tabular}\vspace{-0.5cm}
        \end{center}
        \caption{\small{Decomposing speckle correlations  by number of bounces in a MC path. We simulate correlations of multiple illumination and viewing sets satisfying $\ind_1-\outd_1-(\ind_2-\outd_2)=\bomg$, and plot the contribution  as a function of two angles $\alpha=\min( \angle(\ind_1,\ind_2),\angle(\outd_1,\outd_2))$, $\beta=\min( \angle(\ind_1,\outd_1),\angle(\ind_2,\outd_2))$. The multiple scattering term drops to zero as soon as one of the above angular differences increases. As a result for every target frequency one wishes to image, it is easy to find illumination and viewing configurations whose correlation accounts for the single scattering component alone.      }  }\label{fig:single-multiple}
\end{figure*}

When simulating covariance using Monte Carlo rendering, we can separate contributions from paths of different numbers of bounces $B$. For example, in \figref{fig:single-multiple}, we show simulations for a cube volume $\V$ of dimensions $100\lambda\times100\lambda\times100\lambda$, and with O.D.=5, resulting in strong multiple scattering. We simulate the covariance for multiple pairs of illumination and imaging sets satisfying $\ind_1-\outd_1-(\ind_2-\outd_2)=\bomg$, for some target 3D vector $\bomg$. In each simulation, we decompose the rendered speckle covariance into two components, one accounting for contributions from paths that scattered once ($B=1$), and another accounting for paths that scattered two or more times ($B\ge 2$). Within each rendered covariance matrix, the bottom left corner corresponds to rendering intensity.

We observe that, for the intensity case, the multiple-scattering component is dominant. By contrast, for cases where the difference between the two illumination or the two viewing directions is more than some small amount, the multiple-scattering component becomes negligible. This happens because, as the angle difference becomes large enough to bring us outside the range of the memory effect, multiply-scattered paths have complex contributions with randomly-varying phase, and therefore average to zero. 

We conclude that, whenever the imaging and illumination conditions are such that we are outside the memory effect range, speckle covariance can be computed using only single scattering. This evaluation can be done in closed-form, avoiding the need for computationally-expensive Monte Carlo rendering. Namely, from a short derivation we can obtain the formula:
\BE\label{eq:singlesct-fourier}
\!\!C^{\ind_1,\ind_2}_{\outd_1,\outd_2}\!=\!\ampf(\ind_1\cdot\outd_1){\ampf(\ind_2\cdot\outd_2)}^*\!\!\!\int_\V \sctCoff(\ptd)e^{ik((\ind_1-\outd_1)-(\ind_2-\outd_2))\cdot\ptd} \eta(\ptd) \ud \ptd,
\EE
where $\eta(\ptd)=\att(\ptd,\ind_1)\att(\ptd,\outd_1)^*\att(\ptd,\ind_2)^*\att(\ptd,\outd_2)$.

The above discussion indicates that, whenever we are outside the memory effect range, we can accelerate the computation of speckle covariance by using the single-scattering approximation, without significant loss in accuracy. This is analogous to the use of the single-scattering approximation for accelerating intensity rendering~\cite{sun2005practical,walter2009single}, with an important difference: In the case of intensity the single-scattering approximation is valid only for very optically-thin volumes~\cite{narasimhan2006acquiring}. By contrast, in the case of covariance, the approximation can be accurate even for optically thick materials, given appropriate illumination and viewing conditions, making it more broadly applicable.
\comment{
\igkiou{I think we can skip the text below. It's quite complicated, and doesn't add much to the discussion as far as I can tell.} For a  homogeneous target,  \equref{eq:singlesct-fourier}  reduces to evaluating the phase function directly. Assume a  homogeneous target which  is relatively wide in the $x,y$ dimensions and has thickness $d$ in the $z$ dimension.   Selecting $\ind_1-\outd_1=\ind_2-\outd_2$ and denoting $\theta=<\ind_1,\outd_1>=<\ind_2,\outd_2>$,    the integral of \equref{eq:singlesct-fourier} reduces to \vspace{-0.1cm}
\BE \label{eq:ss-inf}
C^{\ind_1,\ind_2}_{\outd_1,\outd_2}=\rho(\theta)\frac{\sctCoff}{\extCoff r_z} \left(1-e^{-\extCoff\cdot d \cdot r_z }\right)\EE
with
\BE r_z=\frac12\left(|{\ind_1}(z)|^{-1}+|{\ind_2}(z)|^{-1}+|{\outd_1}(z)|^{-1}+|{\outd_2}(z)|^{-1}\right)
\EE  %
where ${{\ind}_j}(z),{{\outd}_j}(z)$ denotes the $z$ component of the direction vectors.
}

\comment{\equref{eq:ss-inf} implies that the speckle correlation allows us to measure directly the phase function of the material.
Having an  accurate measure of all angles of the phase function is important for particle sizing~\cite{pusey1999suppression,johnson1994laser,pine1990diffusing} and molecular weighting applications, in which one aims to estimate the exact portion of particles with each size in a certain dispersion. As the phase function of a particle varies with its radius and can be computed using Mie theory\cite{BorenScattering}, given an accurate description of the bulk phase function we can try to decompose it as a superposition of a Mie function dictionary, and the mixing weights correspond to the portion of particles from each size. }

\comment{

\subsection{Similarity theory relations}

Similarity theory~\cite{wyman1989similarity,zhao2014high} states  after multiple scattering events where the angular distribution of light is band limited   one can often exchange between scattering coefficient and phase function without alerting the RTE, and any pair of scattering coefficient and HG anisotropy satisfying  \BE\sctCoff^*,(1-g^*)=\sctCoff(1-g)\EE should lead to visually similar results.
 This is often used to accelerate MC intensity rendering as one can use  the parameter configuration within an equivalent class but with the largest mean free path and hence save steps in the MC process. \comment{This ambiguity results from the fact that after multiple scattering events  one can exchange between the scattering coefficient and the phase function without alerting the RTE output.} As the similarity theory ambiguity arise after multiple scattering it seems reasonable to expect it will reduce with covariance measurements that can be described as single scattering.

To show the difference between speckle covariance measurement and intensity measurements we compare them on the  setup  used for intensity-based extraction of scattering parameters in~\cite{InverseRendSA13}.  In this setup a dispersion was placed in a thin glass cell. The target  was illuminated with a narrow beam from a few angles and the scattered intensity  was imaged by a 2D sensor focused at the edge of the glass cell.  We simulated a 2D family of materials, varying the phase function and the densities parameters. We considered Henyey-Greenstein (HG) phase function whose anisotropy parameter $g\in[-1, 1]$, and extinction coefficient values ranging from single scattering to multiple scattering densities, $\extCoff\in[1/(2d),\cdots,1/(0.1d)],$ were $d$ is the  thickness of the target, implying an optical depth (O.D.) between $0.5$ to $10$. 
As in the setup of~\cite{InverseRendSA13}, for each assignment of material parameters we rendered $18$ intensity images corresponding to different combinations of viewing and illumination directions.
In \figref{fig:similarity-tables} we plot the 2D error surface obtained when trying to explain the data with the wrong material parameters, namely
\BE \label{eq:2D-err}Err(\extCoff,g)=\sum_k \|I_k^{\extitionCoefficient,g}-I_k^{\extitionCoefficient^o,g^o}\|^2 \EE where $\extitionCoefficient^o,g^o$ denote the true values and $k$ runs over the $18$ illumination and viewing configurations.
To compare these with correlation measurements we 
rendered $18$ correlations at equivalent illumination and viewing angles. Note that while in the intensity case one illuminates with a narrow beam and measure a 2D focused image, in the covariance case one illuminates with a plane wave and image at the far field, and the correlation output is  a single scalar rather than  a 2D image. Specifically for every scattering angle $\theta$ considered in~\cite{InverseRendSA13} we selected illumination and viewing pairs such that   the angle between $\ind_1,\outd_1$ is approximately $\theta=\theta_i+\theta_v$, and so is the angle  between $\ind_2,\outd_2$, and yet  $\ind_1,\ind_2$ and $\outd_1,\outd_2$ are different enough so that correlation can be modeled as single scattering.  
\comment{With slight adjustment of the angles we got combinations whose mean frequency is $\ind_1-\outd_1-(\ind_2-\outd_2)=[0,0,\omega_z]$ for $5$ $\omega_z$ values between $0$ to $5/d$. (the multiple $\omega_z$ values are required to obtain a bit more information about $\extitionCoefficient$).  }\comment{One such pairs is given by \BE\ind_1=\begin{pmatrix}sin(\frac{\theta}{2}) \\
0 \\
cos(\frac{\theta}{2}) \\
\end{pmatrix},\;\outd_1=\begin{pmatrix}sin(\frac{\theta}{2}) \\
0 \\
-cos(\frac{\theta}{2}) \\
\end{pmatrix},\;\ind_2=\begin{pmatrix}0 \\
sin(\frac{\theta}{2}+\eps) \\
cos(\frac{\theta}{2}+\eps) \\
\end{pmatrix} ,\;\outd_2=\begin{pmatrix}0 \\
sin(\frac{\theta}{2}+\eps) \\
-cos(\frac{\theta}{2}+\eps) \\
\end{pmatrix}\EE
Effectively in this pair  $\ind_2,\outd_2$ are a $90\degree$ 3D rotation of 
$\ind_1,\outd_1$ around their mean vector, but a rotation of another angle 
works as well. $\eps$ is selected such that $2cos(\theta/2)-2cos(\theta/2+\eps)=\omega_z$. As}
{
   For example we can choose vectors     of the form
        \begin{equation}
        \begin{split}
        &\ind_1=[sin(\theta/2),0,cos(\theta/2)],~~~ \outd_1=[sin(\theta/2),0,-cos(\theta/2)],\\&\ind_2=R_\alpha[sin((\theta+\eps)/2),0,cos((\theta+\eps)/2)],\\
        &\outd_2=R_\alpha[sin((\theta+\eps)/2),0,-cos((\theta+\eps)/2)]
        \end{split}\end{equation}
        were $R_\alpha$ is a rotation around the vector $\ind_1-\outd_1$. This rotation is chosen such that $\ind_1,\ind_2$ and $\outd_1,\outd_2$ are different enough so that correlation can be modeled as single scattering.
        $\eps$ is a small angle shift selected such that $\ind_1+\outd_1-(\ind_2+\outd_2)=[0,0,\omega_z]$, where we used $\omega_z$ values between $0$ to $10/d$,  the multiple $\omega_z$ values are required to capture a few law frequencies of the function $exp(-\extCoff  z)$ and obtain a bit more information about $\extitionCoefficient$. }
For simplicity,  as an amplitude function $\ampf$ we used the square root of the 
Henyey-Greenstein (HG) phase function with uniform phase, thought application specific   amplitude functions can be derived from Mie theory. 
Using \equref{eq:singlesct-fourier} this allows us to measure correlation values of the form
\BE
C^{\ind_1,\ind_2}_{\outd_1,\outd_2}=\rho(\theta_i+\theta_v)\sctCoff\int_{z=0}^d e^{\left(\frac{i2\pi}{\lambda}\omega_z +\extCoff r_z\right) z }  dz
\EE
\begin{figure*}[t]
        \centering
        \subfloat{
                \raisebox{1cm}{\rotatebox[origin=t]{90}{Intensity}}}  
                \subfloat{\label{fig:int_setup}\includegraphics[width=0.1\textwidth]
                {figures/setup/Slide2.JPG}}
        \subfloat{\label{fig:int_OD3_g-08}\includegraphics[width=0.145\textwidth]
                {figures/similarity_tables/similarity_int_OD3_g-8}}
        \subfloat{\label{fig:int_OD3_g0}\includegraphics[width=0.145\textwidth]
                {figures/similarity_tables/similarity_int_OD3_g0}}
        \subfloat{\label{fig:int_OD3_g08}\includegraphics[width=0.145\textwidth]
                {figures/similarity_tables/similarity_int_OD3_g8}}
        \subfloat{\label{fig:int_OD9_g-08}\includegraphics[width=0.145\textwidth]
                {figures/similarity_tables/similarity_int_OD9_g-8}}
        \subfloat{\label{fig:int_OD9_g0}\includegraphics[width=0.145\textwidth]
                {figures/similarity_tables/similarity_int_OD9_g0}}
        \subfloat{\label{fig:int_OD9_g08}\includegraphics[width=0.145\textwidth]
                {figures/similarity_tables/similarity_int_OD9_g8}}
        \\\vspace{-0.25cm}
        \subfloat{
        \raisebox{1cm}{\rotatebox[origin=t]{90}{Covariance}}} 
         \subfloat{\label{fig:int_setup}\includegraphics[width=0.1\textwidth]
                {figures/setup/Slide1.JPG}}
   \subfloat{\label{fig:cov_OD3_g-08}\includegraphics[width=0.145\textwidth]
        {figures/similarity_tables/similarity_cov_OD3_g-8}}
   \subfloat{\label{fig:cov_OD3_g0}\includegraphics[width=0.145\textwidth]
        {figures/similarity_tables/similarity_cov_OD3_g0}}
   \subfloat{\label{fig:cov_OD3_g08}\includegraphics[width=0.145\textwidth]
        {figures/similarity_tables/similarity_cov_OD3_g8}}
   \subfloat{\label{fig:cov_OD9_g-08}\includegraphics[width=0.145\textwidth]
        {figures/similarity_tables/similarity_cov_OD9_g-8}}
   \subfloat{\label{fi:cov_OD9_g0}\includegraphics[width=0.145\textwidth]
        {figures/similarity_tables/similarity_cov_OD9_g0}}
   \subfloat{\label{fi:cov_OD9_g08}\includegraphics[width=0.145\textwidth]
        {figures/similarity_tables/similarity_cov_OD9_g8}}  
  \caption{Similarity theory for intensity and covariance. Top row shows the error in intensity between a rendering of a reference pair of scattering parameters (optical depth (O.D). and phase function anisotropy g), and a 2D scattering parameter space considered. The reference pair is indicated in the title of each subfigure. Bottom row shows that this error for covariance.}  
   \label{fig:similarity-tables}        
\end{figure*}

We compute 2D error surfaces with correlation measurements as in \equref{eq:2D-err}.
In \figref{fig:similarity-tables}, we plot the error of explaining the data by the wrong phase function for each type of measurement.
Looking at the intensity errors one can see the similarity theory ambiguity, an uncertainty between $\extitionCoefficient$ and $g$ parameters, seen as a valley of low error values rather than a single minimum. Similar error surfaces were demonstrated  in~\cite{zhao2014high}. 
For positive $g$ values, this ambiguity is relaxed by the covariance measurements, which provide 
a sharper minima around the true solution compared to the intensity measurements. This is not surprising  since the covariance in \equref{eq:ss-inf} measures the phase function directly, while the intensity measurements capture the full random walk of light in the medium, including a non negligible multiple scattering term.   

\comment{
        \subsubsection{Estimating material density} \label{sec:Classifier}
        \Anat{Not sure if we want to keep this section...}
        As a preliminary demonstration of the information captured by speckle correlation we have designed the following experiment attempting to estimate material density. We consider a rectangular slab with a known type of particles and an unknown density. We assume the slab is illuminated with plane waves at a few different orientations $\ind_j$ and measure backscattering speckles on a set of near field lensless sensors $\snsp$, located on a line parallel to the slab, in a setup similar to \figref{fig:ScatteringCvsIC}. The goal is to extract density from speckle statistics.   This task would be simple if the source intensity were known, since denser materials have stronger backscattering. However if the source power is unknown the intensity images alone contain no information (the intensity image one observes under a plane wave illumination whose extent is wider than the test sample is just a flat uniform image). Our preliminary experiments below show that speckle images can help us measure density despite the fact that the intensity images are identical.     
              
        To  demonstrate density estimation  we rendered covariance matrices $C_{\text{MC}}(\extCoff^k)$ for a large discrete set of $\extCoff^k$ values. We also generated test speckle images $u^{\ind_j,O_m}_\snsp$,  by drawing particle instantiations and solving the wave equations. Since we assume the source power is unknown, we normalized both speckle images $u^{O_m}$ and target covariances $C(\extCoff)$ by the mean intensity. 
        
        We use the  speckle images to compute spatial correlations empirically for different spatial shifts
        \BE\label{eq:corr-spt-sft} c^m_{\text{emp}}(\Delta,\ind_1,\ind_2)=\sum_{\snsp} u^{\ind_1,O_m}_\snsp\cdot {u^{\ind_2,O_m}_{\snsp+\Delta}}^* \EE
        %we compere these with the MC predicted correlation
        %\BE
        %c^k_{\text{MC}}(\Delta,\ind_1,\ind_2)=\sum_{\snsp}C^{\ind_1,\ind_2}_{\snsp,\snsp+\Delta}(\extCoff^k)
        %\EE
        For each $O_m$ we pick the  value of $k$ that minimizes the $L_2$ error: 
        \BE \extCoff(u^{O_m})=\arg\min_k \sum_{\ind_1,\ind_2,\Delta}\left|c^k_{\text{MC}}(\Delta,\ind_1,\ind_2)-c^m_{\text{emp}}(\Delta,\ind_1,\ind_2) \right|^2\EE
        \figref{fig:classifier}  plots  estimation error. While the empirical speckles are noisy and do not match the exact correlation, the estimation error is significantly lower  than that of a random guess, which is the best we can hope for with intensity images alone. This demonstrates that  speckles capture information that is not present in the intensities alone.
        \comment{ The estimation is made more accurate when a wider spatial support is available so the empirical correlation is less noisy, but with the $\mu$-diff simulator~\cite{mudiff} the geometries we can simulate with reasonable computational resources are rather limited. To test this we averaged the $c^m_{\text{emp}}$ values over a few particle instantiations $O^m$ sampled from the same density. As illustrated in \figref{fig:classifier}, the estimation accuracy is improved when more samples are available, but even with a single sensor row estimation is better than random.   
}}

}

\section{Discussion and Future Prospects}

We conclude with a discussion of our contributions and possible future directions. We presented a path-integral formulation for the covariance of speckle fields generated by the interaction of coherent light with scattering volumes. Using this formulation, we derived two Monte Carlo rendering algorithms, one for directly estimating covariance, and another for directly generating speckle patterns. As we demonstrated in Section~\ref{sec:results}, our algorithms provide a unique combination of physical accuracy (closely matching solutions of the wave equation, reproducing known physical phenomena such as memory effect and coherent backscattering), computational efficiency (outperforming wave equation solvers by orders of magnitude), and parsimony (using only bulk macroscopic parameters of a volume, instead of requiring knowledge of its microscopic structure). 

%Both of our Monte Carlo rendering algorithms share strong similarities with standard volume Monte Carlo rendering algorithms for intensity. This facilitates integration into popular physically-accurate rendering engines~\cite{Mitsuba,pharr2016physically}, and reusing existing technology for efficient implementations. We note, however, an important difference with the intensity case: At every next-event estimation, it is necessary to perform $J^2$ or $J$, for covariance or speckle rendering, throughput calculations and updates, where $J$ is the number of lighting and viewing conditions we are simulating (e.g., $J$ can be number of pixels). When $J$ is large, e.g., when rendering a Megapixel image, these next-event estimation operations become the major computational bottleneck of our rendering algorithms. Therefore, it will be important to investigate techniques for accelerating these operations.

We note that our path-integral formulation for speckle covariance can potentially provide the foundation for deriving more sophisticated path sampling algorithms for rendering speckles, analogous to how the path-integral formulation for intensity spurred the invention of algorithms such as bidirectional path tracing and Metropolis light transport~\cite{veach1997robust}. In this context, we observe that our formulation is \emph{reciprocal}, and therefore lends itself to the development of bidirectional path sampling algorithms.
\comment{
\Anat{I think the following paragraph can be skipped}
A fascinating future direction is investigating how to extend our theory and algorithms to scattering volumes that do not satisfy the assumptions we made at the start of \secref{sce:speckeles-def}, namely: Discrete random media with large and dense scatterers~\cite{moon2007rendering}; anisotropic media where scattering is not rotation-invariant~\cite{jakob2010radiative}; and non-exponential media where the locations of scatterers are not independent of each other~\cite{bitterli2018radiative}. Our knowledge of speckle statistics in such media is currently very limited, because of the simultaneous lack of comprehensive lab measurements and theoretical tools to study them. For instance, the central limit theorem used to derive the Gaussianity property in \secref{sce:speckeles-def}, does not hold for non-exponential media, as scatterer locations are no longer independent random variables. Creating physically-accurate rendering tools for these types of scattering can help expand our understanding of speckle effects in a large variety of real-world materials.}

In this paper we chose to focus on \emph{spatial} speckle correlations. In doing so, we ignored another important class of second-order speckle statistics, \emph{temporal} correlations resulting from moving scatterers. We hope that our results will motivate the development of analogous theoretical and simulation tools for temporal correlations. For instance, our rendering algorithms allow us to study the memory effect and related applications in cases where common assumptions (diffusion, Fokker-Planck limit) do not hold. Likewise, rendering algorithms for temporal correlations can allow extending related applications such as dynamic light scattering to cases where the common assumption of Brownian motion of scatterers is invalid~\cite{Duncan:08}.

The ability to render physically correct speckles can be highly beneficial for
incorporating machine learning techniques into optics problems, where the collection of training data has been a major burden.

Last but not least, the findings of \secref{sec:bounces} suggest that measuring and rendering speckle covariance holds promise for \emph{inverse rendering} applications. The fact that speckle covariance measurements are dominated by single scattering for a much larger class of materials than intensity measurements can potentially drastically simplify the volumetric inverse rendering problem, e.g., by potentially allowing us to replace the complex differentiable rendering of Gkioulekas et al.~\shortcite{InverseRendSA13,InvRenderingECCV16} with simple analytic algorithms of Narasimhan et al.~\shortcite{narasimhan2006acquiring}. In addition to simplifying computation, it will be interesting to examine whether speckle covariance measurements can be used to relax previously reported ambiguities between scattering parameters~\cite{wyman1989similarity,zhao2014high}.

% Bibliography
\bibliographystyle{ACM-Reference-Format}
\bibliography{biblio,scatterometer,biblioscattering,bibliography,shuangz,proposal,perception,kb}


\begin{thebibliography}{74}

%%% the Web address.

\ifx \showCODEN    \undefined \def \showCODEN     #1{\unskip}     \fi
\ifx \showDOI      \undefined \def \showDOI       #1{#1}\fi
\ifx \showISBNx    \undefined \def \showISBNx     #1{\unskip}     \fi
\ifx \showISBNxiii \undefined \def \showISBNxiii  #1{\unskip}     \fi
\ifx \showISSN     \undefined \def \showISSN      #1{\unskip}     \fi
\ifx \showLCCN     \undefined \def \showLCCN      #1{\unskip}     \fi
\ifx \shownote     \undefined \def \shownote      #1{#1}          \fi
\ifx \showarticletitle \undefined \def \showarticletitle #1{#1}   \fi
\ifx \showURL      \undefined \def \showURL       {\relax}        \fi
% invisible to TeX
\providecommand\bibfield[2]{#2}
\providecommand\bibinfo[2]{#2}
\providecommand\natexlab[1]{#1}
\providecommand\showeprint[2][]{arXiv:#2}

\bibitem[\protect\citeauthoryear{Akkermans and Montambaux}{Akkermans and
  Montambaux}{2007}]%
        {Akkermans07}
\bibfield{author}{\bibinfo{person}{Eric Akkermans} {and}
  \bibinfo{person}{Gilles Montambaux}.} \bibinfo{year}{2007}\natexlab{}.
\newblock \bibinfo{booktitle}{\emph{Mesoscopic Physics of Electrons and
  Photons}}.
\newblock \bibinfo{publisher}{Cambridge University Press}.
\newblock


\bibitem[\protect\citeauthoryear{Batarseh, Sukhov, Shen, Gemar, Rezvani, and
  Dogariu}{Batarseh et~al\mbox{.}}{2018}]%
        {Batarseh2018}
\bibfield{author}{\bibinfo{person}{M. Batarseh}, \bibinfo{person}{S. Sukhov},
  \bibinfo{person}{Z. Shen}, \bibinfo{person}{H. Gemar}, \bibinfo{person}{R.
  Rezvani}, {and} \bibinfo{person}{A. Dogariu}.}
  \bibinfo{year}{2018}\natexlab{}.
\newblock \showarticletitle{Passive sensing around the corner using spatial
  coherence}.
\newblock \bibinfo{journal}{\emph{Nature Communications}}.
\newblock


\bibitem[\protect\citeauthoryear{Baydoun, Baresch, Pierrat, and Derode}{Baydoun
  et~al\mbox{.}}{2016}]%
        {Baydoun2016}
\bibfield{author}{\bibinfo{person}{Ibrahim Baydoun}, \bibinfo{person}{Diego
  Baresch}, \bibinfo{person}{Romain Pierrat}, {and} \bibinfo{person}{Arnaud
  Derode}.} \bibinfo{year}{2016}\natexlab{}.
\newblock \showarticletitle{Radiative transfer of acoustic waves in continuous
  complex media: Beyond the Helmholtz equation}.
\newblock \bibinfo{journal}{\emph{Physical Review E}}.
\newblock


\bibitem[\protect\citeauthoryear{Bergmann, Mohammadikaji, Irgenfried, Worn,
  Beyerer, and Dachsbacher}{Bergmann et~al\mbox{.}}{2016}]%
        {vmv.20161357}
\bibfield{author}{\bibinfo{person}{Stephan Bergmann}, \bibinfo{person}{Mahsa
  Mohammadikaji}, \bibinfo{person}{Stephan Irgenfried}, \bibinfo{person}{Heinz
  Worn}, \bibinfo{person}{Jürgen Beyerer}, {and} \bibinfo{person}{Carsten
  Dachsbacher}.} \bibinfo{year}{2016}\natexlab{}.
\newblock \showarticletitle{{A Phenomenological Approach to Integrating
  Gaussian Beam Properties and Speckle into a Physically-Based Renderer}}. In
  \bibinfo{booktitle}{\emph{Vision, Modeling \& Visualization}}.
\newblock


\bibitem[\protect\citeauthoryear{Berkovits and Feng}{Berkovits and
  Feng}{1994}]%
        {BERKOVITS1994135}
\bibfield{author}{\bibinfo{person}{Richard Berkovits} {and}
  \bibinfo{person}{Shechao Feng}.} \bibinfo{year}{1994}\natexlab{}.
\newblock \showarticletitle{Correlations in coherent multiple scattering}.
\newblock \bibinfo{journal}{\emph{Physics Reports}} (\bibinfo{year}{1994}).
\newblock
\urldef\tempurl%
\url{https://doi.org/10.1016/0370-1573(94)90079-5}
\showDOI{\tempurl}


\bibitem[\protect\citeauthoryear{Bertolotti, van Putten, Blum, Lagendijk, Vos,
  and Mosk}{Bertolotti et~al\mbox{.}}{2012}]%
        {Bertolotti2012}
\bibfield{author}{\bibinfo{person}{Jacopo Bertolotti},
  \bibinfo{person}{Elbert~G. van Putten}, \bibinfo{person}{Christian Blum},
  \bibinfo{person}{Ad Lagendijk}, \bibinfo{person}{Willem~L. Vos}, {and}
  \bibinfo{person}{Allard~P. Mosk}.} \bibinfo{year}{2012}\natexlab{}.
\newblock \showarticletitle{Non-invasive imaging through opaque scattering
  layers}.
\newblock \bibinfo{journal}{\emph{Nature}} \bibinfo{number}{491(7423)},
  \bibinfo{pages}{232}.
\newblock


\bibitem[\protect\citeauthoryear{Bitterli, Ravichandran, M{\"u}ller, Wrenninge,
  Nov{\'a}k, Marschner, and Jarosz}{Bitterli et~al\mbox{.}}{2018}]%
        {bitterli2018radiative}
\bibfield{author}{\bibinfo{person}{Benedikt Bitterli}, \bibinfo{person}{Srinath
  Ravichandran}, \bibinfo{person}{Thomas M{\"u}ller}, \bibinfo{person}{Magnus
  Wrenninge}, \bibinfo{person}{Jan Nov{\'a}k}, \bibinfo{person}{Steve
  Marschner}, {and} \bibinfo{person}{Wojciech Jarosz}.}
  \bibinfo{year}{2018}\natexlab{}.
\newblock \showarticletitle{A radiative transfer framework for non-exponential
  media}. In \bibinfo{booktitle}{\emph{SIGGRAPH Asia}}. ACM,
  \bibinfo{pages}{225}.
\newblock


\bibitem[\protect\citeauthoryear{Boas and Yodh}{Boas and Yodh}{1997}]%
        {Boas:97}
\bibfield{author}{\bibinfo{person}{David~A Boas} {and}
  \bibinfo{person}{Arjun~G. Yodh}.} \bibinfo{year}{1997}\natexlab{}.
\newblock \showarticletitle{Spatially varying dynamical properties of turbid
  media probed with diffusing temporal light correlation}.
\newblock \bibinfo{journal}{\emph{J. Opt. Soc. Am. A}}.
\newblock


\bibitem[\protect\citeauthoryear{Bohren and Huffman}{Bohren and
  Huffman}{1983}]%
        {BorenScattering}
\bibfield{author}{\bibinfo{person}{Craig~F. Bohren} {and}
  \bibinfo{person}{Donald~R. Huffman}.} \bibinfo{year}{1983}\natexlab{}.
\newblock \bibinfo{booktitle}{\emph{Absorption and scattering of light by small
  particle}}.
\newblock \bibinfo{publisher}{John Wiley \& Sons}.
\newblock


\bibitem[\protect\citeauthoryear{Cuypers, Haber, Bekaert, Oh, and
  Raskar}{Cuypers et~al\mbox{.}}{2012}]%
        {Cuypers:2012:RMD:2231816.2231820}
\bibfield{author}{\bibinfo{person}{Tom Cuypers}, \bibinfo{person}{Tom Haber},
  \bibinfo{person}{Philippe Bekaert}, \bibinfo{person}{Se~Baek Oh}, {and}
  \bibinfo{person}{Ramesh Raskar}.} \bibinfo{year}{2012}\natexlab{}.
\newblock \showarticletitle{Reflectance Model for Diffraction}.
\newblock \bibinfo{journal}{\emph{ACM Trans. Graph.}} \bibinfo{volume}{31},
  \bibinfo{number}{5}, Article \bibinfo{articleno}{122},
  \bibinfo{numpages}{11}~pages.
\newblock
\showISSN{0730-0301}
\urldef\tempurl%
\url{https://doi.org/10.1145/2231816.2231820}
\showDOI{\tempurl}


\bibitem[\protect\citeauthoryear{d'Eon}{d'Eon}{2018a}]%
        {d2018reciprocalpt2}
\bibfield{author}{\bibinfo{person}{Eugene d'Eon}.}
  \bibinfo{year}{2018}\natexlab{a}.
\newblock \showarticletitle{A Reciprocal Formulation of Non-Exponential
  Radiative Transfer. 2: Monte Carlo Estimation and Diffusion Approximation}.
\newblock \bibinfo{journal}{\emph{arXiv preprint arXiv:1809.05881}}.
\newblock


\bibitem[\protect\citeauthoryear{d'Eon}{d'Eon}{2018b}]%
        {d2018reciprocal}
\bibfield{author}{\bibinfo{person}{Eugene d'Eon}.}
  \bibinfo{year}{2018}\natexlab{b}.
\newblock \showarticletitle{A reciprocal formulation of non-exponential
  radiative transfer with uncorrelated sources, detectors and boundaries. 1:
  Sketch and motivation}.
\newblock \bibinfo{journal}{\emph{arXiv preprint arXiv:1803.03259}}.
\newblock


\bibitem[\protect\citeauthoryear{Dougherty, Ackerson, Reguigui,
  Dorri-Nowkoorani, and Nobbmann}{Dougherty et~al\mbox{.}}{1994}]%
        {DOUGHERTY94}
\bibfield{author}{\bibinfo{person}{Ronald~L. Dougherty},
  \bibinfo{person}{Bruce~J. Ackerson}, \bibinfo{person}{N.M. Reguigui},
  \bibinfo{person}{F. Dorri-Nowkoorani}, {and} \bibinfo{person}{Ulf Nobbmann}.}
  \bibinfo{year}{1994}\natexlab{}.
\newblock \showarticletitle{Correlation transfer: Development and application}.
\newblock \bibinfo{journal}{\emph{J. of Quantitative Spectroscopy and Radiative
  Transfer}}.
\newblock


\bibitem[\protect\citeauthoryear{Duncan and Kirkpatrick}{Duncan and
  Kirkpatrick}{2008}]%
        {Duncan:08}
\bibfield{author}{\bibinfo{person}{Donald~D. Duncan} {and}
  \bibinfo{person}{Sean~J. Kirkpatrick}.} \bibinfo{year}{2008}\natexlab{}.
\newblock \showarticletitle{Can laser speckle flowmetry be made a quantitative
  tool?}
\newblock \bibinfo{journal}{\emph{J. Opt. Soc. Am. A}} \bibinfo{volume}{25},
  \bibinfo{number}{8}, \bibinfo{pages}{2088--2094}.
\newblock
\urldef\tempurl%
\url{https://doi.org/10.1364/JOSAA.25.002088}
\showDOI{\tempurl}


\bibitem[\protect\citeauthoryear{Durduran, Choe, Baker, and Yodh}{Durduran
  et~al\mbox{.}}{2010}]%
        {Durduran2010}
\bibfield{author}{\bibinfo{person}{Turgut Durduran}, \bibinfo{person}{Regine
  Choe}, \bibinfo{person}{Wesley~B. Baker}, {and} \bibinfo{person}{Arjun~G.
  Yodh}.} \bibinfo{year}{2010}\natexlab{}.
\newblock \showarticletitle{Diffuse optics for tissue monitoring and
  tomography}.
\newblock \bibinfo{journal}{\emph{Reports on Progress in Physics}}.
\newblock


\bibitem[\protect\citeauthoryear{Dutr{\'e}, Bala, and Bekaert}{Dutr{\'e}
  et~al\mbox{.}}{2006}]%
        {AGI}
\bibfield{author}{\bibinfo{person}{Philip Dutr{\'e}}, \bibinfo{person}{Kavita
  Bala}, {and} \bibinfo{person}{Philippe Bekaert}.}
  \bibinfo{year}{2006}\natexlab{}.
\newblock \bibinfo{booktitle}{\emph{Advanced global illumination}}.
\newblock \bibinfo{publisher}{AK Peters, Ltd.}
\newblock


\bibitem[\protect\citeauthoryear{Erf}{Erf}{1978}]%
        {SpeckleMetrology}
\bibfield{author}{\bibinfo{person}{Robert Erf}.}
  \bibinfo{year}{1978}\natexlab{}.
\newblock \showarticletitle{Speckle Metrology}.
\newblock \bibinfo{publisher}{Elsevier}.
\newblock


\bibitem[\protect\citeauthoryear{Feng, Kane, Lee, and Stone}{Feng
  et~al\mbox{.}}{1988}]%
        {feng1988correlations}
\bibfield{author}{\bibinfo{person}{Shechao Feng}, \bibinfo{person}{Charles
  Kane}, \bibinfo{person}{Patrick~A Lee}, {and} \bibinfo{person}{A~Douglas
  Stone}.} \bibinfo{year}{1988}\natexlab{}.
\newblock \showarticletitle{Correlations and fluctuations of coherent wave
  transmission through disordered media}.
\newblock \bibinfo{journal}{\emph{Physical review letters}}
  \bibinfo{volume}{61}, \bibinfo{number}{7}, \bibinfo{pages}{834}.
\newblock


\bibitem[\protect\citeauthoryear{Fienup}{Fienup}{1982}]%
        {Fienup:82}
\bibfield{author}{\bibinfo{person}{James~R. Fienup}.}
  \bibinfo{year}{1982}\natexlab{}.
\newblock \showarticletitle{Phase retrieval algorithms: a comparison}.
\newblock \bibinfo{journal}{\emph{Appl. Opt.}} \bibinfo{volume}{21},
  \bibinfo{number}{15}, \bibinfo{pages}{2758--2769}.
\newblock


\bibitem[\protect\citeauthoryear{Freund}{Freund}{1990}]%
        {FREUND199049}
\bibfield{author}{\bibinfo{person}{Isaac Freund}.}
  \bibinfo{year}{1990}\natexlab{}.
\newblock \showarticletitle{Looking through walls and around corners}.
\newblock \bibinfo{journal}{\emph{Physica: Statistical Mechanics and its App.}}
\newblock


\bibitem[\protect\citeauthoryear{Freund and Eliyahu}{Freund and
  Eliyahu}{1992}]%
        {Freund92}
\bibfield{author}{\bibinfo{person}{Isaac Freund} {and} \bibinfo{person}{Danny
  Eliyahu}.} \bibinfo{year}{1992}\natexlab{}.
\newblock \showarticletitle{Surface correlations in multiple-scattering media}.
\newblock \bibinfo{journal}{\emph{Phys Rev A}} (\bibinfo{year}{1992}).
\newblock


\bibitem[\protect\citeauthoryear{Freund, Rosenbluh, and Feng}{Freund
  et~al\mbox{.}}{1988}]%
        {PhysRevLett.61.2328}
\bibfield{author}{\bibinfo{person}{Isaac Freund}, \bibinfo{person}{Michael
  Rosenbluh}, {and} \bibinfo{person}{Shechao. Feng}.}
  \bibinfo{year}{1988}\natexlab{}.
\newblock \showarticletitle{Memory Effects in Propagation of Optical Waves
  through Disordered Media}.
\newblock \bibinfo{journal}{\emph{Phys. Rev. Lett.}}  \bibinfo{volume}{61},
  \bibinfo{pages}{2328--2331}.
\newblock
Issue 20.
\urldef\tempurl%
\url{https://doi.org/10.1103/PhysRevLett.61.2328}
\showDOI{\tempurl}


\bibitem[\protect\citeauthoryear{Fried}{Fried}{1982}]%
        {Fried:82}
\bibfield{author}{\bibinfo{person}{David~L. Fried}.}
  \bibinfo{year}{1982}\natexlab{}.
\newblock \showarticletitle{Anisoplanatism in adaptive optics}.
\newblock \bibinfo{journal}{\emph{J. Opt. Soc. Am.}} \bibinfo{volume}{72},
  \bibinfo{number}{1}, \bibinfo{pages}{52--61}.
\newblock
\urldef\tempurl%
\url{https://doi.org/10.1364/JOSA.72.000052}
\showDOI{\tempurl}


\bibitem[\protect\citeauthoryear{Frisvad, Christensen, and Jensen}{Frisvad
  et~al\mbox{.}}{2007}]%
        {FCJ07}
\bibfield{author}{\bibinfo{person}{Jeppe~Revall Frisvad},
  \bibinfo{person}{Niels~Jørgen Christensen}, {and}
  \bibinfo{person}{Henrik~Wann Jensen}.} \bibinfo{year}{2007}\natexlab{}.
\newblock \showarticletitle{Computing the scattering properties of
  participating media using {L}orenz-{M}ie theory}.
\newblock \bibinfo{journal}{\emph{SIGGRAPH}}.
\newblock


\bibitem[\protect\citeauthoryear{Gkioulekas, Levin, and Zickler}{Gkioulekas
  et~al\mbox{.}}{2016}]%
        {InvRenderingECCV16}
\bibfield{author}{\bibinfo{person}{Ioannis Gkioulekas}, \bibinfo{person}{Anat
  Levin}, {and} \bibinfo{person}{Todd Zickler}.}
  \bibinfo{year}{2016}\natexlab{}.
\newblock \bibinfo{title}{An Evaluation of Computational Imaging Techniques for
  Heterogeneous Inverse Scattering}.
\newblock
\newblock


\bibitem[\protect\citeauthoryear{Gkioulekas, Zhao, Bala, Zickler, and
  Levin}{Gkioulekas et~al\mbox{.}}{2013}]%
        {InverseRendSA13}
\bibfield{author}{\bibinfo{person}{I. Gkioulekas}, \bibinfo{person}{S. Zhao},
  \bibinfo{person}{K. Bala}, \bibinfo{person}{T. Zickler}, {and}
  \bibinfo{person}{A. Levin}.} \bibinfo{year}{2013}\natexlab{}.
\newblock \showarticletitle{Inverse Volume Rendering with Material
  Dictionaries}.
\newblock \bibinfo{journal}{\emph{ACM Transactions on Graphics (Proc. ACM
  SIGGRAPH Asia)}} (\bibinfo{year}{2013}).
\newblock


\bibitem[\protect\citeauthoryear{Goldburg}{Goldburg}{1999}]%
        {DynamicLightScattering}
\bibfield{author}{\bibinfo{person}{W.~I. Goldburg}.}
  \bibinfo{year}{1999}\natexlab{}.
\newblock \showarticletitle{Dynamic light scattering}.
\newblock \bibinfo{journal}{\emph{American Journal of Physics}}
  (\bibinfo{year}{1999}).
\newblock


\bibitem[\protect\citeauthoryear{Goodman}{Goodman}{2007}]%
        {GoodmanSpeckle}
\bibfield{author}{\bibinfo{person}{Goodman}.} \bibinfo{year}{2007}\natexlab{}.
\newblock \bibinfo{booktitle}{\emph{Speckle Phenomena in Optics: Theory and
  Applications}}.
\newblock \bibinfo{publisher}{Roberts and Company Pub.}
\newblock


\bibitem[\protect\citeauthoryear{Holodovski, Schechner, Levin, Levis, and
  Aides}{Holodovski et~al\mbox{.}}{2016}]%
        {VadimICCP}
\bibfield{author}{\bibinfo{person}{Vadim Holodovski}, \bibinfo{person}{Yoav~Y.
  Schechner}, \bibinfo{person}{Anat Levin}, \bibinfo{person}{Aviad Levis},
  {and} \bibinfo{person}{Amit Aides}.} \bibinfo{year}{2016}\natexlab{}.
\newblock \showarticletitle{In-situ multi-view multi-scattering stochastic
  tomography}. In \bibinfo{booktitle}{\emph{ICCP}}.
\newblock


\bibitem[\protect\citeauthoryear{Ilyushin}{Ilyushin}{2012}]%
        {Ilyushin2012}
\bibfield{author}{\bibinfo{person}{Y.A. Ilyushin}.}
  \bibinfo{year}{2012}\natexlab{}.
\newblock \showarticletitle{Coherent backscattering enhancement in highly
  anisotropically scattering media: Numerical solution}.
\newblock \bibinfo{journal}{\emph{Journal of Quantitative Spectroscopy and
  Radiative Transfer}} (\bibinfo{year}{2012}).
\newblock


\bibitem[\protect\citeauthoryear{Ishimaru}{Ishimaru}{1999}]%
        {ishimaru1999wave}
\bibfield{author}{\bibinfo{person}{Akira Ishimaru}.}
  \bibinfo{year}{1999}\natexlab{}.
\newblock \bibinfo{booktitle}{\emph{Wave propagation and scattering in random
  media}}. Vol.~\bibinfo{volume}{12}.
\newblock \bibinfo{publisher}{John Wiley \& Sons}.
\newblock


\bibitem[\protect\citeauthoryear{Jacquot and Fournier}{Jacquot and
  Fournier}{2000}]%
        {IntSpeckleLight}
\bibfield{author}{\bibinfo{person}{P. Jacquot} {and} \bibinfo{person}{J.~M.
  Fournier}.} \bibinfo{year}{2000}\natexlab{}.
\newblock \showarticletitle{Interferometry in Speckle Light}.
\newblock \bibinfo{publisher}{Springer}.
\newblock


\bibitem[\protect\citeauthoryear{Jacquot and Rastogi}{Jacquot and
  Rastogi}{1979}]%
        {Jacquot:79}
\bibfield{author}{\bibinfo{person}{Pierre Jacquot} {and}
  \bibinfo{person}{Pramod~K. Rastogi}.} \bibinfo{year}{1979}\natexlab{}.
\newblock \showarticletitle{Speckle motions induced by rigid-body movements in
  free-space geometry: an explicit investigation and extension to new cases}.
\newblock \bibinfo{journal}{\emph{Appl. Opt.}} (\bibinfo{year}{1979}).
\newblock


\bibitem[\protect\citeauthoryear{Jakobsen, Yura, and Hanson}{Jakobsen
  et~al\mbox{.}}{2012}]%
        {Jakobsen2012}
\bibfield{author}{\bibinfo{person}{M.~L. Jakobsen}, \bibinfo{person}{H.~T.
  Yura}, {and} \bibinfo{person}{S.~G. Hanson}.}
  \bibinfo{year}{2012}\natexlab{}.
\newblock \showarticletitle{Spatial filtering velocimetry of objective speckles
  for measuring out-of-plane motion}.
\newblock \bibinfo{journal}{\emph{Appl. Opt.}} (\bibinfo{year}{2012}).
\newblock


\bibitem[\protect\citeauthoryear{Jarabo, Aliaga, and Gutierrez}{Jarabo
  et~al\mbox{.}}{2018}]%
        {Jarabo2018radiative}
\bibfield{author}{\bibinfo{person}{Adrian Jarabo}, \bibinfo{person}{Carlos
  Aliaga}, {and} \bibinfo{person}{Diego Gutierrez}.}
  \bibinfo{year}{2018}\natexlab{}.
\newblock \showarticletitle{A Radiative Transfer Framework for
  Spatially-Correlated Materials}.
\newblock \bibinfo{journal}{\emph{ACM Transactions on Graphics}}
  \bibinfo{volume}{37}, \bibinfo{number}{4} (\bibinfo{year}{2018}).
\newblock


\bibitem[\protect\citeauthoryear{Katz, Heidmann, Fink, and Gigan}{Katz
  et~al\mbox{.}}{2014}]%
        {Katz2014}
\bibfield{author}{\bibinfo{person}{O. Katz}, \bibinfo{person}{P. Heidmann},
  \bibinfo{person}{M. Fink}, {and} \bibinfo{person}{S. Gigan}.}
  \bibinfo{year}{2014}\natexlab{}.
\newblock \showarticletitle{Non-invasive single-shot imaging through scattering
  layers and around corners via speckle correlation}.
\newblock \bibinfo{journal}{\emph{Nat. Photonics}} (\bibinfo{year}{2014}).
\newblock


\bibitem[\protect\citeauthoryear{Katz, Small, and Silberberg}{Katz
  et~al\mbox{.}}{2012}]%
        {Katz2012}
\bibfield{author}{\bibinfo{person}{O. Katz}, \bibinfo{person}{E. Small}, {and}
  \bibinfo{person}{Y. Silberberg}.} \bibinfo{year}{2012}\natexlab{}.
\newblock \showarticletitle{Looking around corners and through thin turbid
  layers in real time with scattered incoherent light}.
\newblock \bibinfo{journal}{\emph{Nature}} (\bibinfo{year}{2012}).
\newblock


\bibitem[\protect\citeauthoryear{Kaufmann}{Kaufmann}{2011}]%
        {AdvSpeckleMetrology}
\bibfield{author}{\bibinfo{person}{Guillermo~H. Kaufmann}.}
  \bibinfo{year}{2011}\natexlab{}.
\newblock \showarticletitle{Advances in Speckle Metrology and Related
  Techniques}.
\newblock \bibinfo{publisher}{Wiley}.
\newblock


\bibitem[\protect\citeauthoryear{Levis, Schechner, Aides, and Davis}{Levis
  et~al\mbox{.}}{2015}]%
        {AviadICCV}
\bibfield{author}{\bibinfo{person}{Aviad Levis}, \bibinfo{person}{Yoav~Y.
  Schechner}, \bibinfo{person}{Amit Aides}, {and} \bibinfo{person}{Anthony~B.
  Davis}.} \bibinfo{year}{2015}\natexlab{}.
\newblock \showarticletitle{Airborne three-dimensional cloud tomography}. In
  \bibinfo{booktitle}{\emph{ICCV}}.
\newblock


\bibitem[\protect\citeauthoryear{Li and Genack}{Li and Genack}{1994}]%
        {PhysRevE.49.4530}
\bibfield{author}{\bibinfo{person}{J.~H. Li} {and} \bibinfo{person}{A.~Z.
  Genack}.} \bibinfo{year}{1994}\natexlab{}.
\newblock \showarticletitle{Correlation in laser speckle}.
\newblock \bibinfo{journal}{\emph{Phys. Rev. E}}  \bibinfo{volume}{49}
  (\bibinfo{date}{May} \bibinfo{year}{1994}), \bibinfo{pages}{4530--4533}.
\newblock
Issue 5.
\urldef\tempurl%
\url{https://doi.org/10.1103/PhysRevE.49.4530}
\showDOI{\tempurl}


\bibitem[\protect\citeauthoryear{Lu, Gan, Gu, and Luo}{Lu
  et~al\mbox{.}}{2004}]%
        {lu2004monte}
\bibfield{author}{\bibinfo{person}{Qiang Lu}, \bibinfo{person}{Xiaosong Gan},
  \bibinfo{person}{Min Gu}, {and} \bibinfo{person}{Qingming Luo}.}
  \bibinfo{year}{2004}\natexlab{}.
\newblock \showarticletitle{Monte Carlo modeling of optical coherence
  tomography imaging through turbid media}.
\newblock \bibinfo{journal}{\emph{Applied optics}} \bibinfo{volume}{43},
  \bibinfo{number}{8} (\bibinfo{year}{2004}), \bibinfo{pages}{1628--1637}.
\newblock


\bibitem[\protect\citeauthoryear{Meng, Papas, Habel, Dachsbacher, Marschner,
  Gross, and Jarosz}{Meng et~al\mbox{.}}{2015}]%
        {meng2015multi}
\bibfield{author}{\bibinfo{person}{Johannes Meng}, \bibinfo{person}{Marios
  Papas}, \bibinfo{person}{Ralf Habel}, \bibinfo{person}{Carsten Dachsbacher},
  \bibinfo{person}{Steve Marschner}, \bibinfo{person}{Markus~H Gross}, {and}
  \bibinfo{person}{Wojciech Jarosz}.} \bibinfo{year}{2015}\natexlab{}.
\newblock \showarticletitle{Multi-scale modeling and rendering of granular
  materials.}
\newblock \bibinfo{journal}{\emph{ACM Trans. Graph.}} \bibinfo{volume}{34},
  \bibinfo{number}{4} (\bibinfo{year}{2015}), \bibinfo{pages}{49--1}.
\newblock


\bibitem[\protect\citeauthoryear{Mesradi, Genoux, Cuplov, Haidar, Jan, Buvat,
  and Pain}{Mesradi et~al\mbox{.}}{2013}]%
        {mesradi:hal-01316109}
\bibfield{author}{\bibinfo{person}{M. Mesradi}, \bibinfo{person}{A. Genoux},
  \bibinfo{person}{V. Cuplov}, \bibinfo{person}{D.~Abi Haidar},
  \bibinfo{person}{S. Jan}, \bibinfo{person}{I. Buvat}, {and}
  \bibinfo{person}{F. Pain}.} \bibinfo{year}{2013}\natexlab{}.
\newblock \showarticletitle{{Experimental and analytical comparative study of
  optical coefficient of fresh and frozen rat tissues}}.
\newblock \bibinfo{journal}{\emph{{Journal of Biomedical Optics}}}
  (\bibinfo{date}{Nov.} \bibinfo{year}{2013}).
\newblock
\urldef\tempurl%
\url{https://doi.org/10.1117/1.JBO.18.11.117010.}
\showDOI{\tempurl}


\bibitem[\protect\citeauthoryear{Mishchenko, Travis, and Lacis}{Mishchenko
  et~al\mbox{.}}{2006}]%
        {mishchenko2006multiple}
\bibfield{author}{\bibinfo{person}{M.I. Mishchenko}, \bibinfo{person}{L.D.
  Travis}, {and} \bibinfo{person}{A.A. Lacis}.}
  \bibinfo{year}{2006}\natexlab{}.
\newblock \bibinfo{booktitle}{\emph{Multiple scattering of light by particles:
  radiative transfer and coherent backscattering}}.
\newblock \bibinfo{publisher}{Cambridge Univ Pr}.
\newblock


\bibitem[\protect\citeauthoryear{Moon, Walter, and Marschner}{Moon
  et~al\mbox{.}}{2007}]%
        {moon2007rendering}
\bibfield{author}{\bibinfo{person}{Jonathan~T Moon}, \bibinfo{person}{Bruce
  Walter}, {and} \bibinfo{person}{Stephen~R Marschner}.}
  \bibinfo{year}{2007}\natexlab{}.
\newblock \showarticletitle{Rendering discrete random media using precomputed
  scattering solutions}. In \bibinfo{booktitle}{\emph{Proceedings of the 18th
  Eurographics conference on Rendering Techniques}}. Eurographics Association,
  \bibinfo{pages}{231--242}.
\newblock


\bibitem[\protect\citeauthoryear{Mosk, Lagendijk, Lerosey, and Fink}{Mosk
  et~al\mbox{.}}{2013}]%
        {Mosk2012}
\bibfield{author}{\bibinfo{person}{Allard~P. Mosk}, \bibinfo{person}{Ad
  Lagendijk}, \bibinfo{person}{Geoffroy Lerosey}, {and}
  \bibinfo{person}{Mathias Fink}.} \bibinfo{year}{2013}\natexlab{}.
\newblock \showarticletitle{Controlling waves in space and time for imaging and
  focusing in complex media}.
\newblock \bibinfo{journal}{\emph{Nat. Photonics}} (\bibinfo{year}{2013}).
\newblock


\bibitem[\protect\citeauthoryear{M{\"u}ller, Papas, Gross, Jarosz, and
  Nov{\'a}k}{M{\"u}ller et~al\mbox{.}}{2016}]%
        {muller2016efficient}
\bibfield{author}{\bibinfo{person}{Thomas M{\"u}ller}, \bibinfo{person}{Marios
  Papas}, \bibinfo{person}{Markus~H Gross}, \bibinfo{person}{Wojciech Jarosz},
  {and} \bibinfo{person}{Jan Nov{\'a}k}.} \bibinfo{year}{2016}\natexlab{}.
\newblock \showarticletitle{Efficient rendering of heterogeneous polydisperse
  granular media.}
\newblock \bibinfo{journal}{\emph{ACM Trans. Graph.}} \bibinfo{volume}{35},
  \bibinfo{number}{6} (\bibinfo{year}{2016}), \bibinfo{pages}{168--1}.
\newblock


\bibitem[\protect\citeauthoryear{Narasimhan, Gupta, Donner, Ramamoorthi, Nayar,
  and Jensen}{Narasimhan et~al\mbox{.}}{2006}]%
        {narasimhan2006acquiring}
\bibfield{author}{\bibinfo{person}{S.G. Narasimhan}, \bibinfo{person}{M.
  Gupta}, \bibinfo{person}{C. Donner}, \bibinfo{person}{R. Ramamoorthi},
  \bibinfo{person}{S.K. Nayar}, {and} \bibinfo{person}{H.W. Jensen}.}
  \bibinfo{year}{2006}\natexlab{}.
\newblock \showarticletitle{Acquiring scattering properties of participating
  media by dilution}.
\newblock \bibinfo{journal}{\emph{ACM Trans. Graph.}} \bibinfo{volume}{25},
  \bibinfo{number}{3} (\bibinfo{year}{2006}).
\newblock


\bibitem[\protect\citeauthoryear{Nixon, Katz, Small, Bromberg, Friesem,
  Silberberg, and Davidson}{Nixon et~al\mbox{.}}{2013}]%
        {Nixon2013}
\bibfield{author}{\bibinfo{person}{Micha Nixon}, \bibinfo{person}{Ori Katz},
  \bibinfo{person}{Eran Small}, \bibinfo{person}{Yaron Bromberg},
  \bibinfo{person}{Asher~A. Friesem}, \bibinfo{person}{Yaron Silberberg}, {and}
  \bibinfo{person}{Nir Davidson}.} \bibinfo{year}{2013}\natexlab{}.
\newblock \showarticletitle{Real-time wavefront shaping through scattering
  media by all-optical feedback}.
\newblock \bibinfo{journal}{\emph{Nat. Photonics}} (\bibinfo{year}{2013}).
\newblock


\bibitem[\protect\citeauthoryear{Osnabrugge, Horstmeyer, Papadopoulos,
  Judkewitz, and Vellekoop}{Osnabrugge et~al\mbox{.}}{2017}]%
        {GenOptMemory17}
\bibfield{author}{\bibinfo{person}{Gerwin Osnabrugge}, \bibinfo{person}{Roarke
  Horstmeyer}, \bibinfo{person}{Ioannis~N. Papadopoulos},
  \bibinfo{person}{Benjamin Judkewitz}, {and} \bibinfo{person}{Ivo~M.
  Vellekoop}.} \bibinfo{year}{2017}\natexlab{}.
\newblock \showarticletitle{Generalized optical memory effect}.
\newblock \bibinfo{journal}{\emph{Optica}} (\bibinfo{year}{2017}).
\newblock


\bibitem[\protect\citeauthoryear{Pan, Birngruber, Rosperich, and
  Engelhardt}{Pan et~al\mbox{.}}{1995}]%
        {pan1995low}
\bibfield{author}{\bibinfo{person}{Yingtian Pan}, \bibinfo{person}{Reginald
  Birngruber}, \bibinfo{person}{J{\"u}rgen Rosperich}, {and}
  \bibinfo{person}{Ralf Engelhardt}.} \bibinfo{year}{1995}\natexlab{}.
\newblock \showarticletitle{Low-coherence optical tomography in turbid tissue:
  theoretical analysis}.
\newblock \bibinfo{journal}{\emph{Applied optics}} \bibinfo{volume}{34},
  \bibinfo{number}{28} (\bibinfo{year}{1995}), \bibinfo{pages}{6564--6574}.
\newblock


\bibitem[\protect\citeauthoryear{Pierrat, Greffet, Carminati, and
  Elaloufi}{Pierrat et~al\mbox{.}}{2005}]%
        {Pierrat:05}
\bibfield{author}{\bibinfo{person}{Romain Pierrat},
  \bibinfo{person}{Jean-Jacques Greffet}, \bibinfo{person}{R\'{e}mi Carminati},
  {and} \bibinfo{person}{Rachid Elaloufi}.} \bibinfo{year}{2005}\natexlab{}.
\newblock \showarticletitle{Spatial coherence in strongly scattering media}.
\newblock \bibinfo{journal}{\emph{J. Opt. Soc. Am. A}} \bibinfo{volume}{22},
  \bibinfo{number}{11} (\bibinfo{date}{Nov} \bibinfo{year}{2005}),
  \bibinfo{pages}{2329--2337}.
\newblock


\bibitem[\protect\citeauthoryear{Pine, Weitz, Chaikin, and Herbolzheimer}{Pine
  et~al\mbox{.}}{1988}]%
        {pine1988diffusing}
\bibfield{author}{\bibinfo{person}{DJ Pine}, \bibinfo{person}{DA Weitz},
  \bibinfo{person}{PM Chaikin}, {and} \bibinfo{person}{E Herbolzheimer}.}
  \bibinfo{year}{1988}\natexlab{}.
\newblock \showarticletitle{Diffusing wave spectroscopy}.
\newblock \bibinfo{journal}{\emph{Physical review letters}}
  \bibinfo{volume}{60}, \bibinfo{number}{12} (\bibinfo{year}{1988}),
  \bibinfo{pages}{1134}.
\newblock


\bibitem[\protect\citeauthoryear{Sawicki, Kastor, and Xu}{Sawicki
  et~al\mbox{.}}{2008}]%
        {Sawicki:08}
\bibfield{author}{\bibinfo{person}{John Sawicki}, \bibinfo{person}{Nikolas
  Kastor}, {and} \bibinfo{person}{Min Xu}.} \bibinfo{year}{2008}\natexlab{}.
\newblock \showarticletitle{Electric field Monte Carlo simulation of coherent
  backscattering of polarized light by a turbid medium containing Mie
  scatterers}.
\newblock \bibinfo{journal}{\emph{Opt. Express}} \bibinfo{volume}{16},
  \bibinfo{number}{8} (\bibinfo{date}{Apr} \bibinfo{year}{2008}),
  \bibinfo{pages}{5728--5738}.
\newblock


\bibitem[\protect\citeauthoryear{Schmitt and Kn{\"u}ttel}{Schmitt and
  Kn{\"u}ttel}{1997}]%
        {schmitt1997model}
\bibfield{author}{\bibinfo{person}{Joseph~M Schmitt} {and} \bibinfo{person}{A
  Kn{\"u}ttel}.} \bibinfo{year}{1997}\natexlab{}.
\newblock \showarticletitle{Model of optical coherence tomography of
  heterogeneous tissue}.
\newblock \bibinfo{journal}{\emph{JOSA A}} \bibinfo{volume}{14},
  \bibinfo{number}{6} (\bibinfo{year}{1997}), \bibinfo{pages}{1231--1242}.
\newblock


\bibitem[\protect\citeauthoryear{Schott, Bertolotti, L\'{e}ger, Bourdieu, and
  Gigan}{Schott et~al\mbox{.}}{2015}]%
        {Schott:15}
\bibfield{author}{\bibinfo{person}{Schott}, \bibinfo{person}{Bertolotti},
  \bibinfo{person}{L\'{e}ger}, \bibinfo{person}{Bourdieu}, {and}
  \bibinfo{person}{Gigan}.} \bibinfo{year}{2015}\natexlab{}.
\newblock \showarticletitle{Characterization of the angular memory effect of
  scattered light in biological tissues}.
\newblock \bibinfo{journal}{\emph{Opt. Express}} (\bibinfo{year}{2015}).
\newblock


\bibitem[\protect\citeauthoryear{Shen, Sukhov, and Dogariu}{Shen
  et~al\mbox{.}}{2017}]%
        {Shen:17}
\bibfield{author}{\bibinfo{person}{Zhean Shen}, \bibinfo{person}{Sergey
  Sukhov}, {and} \bibinfo{person}{Aristide Dogariu}.}
  \bibinfo{year}{2017}\natexlab{}.
\newblock \showarticletitle{Monte Carlo method to model optical coherence
  propagation in random media}.
\newblock \bibinfo{journal}{\emph{J. Opt. Soc. Am. A}} \bibinfo{volume}{34},
  \bibinfo{number}{12} (\bibinfo{date}{Dec} \bibinfo{year}{2017}),
  \bibinfo{pages}{2189--2193}.
\newblock
\urldef\tempurl%
\url{https://doi.org/10.1364/JOSAA.34.002189}
\showDOI{\tempurl}


\bibitem[\protect\citeauthoryear{Smith, Desai, Agarwal, and Gupta}{Smith
  et~al\mbox{.}}{2017}]%
        {Smith:2017:Speckle}
\bibfield{author}{\bibinfo{person}{Brandon~M. Smith}, \bibinfo{person}{Pratham
  Desai}, \bibinfo{person}{Vishal Agarwal}, {and} \bibinfo{person}{Mohit
  Gupta}.} \bibinfo{year}{2017}\natexlab{}.
\newblock \showarticletitle{CoLux: Multi-object 3D Micro-motion Analysis Using
  Speckle Imaging}.
\newblock \bibinfo{journal}{\emph{ACM Trans. Graph.}} (\bibinfo{year}{2017}).
\newblock


\bibitem[\protect\citeauthoryear{Stam}{Stam}{1999}]%
        {Stam:1999:DS}
\bibfield{author}{\bibinfo{person}{J. Stam}.} \bibinfo{year}{1999}\natexlab{}.
\newblock \showarticletitle{Diffraction shaders}. In
  \bibinfo{booktitle}{\emph{SIGGRAPH}}.
\newblock


\bibitem[\protect\citeauthoryear{Sun, Ramamoorthi, Narasimhan, and Nayar}{Sun
  et~al\mbox{.}}{2005}]%
        {sun2005practical}
\bibfield{author}{\bibinfo{person}{Bo Sun}, \bibinfo{person}{Ravi Ramamoorthi},
  \bibinfo{person}{Srinivasa~G Narasimhan}, {and} \bibinfo{person}{Shree~K
  Nayar}.} \bibinfo{year}{2005}\natexlab{}.
\newblock \showarticletitle{A practical analytic single scattering model for
  real time rendering}. In \bibinfo{booktitle}{\emph{ACM Transactions on
  Graphics (TOG)}}, Vol.~\bibinfo{volume}{24}. ACM,
  \bibinfo{pages}{1040--1049}.
\newblock


\bibitem[\protect\citeauthoryear{Thierry, Antoine, Chniti, and
  Alzubaidi}{Thierry et~al\mbox{.}}{2015}]%
        {mudiff}
\bibfield{author}{\bibinfo{person}{B. Thierry}, \bibinfo{person}{X. Antoine},
  \bibinfo{person}{C. Chniti}, {and} \bibinfo{person}{H. Alzubaidi}.}
  \bibinfo{year}{2015}\natexlab{}.
\newblock \showarticletitle{$\mu$-diff: An open-source Matlab toolbox for
  computing multiple scattering problems by disks}.
\newblock \bibinfo{journal}{\emph{Computer Physics Communications}}
  \bibinfo{volume}{192} (\bibinfo{year}{2015}), \bibinfo{pages}{348 -- 362}.
\newblock
\showISSN{0010-4655}
\urldef\tempurl%
\url{https://doi.org/10.1016/j.cpc.2015.03.013}
\showDOI{\tempurl}


\bibitem[\protect\citeauthoryear{Treeby and Cox.}{Treeby and Cox.}{2010}]%
        {kWaves}
\bibfield{author}{\bibinfo{person}{B.~E. Treeby} {and} \bibinfo{person}{B.~T.
  Cox.}} \bibinfo{year}{2010}\natexlab{}.
\newblock \showarticletitle{k-Wave: MATLAB toolbox for the simulation and
  reconstruction of photoacoustic wave-fields,}.
\newblock \bibinfo{journal}{\emph{J. Biomed. Opt.}} (\bibinfo{year}{2010}).
\newblock


\bibitem[\protect\citeauthoryear{Twersky}{Twersky}{1964}]%
        {Twersky64}
\bibfield{author}{\bibinfo{person}{V. Twersky}.}
  \bibinfo{year}{1964}\natexlab{}.
\newblock \showarticletitle{On propagation in random media of discrete
  scatterers}.
\newblock \bibinfo{journal}{\emph{Am. Math. Sot. Symp. Stochastic Processes in
  Mathematical Physics and Engineering, Vol. 16, p. 84}}
  (\bibinfo{year}{1964}).
\newblock


\bibitem[\protect\citeauthoryear{Veach}{Veach}{1997}]%
        {veach1997robust}
\bibfield{author}{\bibinfo{person}{E. Veach}.} \bibinfo{year}{1997}\natexlab{}.
\newblock \emph{\bibinfo{title}{Robust Monte Carlo methods for light transport
  simulation}}.
\newblock \bibinfo{thesistype}{Ph.D. Dissertation}. \bibinfo{school}{PhD
  thesis, Stanford University}.
\newblock


\bibitem[\protect\citeauthoryear{Vellekoop and Aegerter}{Vellekoop and
  Aegerter}{2010}]%
        {Vellekoop:10}
\bibfield{author}{\bibinfo{person}{Ivo~M. Vellekoop} {and}
  \bibinfo{person}{Christof~M. Aegerter}.} \bibinfo{year}{2010}\natexlab{}.
\newblock \showarticletitle{Scattered light fluorescence microscopy: imaging
  through turbid layers}.
\newblock \bibinfo{journal}{\emph{Opt. Lett.}} \bibinfo{volume}{35},
  \bibinfo{number}{8} (\bibinfo{date}{Apr} \bibinfo{year}{2010}),
  \bibinfo{pages}{1245--1247}.
\newblock
\urldef\tempurl%
\url{https://doi.org/10.1364/OL.35.001245}
\showDOI{\tempurl}


\bibitem[\protect\citeauthoryear{Walter, Zhao, Holzschuch, and Bala}{Walter
  et~al\mbox{.}}{2009}]%
        {walter2009single}
\bibfield{author}{\bibinfo{person}{Bruce Walter}, \bibinfo{person}{Shuang
  Zhao}, \bibinfo{person}{Nicolas Holzschuch}, {and} \bibinfo{person}{Kavita
  Bala}.} \bibinfo{year}{2009}\natexlab{}.
\newblock \showarticletitle{Single scattering in refractive media with triangle
  mesh boundaries}. In \bibinfo{booktitle}{\emph{ACM Transactions on Graphics
  (TOG)}}, Vol.~\bibinfo{volume}{28}. ACM, \bibinfo{pages}{92}.
\newblock


\bibitem[\protect\citeauthoryear{Werner, Velinov, Jakob, and Hullin}{Werner
  et~al\mbox{.}}{2017}]%
        {Werner17}
\bibfield{author}{\bibinfo{person}{Sebastian Werner}, \bibinfo{person}{Zdravko
  Velinov}, \bibinfo{person}{Wenzel Jakob}, {and} \bibinfo{person}{Matthias~B.
  Hullin}.} \bibinfo{year}{2017}\natexlab{}.
\newblock \showarticletitle{Scratch Iridescence: Wave-Optical Rendering of
  Diffractive Surface Structure}.
\newblock \bibinfo{journal}{\emph{ACM SIGGRAPH Asia}} (\bibinfo{year}{2017}).
\newblock


\bibitem[\protect\citeauthoryear{Wyman, Patterson, and Wilson}{Wyman
  et~al\mbox{.}}{1989}]%
        {wyman1989similarity}
\bibfield{author}{\bibinfo{person}{Douglas~R Wyman}, \bibinfo{person}{Michael~S
  Patterson}, {and} \bibinfo{person}{Brian~C Wilson}.}
  \bibinfo{year}{1989}\natexlab{}.
\newblock \showarticletitle{Similarity relations for anisotropic scattering in
  Monte Carlo simulations of deeply penetrating neutral particles}.
\newblock \bibinfo{journal}{\emph{J. Comput. Phys.}} \bibinfo{volume}{81},
  \bibinfo{number}{1} (\bibinfo{year}{1989}), \bibinfo{pages}{137--150}.
\newblock


\bibitem[\protect\citeauthoryear{Xu}{Xu}{2004}]%
        {Xu:04}
\bibfield{author}{\bibinfo{person}{Min Xu}.} \bibinfo{year}{2004}\natexlab{}.
\newblock \showarticletitle{Electric field Monte Carlo simulation of polarized
  light propagation in turbid media}.
\newblock \bibinfo{journal}{\emph{Opt. Express}} \bibinfo{volume}{12},
  \bibinfo{number}{26} (\bibinfo{date}{Dec} \bibinfo{year}{2004}),
  \bibinfo{pages}{6530--6539}.
\newblock


\bibitem[\protect\citeauthoryear{Yan, Ha{\v{s}}an, Walter, Marschner, and
  Ramamoorthi}{Yan et~al\mbox{.}}{2018}]%
        {yan2018rendering}
\bibfield{author}{\bibinfo{person}{Ling-Qi Yan}, \bibinfo{person}{Milo{\v{s}}
  Ha{\v{s}}an}, \bibinfo{person}{Bruce Walter}, \bibinfo{person}{Steve
  Marschner}, {and} \bibinfo{person}{Ravi Ramamoorthi}.}
  \bibinfo{year}{2018}\natexlab{}.
\newblock \showarticletitle{Rendering specular microgeometry with wave optics}.
\newblock \bibinfo{journal}{\emph{ACM Transactions on Graphics (TOG)}}
  \bibinfo{volume}{37}, \bibinfo{number}{4} (\bibinfo{year}{2018}),
  \bibinfo{pages}{75}.
\newblock


\bibitem[\protect\citeauthoryear{Yang, Pu, and Psaltis}{Yang
  et~al\mbox{.}}{2014}]%
        {Yang:14}
\bibfield{author}{\bibinfo{person}{Xin Yang}, \bibinfo{person}{Ye Pu}, {and}
  \bibinfo{person}{Demetri Psaltis}.} \bibinfo{year}{2014}\natexlab{}.
\newblock \showarticletitle{Imaging blood cells through scattering biological
  tissue using speckle scanning microscopy}.
\newblock \bibinfo{journal}{\emph{Opt. Express}} \bibinfo{volume}{22},
  \bibinfo{number}{3} (\bibinfo{date}{Feb} \bibinfo{year}{2014}),
  \bibinfo{pages}{3405--3413}.
\newblock
\urldef\tempurl%
\url{https://doi.org/10.1364/OE.22.003405}
\showDOI{\tempurl}


\bibitem[\protect\citeauthoryear{Yee}{Yee}{1966}]%
        {FDTDYee66}
\bibfield{author}{\bibinfo{person}{K. Yee}.} \bibinfo{year}{1966}\natexlab{}.
\newblock \showarticletitle{Numerical solution of initial boundary value
  problems involving Maxwell's equations in isotropic media}.
\newblock \bibinfo{journal}{\emph{EEE Trans. on Antennas and Propagation}}
  (\bibinfo{year}{1966}).
\newblock


\bibitem[\protect\citeauthoryear{Yeh, Mehra, Ren, Antani, Manocha, and Lin}{Yeh
  et~al\mbox{.}}{2013}]%
        {Yeh:2013:WCI:2508363.2508420}
\bibfield{author}{\bibinfo{person}{Hengchin Yeh}, \bibinfo{person}{Ravish
  Mehra}, \bibinfo{person}{Zhimin Ren}, \bibinfo{person}{Lakulish Antani},
  \bibinfo{person}{Dinesh Manocha}, {and} \bibinfo{person}{Ming Lin}.}
  \bibinfo{year}{2013}\natexlab{}.
\newblock \showarticletitle{Wave-ray Coupling for Interactive Sound Propagation
  in Large Complex Scenes}.
\newblock \bibinfo{journal}{\emph{ACM Trans. Graph.}} (\bibinfo{year}{2013}).
\newblock


\bibitem[\protect\citeauthoryear{Zhao, Ramamoorthi, and Bala}{Zhao
  et~al\mbox{.}}{2014}]%
        {zhao2014high}
\bibfield{author}{\bibinfo{person}{Shuang Zhao}, \bibinfo{person}{Ravi
  Ramamoorthi}, {and} \bibinfo{person}{Kavita Bala}.}
  \bibinfo{year}{2014}\natexlab{}.
\newblock \showarticletitle{High-order similarity relations in radiative
  transfer}.
\newblock \bibinfo{journal}{\emph{ACM Transactions on Graphics (TOG)}}
  \bibinfo{volume}{33}, \bibinfo{number}{4} (\bibinfo{year}{2014}),
  \bibinfo{pages}{104}.
\newblock


\end{thebibliography}
\section{Appendix}
\subsection{Bulk parameters with multiple particle types  }\label{sec:bulkandmean}
For the simplicity of the exposition  our algorithm was derived in the main paper assuming particles of a single type. Here we show that with small adjustments we can handle any mixture of particle types.
We index the particle type with a subscript  $\ptctype$.  We assume we are given as input a possibly spatially varying distribution of the density of each type of particle, denoted by $\materialDensity_{\ptctype}(\ptd)$.
  For each particle type $\ptctype$ we denote its  scattering and absorption  cross sections by $\scatteringCrossSectionPtc,\atnCSPtc$ its radius by $r_\ptctype$, and its  normalized amplitude function by $ \ampf_\ptctype$.\comment{, with $\ampf_\ptctype={\sqrt{\scatteringCrossSectionPtc}}^{-1}\unampf_\ptctype$.} 
  
  We start by defining the  bulk parameters that summarize the density and scattering amplitude of all particle  types using some mean statistics.  
Then we  revisit some equations in the main text that should be readjusted when more than one type of particles is involved.

First let us denote by $\Nmean(\ptd)$ the mean number of particles from type $\ptctype$ in a unit volume. For spherical particles, this can be computed by dividing the density by the volume of a particle with radius $r_\ptctype$
\BE
\Nmean(\ptd)=\frac{\materialDensity_\ptctype(\ptd)}{4/3\pi r_\ptctype^3}
\EE 

 To account for the fact that we  work with normalized  amplitude and phase functions, we define the scattering and absorption coefficients  $\sctCoff,\atnCoff$ as the expected energy scattered/absorbed from a unit sized volume, namely the expected number of particles in a unit volume times the cross section
 \BE
 \sctCoff(\ptd)=\sum_\ptctype \Nmean(\ptd)\scatteringCrossSectionPtc,\;\;\atnCoff(\ptd)=\sum_\ptctype \Nmean(\ptd)\atnCSPtc+ \atnCoff^{\text{med}}
 \EE
 where $\atnCoff^{\text{med}}$ is the attenuation coefficient of the leading medium.
 extinction coefficient $\extCoff$ is defined via $\extCoff=\sctCoff+\atnCoff$.
 
 We define the bulk  outer product of  amplitude functions as
 \BE\label{eq:outprod-S} 
 \ampF(\theta_1,\theta_2)={\sum_\ptctype  \ampm_{\ptctype}  \ampf_{\ptctype}(\theta_1) { \ampf_{\ptctype}(\theta_2)}^{*}} \;\;\text{~~~~with} \;\;\text{~~~~with}\;\; \ampm_\ptctype=\frac{\scatteringCrossSectionPtc \Nmean}  {\sum_\ptctype \scatteringCrossSectionPtc \Nmean}
 \EE
 and as a special case of the above, define the phase function as $\rho(\theta)= \ampF(\theta,\theta)$.
 
 We will also define a bulk amplitude function as
 \BE
 \ampf(\theta)= \sum_\ptctype  \ampm_{\ptctype} \frac{1}{\sqrt{\scatteringCrossSectionPtc}} \ampf_{\ptctype}(\theta)
 \EE
 However, in most cases the product of scattering amplitude at two angles is defined via \equref{eq:outprod-S} and  not as $\ampf(\theta_1)\ampf(\theta_2)$, since two paths scattering at a single point also scatter at a particle of the same type.
 
To see how this quantity folds into the speckle covariance derivation, let us start with the speckle  mean. 

%Here our goal is to prove \equref{eq:mean-ff}, expressing the mean of all %complex speckle fields we can get over all particle positions sampled from %the same density.

\subsection{The speckle mean}\label{sec:app-mean}
  \equref{eq:int-paths-1} for the mean of all paths from a near field point $\ptd_1$ to a point $\ptd_2$ (near field or far field) should be conditioned on the type of particle at $\ptd_1$, and should be expressed as:
\BE\label{eq:mean-nf-ptct}
\int_{\Path^{\inp}_{\snsp}}p(\pathseq)\sigsct(\pathseq)\ud\pathseq =\sqrt{\scatteringCrossSectionPtc}\mulsct_\ptctype(\omgv\rightarrow \ptd_1 \rightarrow \ptd_2)
\EE
where we scale by $\sqrt{\scatteringCrossSectionPtc}$ because, while $\mulsct_\ptctype$ is defined  using the normalized scattering amplitude function, the total energy scattered  by a particle is a function of its size. 

If either the source or the sensor are at the near field (i.e. points rather than directions), the mean of all speckle fields given a source at $\inp$ and a sensor at $\snsp$  is a special case of \equref{eq:mean-nf-ptct},  omitting the  $\ampf$ term representing the direction change
\BE \meanspk^{\inp}_{\snsp}=\mulsct(\inp\sto\snsp)=
\att(\inp,\snsp)\sigsct(\inp\sto \snsp),
\EE
compare with \equref{eq:sigsct-0-def} and \equref{eq:sigsct-def} for the definition of $\sigsct$. 

When both source and sensor are at the far field, we need to integrate \equref{eq:mean-nf-ptct} through all possible scattering points in space.

\begin{claim}  The mean of all paths from a far field source $\ind$ to a far field sensor $\outd$ is
\BE\label{eq:mean-ff-ptct}
\meanspk_{\outd}^{\ind}=\int \sum_{\ptctype} {\Nmean}{\sqrt{\scatteringCrossSectionPtc(\ptd)}} \sigsct_\ptctype(\ind\sto\ptd)\mulsct_\ptctype(\ind\rightarrow\ptd\rightarrow\outd)\ud\ptd
\EE
\end{claim}
As this integral involves the phase of the paths from $\ind$ to $\ptd$ and from $\ptd$ to $\outd$, it reduces quickly if paths with multiple phases are involved and it is non zero mostly for the forward scattering direction $\ind\approx \outd$. 
%In practice for $\ind= \outd$ this is equivalent to the attenuated direct %component and $\meanspk_{\ind}^{\ind}=\att(\ind,\ind)$.\Anat{this requires %a formula with $\mulsct$ rather than $\sigsct$...}
\proof
To justify this equation let us divide the space of all paths from $\ind$ to $\outd$ to sets $L(\ptd)$, defined as the set of all paths whose first scattering event happens at point $\ptd$. To average the contribution of all paths in this group, let us assume that we know point $\ptd$ includes a particle of type $\ptctype$.  Note that the path phase from the beginning at $\ind$ to $\ptd$ is given by $\sigsct_\ptctype(\ind\sto\ptd)$. From $\ptd$ to $\outd$ there are many paths,   since they all start at the same point we can use \equref{eq:mean-nf-ptct} implying that their average contribution is $\sqrt{\scatteringCrossSectionPtc}\mulsct_\ptctype(\ind\rightarrow\ptd\rightarrow\outd)$. Thus the contribution of all paths in the set $L(\ptd)$  integrates to \BE\label{eq:mean-ff-ptcsng}\sqrt{\scatteringCrossSectionPtc}\sigsct_\ptctype(\ind\sto\ptd)\mulsct_\ptctype(\ind\rightarrow\ptd\rightarrow\outd).\EE To get from here to \equref{eq:mean-ff-ptct} we need to weight \equref{eq:mean-ff-ptcsng} by the expected number of particles in a unit volume.   \eop

\subsection{Integrals in path space}\label{sec:path-integrals}

Having derived the mean let us move to the covariance. Our goal here is to derive expectations of path contributions and justify \equref{eq:path-cont-markov}. For this we introduce the notation
 \begin{align}
 &\Pmulsct((\omgv_1,\omgv_2)\rightarrow\ptd_o\rightarrow(\ptd_1,\ptd_2))=\nonumber\\
&\sum_\ptctype \ampm_\ptctype \mulsct_\ptctype(\omgv_1\rightarrow \ptd_o\rightarrow\ptd_1){\mulsct_\ptctype(\omgv_2\rightarrow \ptd_o\rightarrow \ptd_2)}^*,
 \end{align}
%=======
%Having derived the mean let us move to the covariance. Our goal here is to derive expectations of path contributions and justify \equref{eq:path-cont-markov}.

%For this we introduce the notation
%\begin{multline}
% \Pmulsct((\omgv_1,\omgv_2)\to\ptd_o\to(\ptd_1,\ptd_2))=\\ \sum_\ptctype \ampm_\ptctype \mulsct_\ptctype(\omgv_1\to \ptd_o\to\ptd_1){\mulsct_\ptctype(\omgv_2\to \ptd_o\to \ptd_2)}^* \end{multline}
%>>>>>>> 1cd6f0a7a9a111294976de0e9847531ac40a0e90
 with $\ampm_\ptctype$ defined in \equref{eq:outprod-S}, and where $\mulsct_\ptctype$ is the equivalent of $\mulsct$ with an amplitude function of particle type $\ptctype$:
\BE
\mulsct_\ptctype(\omgv_j\sto\ptd_o\sto\ptd_{j})=
\att(\ptd_{o},\ptd_{j})
\phase(\ptd_{o}{\sto\ptd_{j}})\ampf_\ptctype(\omgv_j\cdot\dir{\ptd_{o},{\ptd_{j}}}).
\EE
The term $\Pmulsct((\omgv_1,\omgv_2)\sto\ptd_o\sto(\ptd_1,\ptd_2))$ should replace all terms of the form $\mulsct(\omgv_1\rightarrow\ptd_o\rightarrow\ptd_1){\mulsct(\omgv_2\rightarrow\ptd_o\rightarrow\ptd_2)}^*$ in the definition of the speckle covariance, and the resulting changes to the MC process are summarised in Alg \ref{alg:MCpair-app}. 
%=======
%The term $\Pmulsct((\omgv_1,\omgv_2)\to\ptd_o\to(\ptd_1,\ptd_2))$ should replace all terms of the form $\mulsct(\omgv_1\to\ptd_o\to\ptd_1){\mulsct(\omgv_2\to\ptd_o\to\ptd_2)}^*$ in the definition of the speckle covariance. 
%>>>>>>> 1cd6f0a7a9a111294976de0e9847531ac40a0e90
 Effectively this is encoding the fact that when two paths scatter at the same particle they scatter at a particle from the same type, so the same $\ampf_\ptctype$ should apply to both.

%Using the result of \equref{eq:int-paths-1} we are ready to move path pairs. 

To analyze the path contributions we will divide the space of all path pairs $\pathseq_1,\pathseq_2$ from $\inp_1,\inp_2$ to $\snsp_1,\snsp_2$ to sets defined by the nodes they have in common. Let $\subpath=\{\ptd_1,\ldots,\ptd_B\}$ denote a set of nodes and $\pathseq^{s,P_1},\pathseq^{s,P_2}$ two possibly different permutations of these nodes. We look at the set of all paths that share exactly the nodes in $\pathseq^s$  in the $P_1,P_2$ orders:
\begin{multline}\label{eq:l1l2-disj-seg}
L(\pathseq^{s,P_1},\pathseq^{s,P_2})\!=\\\;\;\;\;\;\;\;\left\{\!\!\begin{array}{@{~}c|c@{}}(\pathseq^1,\pathseq^2)\!\!&\!\!\pathseq^{1}\!=\!\{\inp_1\sto\ldots\sto\ptd_{P_1(1)}\sto\ldots\sto \ptd_{P_1(B)}\sto\ldots\sto\snsp_1\}\\&\!\!\pathseq^2\!=\!\{\inp_2\sto\ldots\sto\ptd_{P_2(1)}\sto\ldots \sto\ptd_{P_2(B)}\sto\ldots\sto\snsp_2\}\end{array}\!\right\}
\end{multline}
where any occurrence of $\ldots$ in \equref{eq:l1l2-disj-seg} can be replaced with  any sequence of nodes, as long as they are different from each other. With this definition we can divide the space of all paths $\pathseq^1,\pathseq^2$ to disjoint sets. Below we argue that the throughput contribution from each  set $L(\pathseq^{s,P_1},\pathseq^{s,P_2})$ averages to the volumetric throughput contribution of the direct paths $\pathseq^{s,P_1},\pathseq^{s,P_2}$.

In the following we use the notation
$b_1^-=P_1^{-1}(b)-1$, $b_1^+=P_1^{-1}(b)+1$ for the nodes before and after $\ptd_b$ in the permuted sequence $P_1$, and similarly for $P_2$,  $b_2^-=P_2^{-1}(b)-1$, $b_2^+=P_2^{-1}(b)+1$ . 
\begin{claim}\label{claim:pathsetsint}
\begin{multline}\label{eq:-path-int-2-ord-sets}
 \int_{(\pathseq^1,\pathseq^2)\in L(\pathseq^{s,P_1},\pathseq^{s,P_2})} p(\pathseq^1,\pathseq^2)\sigsct(\pathseq^1){\sigsct(\pathseq^2)}^*=\\
\prod_{b=0}^{B}\Pmulsct_b(\pathseq^{1,P_1},\pathseq^{1,P_2})\prod_{b=1}^{B}\sctCoff(\ptd_b)\end{multline}  
with        
\BE
\Pmulsct_b(\pathseq^{1,P_1},\pathseq^{2,P_2})=\Pmulsct((\ptd_{b_1^-},\ptd_{b_2^-})\to\ptd_{b}\to(\ptd_{b_1^+},\ptd_{b_2^+}))
\EE  
\end{claim}
\proof
Let us start by drawing an independent set of $B$ nodes $\ptd_1,\ldots,\ptd_B$. According to the target density, the probability for these particles is  the last term of \equref{eq:-path-int-2-ord-sets}, $\prod_{b=1}^{B}\sctCoff(\ptd_b)$.
For each position $\ptd_b$ we draw a particle type  $\ptctype(b)\sim\ampm_\ptctype$.
Given the type of all particles on the paths  we decompose the path probabilities.

Let $L_b$ denote the set of all disjoint paths $(\pathseq^{1,b},\pathseq^{2,b})$  from $\ptd_{P_1(b)}$ to $\ptd_{P_1(b+1)}$ and from  $\ptd_{P_2(b)}$ to $\ptd_{P_2(b+1)}$, and let $\omega_1^b,\omega_2^b$ denote the end direction of $\pathseq^{1,b},\pathseq^{2,b}$ (i.e. the direction at which the last segment is entering $\ptd_{P_1(b+1)}$ or $\ptd_{P_2(b+1)}$). 
While the only constraint on $\pathseq^{1,b},\pathseq^{2,b}$  is that they are disjoint, we will make the approximation that they are independent. \cite{mishchenko2006multiple} shows that the error introduced by this approximation is $o(1/\Nomean)$ where $\Nomean$ is the expected number of particles in the medium. Thus we can write

\BE
p(\pathseq^1,\pathseq^2)=p(\pathseq^{1,0})p(\pathseq^{2,0})\prod_{b=1}^{B} p(\pathseq^{1,b}|\omega^{1,b-1})p(\pathseq^{2,b}|\omega_1^{b-1})
\EE

Using \equref{eq:int-paths-1}:
\BEA\label{eq:path-int-sing-seg}
\int_{L_0}\!\!p(\pathseq^{1,0},\pathseq^{2,0})\sigsct(\pathseq^{1,0}){\sigsct(\pathseq^{2,0})}^*\!\!\!\!\!\!&\!=\!\!\!\!&\!\!\!\!\int_{\pathseq^{1,0}} \!p(\pathseq^{1,0})\sigsct(\pathseq^{1,0})\cdot\!\!{\int_{\pathseq^{1,0}} \!\!p(\pathseq^{2,0})\sigsct(\pathseq^{2,0})}^*\nonumber\\\!\!\!\!&\!=\!&\!\!\! \! \mulsct(\inp_1\to\ptd_{P_1(1)}) {\mulsct(\inp_2\to\ptd_{P_2(1)}) }^*
\EEA
Since all paths in the set $L_0$ integrate to the direct path, we know that the end directions when entering  $\ptd_{P_1(1)},\ptd_{P_2(1)}$ are $\omega_0^1=\dir{\inp_1\ptd_{P_1(1)}}$, $\omega_0^2=\dir{\inp_2\ptd_{P_2(1)}}$. Given the end direction of the first segment we can apply \equref{eq:int-paths-1} to the second segment, and in a similar way, to all successive segments:
%\BEA\label{eq:path-int-sing-seg-2}
\begin{multline}
\label{eq:path-int-sing-seg-2}
\int_{L_b}p(\pathseq^{1,b},\pathseq^{2,b}|\omgv^{1}_{b-1},\omgv^{2}_{b-1})\sigsct(\pathseq^{1,b}){\sigsct(\pathseq^{2,b})}^*=\\\mulsct_{\ptctype(P_1(b))}(\omgv^1_{b-1}\to\ptd_{P_1(b)}\to\ptd_{P_1(b+1)})\\{\mulsct_{\ptctype(P2(b))}(\omgv^2_{b-1}\to\ptd_{P_2(b)}\to\ptd_{P_2(b+1)} ) }^*
%\EEA
\end{multline}
%Applying \equref{eq:path-int-sing-seg} to all segments in the sequence result in \equref{eq:-path-int-2-ord-sets}. 

Concatenating \equpref{eq:path-int-sing-seg}{eq:path-int-sing-seg-2}  assuming the particle position and type is given, and permuting the order of nodes accordingly, provides
\BE\label{eq:ms-prod-cond-pt}
\prod_{b=0}^B \mulsct_{\ptctype(b)}(\ptd_{b_1^-}\to\ptd_{b}\to\ptd_{b_1^+}) { \mulsct_{\ptctype(b)}(\ptd_{b_2^-}\to\ptd_{b}\to,\ptd_{b_2^+})}^* 
\EE
If we now sum \equref{eq:ms-prod-cond-pt} for all all possible assignment  of particle types and consider also the probability of sampling the nodes themselves, we get \equref{eq:-path-int-2-ord-sets}.

\eop

\subsection{Path permutations}\label{sec:permutations}
As mentioned in \secref{cov-path-def}, Claim \ref{claim:pathsetsint} significantly simplifies the path sampling algorithm, since the fact that all paths collapse to their joint nodes allows us to largely reduce the sampling space. However, Claim \ref{claim:pathsetsint} does not imply that the joint paths appear at the same order.
Since every particle instantiation that contains a set of nodes $\ptd_1,\ldots,\ptd_B$ contains all its permutations, paths pairs tracing the same set of nodes at different orders are not independent. 
If we want to derive a MC algorithm which accounts for all permutations, we can use the update rule defined below.

  Let $\pathseq^s=\ptd_1\to\ldots\to\ptd_B$ be a set of $B$ nodes sampled according to some path probability $q$ . 
%Let $\pathseq_1,\pathseq_2$ denote the an extension of $\pathseq_o$ to the start and %end nodes.
Let  $\Lambda=\{P_1,P_2\ldots\}$ be a set of permutations on $B$ entries, and let $\pathseq^{j,P}$ (for $j=1,2$) denote a permuted version of $\pathseq^s$ connected to the start and end nodes
\BE
\pathseq^{j,P}=\{\inp_j\to\ptd_{P(1)}\to\ldots\to \ptd_{P(B)}\to\snsp_j\}
\EE
To account for all permutations, we redefine $c(\subpath)$ from \equref{eq:covariance-contrib} as
\BE c(\subpath)=\sum_{(P_{n_1},P_{n_2})\in \Lambda}c_{\pathseq^{1,P_{n_1}},\pathseq^{2,P_{n_2}}}
\EE and the next event estimation should update the correlation by
\comment{ \BE\label{eq:all-perm}
\frac{\frac{1}{|\Lambda|^2} \sum_{(P_{n_1},P_{n_2})\in \Lambda}c_{\pathseq_1^{P_{n_1}},\pathseq_2^{P_{n_2}}}}{\sum_{P_n\in\Lambda}q(\pathseq_o^{P_n})}
 \EE}
 \BE\label{eq:all-perm}
\frac{c(\subpath)}{\sum_{P_n\in\Lambda}q(\pathseq^{s,P_n})}
 \EE
 where the numerator sums over all path pairs in all permutations, and in the denominator we divide by the probability $q$ of sampling these paths, since for every path we add all permutations, there are multiple ways to sample it. 
In practice the update rules of \equref{eq:cov-app-fb} is a special case of \equref{eq:all-perm} when the permutation set $\Lambda$ was taken to include the identity and reversed permutations.

For materials whose optical depth is low and the average number of scattering events in a path is not high, we could implement the accurate update rule of \equref{eq:all-perm}. For large path lengths $B$, the $B!$ factor is computationally prohibiting. 

As mentioned in the main text, our simulations show that considering the forward and reversed permutations only is accurate enough, as it agrees very well with the covariances produced by the exact wave solver.

\begin{figure*}[!t]
\begin{center}\begin{tabular}{c@{}c@{~~~~~~}c@{~~~~~~}c@{~~~~~~}c@{}}
&\tiny$\pathseq^1=\ind\to\ptd_1\to \ptd_2\to\ptd_3\to\outd$
&\tiny$\pathseq^1=\ind\to\ptd_1\to \ptd_2\to\ptd_3\to\outd$
&\tiny$\pathseq^1=\ind\to\ptd_1\to \ptd_2\to\ptd_3\to\outd$
&\tiny$\pathseq^1=\ind\to\ptd_1\to \ptd_2\to\ptd_3\to\outd$\\
&\tiny$\pathseq^2=\ind\to\ptd_1\to \ptd_2\to\ptd_3\to\outd$&
\tiny$\pathseq^2=\ind\to\ptd_3\to \ptd_2\to\ptd_1\to\outd$&
\tiny$\pathseq^2=\ind\to\ptd_3\to \ptd_1\to\ptd_2\to\outd$&
\tiny$\pathseq^2=\ind\to\ptd_1\to \ptd_3\to\ptd_2\to\outd$
{\vspace{-0.0cm}}\\
        \includegraphics[width=0.13\textwidth,height=0.1\textwidth]{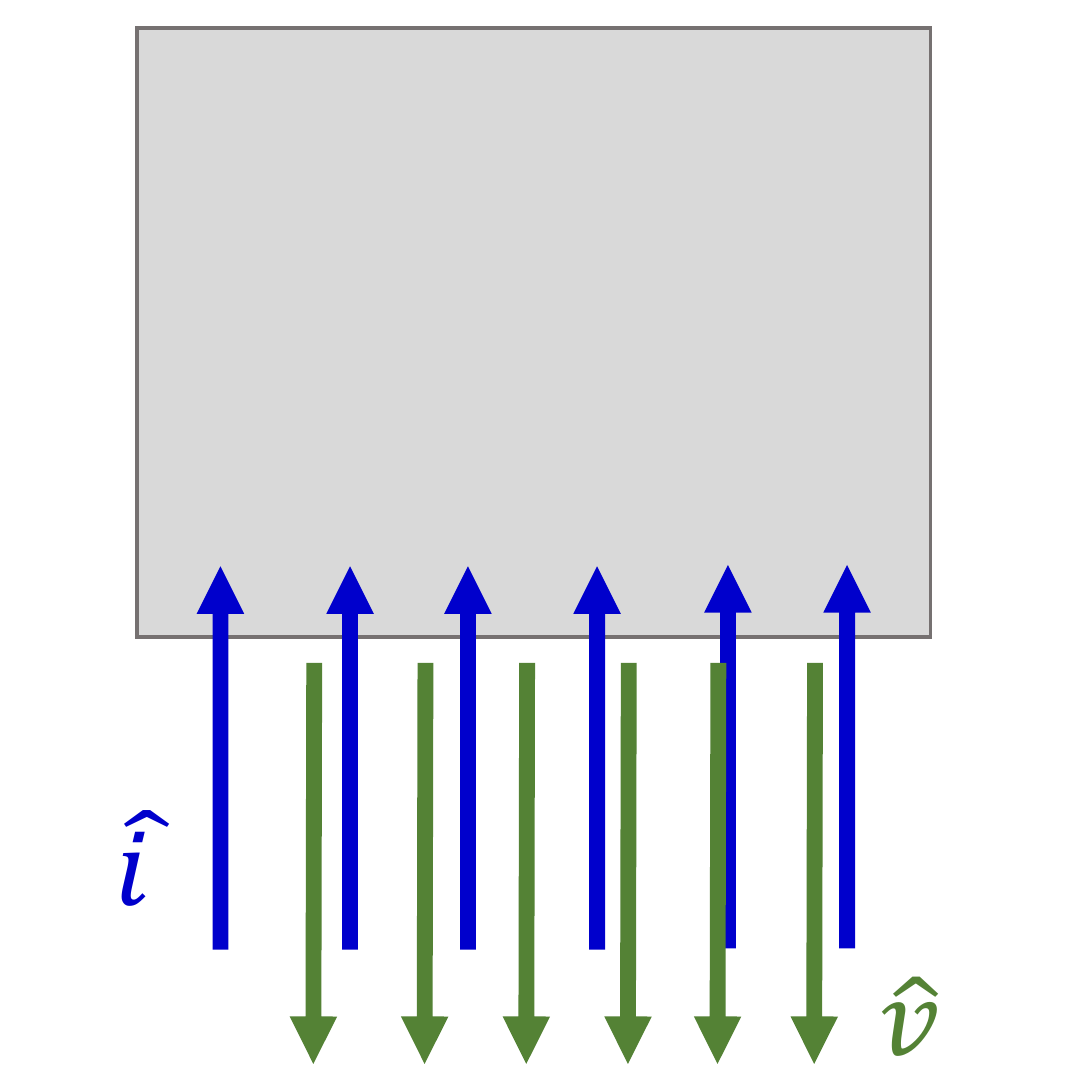}&
        \includegraphics[width=0.15\textwidth]{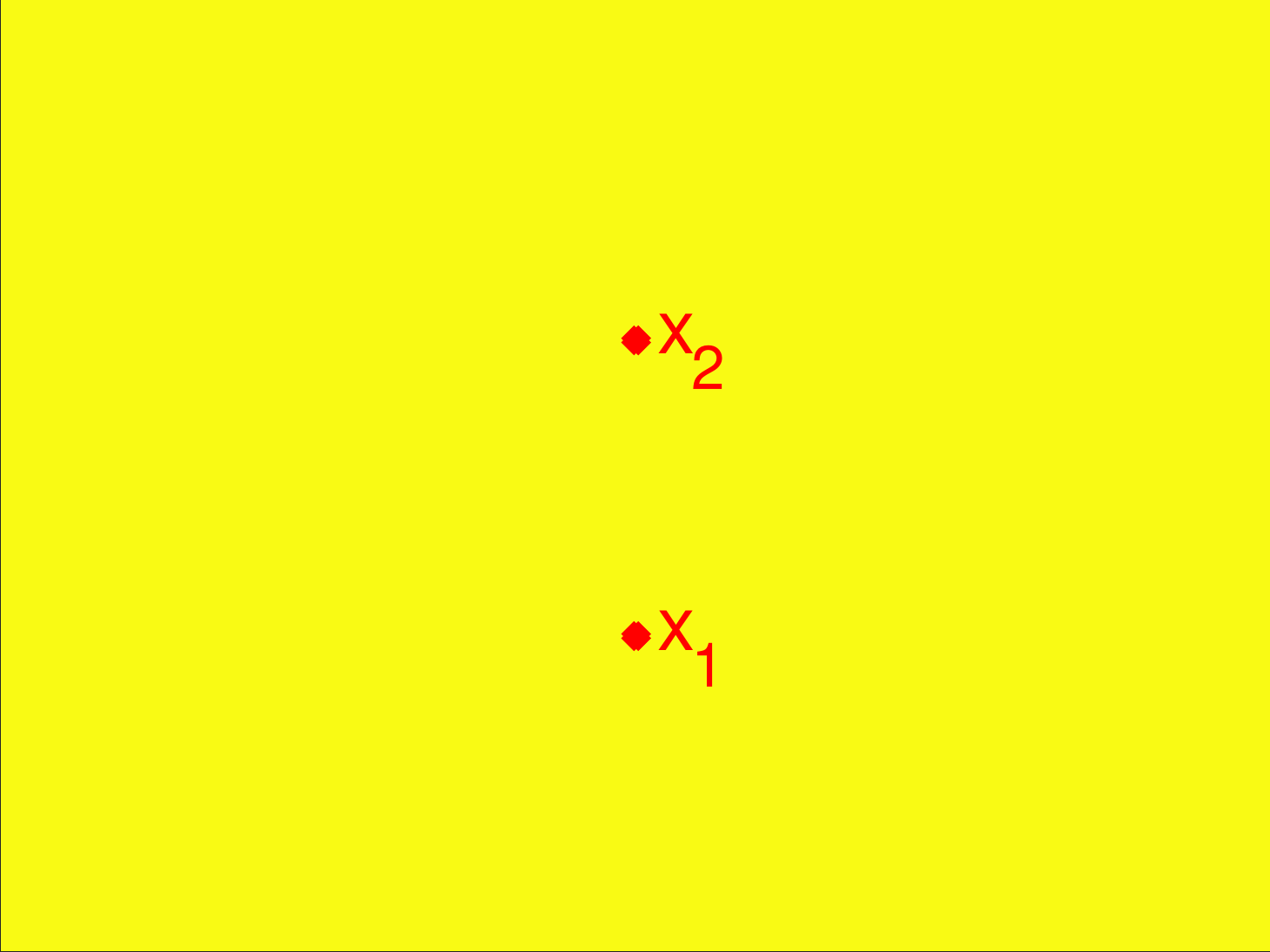}&
    \includegraphics[width=0.15\textwidth]{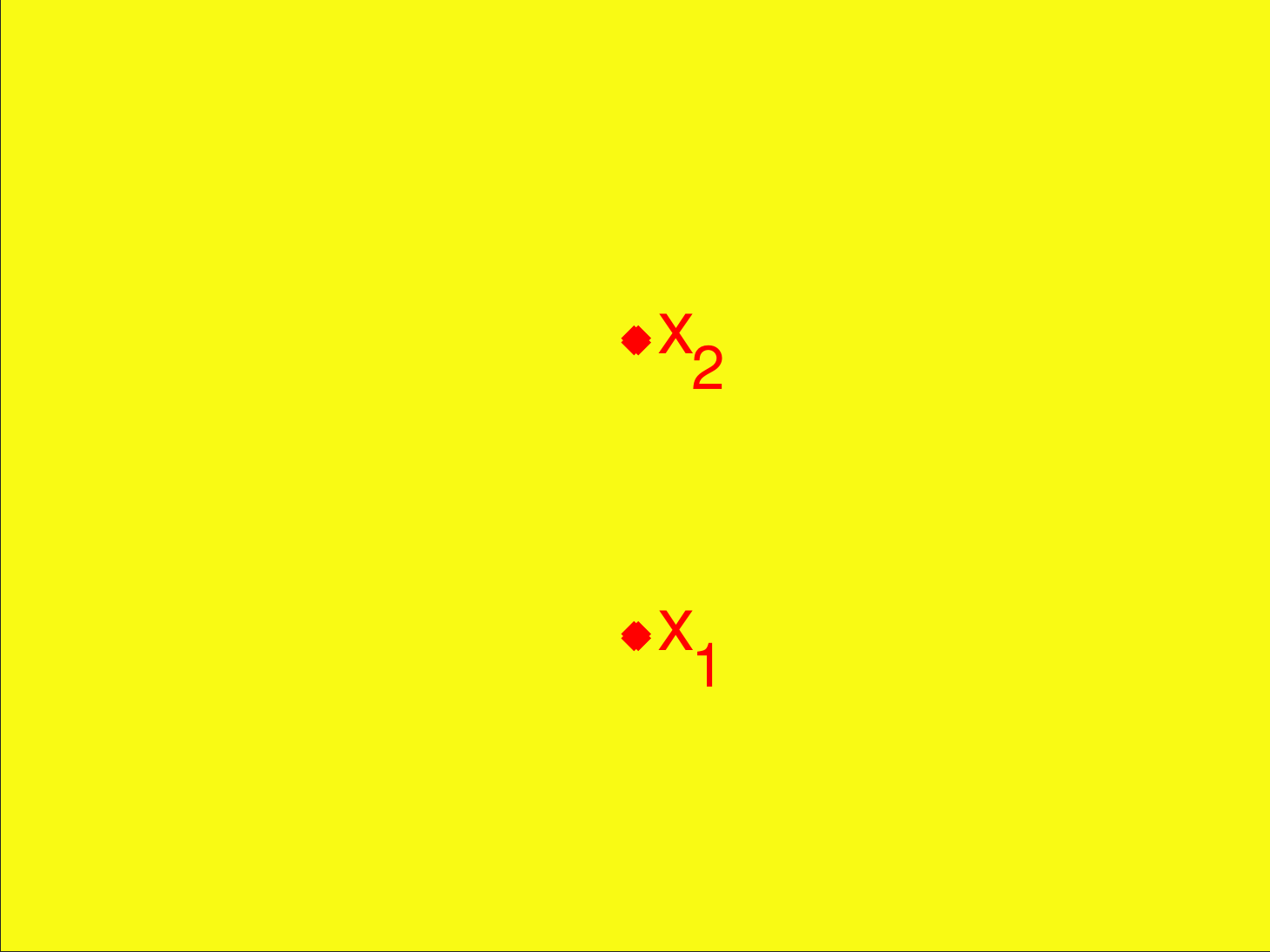}&
        \includegraphics[width=0.15\textwidth]{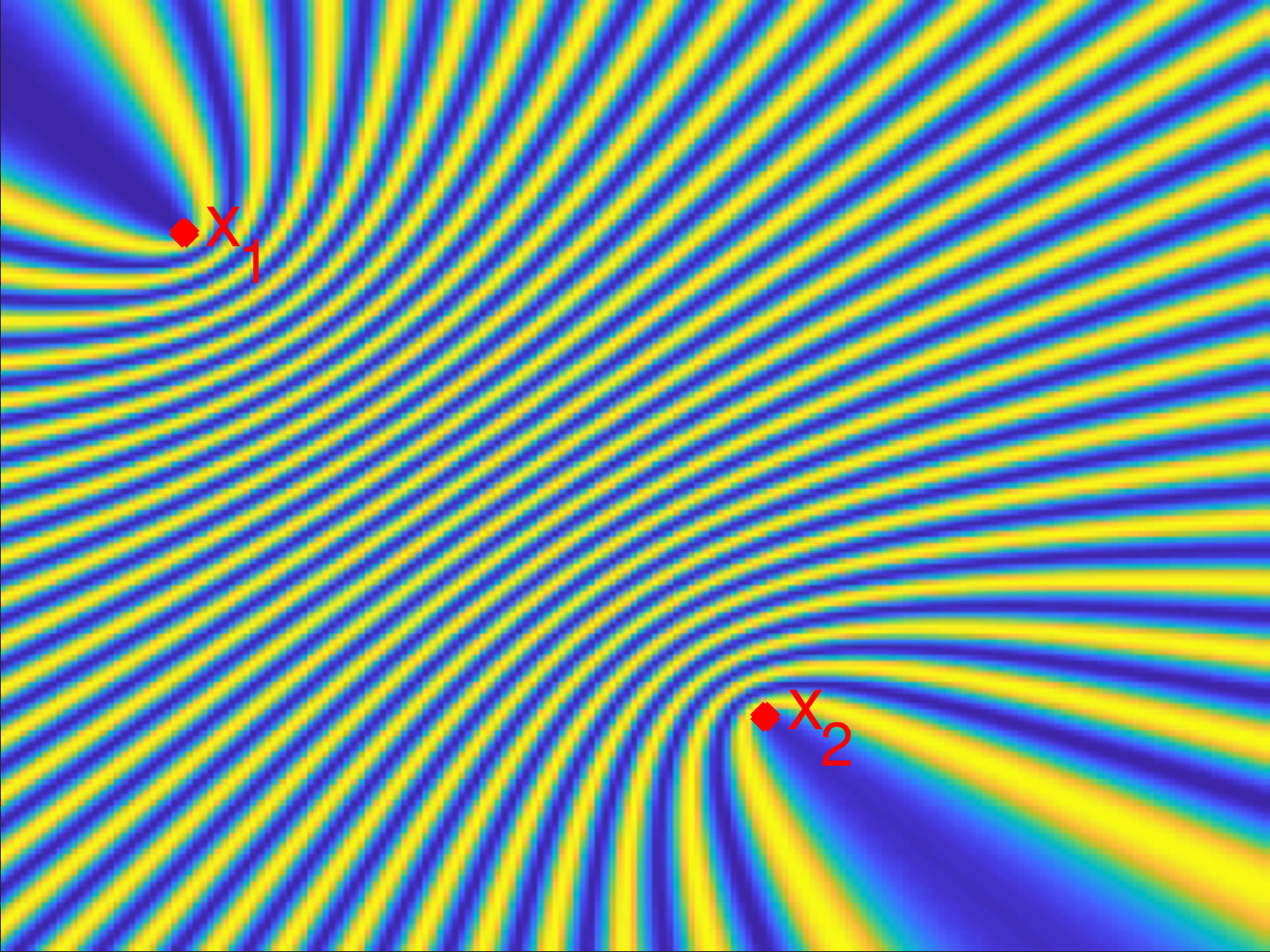}&
           \includegraphics[width=0.15\textwidth]{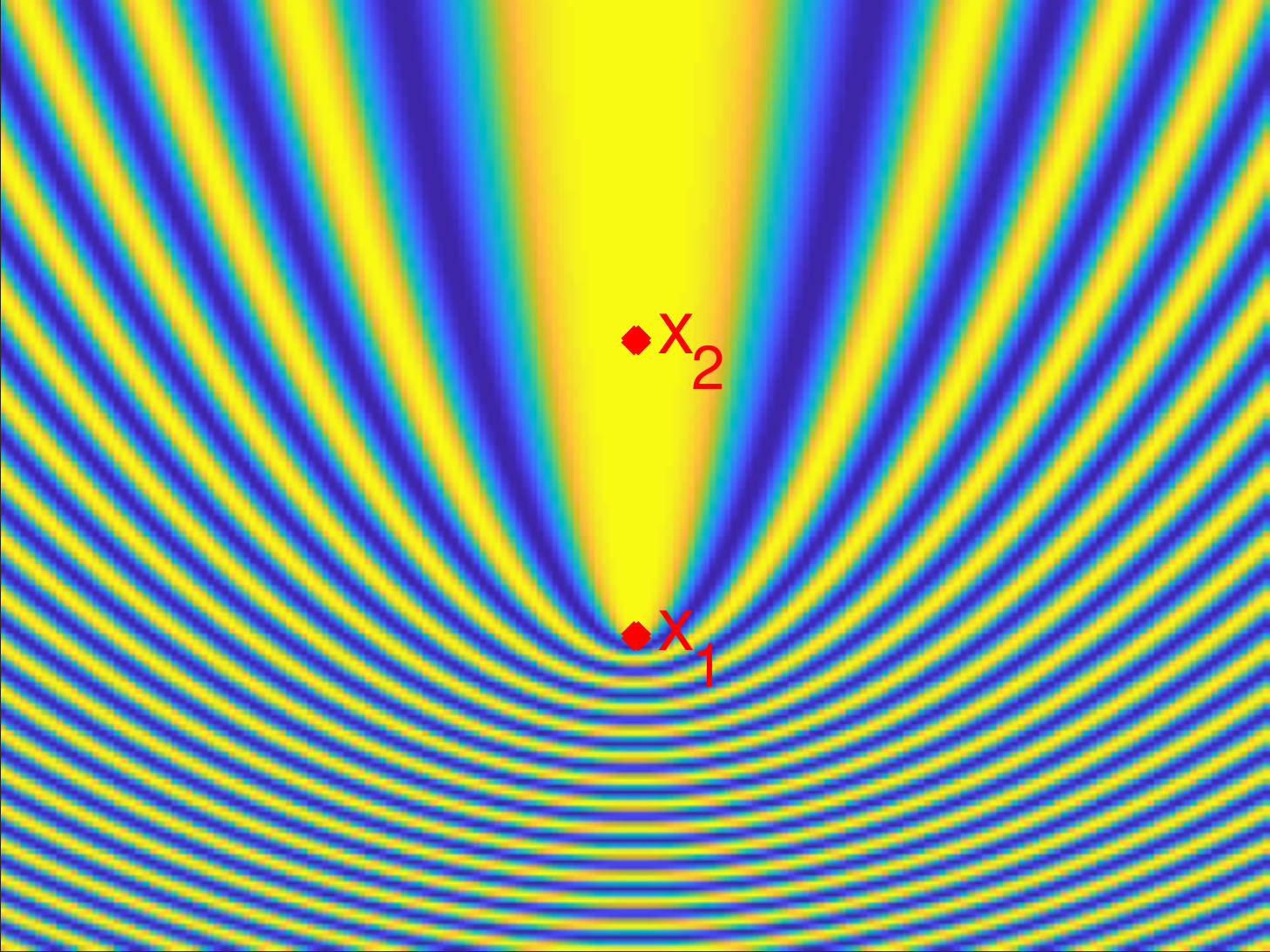}
\\
        \includegraphics[width=0.13\textwidth,height=0.1\textwidth]{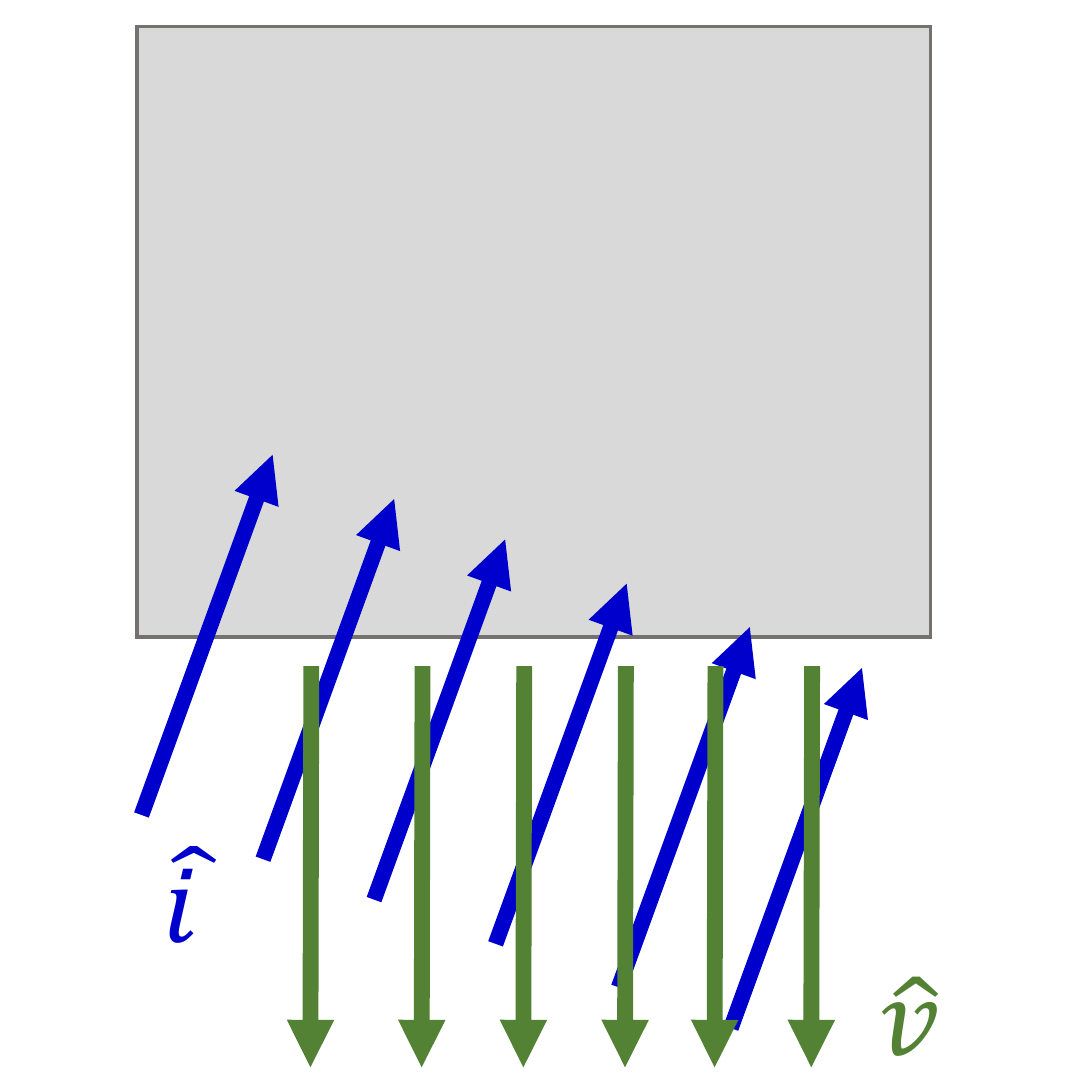}&
        \includegraphics[width=0.15\textwidth]{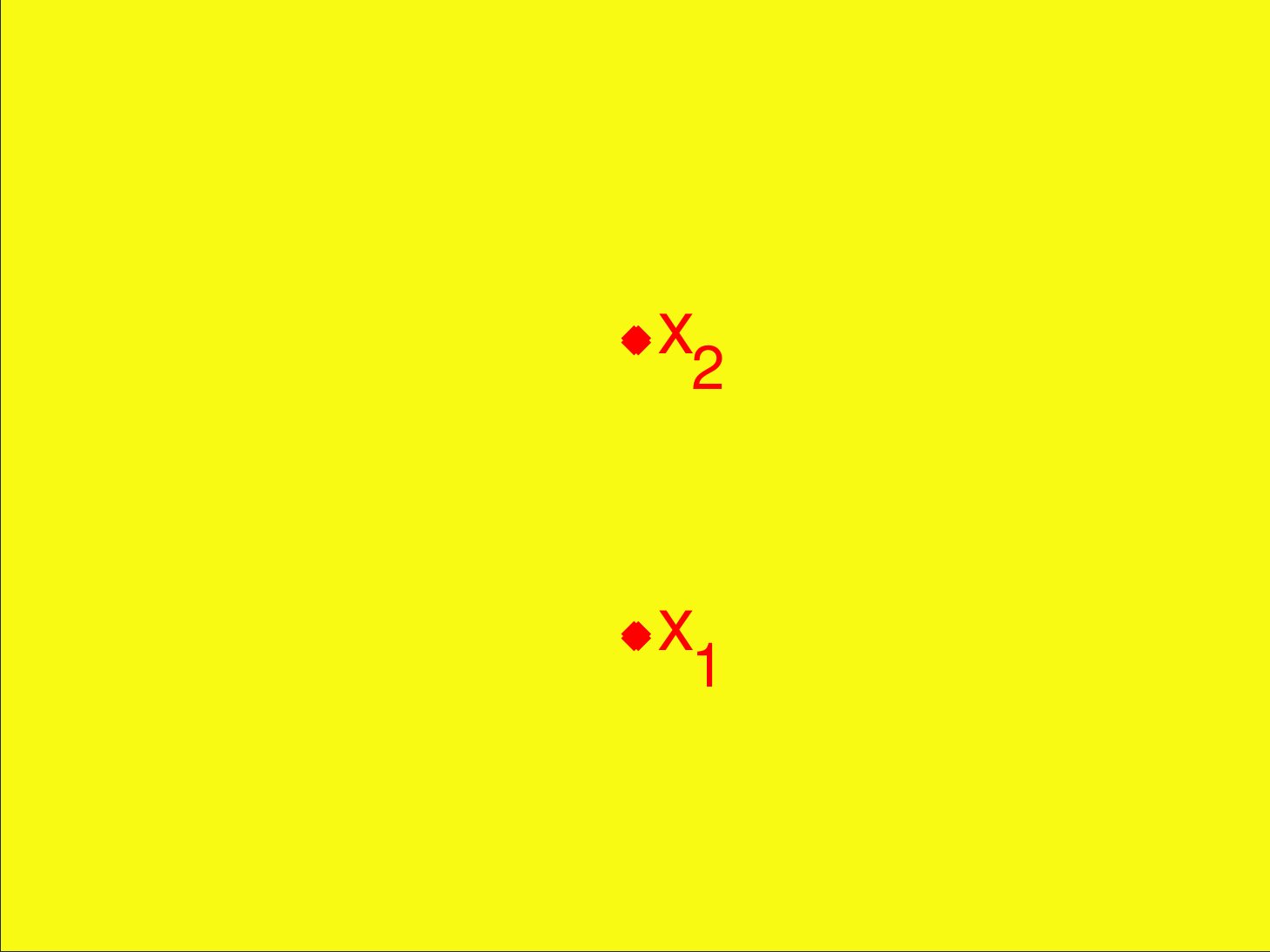}&
     \includegraphics[width=0.15\textwidth]{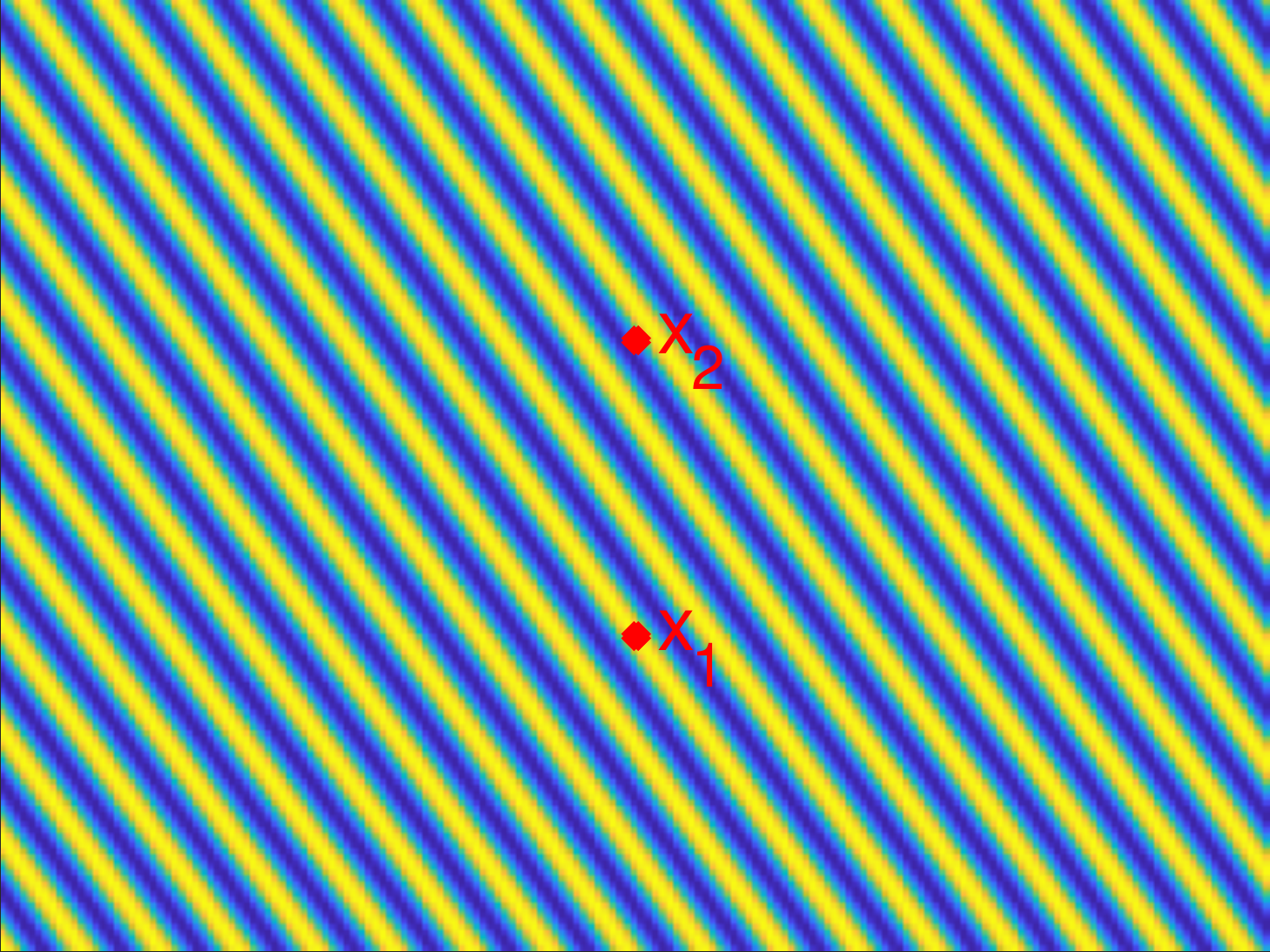}&
        \includegraphics[width=0.15\textwidth]{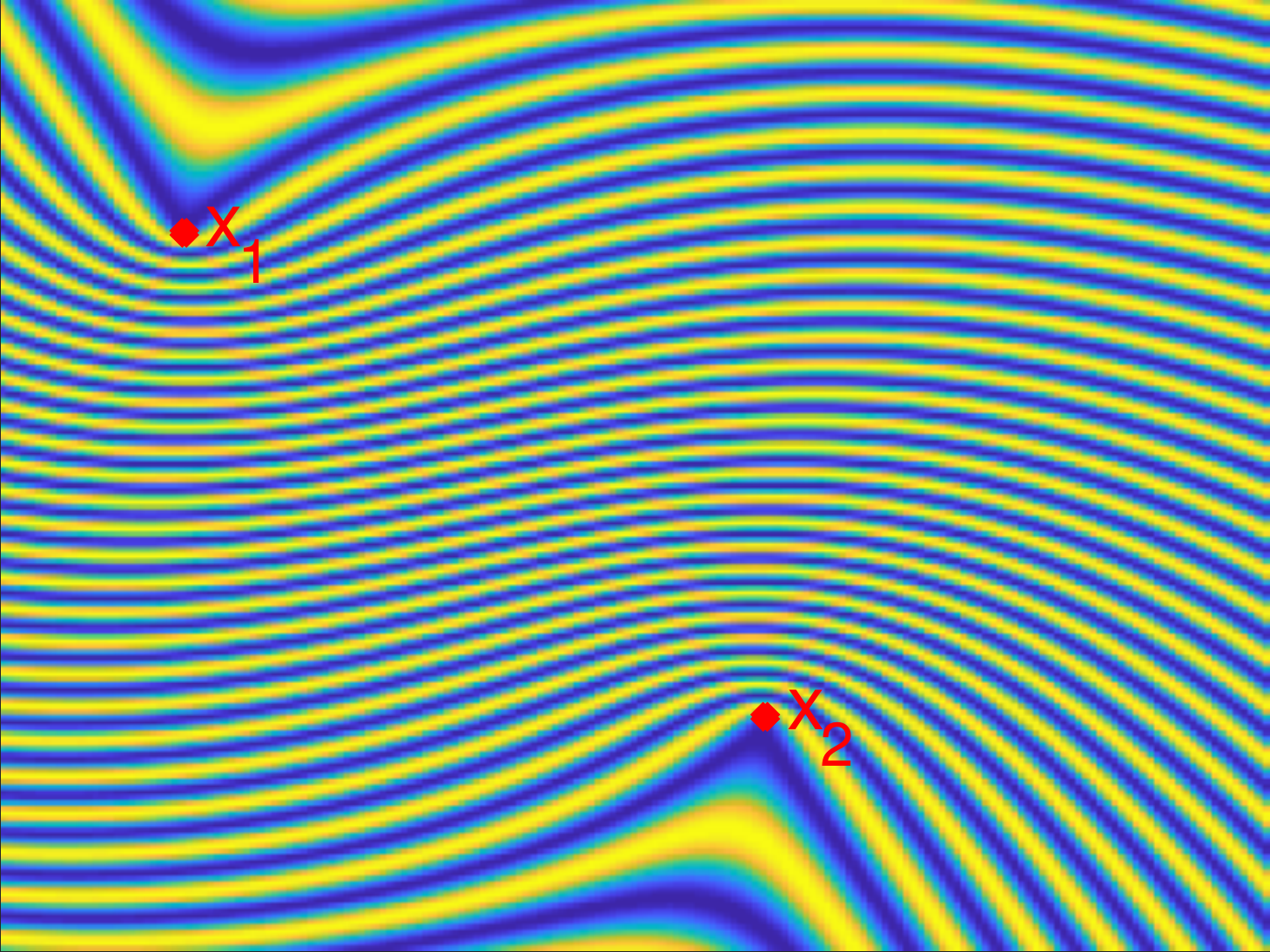}&
        \includegraphics[width=0.15\textwidth]{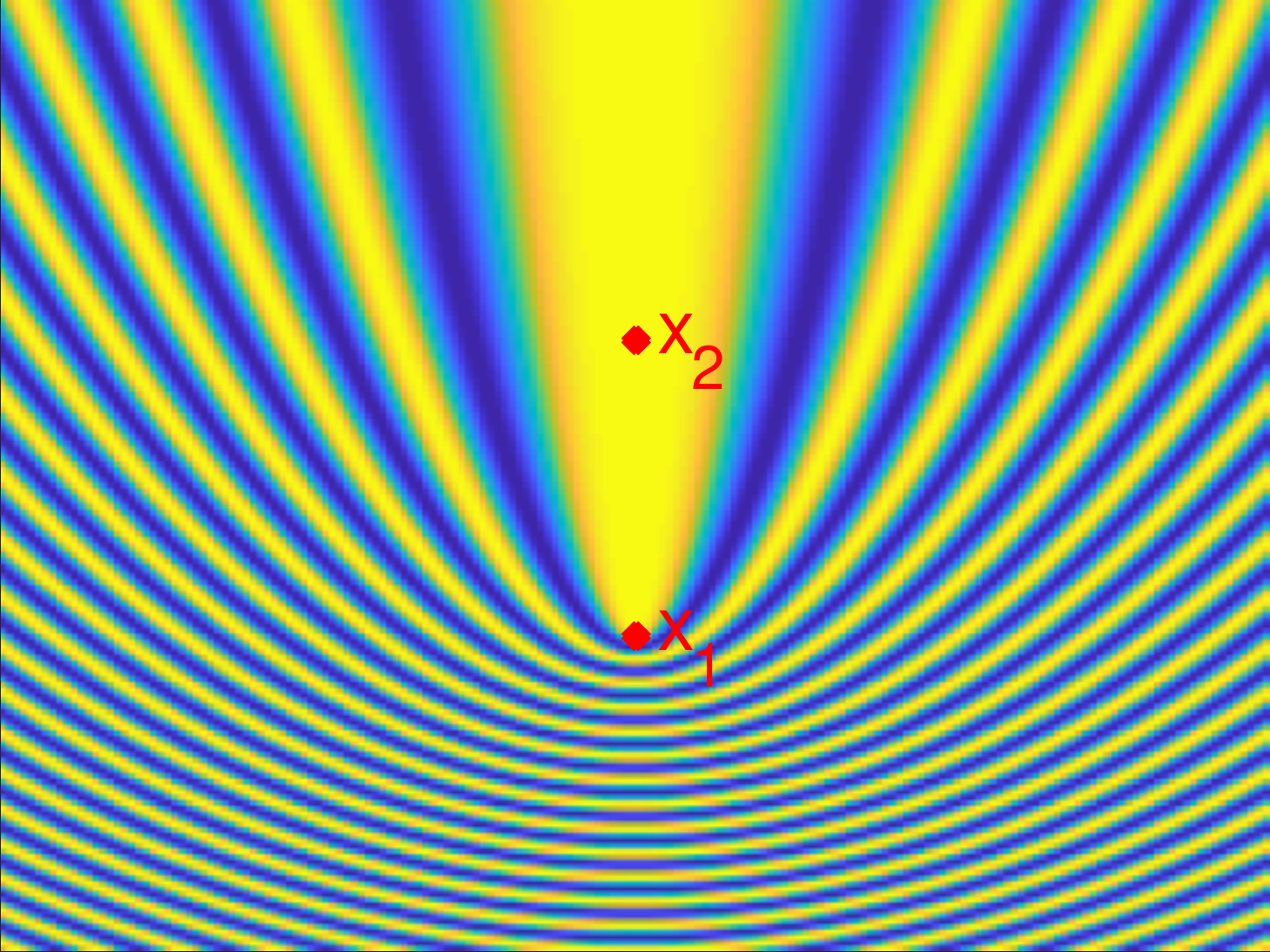}
\\&(a) Forward\;\;\;\;\;\;\;\; &(b) Reversed\;\;\;\;\;\;&(c) Permutation 1\;\;\;&(d) Permutation 2
\end{tabular}
\caption{The phase of pairwise path throughput as a function of the position of one of the points, for different node permutations. For the forward permutation the phase is constant. For the reversed permutation a constant phase is achieved only at the backscattering direction. Other permutations result in spatially varying phase, thus cancel out   in spatial integration.   }  
\label{fig:perms-plots}
\end{center}        
\end{figure*}

\begin{figure}[!t]
\begin{center}\begin{tabular}{cc}
       \subfloat[Mean throughput]{ \includegraphics[width=0.22\textwidth]{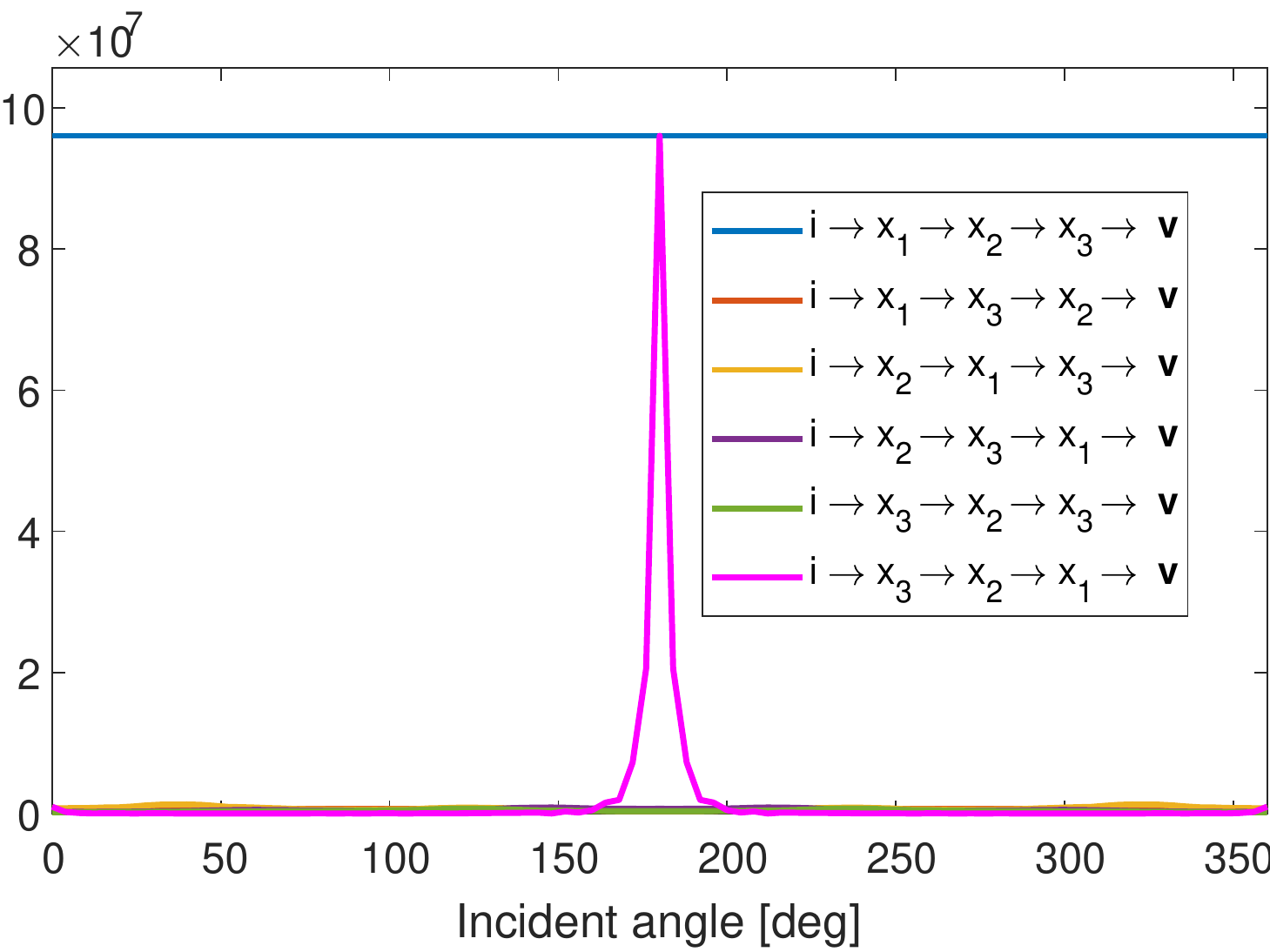}}&
        \subfloat[Log10 mean throughput]{\includegraphics[width=0.21\textwidth]{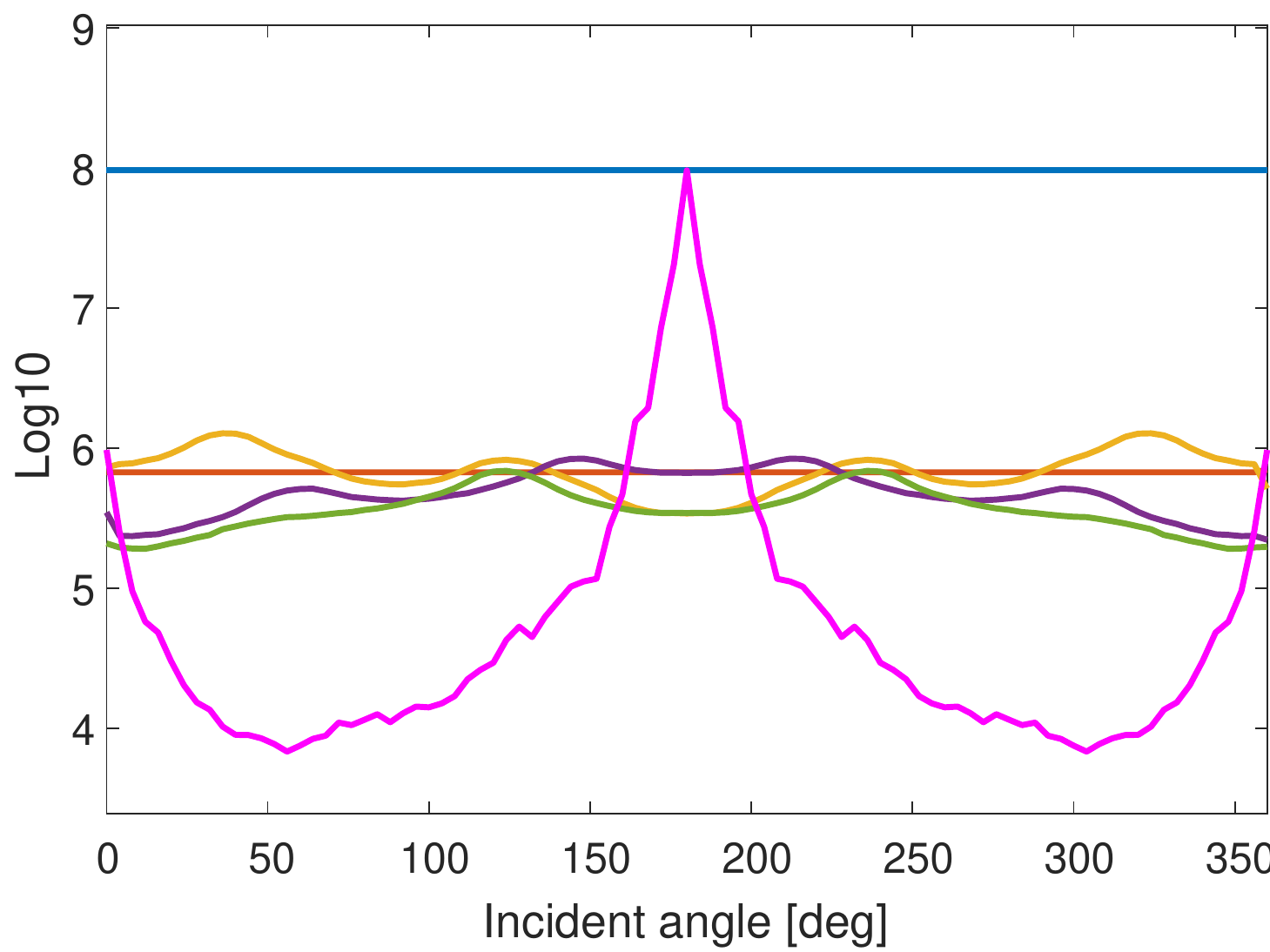}}
\end{tabular}
\caption{The mean permutation throughput after integrating  spatial shifts of $\ptd_2,\ptd_3$. As seen in the log plot (b), the mean contribution of the neglected permutations is about two order of magnitudes lower than the forward and reversed permutations.}  
\label{fig:path-perms-plots}
\end{center}        
\end{figure}
 
The  reason most permutations can be ignored is that the phase of the path  throughput is equivalent to the path length and for permutations that do not trace the nodes in the same order the segment lengths are different (\figref{fig:paths-permute}), hence we are summing complex numbers with different phases that quickly integrate to zero. To see this we considered paths of 3 nodes $\ptd_1,\ptd_2,\ptd_3$, which is the smallest path length with non trivial permutations. In   \figref{fig:perms-plots} we used $\ind_1=\ind_2,\outd_1=\outd_2$,  fixed the nodes $\ptd_1,\ptd_2$ while varying the third node $\ptd_3$ over a $20\lambda\times 20\lambda$ area. We evaluated the path  throughput contributions $\sigsct(\pathseq^1){\sigsct(\pathseq^{2,P})}^*$ for various permutations $P$ on 3 nodes. When $P$ is the identity permutations $\pathseq^1$ and $\pathseq^{2,P}$ have the same length, hence  $\sigsct(\pathseq^1){\sigsct(\pathseq^{2,P})}^*$ is always a positive number. If $P$ is the reversed permutation leading to the path $\pathseq_2=\ind\to\ptd_3\to \ptd_2\to\ptd_1\to\outd$ we get a fixed phase only for the backscattering direction $\outd\sim-\ind$. For other  directions one can see in \figref{fig:perms-plots}b lower row that perturbing the position of $\ptd_3$ changes the phase, hence it is clear that by averaging the pairwise path throughput over all positions of $\ptd_3$ averages to zero
\BE
\int\sigsct(\pathseq^1){\sigsct(\pathseq^{2,P})}^*d{\ptd_3}\approx0
\EE
For all other permutations, there is not even a single configuration of illumination and viewing directions leading to a fixed phase, and as can be seen in \figref{fig:perms-plots}c, varying the position of one of the nodes locally quickly changes the phase, hence averaging different path contributions over a local window integrates to zero. There are some rare path selections leading to a locally stationary phase, as can be seen in \figref{fig:perms-plots}d.\comment{Some of these stationary points are stationary only with respect to $\ptd_3$ and a small shift of $\ptd_2$ will shift their phase as in \figref{}d. In other cases (\figref{}ef) this is really a stationary point that do not change phase if we shift $\ptd_2$ (\figref{}f).  For $B=3$ this happens only when all 3 points lie on a line in the direction of $\outd$, in which case the permutation $P:\ind\to\ptd_1\to \ptd_3\to\ptd_2\to\outd$ has a length equivalent to $\ind\to\ptd_1\to \ptd_2\to\ptd_3\to\outd$ for any   local change in the position of the scatterers.} However, the probability of selecting such paths is low, and therefore the contribution to the overall covariance is negligible. In \figref{fig:path-perms-plots} we numerically evaluated  the integral of all 6 permutations of 3 numbers when varying two of the scatterers positions through a 2D square  \BE
\iint\sigsct(\pathseq^1){\sigsct(\pathseq^{2,P})}^*d{\ptd_2}d{\ptd_3}
\EE
One can see that except of the forward and reversed permutations the throughput of all permutations integrates to a contribution that is about two order of magnitude lower than the forward contribution. 

 \begin{algorithm}[!t]
         \mbox{}\hfill {\stcom\it{Initialize covariance estimate.}}\\
        Set $C=0$. \\
        \For{itr=1:N}{\mbox{}\hfill {\stcom\it{Sample first vertex of subpath.}}\\ Sample point $\ptd_1\sim q_{o}(\ptd_1)$  inside medium.\\ \mbox{}\hfill {\stcom\it{Update covariance by single scattering path}}\\
                Sample uniform direction  $\omgv_1$.\\
                $C+=V\mulsct(\inp_1\sto\ptd_1){\mulsct(\inp_2\sto\ptd_1)}^*{\Pmulsct((\inp_1,\inp_1)\sto \ptd_1\sto(\snsp_1,\snsp_2))}$ \\ \mbox{}\hfill {\stcom\it{Continue tracing the subpath.}}\\
                \mbox{}\hfill {\stcom\it{Sample second vertex of subpath.}}\\
                Sample $d\sim \sctCoff(\ptd_1)|\att(\ptd_1,\ptd_{1}+d\omgv_{1})|^2$\\ Set $\ptd_2=\ptd_{1}+d\omgv_{1}$\\$b=2$\\
                \While{ $\ptd_b$ inside medium}{\mbox{}\hfill {\stcom\it{Perform next-event estimation.}}\\{\textcolor{blue}{Forward version:}}\\ {\textcolor{blue}{$C+=\frac{V}{2}\Pmulsct(\ptd_2\sto\ptd_1\sto(\inp_1,\inp_2))\Pmulsct(\ptd_{b-1}\sto\ptd_b\sto(\snsp_1,\snsp_2)) $}}\\ 
                        {\textcolor{red}{Or, forward+backward version:}}\\{\textcolor{red}{$C+=\frac{V}{2}\Big(\Pmulsct(\ptd_2\sto\ptd_1\sto(\inp_1,\inp_2))\Pmulsct(\ptd_{b-1}\sto\ptd_b\sto(\snsp_1,\snsp_2))+$\\\quad\quad\quad\quad$\Pmulsct(\ptd_2\sto\ptd_1\sto(\inp_1,\snsp_2))\Pmulsct(\ptd_{b-1}\sto\ptd_b\sto(\snsp_1,\inp_2))+$\\\quad\quad\quad\quad$\Pmulsct(\ptd_2\sto\ptd_1\sto(\snsp_1,\inp_2))\Pmulsct(\ptd_{b-1}\sto\ptd_b\sto(\inp_1,\snsp_2))+$\\\quad\quad\quad\quad$\Pmulsct(\ptd_2\sto\ptd_1\sto(\snsp_1,\snsp_2))\Pmulsct(\ptd_{b-1}\sto\ptd_b\sto(\inp_1,\inp_2))\;\Big)$}}\\$\omgv_b\sim \rho(\omgv_{b-1},\omgv_b)$\\\mbox{}\hfill {\stcom\it{Sample next vertex of subpath}}\\Sample $d\sim\sctCoff(\ptd_b) |\att(\ptd_{b},\ptd_{b}+d\omgv_{b})|^2$\\ 
                        Set $\ptd_{b+1}=\ptd_{b}+d\omgv_{b}$\\\mbox{}\hfill {\stcom\it{Account for absorption}}\\ Sample a uniform random number $a\sim[0,1]$\\\If {$a>\sctCoff(\ptd_{b+1})/\extCoff(\ptd_{b+1})$}{ \mbox{}\hfill {\stcom\it{Terminate subpath at absorption event.}}\\ break\\ }$b=b+1$}
        }  \mbox{}\hfill {\stcom\it{Produce final covariance estimate.}}\\ 
        $C=\frac{1}{N}C$\\
        \Return{$C$}
        \caption{MC covariance $C^{\inp_1,\inp_2}_{\snsp_1,\snsp_2}$}
        \label{alg:MCpair-app}
 \end{algorithm}

 \begin{algorithm}[t!]\mbox{}\hfill {\stcom\it{Initialize field estimate.}}\\
        Set $u(j)=0$ \\
        \For{itr=1:N}{ \mbox{}\hfill {\stcom\it{Sample first vertex of subpath.}}\\ Sample $\ptd_1\sim q_o(\ptd_1)$  inside medium.\\Sample random phase $\zeta\sim  \text{Unif}[0,1]$, $z=e^{2\pi i \zeta}$\\
                Sample $\ptctype\sim \ampm_\ptctype$.\\ \mbox{}\hfill {\stcom\it{Update field with single scattering path}}\\
                $\forall j,~ u(j)+= z\cdot\sqrt{V}\cdot\mulsct_\ptctype(\inp_j\sto\ptd_1)\mulsct_\ptctype(\inp_j\sto\ptd_1\sto\snsp_j)$ \\  \mbox{}\hfill {\stcom\it{Continue tracing the subpath.}}\\
                \mbox{}\hfill {\stcom\it{Sample second vertex of subpath.}}\\ Sample uniform direction  $\omgv_1$.\\ 
                Sample $d\sim \sctCoff(\ptd_1)|\att(\ptd_1,\ptd_{1}+d\omgv_{1})|^2$\\ Set $\ptd_2=\ptd_{1}+d\omgv_{1}$\\$k=2$\\
                \While{ $\ptd_b$ inside medium}{Sample random phase $\zeta\sim  \text{Unif}[0,1]$, $z=e^{2\pi i \zeta}$.\\
                        Sample $\ptctype\sim \ampm_\ptctype$.\\ \mbox{}\hfill {\stcom\it{Update field with next-event estimation}}\\{\textcolor{blue}{Forward version:}}\\ {\textcolor{blue}{$\forall j,~u(j)+= z\cdot\sqrt{\frac{V}{2}}\mulsct_{\ptctype}(\ptd_2\sto\ptd_1\sto\inp_j)\mulsct_{\ptctype}(\ptd_{k-1}\sto\ptd_k\sto\snsp_j) $}}\\ 
                        {\textcolor{red}{Or, forward+backward version:}}\\ {\textcolor{red}{$\forall j,~u(j)+= z\cdot\sqrt{\frac{V}{2}}\Big(\mulsct_{\ptctype}(\ptd_2\sto\ptd_1\sto\inp_j)\mulsct_{\ptctype}(\ptd_{b-1}\sto\ptd_b\sto\snsp_j)$\\\quad\quad\quad\quad\quad\quad\quad\quad\quad$+\mulsct_{\ptctype}(\ptd_2\sto\ptd_1\sto\snsp_j)\mulsct_{\ptctype}(\ptd_{b-1}\sto\ptd_b\sto\inp_j)\Big)$}}\\\mbox{}\hfill {\stcom\it{Sample next node}}\\$\omgv_b\sim \rho(\omgv_{b-1},\omgv_b)$\\Sample $d\sim\sctCoff(\ptd_b) |\att(\ptd_{b},\ptd_{b}+d\omgv_{b})|^2$\\
                        Set $\ptd_{b+1}=\ptd_{b}+d\omgv_{b}$\\\mbox{}\hfill {\stcom\it{Account for absorption}}\\ Sample a uniform random number $a\sim[0,1]$\\\If {$a>\sctCoff(\ptd_{b+1})/\extCoff(\ptd_{b+1})$}{  \mbox{}\hfill {\stcom\it{Terminate subpath at absorption event.}}\\ break\\ }$b=b+1$}
        } \mbox{}\hfill {\stcom\it{Produce final field with correct mean.}}\\
        $u(j)=\meanspk^{\inp_j}_{\snsp_j}+\sqrt{\frac{1}{{N}}}u(j)$\\
        \Return{$u$}
        \caption{MC field $u^{\inp_j}_{\snsp_j}$}
        \label{alg:MCfieldPT}
\end{algorithm}

\subsection{The correlation transfer equation}\label{sec:CTE}

MC algorithms evaluating scattering intensity were historically derived in computer graphics based on the radiative transfer equation (RTE), which is an integral equation expressing the intensity at one point using the intensity at other points in space. A MC is then defined as a recursive evaluation of the RTE.
One of the main results in the speckle correlation literature is an analogues {\em correlation transfer equation (CTE)}~\cite{ishimaru1999wave,Twersky64}. 
This  is an integral equation that expresses the correlation through the light field intensity at other points in space. In contrary to the way this was developed in computer graphics, textbooks like~\cite{mishchenko2006multiple} derive the CTE and RTE starting from electro-magnatic equations and express their solution as a summation of path-pairs contributions.

The CTE~\cite{ishimaru1999wave,Twersky64}  considers speckles at different spatial points under the {\em same} illumination direction, and express their second order moments. 
In our notations $C^{\inp}_{\snsp_1,\snsp_2}$ is the speckle covariance that relates to the second order moments as: $E[u^{\inp,O}_{\snsp_1}\cdot {u^{\inp,O}_{\snsp_2}}^*]=C^{\inp}_{\snsp_1,\snsp_2}+\meanspk^{\inp}_{\snsp_1}{\meanspk^{\inp}_{\snsp_2}}^*$. 
The CTE states that:
\begin{multline}\label{eq:CTE-points}
E[u^{\inp,O}_{\snsp_1}\cdot {u^{\inp,O}_{\snsp_2}}^*]= \meanspk^{\inp}_{\snsp_1}{\meanspk^{\inp}_{\snsp_2}}^*+\\\negthickspace\negthickspace\int_{\ptd }\sctCoff(\ptd)\negthickspace\int_{\omgv}\negthickspace\negthickspace\mulsct(\omgv\to\ptd\to\snsp_1)\mulsct(\omgv\to\ptd\to\snsp_2)^*L^{\inp}_{\ptd,\omgv}
\end{multline}
Where $L^{\inp}_{\ptd,\omgv}$ is the ``light field'' as used traditionally in computer graphics, namely the intensity arriving point $\ptd$ from direction $\omgv$.

The important observation made by the CTE is that to compute correlations between the fields at sensor points ${\snsp}_1,{\snsp}_2$, 
we need to integrate only {\em intensity} from other space points, and there is no need to memorize any other correlations.
The intensity at other space points is weighted by the volumetric throughput $\mulsct$, namely the probability and phase of making a ``single scattering'' step from $\ptd$ to $\snsp_1$ and $\snsp_2$.
For the case $\snsp_1=\snsp_2$ the covariance reduces to intensity and indeed \equref{eq:CTE-points} reduces to the integral version of the RTE assuming zero emission inside the target. The $\mulsct(\ptd)$ is basically the attenuation and phase function part of the RTE, and the term $\meanspk$ is the attenuation of the direct source.

It is not hard to show that for the case $\inp_1=\inp_2$ a forward-only MC version summarized in Algorithm \ref{alg:MCpair-app} is basically a recursive evaluation of this CTE integral. 
This variant is derived by 
approximating the covariance as
\BE\label{eq:app-cov-imps-path-app}
C^{\ind_1,\ind_2}_{\snsp_1,\snsp_2}\approx\frac1N\sum_{\pathseq^n_0\sim q}\frac{c_{\pathseq^{1,n},\pathseq^{2,n}}}{q(\subpathn)}
\EE
 Rather than as\BE\label{eq:cov-app-fb-app}
C^{\ind_1,\ind_2}_{\snsp_1,\snsp_2}\approx\frac1N\sum_{\pathseq^n_0\sim q}\frac{c_{\pathseq^{1,n},\pathseq^{2,n}}+c_{\pathseq^{1,n}\pathseq^{2,n,r}}+c_{\pathseq^{1,n,r},\pathseq^{2,n}}+c_{\pathseq^{1,n,r},\pathseq^{2,n,r}}}{q(\subpathn)+q(\rsubpathn)}
\EE

\comment{
We update the covariance at each step of the MC algorithm as 
 \BEA\label{eq:cov-ratio-fo-app}
 \frac{\cll}{q(\subpath)}&=&V\cdot\mulsct(\ptd_2\sto\ptd_1\sto\inp_1)\mulsct(\ptd_{B-1}\sto\ptd_B\sto\snsp_1)\nonumber\\
&&\cdot{\mulsct(\ptd_2\sto\ptd_1\sto\inp_2)}^*{\mulsct(\ptd_{B-1}\sto\ptd_B\sto\snsp_2)}^* \label{eq:update-f-paths}\EEA
Rather than as
\begin{multline}\label{eq:update-fb-paths-app}
 \frac{\cll+c_{\pathseq^1\pathseq^{2,r}}+c_{\pathseq^{1,r},\pathseq^2}+c_{\pathseq^{1,r},\pathseq^{2,r}}}{2q(\subpath)}=\\\;\;\;\;\frac{V}{2}\Big(\mulsct(\ptd_2\sto\ptd_1\sto\inp_1)\mulsct(\ptd_{B-1}\sto\ptd_B\sto\snsp_1)\\\;\;\;\;\;\;\;\;\;+\mulsct(\ptd_2\sto\ptd_1\sto\snsp_1)\mulsct(\ptd_{B-1}\sto\ptd_B\sto\inp_1)\Big)\\\;\;\;\;\;\;\cdot\Big(\mulsct(\ptd_2\sto\ptd_1\sto\inp_2)\mulsct(\ptd_{B-1}\sto\ptd_B\sto\snsp_2)\\{\!\!+\mulsct(\ptd_2\sto\ptd_1\sto\snsp_2)\mulsct(\ptd_{B-1}\sto\ptd_B\sto\inp_2)\Big)}^*
%\end{array}
\end{multline}
}

The derivation of the CTE e.g. in~\cite{mishchenko2006multiple} follows an expression of the solution to the wave equation as a sum of path contributions as we have presented in the previous sections. Making the simplifying assumption that only forward path pairs need to be considered they reorganize all the paths in the summation in a more compact recursive formula which is essentially the CTE, or the RTE in the $\snsp_1=\snsp_2$ case. 
The fact that only forward paths are considered is an inherent assumption necessary for the compactness of the CTE, namely that one needs to memorize only the last node on a path and the rest of its history does not matter. However, this is also its main drawback, as it does not allow it to explain physical phenomena such as coherent backscattering, which is an interference effect generated by the full path and not only by the last event.
Due to this problem we chose to derive our covariance rendering directly from a path space formulation and not through an integral equation.

\subsection{Sampling a speckle image with multiple particle types}
Before concluding  we note that sampling a speckle image as defined in Alg. \ref{alg:MCfield} should also adjust when multiple types of particles are available. The covariance in the multiple type case is expressed as:

\begin{equation}\label{eq:covariance_rank1_sum_app}
\covmat=\int_{\Path} f(\subpath) \sum_{\ptctype_1,\ptctype_2}\! \ampm_{\ptctype_1}\ampm_{\ptctype_2}\cdot \avec_{\ptctype_1,\ptctype_2}(\subpath)\cdot\avec^*_{\ptctype_1,\ptctype_2}(\subpath) \ud \subpath ,
\end{equation}

With
\begin{align}
\avec_{\ptctype,j}(\subpath)=&\Big(\mulsct_\ptctype(\ptd_2\rightarrow\ptd_1\rightarrow\inp_j)\mulsct_\ptctype(\ptd_{B-1}\rightarrow\ptd_B\rightarrow\snsp_j)\nonumber\\
&+\mulsct_\ptctype(\ptd_{B-1}\rightarrow\ptd_B\rightarrow\inp_j)\mulsct_\ptctype(\ptd_2\rightarrow\ptd_1\rightarrow\snsp_j)\Big).
\end{align}

Therefore, for every node sample we should also sample a particle type  $\ptctype\sim \ampm_\ptctype$. We summarize the changes in Alg. \ref{alg:MCfieldPT}.

\end{document}